\documentclass[12pt] {article} 
\usepackage{psfrag}
\usepackage{graphicx} 
\usepackage{latexsym,amsfonts}
\usepackage{amssymb}
\begin{document}
\setcounter{page}{1}
\def\theequation{\arabic{section}.\arabic{equation}}
\def\theequation{\thesection.\arabic{equation}}
\setcounter{section}{0}

\title{~\rightline{\normalsize {\rm IK-TUW-Preprint 0111401}}\\
[0.2in]On the solution of the massless Thirring model with fermion
fields quantized in the chiral symmetric phase}

\author{M. Faber\thanks{E--mail: faber@kph.tuwien.ac.at, Tel.:
+43--1--58801--14261, Fax: +43--1--5864203} ~~and~
A. N. Ivanov\thanks{E--mail: ivanov@kph.tuwien.ac.at,  Tel.:
+43--1--58801--14261, Fax: +43--1--5864203}~\thanks{Permanent Address:
State Technical University, Department of Nuclear Physics, 195251
St. Petersburg, Russian Federation}}

\date{\today}

\maketitle
\vspace{-0.5in}
\begin{center}
{\it Atominstitut der \"Osterreichischen Universit\"aten,
Arbeitsbereich Kernphysik und Nukleare Astrophysik, Technische
Universit\"at Wien, Wiedner Hauptstr. 8-10, A-1040 Wien,
\"Osterreich }
\end{center}

\begin{center}
\begin{abstract}
Correlation functions of fermionic fields described by the massless
Thirring model are analysed within the operator formalism developed by
Klaiber and the path--integral approach with massless fermions
quantized in the chiral symmetric phase. We notice that Klaiber's
composite fermion operators possess non--standard properties under
parity transformations and construct operators with standard parity
properties.  We find that Klaiber's parameterization of a
one--parameter family of solutions of the massless Thirring model is
not well defined, since it is not consistent with the requirement of
chiral symmetry. We show that the dynamical dimensions of correlation
functions depend on an arbitrary parameter induced by ambiguities of
the evaluation of the chiral Jacobian. A non--perturbative
renormalization of the massless Thirring model is discussed. We
demonstrate that the infrared divergences of Klaiber's correlation
functions can be transferred into ultra--violet divergences by
renormalization of the wave function of fermionic fields. This makes
Klaiber's correlation functions non--singular in the infrared
limit. We show that the requirement of non--perturbative
renormalizability of the massless Thirring model fixes a free
parameter of the path--integral approach. In turn, the operator
formalism is inconsistent with non--perturbative renormalizability of
the massless Thirring model. We carry out a non--perturbative
renormalization of the sine--Gordon model and show that it is not an
asymptotically free theory as well as the massless Thirring model. We
calculate the fermion condensate by using the Fourier transform of the
two--point Green function of massless Thirring fermion fields
quantized in the chiral symmetric phase.
\end{abstract}
\end{center}

\newpage

\section{Introduction}
\setcounter{equation}{0}

\hspace{0.2in} The massless Thirring model [1] is a theory of a
self--coupled Dirac field $\psi(x)$
\begin{eqnarray}\label{label1.1}
{\cal L}_{\rm Th}(x) = \bar{\psi}(x)i\gamma^{\mu}\partial_{\mu}\psi(x) -
\frac{1}{2}\,g\,\bar{\psi}(x)\gamma^{\mu}\psi(x)\bar{\psi}(x)
\gamma_{\mu}\psi(x),
\end{eqnarray}
where $g$ is a dimensionless coupling constant that can be both
positive and negative as well. The field $\psi(x)$ is a spinor field
with two components $\psi_1(x)$ and $\psi_2(x)$, $x$ is a 2--vector
$x^{\mu} = (x^0, x^1)$, where $x^0$ and $x^1$ are time and spatial
components. The $\gamma$--matrices are defined in terms of the
well--known $2\times 2$ Pauli matrices $\sigma_1$, $\sigma_2$ and
$\sigma_3$
\begin{eqnarray}\label{label1.2}
\gamma^0 = \sigma_1 = \left(\begin{array}{cc}
0 & 1\\
1 & 0 
\end{array}\right)\,,\,\gamma^1 = - i\sigma_2 = \left(\begin{array}{cc}
0 & - 1\\
1 & 0 
\end{array}\right)\,,\,\gamma^5 = \gamma^0\gamma^1 = \sigma_3 = 
\left(\begin{array}{cc} 1 & 0 \\ 0 & -1
\end{array}\right).
\end{eqnarray}
These $\gamma$--matrices obey the standard relations
\begin{eqnarray}\label{label1.3}
\gamma^{\mu}\gamma^{\nu}&+& \gamma^{\nu}\gamma^{\mu} = 2
g^{\mu\nu},\nonumber\\ 
\gamma^{\mu}\gamma^5&+& \gamma^5\gamma^{\mu} = 0.
\end{eqnarray}
We use the metric tensor $g^{\mu\nu}$ defined by $g^{00} = - g^{11} =
1$ and $g^{01} = g^{10} = 0$. The axial--vector product
$\gamma^{\mu}\gamma^5$ can be expressed in terms of $\gamma^{\nu}$
\begin{eqnarray}\label{label1.4}
\gamma^{\mu}\gamma^5 &=& - \epsilon^{\mu\nu}\gamma_{\nu},\
\end{eqnarray}
where $\epsilon^{\mu\nu}$ is the anti--symmetric tensor defined by
$\epsilon^{01} = - \epsilon^{10} = 1$. Further, we also use the
relation $\gamma^{\mu}\gamma^{\nu} = g^{\mu\nu} +
\epsilon^{\mu\nu}\gamma^5$.

The Lagrangian (\ref{label1.1}) is obviously invariant under the
chiral group $U_{\rm V}(1)\times U_{\rm A}(1)$
\begin{eqnarray}\label{label1.5}
\psi(x) \stackrel{\rm V}{\longrightarrow} \psi^{\prime}(x) &=&
e^{\textstyle i\alpha_{\rm V}}\psi(x) ,\nonumber\\ \psi(x)
\stackrel{\rm A}{\longrightarrow} \psi^{\prime}(x) &=& e^{\textstyle
i\alpha_{\rm A}\gamma^5}\psi(x),
\end{eqnarray}
where $\alpha_{\rm V}$ and $\alpha_{\rm A}$ are real parameters
defining global rotations.

Due to invariance under chiral group $U_{\rm V}(1)\times U_{\rm A}(1)$
the vector and axial--vector current $j^{\mu}(x)$ and $j^{\mu}_5(x)$,
induced by vector (V) and axial--vector (A) rotations and defined by
\begin{eqnarray}\label{label1.6}
j^{\mu}(x) &=&
\bar{\psi}(x)\gamma^{\mu}\psi(x),\nonumber\\
j^{\mu}_5(x) &=&
\bar{\psi}(x)\gamma^{\mu}\gamma^5\psi(x),
\end{eqnarray}
are conserved
\begin{eqnarray}\label{label1.7}
\partial_{\mu}j^{\mu}(x) = \partial_{\mu}j^{\mu}_5(x) = 0.
\end{eqnarray}
Recall, that in 1+1--dimensional field theories the vector and
axial--vector currents are related by $j^{\mu}_5(x) = -
\varepsilon^{\mu\nu}j_{\nu}(x)$ due to the properties of Dirac
matrices.

The massless Thirring model being a local quantum field theoretic
model is invariant under $C$, $P$ and $T$ transformations
separately. Since below we are concerned with the properties of
fermion fields under parity transformations, we remind that
invariance of the massless Thirring model under parity transformation
imposes the constraints [2,3] (see also Appendix A of this paper)
\begin{eqnarray}\label{label1.8}
{\cal P}\,{\cal L}(x^0,x^1)\,{\cal P}^{\dagger} &=& + {\cal L}(x^0,
-x^1),\nonumber\\ {\cal P}\,j^{\mu}(x^0,x^1)\,{\cal P}^{\dagger} &=& +
j_{\mu}(x^0, -x^1),\nonumber\\ {\cal P}\,j^{\mu}_5(x^0,x^1)\,{\cal
P}^{\dagger} &=&- j_{5\mu}(x^0, -x^1).
\end{eqnarray}
These constraints fix the transformation of the fermionic field [2,3]
\begin{eqnarray}\label{label1.9}
{\cal P}\,\psi(x^0,x^1)\,{\cal P}^{\dagger} = \gamma^0\psi(x^0, - x^1)
\end{eqnarray}
up to an insignificant phase factor [2,3] which we have dropped. 

The relation (\ref{label1.9}) will be important for the discussion of
solutions of the Dirac equation for a free massless fermion field. We
will show below that for a free massless fermionic field in
1+1--dimensional space--time the Dirac equation has two solutions. One
of them possesses non--standard properties under parity
transformations and does not obey the constraint (\ref{label1.9}). For
the first time this solution has been obtained by Thirring in his
seminal paper [1], though, without reference to non--standard
properties under parity transformation.

As has been shown by Thirring [1] the model described by the
Lagrangian (\ref{label1.1}) can be solved exactly. Thirring
constructed the eigenstates of the Hamiltonian and calculated some
observable quantities and found them to be finite after
renormalization.

Of course, the exact solution of the massless Thirring model means
also the possibility of an explicit evaluation of any correlation
function of fermions. For the first time an explicit evaluation of any
correlation function of the massless Thirring model has been carried
out by Klaiber in the traditional quantum field theoretic way within
the operator technique with fermion fields quantized in the chiral
symmetric phase relative to the perturbative chiral symmetric vacuum
[4]\,\footnote{The same technique has been used then by Lowenstein [5]
who studied the short--distance behaviour of products of fermion
fields in the massless Thirring model and the validity of Wilson's
expansion hypothesis.}. According to Klaiber's statement the massless
Thirring model can be reduced to a quantum field theory of a massless
free fermionic field $\Psi$ by a corresponding canonical transformation
of the self--coupled fermionic field $\psi \to \Psi$. The two--point
Green function $S_F(x-y)$ of the free massless fermionic field $\Psi$ is
defined by
\begin{eqnarray}\label{label1.10}
S_F(x-y) = i\langle 0|{\rm T}(\Psi(x)\bar{\Psi}(y))|0\rangle =
\frac{1}{2\pi}\,\frac{\hat{x} - \hat{y}}{(x-y)^2 -\,i\,0},
\end{eqnarray}
where $\hat{x} - \hat{y} =\gamma^{\mu}(x-y)_{\mu}$.

The main implicit point of Klaiber's approach for the evaluation of
correlation functions in the massless Thirring model is the
quantization of the self--coupled fermionic field $\psi$ and the free
field $\Psi$ in the chiral symmetric phase [6]. As we have shown in
Ref.[6] the massless Thirring model with an attractive coupling $g >
0$ is unstable under spontaneous breaking of chiral symmetry and
energetic the chirally broken phase is preferable with respect to
the chiral symmetric one. This implies that for $g >0$ the
self--coupled fermion fields $\psi$ should be quantized in the
chirally broken phase. The wave function of the non--perturbative
vacuum coincides with the wave function of the ground state of the
superconducting phase in the Bardeen, Cooper and Schrieffer theory of
superconductivity (the BCS--theory) [6].

The problem of an exact solution of the massless Thirring model has
been revised within the path--integral approach by Furuya, Gamboa
Saravi and Schaposnik [7] and Na${\acute{\rm o}}$n [8]. These authors
used the technique of auxiliary vector fields where the four--fermion
interaction in the Lagrangian (\ref{label1.1}) is linearized by the
inclusion of a local vector field $V_{\mu}$. The main object of the
path--integral approach is the generating functional of Green
functions
\begin{eqnarray}\label{label1.11}
Z_{\rm Th}[J,\bar{J}] &=& \int {\cal D}\psi{\cal
D}\bar{\psi}\,\exp\,i\int
d^2x\,\Big[\bar{\psi}(x)i\gamma^{\mu}\partial_{\mu}\psi(x) -
\frac{1}{2}\,g\,\bar{\psi}(x)\gamma^{\mu}\psi(x)\bar{\psi}(x)
\gamma_{\mu}\psi(x)\nonumber\\ && + \bar{\psi}(x)J(x) +
\bar{J}(x)\psi(x)\Big],
\end{eqnarray}
where $\bar{J}(x)$ and $J(x)$ are external sources of the $\psi(x)$
and $\bar{\psi}(x)$ fields. 

The exponential of the four--fermion interaction can be replaced by the
Gaussian integral over a vector field $V_{\mu}$:
\begin{eqnarray}\label{label1.12}
&&\exp\,i\int d^2x\,\Big[-\frac{1}{2}\,g\,\bar{\psi}(x)
\gamma^{\mu}\psi(x)\bar{\psi}(x) \gamma_{\mu}\psi(x)\Big] =
\nonumber\\ &&=\int {\cal D}V^{\mu}\,\exp\,i\int
d^2x\,\Big[\frac{1}{2g}\, V_{\mu}(x)V^{\mu}(x) +
\bar{\psi}(x)\gamma^{\mu}\psi(x)\,V_{\mu}(x)\Big].
\end{eqnarray}
Using the path--integral representation (\ref{label1.12}) the generating
functional (\ref{label1.11}) can be recast into the form
\begin{eqnarray}\label{label1.13}
Z_{\rm Th}[J,\bar{J}] &=& \int {\cal D}\psi{\cal D}\bar{\psi}{\cal
D}V^{\mu}\,\exp\,i\int
d^2x\,\Big[\bar{\psi}(x)i\gamma^{\mu}(\partial_{\mu} - i
V_{\mu}(x))\psi(x) + \frac{1}{2g}\, V_{\mu}(x)V^{\mu}(x)\nonumber\\ &&
+ \bar{\psi}(x)J(x) + \bar{J}(x)\psi(x)\Big].
\end{eqnarray}
The first step of the path--integral approach to the solution of the
massless Thirring model suggested in Refs.[7,8] is in the use of the
Hodge decomposition of the vector field $V_{\mu}(x)$
\begin{eqnarray}\label{label1.14}
V_{\mu}(x) = \partial_{\mu}\eta(x) +
\varepsilon_{\mu\nu}\partial^{\nu}\xi(x),
\end{eqnarray}
where $\eta(x)$ and $\xi(x)$ are a scalar and a pseudoscalar field,
respectively.

Substituting the decomposition (\ref{label1.14}) in (\ref{label1.13})
one reduces the generating functional to the form
\begin{eqnarray}\label{label1.15}
Z_{\rm Th}[J,\bar{J}] &=& \int {\cal D}\psi{\cal D}\bar{\psi}{\cal
D}\eta\,{\cal D}\xi\,\exp\,i\int
d^2x\,\Big[\bar{\psi}(x)i\gamma^{\mu}(\partial_{\mu} -
i\partial_{\mu}\eta(x) -
i\varepsilon_{\mu\nu}\partial^{\nu}\xi(x))\psi(x)\nonumber\\ && +
\frac{1}{2g}\,\partial_{\mu}\eta(x)\partial^{\mu}\eta(x) -
\frac{1}{2g}\,\partial_{\mu}\xi(x)\partial^{\mu}\xi(x) +
\bar{\psi}(x)J(x) + \bar{J}(x)\psi(x)\Big],
\end{eqnarray}
where we have used the relation $\varepsilon_{\mu\nu}
\varepsilon^{\mu\alpha} = - g^{\alpha}_{\nu}$.

The second step of the path--integral approach to the solution of the
massless Thirring model suggested in Refs.[7,8] is in the
transformation of the fermion field
\begin{eqnarray}\label{label1.16}
\psi(x) &=& e^{\textstyle i\eta(x) + i\gamma^5\xi(x)
}\,\Psi(x),\nonumber\\ \bar{\psi}(x) &=& \bar{\Psi}(x)\,e^{\textstyle
- i\eta(x) + i\gamma^5\xi(x)}.
\end{eqnarray}
This is an analogy of the canonical transformation suggested by
Klaiber reducing the massless Thirring model of the self--coupled
fermionic field $\psi$ to a quantum field theory of a free massless
fermionic field $\Psi$ [4].

As a result the generating functional of the Green functions reads
\begin{eqnarray}\label{label1.17}
&&Z_{\rm Th}[J,\bar{J}] =\int {\cal D}\Psi{\cal D}\bar{\Psi}{\cal
D}\eta\,{\cal D}\xi\,J[\xi]\,\nonumber\\ &&\times\,
\exp\,i\int d^2x\,\Big[\bar{\Psi}(x)i\gamma^{\mu}\partial_{\mu}\Psi(x)
+ \frac{1}{2g}\,\partial_{\mu}\eta(x)\partial^{\mu}\eta(x) -
\frac{1}{2g}\,\partial_{\mu}\xi(x)\partial^{\mu}
\xi(x)\nonumber\\ && + \bar{\Psi}(x)e^{\textstyle - i\eta(x) +
i\gamma^5\xi(x)}J(x) + \bar{J}(x)e^{\textstyle i\eta(x) +
i\gamma^5\xi(x) }\Psi(x)\Big],
\end{eqnarray}
where $J[\xi]$ is the Jacobian caused by the chiral part of the
transformation (\ref{label1.16}).

In Appendix B we show that following the procedure developed in
[9--14] the chiral Jacobian $J[\xi]$ can be given by the expression
\begin{eqnarray}\label{label1.18}
J[\xi] = \exp\,\Big[-\,i\int d^2x\,\frac{\alpha_J}{2\pi}\,
\partial_{\mu}\xi(x)\partial^{\mu}\xi(x)\Big],
\end{eqnarray}
where an arbitrary real parameter $\alpha_J$ reflects ambiguities of
the evaluation of chiral Jacobians induced by different regularization
procedures [11--13]\,\footnote{In Refs.[13] ambiguities of fermionic
determinants are investigated in the massless Schwinger model.}.

Our expression for the chiral Jacobian (\ref{label1.18}) taken at
$\alpha_J = 1$ coincides with the chiral Jacobian calculated by
Damgaard, Nielsen and Sollacher [15]. Unlike Refs.[7,8] Damgaard,
Nielsen and Sollacher [15] analysed quantum field theories in
1+1--dimensional Minkowski space--time as we are. The agreement can be
reached setting $V^{\mu}(x) = \varepsilon^{\mu\nu}\partial_{\nu}
\theta(x)$ in Eq.(14) of Ref.[15]. This change corresponds to a
decoupling of the vector field $V^{\mu}(x)$ from massless fermion
fields as suggested in Refs.[7,8]\,\footnote{Nice reviews of quantum
field theories in 1+1--dimensional space--time one can find in books
[16] and [17].}.

Due to (\ref{label1.18}) the generating functional (\ref{label1.17})
can be rewritten as follows
\begin{eqnarray}\label{label1.19}
\hspace{-0.3in}&&Z_{\rm Th}[J,\bar{J}] = \int {\cal D}\Psi{\cal D}
\bar{\Psi}{\cal D}\eta\,{\cal D}\xi\,\nonumber\\
\hspace{-0.3in}&&\times \exp\,i\int
d^2x\,\Big[\bar{\Psi}(x)i\gamma^{\mu}\partial_{\mu}\Psi(x) +
\frac{1}{2g}\,\partial_{\mu}\eta(x)\partial^{\mu}\eta(x) -
\frac{1}{2g}\,\Big(1 + \alpha_J\,\frac{g}{\pi}\Big)\,
\partial_{\mu}\xi(x)\partial^{\mu} \xi(x)\nonumber\\
\hspace{-0.3in}&& + \bar{\Psi}(x)e^{\textstyle - i\eta(x) +
i\gamma^5\xi(x)}J(x) + \bar{J}(x)e^{\textstyle i\eta(x) +
i\gamma^5\xi(x) }\Psi(x)\Big].
\end{eqnarray}
Using the identity
\begin{eqnarray}\label{label1.20}
e^{\textstyle i\gamma^5\xi(x)} =
\Big(\frac{1+\gamma^5}{2}\Big)\,e^{\textstyle i\,\xi(x)} +
\Big(\frac{1-\gamma^5}{2}\Big)\,e^{\textstyle -\,i\,\xi(x)}
\end{eqnarray}
one can rewrite the generating functional (\ref{label1.19}) in the
more convenient form
\begin{eqnarray}\label{label1.21}
\hspace{-0.3in}&&Z_{\rm Th}[J,\bar{J}] = \int {\cal D}\Psi{\cal D}
\bar{\Psi}{\cal D}\eta\,{\cal D}\xi\,\nonumber\\
\hspace{-0.3in}&&\times \exp\,i\int
d^2x\,\Big\{\bar{\Psi}(x)i\gamma^{\mu}\partial_{\mu}\Psi(x) +
\frac{1}{2g}\,\partial_{\mu}\eta(x)\partial^{\mu}\eta(x) -
\frac{1}{2g}\,\Big(1 +
\alpha_J\,\frac{g}{\pi}\Big)\,\partial_{\mu}\xi(x)\partial^{\mu}
\xi(x)\nonumber\\
\hspace{-0.3in}&& + \bar{\Psi}(x)\Big[\Big(\frac{1+\gamma^5}{2}\Big)
e^{\textstyle - i\eta(x) + i\xi(x)} + \Big(\frac{1-\gamma^5}{2}\Big)
e^{\textstyle - i\eta(x) - i\xi(x)}\Big]J(x)\nonumber\\
\hspace{-0.3in}&& + \bar{J}(x)\Big[\Big(\frac{1+\gamma^5}{2}\Big)
e^{\textstyle i\eta(x) + i\xi(x) } +
\Big(\frac{1-\gamma^5}{2}\Big)e^{\textstyle i\eta(x) - i\xi(x)
}\Big]\Psi(x)\Big\}.
\end{eqnarray}
The last step in the transformation of the generating functional
$Z_{\rm Th}[J,\bar{J}]$ is to rescale the fields $\eta$ and $\xi$:
$\eta \to \sqrt{g}\,\eta$ and $\xi \to \lambda(\alpha_J)\,\xi$, where
$\lambda(\alpha_J)\,$ is determined by
\begin{eqnarray}\label{label1.22}
\lambda(\alpha_J) =\sqrt{\frac{g}{\displaystyle 1 + \alpha_J\,
\frac{g}{\pi}}}
\end{eqnarray}
As a result the generating functional of Green functions of the
massless Thirring model $Z_{\rm Th}[J,\bar{J}]$ is defined by the path
integral
\begin{eqnarray}\label{label1.23}
\hspace{-0.5in}&&Z_{\rm Th}[J,\bar{J}] = \int {\cal D}\Psi{\cal D}
\bar{\Psi}{\cal D}\eta\,{\cal D}\xi\,\nonumber\\
\hspace{-0.5in}&&\times \exp\,i\int
d^2x\,\Big\{\bar{\Psi}(x)i\gamma^{\mu}\partial_{\mu}\Psi(x) +
\frac{1}{2}\,\partial_{\mu}\eta(x)\partial^{\mu}\eta(x) -
\frac{1}{2}\,\partial_{\mu}\xi(x)\partial^{\mu}
\xi(x)\nonumber\\
\hspace{-0.5in}&& + \bar{\Psi}(x)\Big[\Big(\frac{1+\gamma^5}{2}\Big)
e^{\textstyle - i\sqrt{g}\,\eta(x) + i \lambda(\alpha_J)\,\xi(x)} +
\Big(\frac{1-\gamma^5}{2}\Big) e^{\textstyle - i\sqrt{g}\,\eta(x) - i
\lambda(\alpha_J)\, \xi(x)}\Big]J(x)\nonumber\\
\hspace{-0.5in}&& + \bar{J}(x)\Big[\Big(\frac{1+\gamma^5}{2}\Big)
e^{\textstyle i\sqrt{g}\,\eta(x) + i \lambda(\alpha_J)\, \xi(x)} +
\Big(\frac{1-\gamma^5}{2}\Big)e^{\textstyle i \sqrt{g}\,\eta(x) - i
\lambda(\alpha_J)\, \xi(x)}\Big]\Psi(x)\Big\}.\nonumber\\
\hspace{-0.5in}&&
\end{eqnarray}
It is important to emphasize that the kinetic term of the pseudoscalar
field $\xi$ enters with opposite sign. The free Lagrangian
of the scalar and pseudoscalar fields $\eta(x)$ and $\xi(x)$ is
invariant under global $O(1,1)$ rotations
\begin{eqnarray}\label{label1.24}
\eta(x) \to \eta\,'(x) &=& \cosh\gamma\,\eta(x) +
\sinh\gamma\,\xi(x),\nonumber\\ \xi(x) \to \xi\,'(x) &=&
\sinh\gamma\,\eta(x) + \cosh\gamma\,\xi(x).
\end{eqnarray}
In parallel to the generating functional of Green functions
(\ref{label1.11}) one can consider the generating functional of
correlation functions of the scalar fermionic field density
$\bar{\psi}(x)\psi(x)$ determined by
\begin{eqnarray}\label{label1.25}
\hspace{-0.5in}&&Z_{\rm Th}[\sigma] = \int {\cal D}\psi{\cal
D}\bar{\psi}\nonumber\\
\hspace{-0.5in}&&\times\,\exp\,i\int
d^2x\,\Big[\bar{\psi}(x)i\gamma^{\mu}\partial_{\mu}\psi(x) -
\frac{1}{2}\,g\,\bar{\psi}(x)\gamma^{\mu}\psi(x)\bar{\psi}(x)
\gamma_{\mu}\psi(x) + \bar{\psi}(x)\psi(x)\sigma(x)\Big],
\end{eqnarray}
where $\sigma(x)$ is an external source of $\bar{\psi}(x)\psi(x)$. Due
to the transformations expounded above the generating functional
$Z_{\rm Th}[\sigma]$ can be reduced to the form of the following path
integral
\begin{eqnarray}\label{label1.26}
\hspace{-0.5in}&&Z_{\rm Th}[\sigma] =\nonumber\\
\hspace{-0.5in}&&= \int {\cal D}\Psi{\cal
D}\bar{\Psi}{\cal D}\xi\,\exp\,i\int
d^2x\,\Bigg[\bar{\Psi}(x)i\gamma^{\mu}\partial_{\mu}\Psi(x) -
\frac{1}{2}\,\partial_{\mu}\xi(x)\partial^{\mu}\xi(x)\nonumber\\
\hspace{-0.5in}&&+ \bar{\Psi}(x)\,e^{\textstyle
2i\lambda(\alpha_J)\,\gamma^5\,\xi(x)}\,\Psi(x)\sigma(x)\Bigg]
=\nonumber\\
\hspace{-0.5in}&&= \int {\cal D}\Psi{\cal D}\bar{\Psi}{\cal
D}\xi\,\exp\,i\int
d^2x\,\Bigg\{\bar{\Psi}(x)i\gamma^{\mu}\partial_{\mu}\Psi(x) -
\frac{1}{2}\,\partial_{\mu}\xi(x)\partial^{\mu}\xi(x)\nonumber\\
\hspace{-0.5in}&&+ \bar{\Psi}(x)\,\Bigg[\Bigg(\frac{1 +
\gamma^5}{2}\Bigg)\,e^{\textstyle 2i\lambda(\alpha_J)\,\xi(x)} +
\Bigg(\frac{1 - \gamma^5}{2}\Bigg)\,e^{\textstyle -
2i\lambda(\alpha_J)\,\xi(x)}\Bigg]\,\Psi(x)\sigma(x)\Bigg\}.
\end{eqnarray}
We emphasize that the $\eta$--field does not contribute to the
correlation functions of the scalar fermion density
$\bar{\psi}(x)\psi(x)$.

In order to consider the correlation functions, treated by Coleman
[18] for a proof of equivalence between the massive Thirring model and
the sine--Gordon, one can introduce the following generating
functional
\begin{eqnarray}\label{label1.27}
\hspace{-0.5in}&&Z_{\rm Th}[\sigma,\varphi] = \int {\cal D}\psi{\cal
D}\bar{\psi}\,\exp\,i\int
d^2x\,\Big[\bar{\psi}(x)i\gamma^{\mu}\partial_{\mu}\psi(x) -
\frac{1}{2}\,g\,\bar{\psi}(x)\gamma^{\mu}\psi(x)\bar{\psi}(x)
\gamma_{\mu}\psi(x)\nonumber\\
\hspace{-0.5in}&& + \bar{\psi}(x)\psi(x)\sigma(x) +
\bar{\psi}(x)i\gamma^5\psi(x)\varphi(x)\Big],
\end{eqnarray}
where $\sigma(x)$ and $\varphi(x)$ are external sources of scalar
$\bar{\psi}(x)\psi(x)$ and pseudoscalar
$\bar{\psi}(x)i\gamma^5\psi(x)$ fermion densities, respectively.

Recall that Klaiber's expressions for $2n$--point correlation
functions of massless fermion fields have been used by Coleman [18]
for his proof of equivalence between the massive Thirring model and
the sine--Gordon model which are described by the Lagrangians [18]
\begin{eqnarray}\label{label1.28}
{\cal L}_{\rm Th}(x) = \bar{\psi}(x) (i\gamma^{\mu}\partial_{\mu} -
m)\psi(x) - \frac{1}{2}\,g\,\bar{\psi}(x)\gamma^{\mu}\psi(x)
\bar{\psi}(x) \gamma_{\mu}\psi(x)
\end{eqnarray}
and 
\begin{eqnarray}\label{label1.29}
{\cal L}_{\rm SG}(x) =
\frac{1}{2}\partial_{\mu}\vartheta(x)\partial^{\mu}\vartheta(x) +
\frac{\bar{\alpha}}{\bar{\beta}^2}\,(\cos\bar{\beta}\vartheta(x) - 1),
\end{eqnarray}
where $m$ is the mass of the Thirring fermion fields, $\vartheta(x)$
is a sine--Gordon model pseudoscalar (or scalar) field, and
$\bar{\alpha}$ and $\bar{\beta}$ are real positive parameters [18].

The most interesting property of the sine--Gordon model is the
existence of classical, stable solutions of the equations of motion --
solitons and anti--solitons. Many--soliton solutions obey Pauli's
exclusion principle and can be interpreted as a fermion--like
behaviour.

The equivalence between the massive Thirring model and the
sine--Gordon model Coleman proved perturbatively developing a
perturbation theory with respect to the parameters $m$ and $\alpha$
and quantizing fermion and boson field relative to the perturbative
chiral symmetric vacuum. He showed that $n$--point correlation
functions of the products of massless Thirring fermion field densities
$\bar{\psi}(x)(1 \pm \gamma^5)\psi(x)$ and massless pseudoscalar boson
sine--Gordon fields coincide if (i) the coupling constants $g$ and
$\beta$ are related by
\begin{eqnarray}\label{label1.30}
\frac{4\pi^2}{\beta^2}= 1 + \frac{g}{\pi}
\end{eqnarray}
and (ii) there exist the bosonization rules (so--called Abelian
bosonization rules)
\begin{eqnarray}\label{label1.31}
\bar{\psi}(x)\gamma^{\mu}\psi(x) &=&
\frac{\beta}{2\pi}\,\varepsilon^{\mu\nu}\partial_{\nu}\vartheta(x),
\nonumber\\ Z\,\bar{\psi}(x)\Big(\frac{1\pm \gamma^5}{2}\,\Big)\psi(x)
&=& - \frac{\alpha}{\beta^2}\,e^{\textstyle \mp i\beta\,\vartheta(x)},
\end{eqnarray}
where $Z$ is a constant depending on the regularization procedure
[18].

The generating functional (\ref{label1.27}) reduces itself to the path
integral
\begin{eqnarray}\label{label1.32}
\hspace{-0.5in}&&Z_{\rm Th}[\sigma,\varphi] = Z_{\rm
Th}[\sigma_-,\sigma_+] = \nonumber\\
\hspace{-0.5in}&& =\int {\cal D}\Psi{\cal D}\bar{\Psi}{\cal
D}\xi\,\exp\,i\int
d^2x\,\Bigg\{\bar{\Psi}(x)i\gamma^{\mu}\partial_{\mu}\Psi(x) -
\frac{1}{2}\,\partial_{\mu}\xi(x)\partial^{\mu}\xi(x)\nonumber\\
\hspace{-0.5in}&& + \bar{\Psi}(x)\,\Bigg(\frac{1 +
\gamma^5}{2}\Bigg)\,e^{\textstyle +
2i\lambda(\alpha_J)\,\xi(x)}\Psi(x)\,\sigma_-(x)\nonumber\\
\hspace{-0.5in}&& + \bar{\Psi}(x)\,\Bigg(\frac{1 -
\gamma^5}{2}\Bigg)\,e^{\textstyle -
2i\lambda(\alpha_J)\,\xi(x)}\,\Psi(x)\,\sigma_+(x)\Bigg\},
\end{eqnarray}
where $\sigma_{\mp}(x) = \sigma(x)\mp i\varphi(x)$ are external
sources of the lefthanded and righthanded fermionic field density,
respectively.

The evaluation of correlation functions of Thirring fermions by
means of the generating functionals evidences an exact solution of the
massless Thirring model.  

In this paper we discuss the problems of the uniqueness of the
solution of the massless Thirring model with fermion fields quantized
in the chiral symmetric phase in terms of correlation functions.  We
analyse the evaluation of $n$--point correlation functions of
left(right)handed fermion densities $\bar{\psi}(x)(1 \pm
\gamma^5)\psi(x)$ and the $2n$--point Green functions and the
dependence of these quantities on arbitrary parameters.  In section 2
we discuss the operator formalism suggested by Klaiber for the
solution of the massless Thirring model. We notice that Klaiber's
operator approach to the solution of the massless Thirring model
possesses some shortcomings. One of them concerns the non--standard
parity properties.  In order to show that the fermion vector current
is proportional to the gradient of a scalar density Klaiber used the
solution for a free massless fermionic field contradicting to the
standard properties under parity transformation. This has led to
composite field operators $c(k^1)$ and $c^{\dagger}(k^1)$ which do not
have definite parity when compared with the standard definition
[2,3]. In section 3 we give a representation of the vector current of
the massless Thirring model using the solution of a free massless
fermion field having standard parity properties. In section 4 we
discuss canonical properties of the composite operators $c(k^1)$ and
$c^{\dagger}(k^1)$ and construct new canonical composite operators
$C(k^1)$ and $C^{\dagger}(k^1)$ behaving as scalars under parity
transformations. We show that the transition to these new operators
$C(k^1)$ and $C^{\dagger}(k^1)$ does not affect Klaiber's result for
the evaluation of correlation functions and the statement on the
existence of a one--parameter family of solutions of the massless
Thirring model.  Canonical properties of the operators $c(k^1)$ and
$c^{\dagger}(k^1)$ and $C(k^1)$ and $C^{\dagger}(k^1)$ are discussed
in connection with the derivation of the Schwinger terms. In section 5
within Klaiber's operator formalism we consider quantization of the
massless Thirring model in the chiral symmetric phase.  We derive the
quantum equation of motion for the massless Thirring model and confirm
the existence of a one--parameter family of solutions for quantized
fermionic fields. We show that Klaiber's parameterization of a
one--parameter family of solutions is incorrect.  We demonstrate that
Klaiber's parameterization is based on the definition of the quantum
vector current of the massless Thirring model breaking chiral symmetry
explicitly. In fact, the divergence of the quantum axial--vector
current $j^{\mu}_5(x)$ determined according to Klaiber's definition of
the quantum vector current $j^{\mu}(x)$ does not vanish due to the
contribution of non--covariant terms. We show that if one considers
simultaneously a definition of $j^{\mu}_5(x)$ and requires the
fulfillment of the standard relation $j^{\mu}_5(x) =
-\varepsilon^{\mu\nu}j_{\nu}(x)$ all parameters become fixed and there
is no one--parameter family of solutions of the massless Thirring
model within Klaiber's operator formalism. We give (i) a definition of
$j^{\mu}(x)$ and $j^{\mu}_5(x)$ consistent with chiral symmetry and
the standard relation, $j^{\mu}_5(x) =
-\varepsilon^{\mu\nu}j_{\nu}(x)$, definitions of quantum vector and
axial--vector currents, (ii) a new definition of Klaiber's parameters
and discuss (iii) a possibility to retain a one--parameter family of
solutions of the massless Thirring model with quantized fermionic
fields. In section 6 we replicate Na${\acute{\rm o}}$n's results for
the $n$--point correlation functions of the left(right)handed fermion
densities $\bar{\psi}(x)(1 \pm \gamma^5)\psi(x)$ which have been
treated by Coleman [18] in connection with his proof of equivalence
between the massive Thirring model and the sine--Gordon model
quantizing the Thirring fermion fields in the chiral symmetric phase.
We show that the dynamical dimensions of $\bar{\psi}(x)(1 \pm
\gamma^5)\psi(x)$ depend on an arbitrary parameter $\alpha_J$
reflecting the ambiguities of the evaluation of chiral Jacobian by
different regularization procedures. In section 7 we perform a
complete bosonization of the generating functionals (\ref{label1.32})
and (\ref{label1.26}). We show that the generating functional
(\ref{label1.26}) in the bosonized form coincides with the partition
function of the sine--Gordon model at $\sigma(x) = - m$, where $m$ can
be treated as a Thirring fermion mass.  In section 8 we evaluate the
two--point fermion Green function by using the generating functional
(\ref{label1.23}). We find again a dependence of the dynamical
dimension of the massless fermionic field on the parameter $\alpha_J$.
The dependence of dynamical dimensions of correlation function on
$\alpha_J$ agrees with the existence of a one--parameter family of
solutions of the massless Thirring model. In section 9 we evaluate
$2n$--point Green functions of massless fermion fields and compare the
obtained results with Klaibers' [4].  We show that Klaiber's
correlation functions are singular in the infrared limit. This implies
the necessity of an infrared renormalization of Klaiber's correlation
functions. In section 10 we consider two--point correlation functions
of left(right)handed fermion densities. We treat this correlation
function as the two--point limit of the four--point correlation
function. The former is evaluated explicitly both with Klaiber's
operator formalism and the path--integral approach. We argue the
necessity of the non--perturbative infrared renormalization of
Klaiber's expression to get the result depending on the ultra--violet
cut--off as we have in the path--integral approach.  In section 11 we
investigate the non--perturbative renormalization of the massless
Thirring model. We find that (i) the coupling constant $g$ is
unrenormalized giving a vanishing Gell--Mann--Low function $\beta(g) =
0$ and (ii) the requirement of renormalizability fixes the parameter
$\alpha_J$ to the value $\alpha_J = -2\pi/g$. We notice that our
result concerning the vanishing of the Gell--Mann--Low function or,
correspondingly, the absence of a coupling constant renormalization is
supported by Mueller and Trueman [19] and Gomes and Lowenstein [20]
who observed the same result within the massive Thirring model. In the
Conclusion we discuss the obtained results. In Appendix A we discuss
the properties of fermion fields under parity transformations. In
Appendix B we evaluate the chiral Jacobian induced by local fermionic
field transformations (\ref{label1.16}) and demonstrate, following
Christos [11], the dependence of the chiral Jacobian on an arbitrary
parameter $\alpha_J$. In Appendix C we calculate the two--point
correlation functions defining commutators of scalar and pseudoscalar
fermionic field densities. In Appendix D we make a justification of
Klaiber's ansatz for the definition of the quantum fermionic vector
current $K^{\mu}(x)$ and define in analogous way the quantum fermionic
axial--vector current $K^{\mu}_5(x)$. In Appendix E we calculate a
Fourier transform of the two--point Green function of the massless
Thirring fermion fields. In Appendix F we analyse the
non--perturbative renormalizability of the sine--Gordon model and show
that it is not an asymptotically free quantum field theory as well as
the massless Thirring model.

\section{Proof of Klaiber's basic assumption and  
unusual parity properties of fermion fields} 
\setcounter{equation}{0}

\hspace{0.2in} In this section we discuss Klaiber's operator approach
[4] to the solution of the massless Thirring model. We focus our
discussion on parity properties of fermion fields.  Klaiber's operator
formalism is based on the assumption of a proportionality of the
vector fermion current to the gradient of a {\it scalar} fermion
density, the proof of which has been carried out by the employment of
the solution of the Dirac equation for a free massless fermionic field
with non--standard parity properties.

Klaiber's approach to the solution of the massless Thirring model
starts with the analysis of the equation of motion for the massless
Thirring fermion field.  Therefore, following Klaiber we have to start
with the classical equation of motion and conservation
equations. Using the Lagrangian (\ref{label1.1}) the equation of
motion reads
\begin{eqnarray}\label{label2.1}
i\gamma^{\mu}\,\partial_{\mu}\psi(x) =
g\,j^{\mu}(x)\,\gamma_{\mu}\psi(x),
\end{eqnarray}
where the vector current $j^{\mu}(x)$ as well as the axial vector
current $j^{\mu}_5(x)$ are defined by (\ref{label1.6}) and
(\ref{label1.7}). 

As has been stated by Klaiber [4] the vector and axial--vector
currents can be written in the form of gradients
\begin{eqnarray}\label{label2.2}
j_{\mu}(x) = \frac{1}{\sqrt{\pi}}\,\partial_{\mu}j(x)\quad,\quad j_{5
\mu}(x) = \frac{1}{\sqrt{\pi}}\,\partial_{\mu}j_5(x).
\end{eqnarray}
Due to the properties of the vector and axial--vector currents,
$j_{\mu}(x)$ and $j_{5\mu}(x)$, under the parity transformation
(\ref{label1.8}) the quantities $j(x)$ and $j_5(x)$ should be scalar
and pseudoscalar densities [2,3]
\begin{eqnarray}\label{label2.3}
{\cal P}\,j(x^0,x^1)\,{\cal P}^{\dagger} &=& + j(x^0, -
x^1),\nonumber\\ {\cal P}\,j_5(x^0,x^1)\,{\cal P}^{\dagger} &=& -
j_5(x^0, - x^1).
\end{eqnarray}
Assuming the relations (\ref{label2.2}) Klaiber suggested to introduce
a new fermionic field $\Psi(x)$
\begin{eqnarray}\label{label2.4}
\Psi(x) = e^{\textstyle i\,g\,j(x)/\sqrt{\pi}}\,\psi(x)
\end{eqnarray}
and its currents
\begin{eqnarray}\label{label2.5}
J^{\mu}(x) &=& \bar{\Psi}(x)\gamma^{\mu}\Psi(x),\nonumber\\
J^{\mu}_5(x) &=& \bar{\Psi}(x)\gamma^{\mu}\gamma^5\Psi(x),
\end{eqnarray}
In the case of the validity of (\ref{label2.2}) one can easily show
that the fermion field (\ref{label2.4}) and the currents
(\ref{label2.5})\,\footnote{In the case of quantized fermionic fields
the vector and axial--vector currents should be taken in the
normal--ordered form $J^{\mu}(x) =
:\!\bar{\Psi}(x)\gamma^{\mu}\Psi(x)\!:$ and $J^{\mu}_5(x) =
:\!\bar{\Psi}(x)\gamma^{\mu}\gamma^5\Psi(x)\!:$ as well as the
equation of motion (\ref{label2.1}):
$i\gamma^{\mu}\,\partial_{\mu}\psi(x) =
g\,:\!j^{\mu}(x)\,\gamma_{\mu}\psi(x)\!:$.} satisfy the relations [4]
\begin{eqnarray}\label{label2.6}
i\gamma^{\mu}\partial_{\mu}\Psi(x) &=& 0,\nonumber\\ J^{\mu}(x) &=&
j^{\mu}(x),\nonumber\\ J^{\mu}_5(x) &=& j^{\mu}_5(x).
\end{eqnarray}
This implies that the fermion field $\Psi(x)$ has the same properties
under parity transformations as the field $\psi(x)$
\begin{eqnarray}\label{label2.7}
{\cal P}\,\Psi(x^0,x^1)\,{\cal P}^{\dagger}
&=&\gamma^0\Psi(x^0,-x^1),\nonumber\\ {\cal
P}\,J^{\mu}(x^0,x^1)\,{\cal P}^{\dagger} &=& +
J_{\mu}(x^0,-x^1),\nonumber\\ {\cal P}\,J^{\mu}_5(x^0,x^1)\,{\cal
P}^{\dagger} &=& - J_{5\mu}(x^0,-x^1).
\end{eqnarray}
Thus, the general solution of the equation of motion (\ref{label2.1})
can be written in the form
\begin{eqnarray}\label{label2.8}
\psi(x) = e^{\textstyle -\,i\,g\,J(x)/\sqrt{\pi}}\,\Psi(x),
\end{eqnarray}
where $\Psi(x)$ obeys the free field equation of motion and $J(x)$ is
{\it its integrated current} [4].

Due to the relation $J^{\mu}_5(x) = -\varepsilon^{\mu\nu}\,J_{\nu}(x)$
Klaiber claimed that the solution (\ref{label2.8}) can be extended and
represented in the more general form of a one--parameter family
\begin{eqnarray}\label{label2.9}
\psi(x) &=& e^{\textstyle -i\,\alpha\,J(x) -
i\,\beta\,\gamma^5\,J_5(x)}\,\Psi(x),\nonumber\\
\bar{\psi}(x)&=& \bar{\Psi}(x)\,e^{\textstyle i\,\alpha\,J(x) -
i\,\beta\,\gamma^5\,J_5(x)},
\end{eqnarray}
where the parameters $\alpha$ and $\beta$ obey the constraint
\begin{eqnarray}\label{label2.10}
\alpha - \beta = \frac{g}{\sqrt{\pi}},
\end{eqnarray}
and the scalar and pseudoscalar densities $J(x)$ and $J_5(x)$ are
related to the vector and axial--vector currents $J_{\mu}(x)$ and
$J_{5\mu}(x)$ as
\begin{eqnarray}\label{label2.11}
J_{\mu}(x) = \bar{\Psi}(x)\gamma_{\mu}\Psi(x) =
\frac{1}{\sqrt{\pi}}\,\partial_{\mu}J(x)\quad,\quad J_{5 \mu}(x) =
\bar{\Psi}(x)\gamma_{\mu}\gamma^5\Psi(x) =
\frac{1}{\sqrt{\pi}}\,\partial_{\mu}J_5(x).
\end{eqnarray}
Under parity transformations the densities $J(x)$ and $J_5(x)$ behave
like $j(x)$ and $j_5(x)$, respectively,
\begin{eqnarray}\label{label2.12}
{\cal P}\,J(x^0,x^1)\,{\cal P}^{\dagger} &=& + J(x^0, -
x^1),\nonumber\\ {\cal P}\,J_5(x^0,x^1)\,{\cal P}^{\dagger} &=& -
J_5(x^0, - x^1).
\end{eqnarray}
In quantizing the model Klaiber assumed that $\Psi(x)$ is a canonical
free fermion field. As a free canonical field a quantized $\Psi(x)$
should have the following expansion into plane waves [6]
\begin{eqnarray}\label{label2.13}
&&\Psi(x) = \int^{\infty}_{-\infty}\frac{dp^1}{\sqrt{2\pi}}\,
\frac{1}{\sqrt{ 2p^0}}\,\Big[u(p^0,p^1)a(p^1)\,e^{\textstyle -ip\cdot
x} + v(p^0,p^1)b^{\dagger}(p^1)\,e^{\textstyle ip\cdot
x}\Big],\nonumber\\
&&\bar{\Psi}(x) = \int^{\infty}_{-\infty}\frac{dp^1}{\sqrt{2\pi}}\,
\frac{1}{\sqrt{2p^0}}\Big[\bar{u}(p^0,p^1)a^{\dagger}(p^1)
\,e^{\textstyle ip\cdot x} + \bar{v}(p^0,p^1)b(p^1)\,e^{\textstyle
-ip\cdot x}\Big],
\end{eqnarray}
where $p\cdot x = p^0x^0 - p^1x^1$. The creation
$a^{\dagger}(p^1)\,(b^{\dagger}(p^1))$ and annihilation
$a(p^1)\,(b(p^1))$ operators of fermions (antifermions) with momentum
$p^1$ and energy $p^0 = |p^1|$ obey the anticommutation
relations
\begin{eqnarray}\label{label2.14}
\{a(p^1), a^{\dagger}(q^1)\} &=& \{b(p^1), b^{\dagger}(q^1)\} =
\delta(p^1-q^1),\nonumber\\
\{a(p^1), a(q^1)\} = \{a^{\dagger}(p^1), a^{\dagger}(q^1)\} &=& 
\{b(p^1), b(q^1)\} =
\{b^{\dagger}(p^1), b^{\dagger}(q^1)\} = 0.
\end{eqnarray}
The spinorial wave functions $u(p^0,p^1)$ and
$v(p^0,p^1)=u(-p^0,-p^1)$ are the solutions of the Dirac equation in
the momentum representation for positive and negative energies,
respectively. They are defined by [6]
\begin{eqnarray}\label{label2.15}
u(p^0,p^1) &=& {\displaystyle
\left(\begin{array}{c}\sqrt{p^0 + p^1} \\ \sqrt{p^0 - p^1}
\end{array}\right)} = \sqrt{2p^0}{\displaystyle
\left(\begin{array}{c}\theta(+p^1) \\ \theta(-p^1)
\end{array}\right)},\nonumber\\
\bar{u}(p^0,p^1) &=& (\sqrt{p^0 - p^1}, \sqrt{p^0 + p^1})=
\sqrt{2p^0}\,(\theta(-p^1), \theta(+p^1)),\nonumber\\ v(p^0,p^1) &=&
{\displaystyle \left(\begin{array}{c}\sqrt{p^0 + p^1} \\ -\sqrt{p^0 -
p^1}
\end{array}\right)} = \sqrt{2p^0}{\displaystyle
\left(\begin{array}{c}\theta(+p^1) \\ -\theta(-p^1)
\end{array}\right)},\nonumber\\
\bar{v}(p^0,p^1) &=& (- \sqrt{p^0 - p^1}, \sqrt{p^0 + p^1}) =
\sqrt{2p^0}\,(-\theta(-p^1), \theta(+p^1))
\end{eqnarray}
at $p^0 = |p^1|$ and normalized to 
\begin{eqnarray}\label{label2.16}
u^{\dagger}(p^0,p^1)u(p^0,p^1) &=& v^{\dagger}(p^0,p^1)v(p^0,p^1) =
2p^0,\nonumber\\ \bar{u}(p^0,p^1)u(p^0,p^1) &=&
\bar{v}(p^0,p^1)v(p^0,p^1) = 0,\nonumber\\ \bar{u}(p^0,p^1)v(p^0,p^1)
&=& \bar{v}(p^0,p^1)u(p^0,p^1) = 0.
\end{eqnarray}
The spinorial wave functions $u(p^0,p^1)$ and $v(p^0,p^1)$ satisfy the
following matrix relations [6]
\begin{eqnarray}\label{label2.17}
u(p^0,p^1)\bar{u}(p^0,p^1) = v(p^0,p^1)\bar{v}(p^0,p^1) = \gamma^0p^0
- \gamma^1p^1 = \hat{p}
\end{eqnarray}
and the identities [2,3] (see Appendix A)
\begin{eqnarray}\label{label2.18}
u(p^0,p^1) = \gamma^0u(p^0,-p^1)\quad,\quad v(p^0,p^1) = - \gamma^0
v(p^0, - p^1).
\end{eqnarray}
These identities are important for the correct behaviour of the
operators of creation and annihilation of fermions and antifermions
under parity transformation (see Appendix A)
\begin{eqnarray}\label{label2.19}
{\cal P}\,a(k^1)\,{\cal P}^{\dagger} &=& + a(-k^1)\quad,\quad {\cal
P}\,a^{\dagger}(k^1)\,{\cal P}^{\dagger} = + a^{\dagger}(-k^1),
\nonumber\\ 
{\cal P}\,b(k^1)\,{\cal P}^{\dagger} &=& - b(-k^1)\quad,
\quad {\cal P}\,b^{\dagger}(k^1)\,{\cal P}^{\dagger} =
- b^{\dagger}(-k^1).
\end{eqnarray}
We would like to emphasize that Klaiber [4] following Thirring used
the expansion of the $\Psi$--field into plane wave, where
\begin{eqnarray}\label{label2.20}
v(p^0, p^1) = u(p^0,p^1) = \sqrt{2p^0}{\displaystyle
\left(\begin{array}{c}\theta(+p^1) \\ \theta(-p^1)
\end{array}\right)}.
\end{eqnarray}
For such a solution all relations (\ref{label2.16}) and
(\ref{label2.17}) are retained. However, as has been shown in [6] this
solution cannot be obtained from the solution for a free massive
fermionic field in the zero--mass limit.  As a result the spinorial wave
function $v(p^0, p^1)$ breaks the identity (\ref{label2.18}) that
entails the violation of the relations (\ref{label2.19}). In fact,
according to the solution (\ref{label2.20}) they read (see Appendix A)
\begin{eqnarray}\label{label2.21}
{\cal P}\,a(k^1){\cal P}^{\dagger} &=& + a(-k^1)\quad,\quad {\cal
P}\,a^{\dagger}(k^1){\cal P}^{\dagger} = +
a^{\dagger}(-k^1)\nonumber\\ {\cal P}\,b(k^1){\cal P}^{\dagger}
&=&\stackrel{\textstyle ?}{+} b(-k^1)\quad,\quad {\cal
P}\,b^{\dagger}(k^1){\cal P}^{\dagger} = \stackrel{\textstyle ?}{+}
b^{\dagger}(-k^1).
\end{eqnarray}
This contradicts to the well--known properties of the fermion creation
and annihilation operators under parity transformations.

Now let us replicate the derivation of the representation of the
vector current $J^{\mu}(x) = :\!\bar{\Psi}(x)\gamma^{\mu}\Psi(x)\!:$
suggested by Klaiber [4]
\begin{eqnarray}\label{label2.22}
J^{\mu}(x) = -\frac{i}{\sqrt{2}\pi}\int^{\infty}_{-\infty}
\frac{dk^1}{\sqrt{2k^0}}\,k^{\mu}\,\Big[c(k^1)\,e^{\textstyle -ik\cdot
x} - c^{\dagger}(k^1)\,e^{\textstyle ik\cdot x}\Big],
\end{eqnarray}
and confirming Klaiber's assumption (\ref{label2.11}).

Following Klaiber and using the expansion of the $\Psi$--field into
plane waves with the spinorial functions related by (\ref{label2.20})
we obtain 
\begin{eqnarray}\label{label2.23}
\hspace{-0.5in}&&J^{\mu}(x) = :\bar{\Psi}(x)\gamma^{\mu}\Psi(x):=
\int^{\infty}_{-\infty}\int^{\infty}_{-\infty}\frac{dp^1dq^1}{2\pi}\,\frac{1}{\sqrt{2p^02q^0}}\,
\bar{u}(p^0,p^1)\gamma^{\mu}u(q^0,q^1)\nonumber\\
\hspace{-0.5in}&&\times\,\Big\{a^{\dagger}(p^1)a(q^1) e^{\textstyle
i(p-q)\cdot x} - b^{\dagger}(q^1)b(p^1)\, e^{\textstyle -i(p-q)\cdot
x}\nonumber\\
\hspace{-0.5in}&& + b(p^1) a(q^1)\, e^{\textstyle -i(p + q)\cdot x} +
a^{\dagger}(p^1) b^{\dagger}(q^1)\,e^{\textstyle i(p+q)\cdot
x}\,\Big\}.
\end{eqnarray}
The analytical expression of the quantity $[\bar{u}(p^0,
p^1)\gamma^{\mu}u(q^0, q^1)]$ suitable for the derivation of the
momentum representation is rather ambiguous and the result depends
fully on the skill and intuition of the investigator. Following
Klaiber's intuition we represent this quantity as follows
\begin{eqnarray}\label{label2.24}
&&\bar{u}(p^0,p^1)\gamma^{\mu}u(q^0,q^1) = \nonumber\\
&&=\frac{p^{\mu} - q^{\mu}}{p^0
- q^0}\,\sqrt{2p^02q^0}\,\theta(q^1(p^1-q^1)) + \frac{q^{\mu} -
p^{\mu}}{q^0 - p^0}\,\sqrt{2p^02q^0}\,\theta(p^1(q^1-p^1))
\end{eqnarray}
for the first and second term in (\ref{label2.23}) and 
\begin{eqnarray}\label{label2.25}
\bar{u}(p^0,p^1)\gamma^{\mu}u(q^0,q^1) = \sqrt{2p^02q^0}\,
\frac{p^{\mu} + q^{\mu}}{p^0 + q^0}\,\theta(p^1q^1)
\end{eqnarray}
for the third and fourth term, respectively.

Using (\ref{label2.24}) and (\ref{label2.25}) we represent
(\ref{label2.23}) in the form
\begin{eqnarray}\label{label2.26}
\hspace{-0.5in}&&J^{\mu}(x)=\int^{\infty}_{-\infty}
\int^{\infty}_{-\infty}\frac{dp^1dq^1}{2\pi}\,\nonumber\\
\hspace{-0.5in}&&\times
\Bigg\{\Bigg[\frac{p^{\mu} - q^{\mu}}{p^0 - q^0}\,\theta(q^1(p^1 -
q^1)) + \frac{q^{\mu} - p^{\mu}}{q^0 - p^0}\,\theta(p^1(q^1 - p^1))
\Bigg]\,a^{\dagger}(p^1)a(q^1)\,e^{\textstyle + i(p-q)\cdot x}\nonumber\\
\hspace{-0.5in}&&~ - \Bigg[\frac{p^{\mu} - q^{\mu}}{p^0 -
q^0}\,\theta(q^1(p^1 - q^1)) + \frac{q^{\mu} - p^{\mu}}{q^0 -
p^0}\,\theta(p^1(q^1 -
p^1))\Bigg]\,b^{\dagger}(q^1)b(p^1)\,e^{\textstyle - i(p-q)\cdot x}
\nonumber\\
\hspace{-0.5in}&&~ + \frac{p^{\mu} + q^{\mu}}{p^0 +
q^0}\,\theta(p^1q^1)\,b(p^1) a(q^1)\, e^{\textstyle -i(p + q)\cdot x}\nonumber\\
\hspace{-0.5in}&&~ + \frac{p^{\mu} + q^{\mu}}{p^0 +
q^0}\,\theta(p^1q^1)\, a^{\dagger}(p^1)b^{\dagger}(q^1)\,e^{\textstyle
i(p+q)\cdot x}\,\Bigg\}.
\end{eqnarray}
Then, let us rewrite (\ref{label2.26}) in an equivalent form and
consider the components of the vector current separately
\begin{eqnarray}\label{label2.27}
\hspace{-0.5in}&&J^0(x)=\int^{\infty}_{-\infty}
\int^{\infty}_{-\infty}\frac{dp^1dq^1}{2\pi}\,\nonumber\\
\hspace{-0.5in}&&\times \Bigg\{\theta(q^1(p^1 - q^1))\,[a^{\dagger}(q^1)a(p^1) -
b^{\dagger}(q^1)b(p^1)]\,e^{\textstyle -i(p-q)\cdot x}\nonumber\\
\hspace{-0.5in}&&~ + \theta(q^1(p^1 - q^1))\,[a^{\dagger}(p^1)a(q^1) -
b^{\dagger}(p^1)b(q^1)]\,e^{\textstyle + i(p-q)\cdot x}\nonumber\\
\hspace{-0.5in}&&~ + \theta(p^1q^1)\,b(p^1) a(q^1)\, e^{\textstyle -i(p + q)\cdot x}\nonumber\\
\hspace{-0.5in}&&~ + \theta(p^1q^1)\, a^{\dagger}(p^1)b^{\dagger}(q^1)\,e^{\textstyle
+ i(p+q)\cdot x}\,\Bigg\},\nonumber\\
\hspace{-0.5in}&&J^1(x)=\int^{\infty}_{-\infty}
\int^{\infty}_{-\infty}\frac{dp^1dq^1}{2\pi}\,\nonumber\\
\hspace{-0.5in}&&\times \Bigg\{\varepsilon(p^1 - q^1)
\theta(q^1(p^1 - q^1))\,[a^{\dagger}(q^1)a(p^1) -
b^{\dagger}(q^1)b(p^1)]\,e^{\textstyle -i(p-q)\cdot x}\nonumber\\
\hspace{-0.5in}&&~ + \varepsilon(p^1 - q^1)\theta(q^1(p^1 - q^1))
\,[a^{\dagger}(p^1)a(q^1) -
b^{\dagger}(p^1)b(q^1)]\,e^{\textstyle + i(p-q)\cdot x}\nonumber\\
\hspace{-0.5in}&&~ + \varepsilon(p^1 + q^1)\theta(p^1q^1)
\,b(p^1) a(q^1)\, e^{\textstyle -i(p + q)\cdot x}\nonumber\\
\hspace{-0.5in}&&~ + \varepsilon(p^1 + q^1)\theta(p^1q^1)\,
a^{\dagger}(p^1)b^{\dagger}(q^1)\,e^{\textstyle + i(p+q)\cdot
x}\,\Bigg\},
\end{eqnarray}
where $\varepsilon(p^1-q^1)$ is a sign function. 

Making a change of variables $p^1 - q^1 \to k^1$ in the first and
second terms and $p^1 + q^1 \to k^1$ in the third and fourth ones we
arrive at the expression
\begin{eqnarray}\label{label2.28}
\hspace{-0.5in}&&J^0(x)=\int^{\infty}_{-\infty}
\int^{\infty}_{-\infty}\frac{dk^1dq^1}{2\pi}\,\nonumber\\
\hspace{-0.5in}&&\times \Bigg\{\theta(k^1q^1)\,[a^{\dagger}(q^1)a(k^1
+ q^1) - b^{\dagger}(q^1)b(k^1 + q^1)]\,e^{\textstyle -i(|k^1 +q^1| -|q^1|)x^0 + ik^1x^1}\nonumber\\
\hspace{-0.5in}&&~ + \theta(k^1q^1)\,[a^{\dagger}(k^1 + q^1)a(q^1) -
b^{\dagger}(k^1 + q^1)b(q^1)]\,e^{\textstyle + i(|k^1+q^1| - |q^1|)x^0
- ik^1x^1}\nonumber\\
\hspace{-0.5in}&&~ + \theta(q^1(k^1-q^1))\,b(k^1 - q^1) a(q^1)\,
e^{\textstyle -i(|k^1 - q^1| + |q^1|)x^0 + ik^1x^1}\nonumber\\
\hspace{-0.5in}&&~ + \theta(q^1(k^1 - q^1))\,
a^{\dagger}(q^1)b^{\dagger}(k^1 - q^1)\, e^{\textstyle + i(|k^1 - q^1|
+ |q^1|)x^0 -i k^1x^1}\,\Bigg\},\nonumber\\
\hspace{-0.5in}&&J^1(x)=\int^{\infty}_{-\infty}
\int^{\infty}_{-\infty}\frac{dk^1dq^1}{2\pi}\,\nonumber\\
\hspace{-0.5in}&&\times \Bigg\{\varepsilon(k^1)\theta(k^1q^1)
\,[a^{\dagger}(q^1)a(k^1
+ q^1) - b^{\dagger}(q^1)b(k^1 + q^1)]\,
e^{\textstyle -i(|k^1 +q^1| -|q^1|)x^0 + ik^1x^1}\nonumber\\
\hspace{-0.5in}&&~ + \varepsilon(k^1)\theta(k^1q^1)
\,[a^{\dagger}(k^1 + q^1)a(q^1) -
b^{\dagger}(k^1 + q^1)b(q^1)]\,
e^{\textstyle + i(|k^1+q^1| - |q^1|)x^0
- ik^1x^1}\nonumber\\
\hspace{-0.5in}&&~ + \varepsilon(k^1)
\theta(q^1(k^1-q^1))\,b(k^1 - q^1) a(q^1)\,
e^{\textstyle -i(|k^1 - q^1| + |q^1|)x^0 + ik^1x^1}\nonumber\\
\hspace{-0.5in}&&~ + \varepsilon(k^1) \theta(q^1(k^1 - q^1))\,
a^{\dagger}(q^1)b^{\dagger}(k^1 - q^1)\, e^{\textstyle + i(|k^1 - q^1|
+ |q^1|)x^0 -i k^1x^1}\,\Bigg\}.
\end{eqnarray}
The relations (\ref{label2.28}) can be represented in the compact form
\begin{eqnarray}\label{label2.29}
\hspace{-0.5in}&&J^{\mu}(x)=\int^{\infty}_{-\infty}
\int^{\infty}_{-\infty}\frac{dk^1dq^1}{2\pi}\,
\frac{k^{\mu}}{k^0}\nonumber\\
\hspace{-0.5in}&&\times \Bigg\{\theta(k^1q^1)\,[a^{\dagger}(q^1)a(k^1
+ q^1) - b^{\dagger}(q^1)b(k^1 + q^1)]\,e^{\textstyle -i(|k^1 +q^1|
-|q^1|)x^0 + ik^1x^1}\nonumber\\
\hspace{-0.5in}&&~ + \theta(k^1q^1)\,[a^{\dagger}(k^1 + q^1)a(q^1) -
b^{\dagger}(k^1 + q^1)b(q^1)]\,e^{\textstyle + i(|k^1+q^1| - |q^1|)x^0
- ik^1x^1}\nonumber\\
\hspace{-0.5in}&&~ + \theta(q^1(k^1-q^1))\,b(k^1 - q^1) a(q^1)\,
e^{\textstyle -i(|k^1 - q^1| + |q^1|)x^0 + ik^1x^1}\nonumber\\
\hspace{-0.5in}&&~ + \theta(q^1(k^1 - q^1))\,
a^{\dagger}(q^1)b^{\dagger}(k^1 - q^1)\, e^{\textstyle + i(|k^1 - q^1|
+ |q^1|)x^0 -i k^1x^1}\,\Bigg\}.
\end{eqnarray}
Due to the Heaviside functions in the meaningful region of momentum
integrals we have $|k^1 +q^1| -|q^1| = |k^1| = k^0$ and $|k^1 - q^1|
-|q^1| = |k^1| = k^0$. This yields
\begin{eqnarray}\label{label2.30}
\hspace{-0.5in}&&J^{\mu}(x)=\int^{\infty}_{-\infty}
\int^{\infty}_{-\infty}\frac{dk^1dq^1}{2\pi}\,\frac{k^{\mu}}{k^0}\nonumber\\
\hspace{-0.5in}&&\times \Bigg\{\theta(k^1q^1)\,[a^{\dagger}(q^1)a(k^1
+ q^1) - b^{\dagger}(q^1)b(k^1 + q^1)]\,e^{\textstyle -ik\cdot
x}\nonumber\\
\hspace{-0.5in}&&~ + \theta(k^1q^1)\,[a^{\dagger}(k^1 + q^1)a(q^1) -
b^{\dagger}(k^1 + q^1)b(q^1)]\,e^{\textstyle + ik\cdot x}\nonumber\\
\hspace{-0.5in}&&~ + \theta(q^1(k^1-q^1))\,b(k^1 - q^1) a(q^1)\,
e^{\textstyle -ik\cdot x}\nonumber\\
\hspace{-0.5in}&&~ + \theta(q^1(k^1 - q^1))\,
a^{\dagger}(q^1)b^{\dagger}(k^1 - q^1)\, e^{\textstyle + ik\cdot
x}\,\Bigg\} = \nonumber\\
\hspace{-0.5in}&&= -\frac{i}{\sqrt{2}\pi}\int^{\infty}_{-\infty}
\frac{dk^1}{\sqrt{2k^0}}\,k^{\mu}\,\Big[c(k^1)\,e^{\textstyle -ik\cdot
x} - c^{\dagger}(k^1)\,e^{\textstyle ik\cdot x}\Big].
\end{eqnarray}
Thus, the vector current can be represented in the form
(\ref{label2.22}), where the operators $c\,(k^1)$ and
$c^{\dagger}(k^1)$ are defined by
\begin{eqnarray}\label{label2.31}
c\,(k^1) &=&\frac{i}{\sqrt{k^0}}\int^{\infty}_{-\infty}dq^1\,
\Big\{\theta(k^1q^1 )\,[a^{\dagger}(q^1)\,a(k^1 + q^1) -
b^{\dagger}(q^1)\,b(k^1 + q^1)]\nonumber\\ && + \theta(q^1(k^1 -
q^1))\,b(k^1 - q^1)\,a(q^1)\Big\},\nonumber\\ c^{\dagger}(k^1)
&=&-\frac{i}{\sqrt{k^0}}\int^{\infty}_{-\infty}dq^1\,\Big\{
\theta(k^1q^1)\,[a^{\dagger}(k^1 + q^1)\,a(q^1) - b^{\dagger}(k^1 +
q^1)\,b(q^1)]\nonumber\\ && + \theta(q^1(k^1 -
q^1))\,a^{\dagger}(q^1)\,b^{\dagger}(k^1 - q^1)\Big\}.
\end{eqnarray}
These operators agree fully with the expressions adduced by Klaiber
(see (II.15) of Ref.[4]).

The vector current given by (\ref{label2.28}) satisfies the relation
(\ref{label2.11}) with the fermion density $J(x)$ equal to
\begin{eqnarray}\label{label2.32}
J(x)= \frac{1}{\sqrt{2\pi}}\int^{\infty}_{-\infty}
\frac{dk^1}{\sqrt{2k^0}}\,\Big[c(k^1)\,e^{\textstyle -ik\cdot x} +
c^{\dagger}(k^1)\,e^{\textstyle ik\cdot x}\Big].
\end{eqnarray}
The axial vector current $J^{\mu}_5(x)
=:\!\bar{\Psi}(x)\gamma^{\mu}\gamma^5\Psi(x)\!:$ and the fermion
density $J_5(x)$ related by $J^{\mu}_5(x) =
\partial^{\mu}J_5(x)/\sqrt{\pi}$ have the following momentum
representations
\begin{eqnarray}\label{label2.33}
\hspace{-0.5in}J^{\mu}_5(x) &=&
-\frac{i}{\sqrt{2}\pi}\int^{\infty}_{-\infty}
\frac{dk^1}{\sqrt{2k^0}}\,\varepsilon(k^1)\,k^{\mu}\,\Big[c(k^1)\,e^{\textstyle
-ik\cdot x} - c^{\dagger}(k^1)\,e^{\textstyle ik\cdot
x}\Big],\nonumber\\ J_5(x) &=&
\frac{1}{\sqrt{2\pi}}\int^{\infty}_{-\infty}
\frac{dk^1}{\sqrt{2k^0}}\,\varepsilon(k^1)\,\Big[c(k^1)\,e^{\textstyle
-ik\cdot x} + c^{\dagger}(k^1)\,e^{\textstyle ik\cdot x}\Big],
\end{eqnarray}
where $\varepsilon(k^1)$ is a sign function.

According to Klaiber's analysis the operators
$c\,(k^1)$ and $c^{\dagger}(k^1)$ should be scalar operators. This
implies that under parity transformations these operators should
transform like
\begin{eqnarray}\label{label2.34}
{\cal P}\,c\,(k^1)\,{\cal P}^{\dagger} &=& c\,(-k^1),\nonumber\\ {\cal
P}\,c^{\dagger}(k^1)\,{\cal P}^{\dagger} &=& c^{\dagger}(-k^1).
\end{eqnarray}
Using the standard properties (\ref{label2.19}) of the creation and
annihilation operators of fermions and antifermions under parity
transformation one finds
\begin{eqnarray}\label{label2.35}
&&{\cal P}\,c\,(k^1)\,{\cal P}^{\dagger} =\nonumber\\
&&=\frac{i}{\sqrt{k^0}}\int^{\infty}_{-\infty}dq^1\,
\Big\{\theta(k^1q^1)\,[a^{\dagger}(-q^1)a(-k^1 - q^1) -
b^{\dagger}(-q^1)b(-k^1 - q^1)]\nonumber\\ && - \theta(q^1(k^1 -
q^1))\,b(- k^1 + q^1) a(-q^1)\Big\}=\nonumber\\
&&=\frac{i}{\sqrt{k^0}}\int^{\infty}_{-\infty}dq^1\,
\Big\{\theta(-k^1q^1)\,[a^{\dagger}(q^1)a(-k^1 + q^1) -
b^{\dagger}(q^1)b(-k^1 + q^1)]\nonumber\\ && - \theta(q^1(- k^1 -
q^1))\,b(- k^1 - q^1) a(q^1)\Big\} \neq c\,(-k^1),\nonumber\\ &&{\cal
P}\,c^{\dagger}(k^1)\,{\cal P}^{\dagger} \neq c^{\dagger}(-k^1).
\end{eqnarray}
Hence, within the standard definition of the properties of fermion and
antifermion operators of creation and annihilation, the operators
$J^{\mu}(x)$ and $J^{\mu}_5(x)$ given by (\ref{label2.22}) and
(\ref{label2.33}) do not possess a definite parity as well as the
fermion densities $J(x)$ and $J_5(x)$.

\section{Fermionic 
fields with standard properties under parity transformation and vector
current decomposition}
\setcounter{equation}{0}

\hspace{0.2in} In this section we derive the decomposition of the
vector current operator $J^{\mu}(x) =
:\!\bar{\Psi}(x)\gamma^{\mu}\Psi(x)\!:$ into scalar and pseudoscalar
fermion densities. We use the solution for a free massless fermionic
field (\ref{label2.15}) with standard properties of creation and
annihilation operators under parity transformations [2,3] (see
Appendix A). In this case (\ref{label2.23}) reads
\begin{eqnarray}\label{label3.1}
\hspace{-0.5in}&&J^{\mu}(x) = :\bar{\Psi}(x)\gamma^{\mu}\Psi(x):=
\int^{\infty}_{-\infty}\int^{\infty}_{-\infty}
\frac{dp^1dq^1}{2\pi}\,\frac{1}{\sqrt{2p^02q^0}}\,\Big\{
[\bar{u}(p^0,p^1)\gamma^{\mu}u(q^0,q^1)]\nonumber\\
\hspace{-0.5in}&&\times\,a^{\dagger}(p^1)a(q^1) e^{\textstyle
i(p-q)\cdot x} - [\bar{v}(p^0,p^1)\gamma^{\mu}v(q^0,q^1)]\,
b^{\dagger}(q^1)b(p^1)\, e^{\textstyle -i(p-q)\cdot
x}\nonumber\\
\hspace{-0.5in}&& + [\bar{v}(p^0,p^1)\gamma^{\mu}u(q^0,q^1)]\,b(p^1)
a(q^1)\, e^{\textstyle -i(p + q)\cdot x}\nonumber\\
\hspace{-0.5in}&& +
[\bar{u}(p^0,p^1)\gamma^{\mu}v(q^0,q^1)]\,a^{\dagger}(p^1)
b^{\dagger}(q^1)\,e^{\textstyle i(p+q)\cdot x}\,\Big\}.
\end{eqnarray}
The products of spinorial wave functions are equal to
\begin{eqnarray}\label{label3.2}
\hspace{-0.5in}&&\bar{u}(p^0,p^1)\gamma^{\mu}u(q^0,q^1) =
[\bar{v}(p^0,p^1)\gamma^{\mu}v(q^0,q^1)] =\nonumber\\ \hspace{-0.5in}&&=\frac{p^{\mu}
- q^{\mu}}{p^0 - q^0}\,\sqrt{2p^02q^0}\,\theta(q^1(p^1-q^1)) +
\frac{q^{\mu} - p^{\mu}}{q^0 -
p^0}\,\sqrt{2p^02q^0}\,\theta(p^1(q^1-p^1)),\nonumber\\
\hspace{-0.5in}&&[\bar{v}(p^0,p^1)\gamma^{\mu}u(q^0,q^1)]
=[\bar{v}(p^0,p^1)\gamma^{\mu}u(q^0,q^1)]=
-\,\sqrt{2p^02q^0}\,{\displaystyle \varepsilon^{\mu\nu}\,\frac{p_{\nu}
+ q_{\nu}}{p^0 + q^0}\,\theta(p^1q^1).}
\end{eqnarray}
Using these expressions we recast the r.h.s. of (\ref{label3.1}) into
the form
\begin{eqnarray}\label{label3.3}
\hspace{-0.5in}&&J^{\mu}(x)=\int^{\infty}_{-\infty}
\int^{\infty}_{-\infty}\frac{dk^1dq^1}{2\pi}\, \frac{k_{\nu}
}{k^0}\nonumber\\
\hspace{-0.5in}&&\times\,\Big\{g^{\mu\nu}\,\theta(k^1q^1)\,[a^{\dagger}(q^1)a(k^1 + q^1) - b^{\dagger}(q^1)b(k^1 +
q^1)]\,e^{\textstyle - ik\cdot x} \nonumber\\
\hspace{-0.5in}&& -\varepsilon^{\mu\nu}\,\theta(q^1(k^1 -
q^1))\,b(k^1 - q^1) a(q^1)\, e^{\textstyle -ik\cdot x} \nonumber\\
\hspace{-0.5in}&&+ g^{\mu\nu}\,\theta(k^1 q^1)\,[a^{\dagger}(k^1 +
q^1)a(q^1) - b^{\dagger}(k^1 + q^1)b(q^1)]\,e^{\textstyle ik\cdot
x}\nonumber\\
\hspace{-0.5in}&&-\varepsilon^{\mu\nu}\,\theta(q^1(k^1 - q^1))\,
a^{\dagger}(q^1)b^{\dagger}(k^1 - q^1)\,e^{\textstyle ik\cdot
x}\,\Big\} =\nonumber\\
\hspace{-0.5in}&& =
\frac{1}{\sqrt{2}\pi}\int^{\infty}_{-\infty}\frac{dk^1}{\sqrt{2k^0}}\,
\Big[ -ik^{\mu}\,d(k^1)\,e^{\textstyle -ik\cdot x} -
ik_{\nu}\,\varepsilon^{\mu\nu}\,d_5(k^1)\,e^{\textstyle -ik\cdot
x}\nonumber\\
\hspace{-0.5in}&&+ ik^{\mu}\,d^{\dagger}(k^1)\,e^{\textstyle ik\cdot
x} + ik_{\nu}\,\varepsilon^{\mu\nu}\,d^{\dagger}_5(k^1)\,e^{\textstyle
+ik\cdot x}\Big].
\end{eqnarray}
This implies that the vector--current $J^{\mu}(x)$ has the following
decomposition
\begin{eqnarray}\label{label3.4}
J^{\mu}(x) = \frac{1}{\sqrt{\pi}}\,\partial^{\mu}J(x) +
\frac{1}{\sqrt{\pi}}\,\varepsilon^{\mu\nu}\, \partial_{\nu}J_5(x)
\end{eqnarray}
in terms of gradients of the scalar and pseudoscalar fermion densities
$J(x)$ and $J_5(x)$ defined by
\begin{eqnarray}\label{label3.5}
J(x) &=&
\frac{1}{\sqrt{2\pi}}\int^{\infty}_{-\infty}\frac{dk^1}{\sqrt{2k^0}}\,
\Big[d(k^1)\,e^{\textstyle -ik\cdot x} +
d^{\dagger}(k^1)\,e^{\textstyle ik\cdot x}\Big],\nonumber\\ J_5(x) &=&
\frac{1}{\sqrt{2\pi}}\int^{\infty}_{-\infty}\frac{dk^1}{\sqrt{2k^0}}\,
\Big[d_5(k^1)\,e^{\textstyle -ik\cdot x} +
d^{\dagger}_5(k^1)\,e^{\textstyle ik\cdot x}\Big].
\end{eqnarray}
The operators $d(k^1)$, $d^{\dagger}(k^1)$, $d_5(k^1)$ and
$d^{\dagger}_5(k^1)$ read
\begin{eqnarray}\label{label3.6}
d(k^1) &=&
\frac{i}{\sqrt{k^0}}\int^{\infty}_{-\infty}dq^1\,\theta(k^1q^1)\,[a^{\dagger}(q^1)a(k^1 + q^1) -
b^{\dagger}(q^1)b(k^1 + q^1)],\nonumber\\
d^{\dagger}(k^1)&=&-\frac{i}{\sqrt{k^0}}\int^{\infty}_{-\infty}dq^1\,
\theta(k^1q^1)\,[a^{\dagger}(k^1 + q^1)a(q^1) -
b^{\dagger}(k^1 + q^1)b(q^1)],\nonumber\\ d_5(k^1) &=&
-\,\frac{i}{\sqrt{k^0}}\int^{\infty}_{-\infty}dq^1\,\theta(q^1(k^1 -
q^1))\,b(k^1 - q^1) a(q^1),\nonumber\\ d^{\dagger}_5(k^1)&=&
\frac{i}{\sqrt{k^0}}\int^{\infty}_{-\infty}dq^1 \,\theta(q^1(k^1 -
q^1))\,a^{\dagger}(q^1)b^{\dagger}(k^1 - q^1).
\end{eqnarray}
They possess correct properties under parity transformation
\begin{eqnarray}\label{label3.7}
{\cal P}\,d(k^1)\,{\cal P}^{\dagger} &=& + d(-k^1),\nonumber\\ {\cal
P}\,d_5(k^1)\,{\cal P}^{\dagger} &=& - d_5(-k^1).
\end{eqnarray}
Since $J^{\mu}(x) = j^{\mu}(x)$, the same decomposition is valid 
for $j^{\mu}(x)$, i.e.
\begin{eqnarray}\label{label3.8}
j^{\mu}(x) = \frac{1}{\sqrt{\pi}}\,\partial^{\mu}j(x) +
\frac{1}{\sqrt{\pi}}\,\varepsilon^{\mu\nu}\, \partial_{\nu}j_5(x).
\end{eqnarray}
Thus, the vector currents $J^{\mu}(x)$ and $j^{\mu}(x)$ have standard
Hodge decompositions.

\section{Canonical commutation relation for $c(k^1)$ and 
$c^{\dagger}(k^1)$, the Schwinger term and a one--parameter 
family of solutions}
\setcounter{equation}{0}

\hspace{0.2in} As has been shown by Klaiber the operators $c(k^1)$ and
$c^{\dagger}(p^1)$ obey the canonical commutation relation
\begin{eqnarray}\label{label4.1}
[c(k^1),c^{\dagger}(p^1)] = \delta(k^1 - p^1).
\end{eqnarray}
This is very important and amazing result of Klaiber's operator
formalism, if to take into account that the commutation relation
(\ref{label4.1}) can be obtained by using canonical anti--commutation
relations for operators of creation and annihilation of fermions and
antifermions (\ref{label2.14}).

Recall, as has been proved by Callan, Dashen, and Sharp [21] 
the commutation relation between time and spatial components of the
vector current $j^{\mu}(x) = \bar{\psi}(x)\gamma^{\mu}\psi(x)$
vanishes
\begin{eqnarray}\label{label4.2}
[j^0(x^0,x^1),j^1(x^0,y^1)] = 0
\end{eqnarray}
using canonical equal--time anticommutation relations for the massless
Thirring fermion fields
\begin{eqnarray}\label{label4.3}
\{\psi(x^0,x^1),\psi^{\dagger}(x^0,y^1)\} = \delta(x^1 - y^1).
\end{eqnarray}
However, according to Schwinger [22] the equal--time commutator
(\ref{label4.2}) should not vanish and reads
\begin{eqnarray}\label{label4.4}
[j^0(x^0,x^1),j^1(x^0,y^1)] = -i\,c\,\frac{\partial}{\partial
x^1}\delta(x^1-y^1),
\end{eqnarray}
where $c$ is the Schwinger term. In the massless Thirring model for
fermion fields quantized in the chiral symmetric phase the Schwinger
term has been calculated by Sommerfield [23]: $c = 1/\pi$. We would
like to emphasize, that the result $c = 1/\pi$ is not a consequence of
canonical anti--commutativity of fermionic field operators but the
result of an evaluation of vacuum expectation values (see also [6]).

The calculation of the equal--time commutator
$[j^0(x^0,x^1),j^1(x^0,y^1)]$ within Klaiber's formalism runs in the
way
\begin{eqnarray}\label{label4.5}
\hspace{-0.5in}&&[j^0(x^0,x^1),j^1(x^0,y^1)]
=[J^0(x^0,x^1),J^1(x^0,y^1)]=
-\frac{1}{\sqrt{\pi}}\frac{\partial}{\partial
y^1}[J^0(x^0,x^1),J(x^0,y^1)] = \nonumber\\
\hspace{-0.5in}&&= \frac{\partial}{\partial
y^1}\frac{i}{2\pi^2}\int^{\infty}_{-\infty}\frac{dk^1}{\sqrt{2k^0}}
\int^{\infty}_{-\infty}\frac{dq^1}{\sqrt{2q^0}}\,k^0\,
\Big\{[c(k^1),c^{\dagger}(q^1)] \,e^{\textstyle -i(k^0-q^0)x^0 +
ik^1x^1 - iq^1y^1}\nonumber\\
\hspace{-0.5in}&&\times + [c(q^1),c^{\dagger}(k^1)]\,e^{\textstyle
i(k^0-q^0)x^0 - ik^1x^1 + iq^1y^1}\Big\} =
i\,\frac{1}{\pi}\,\frac{\partial}{\partial
y^1}\int^{\infty}_{-\infty}\frac{dk^1}{2\pi}\,e^{\textstyle ik^1(x^1 -
y^1)} =\nonumber\\
\hspace{-0.5in}&&= - i\,\frac{1}{\pi}\,\frac{\partial}{\partial
x^1}\delta(x^1-y^1).
\end{eqnarray}
The obtained result agrees completely with (\ref{label4.4}) but
contradicts to (\ref{label4.2}). This should be much more astonishing
due to the fact that $J^0(x^0,x^1)$ and $J^1(x^0,y^1)$ are the
components of the vector current of a free massless fermion field. The
derivation of the equal--time commutator (\ref{label4.5}) in agreement
with Schwinger's ansatz (\ref{label4.4}) testifies that the operators
$c(k^1)$ and $c^{\dagger}(k^1)$ are not real canonical quantum field
operators. We prove this statement in section 5.

Now let us calculate the commutator
$[c(k^1),c^{\dagger}(p^1)]$. Substituting (\ref{label2.31}) and using
the relation
\begin{eqnarray}\label{label4.6}
[AB,CD] = A\{B,C\}D + CA\{B,D\} - C\{A,D\}B - \{A,C\}BD,
\end{eqnarray}
where $A$, $B$, $C$ and $D$ are any operators, we get
\begin{eqnarray}\label{label4.7}
\hspace{-0.5in}&&[c(k^1),c^{\dagger}(p^1)] =
\frac{1}{\sqrt{k^0p^0}}\int^{\infty}_{-\infty}dq^1
\int^{\infty}_{-\infty}d{\ell}^1\,\Big\{\theta(k^1q^1)
\theta(p^1{\ell}^1)\nonumber\\
\hspace{-0.5in}&&\times\,\Big([a^{\dagger}(q^1)a(k^1+q^1),
a^{\dagger}(p^1+{\ell}^1)a({\ell}^1)]+
[b^{\dagger}(q^1)b(k^1+q^1),b^{\dagger}(p^1+{\ell}^1)b({\ell}^1)]\Big)
\nonumber\\
\hspace{-0.5in}&&+\theta(q^1(k^1-q^1))
\theta({\ell}^1(p^1-{\ell}^1))[b(k^1-q^1)a(q^1),
a^{\dagger}({\ell}^1)b^{\dagger}(p^1-{\ell}^1)]\nonumber\\
\hspace{-0.5in}&&+\theta(k^1q^1)\theta({\ell}^1(p^1 -
{\ell}^1))[a^{\dagger}(q^1)a(k^1+q^1)-
b^{\dagger}(q^1)b(k^1+q^1),a^{\dagger}({\ell}^1)
b^{\dagger}(p^1-{\ell}^1)]\nonumber\\
\hspace{-0.5in}&&+\theta(q^1(k^1-q^1))\theta(p^1{\ell}^1)
[b(k^1-q^1)a(q^1),a^{\dagger}(p^1+{\ell}^1)a({\ell}^1)-
b^{\dagger}(p^1+{\ell}^1)b({\ell}^1)]\Big\}=\nonumber\\
\hspace{-0.5in}&&=\frac{1}{\sqrt{k^0p^0}}\int^{\infty}_{-\infty}dq^1\,
\Big\{\delta(k^1-p^1)\,\theta(q^1(k^1-q^1))\theta(q^1(p^1-q^1))
+[\theta(k^1q^1)\theta(p^1(k^1 - p^1 + q^1))\nonumber\\
\hspace{-0.5in}&& -
\theta(k^1(q^1 - p^1))\theta(p^1(q^1 - p^1)) -
\theta(q^1(p^1-q^1))\theta((p^1-q^1)(k^1-p^1+q^1))]\nonumber\\
\hspace{-0.5in}&&\times\,[a^{\dagger}(q^1)a(k^1 - p^1 + q^1) +
b^{\dagger}(q^1)b(k^1 - p^1 + q^1)] +
[\theta(k^1q^1)\theta((k^1+q^1)(p^1-k^1-q^1))\nonumber\\
\hspace{-0.5in}&& - \theta(k^1(p^1-k^1-q^1))\theta(q^1(p^1-q^1))]\,
a^{\dagger}(q^1)b^{\dagger}(p^1-k^1-q^1) +
[\theta(p^1(k^1-p^1-q^1)\nonumber\\
\hspace{-0.5in}&&\times\,\theta(q^1(k^1.q^1) -
\theta(p^1q^1)\theta((p^1+q^1)(k^1-p^1-q^1))]\,
b(q^1)a(k^1-p^1-q^1)\Big\} = \delta(k^1 - p^1).
\end{eqnarray}
This confirms Klaiber's result (\ref{label4.1}).

However, in order to identify the operators $c(k^1)$ and
$c^{\dagger}(k^1)$ with operators of annihilation and creation of a
boson with a momentum $k^1$ it is necessary to show that the
commutators $[c(k^1), c(p^1)]$ and
$[c^{\dagger}(k^1),c^{\dagger}(p^1)]$ vanish. In this case $c(k^1)$
and $c^{\dagger}(k^1)$ would be canonical quantum field operators.
The calculation of the commutator $[c(k^1), c(p^1)]$ is analogous to
$[c(k^1), c^{\dagger}(p^1)]$ and reads
\begin{eqnarray}\label{label4.8}
\hspace{-0.5in}&&[c(k^1),c(p^1)] = -
\frac{1}{\sqrt{k^0p^0}}\int^{\infty}_{-\infty}dq^1
\int^{\infty}_{-\infty}d{\ell}^1\,\Big\{\theta(k^1q^1)
\theta(p^1{\ell}^1)\nonumber\\
\hspace{-0.5in}&&\times\,\Big([a^{\dagger}(q^1)a(k^1+q^1),
a^{\dagger}({\ell}^1)a(p^1+{\ell}^1)]+
[b^{\dagger}(q^1)b(k^1+q^1),b^{\dagger}({\ell}^1)b(p^1+{\ell}^1)]\Big)
\nonumber\\
\hspace{-0.5in}&&+\theta(k^1q^1)\theta({\ell}^1(p^1 -
{\ell}^1))[a^{\dagger}(q^1)a(k^1+q^1)- b^{\dagger}(q^1)b(k^1+q^1),
b(p^1-{\ell}^1)a({\ell}^1)]\nonumber\\
\hspace{-0.5in}&&+\theta(q^1(k^1-q^1))\theta(p^1{\ell}^1)
[b(k^1-q^1)a(q^1),a^{\dagger}({\ell}^1)a(p^1+{\ell}^1)-
b^{\dagger}({\ell}^1)b(p^1+{\ell}^1)]\Big\}\nonumber\\
\hspace{-0.5in}&& = -
\frac{1}{\sqrt{k^0p^0}}\int^{\infty}_{-\infty}dq^1\,\Big\{[\theta(k^1q^1)\theta(p^1(k^1+q^1)) -
\theta(p^1q^1)\theta(k^1(p^1+q^1))]\nonumber\\
\hspace{-0.5in}&&\times\,[a^{\dagger}(q^1)a(k^1+p^1+q^1) +
b^{\dagger}(q^1)b(k^1+p^1+q^1)] + [\theta(k^1(q^1-k^1))
\nonumber\\
\hspace{-0.5in}&&\times\,\theta((q^1-k^1)(k^1+p^1-q^1)) -
\theta(k^1(p^1-q^1))\theta(q^1(p^1-q^1)) +
\theta(q^1(k^1-q^1))\nonumber\\
\hspace{-0.5in}&&\times\,\theta(p^1(k^1-q^1)) -
\theta(p^1(q^1-p^1))\theta((q^1-p^1)(k^1+p^1-q^1))]
\,b(q^1)a(k^1+p^1-q^1)\Big\}.
\end{eqnarray}
Accounting for the properties of the Heaviside functions one can show
that all momentum integrals vanish. This yields
\begin{eqnarray}\label{label4.9}
[c(k^1),c(p^1)] = 0.
\end{eqnarray}
This testifies that from the naive point of view $c(k^1)$ and
$c^{\dagger}(k^1)$ are canonical quantum field operators. 

In turn, the operators $d(k^1)$, $d^{\dagger}(k^1)$, $d_5(k^1)$ and
$d^{\dagger}_5(k^1)$ do not satisfy canonical commutation relations
(\ref{label4.1}) and (\ref{label4.9}). Moreover, the commutators of
these operators are $q$--valued quantities. Hence, the only way to
deal with canonical quantum field operators is to follow Klaiber and
represent the vector fermion current $J^{\mu}(x)$ in the form of a
gradient of the scalar fermionic field density.

For this aim we suggest to turn to (\ref{label3.3}) and rewrite the
product $\varepsilon^{\mu\nu}k_{\nu}$ in the equivalent form
\begin{eqnarray}\label{label4.10}
\varepsilon^{\mu\nu}k_{\nu} = - \varepsilon(k^1)\,k^{\mu}.
\end{eqnarray}
This allows to recast the vector current into the form 
\begin{eqnarray}\label{label4.11}
\hspace{-0.3in}J^{\mu}(x)=
\frac{1}{\sqrt{2}\pi}\int^{\infty}_{-\infty}\frac{dk^1}{\sqrt{2k^0}}\,
\Big[ -ik^{\mu}\,C(k^1)\,e^{\textstyle -ik\cdot x} +
ik^{\mu}\,C^{\dagger}(k^1)\,e^{\textstyle ik\cdot x}\Big]
=\frac{1}{\sqrt{\pi}}\,\partial^{\mu}S(x),
\end{eqnarray}
where $S(x)$ is a real scalar fermionic field density 
\begin{eqnarray}\label{label4.12}
S(x)=
\frac{1}{\sqrt{2\pi}}\int^{\infty}_{-\infty}\frac{dk^1}{\sqrt{2k^0}}\,
\Big[C(k^1)\,e^{\textstyle -ik\cdot x} +
C^{\dagger}(k^1)\,e^{\textstyle ik\cdot x}\Big] .
\end{eqnarray}
The operators $C(k^1)$ and $C^{\dagger}(k^1)$ are
defined by
\begin{eqnarray}\label{label4.13}
\hspace{-0.5in}&&C(k^1) = d(k^1) - \varepsilon(k^1)\,d_5(k^1)
=\frac{i}{\sqrt{k^0}}\int^{\infty}_{-\infty}dq^1\,\Big\{\theta(k^1q^1
)\,[a^{\dagger}(q^1)a(k^1 + q^1)\nonumber\\\hspace{-0.5in} && -
b^{\dagger}(q^1)b(k^1 + q^1)] + \varepsilon(k^1)\,\theta(q^1(k^1 -
q^1))\,b(k^1 - q^1) a(q^1)\Big\},\nonumber\\ \hspace{-0.5in}&&
C^{\dagger}(k^1) = d^{\dagger}(k^1) -
\varepsilon(k^1)\,d^{\dagger}_5(k^1)=-\frac{i}{\sqrt{k^0}}\int^{\infty}_{-\infty}dq^1\,\Big\{
\theta(k^1q^1)\,[a^{\dagger}(k^1 +
q^1)a(q^1)\nonumber\\\hspace{-0.5in} && - b^{\dagger}(k^1 +
q^1)b(q^1)] + \varepsilon(k^1)\,\theta(q^1(k^1 -
q^1))\,a^{\dagger}(q^1)b^{\dagger}(k^1 - q^1)\Big\}.
\end{eqnarray}
The operators $C(k^1)$ and $C^{\dagger}(k^1)$ have
standard properties under parity transformations
\begin{eqnarray}\label{label4.14}
{\cal P}C(k^1){\cal P}^{\dagger} = + C(-k^1)\quad,\quad {\cal
P}C^{\dagger}(k^1){\cal P}^{\dagger} = + C^{\dagger}(-k^1).
\end{eqnarray}
The axial--vector current $J^{\mu}_5(x)$ can be written in complete
analogy with the vector current
\begin{eqnarray}\label{label4.15}
J^{\mu}_5(x) = \frac{1}{\sqrt{\pi}}\,\partial^{\mu}P(x),
\end{eqnarray}
where $P(x)$ is a pseudoscalar fermionic field density defined by
\begin{eqnarray}\label{label4.16}
P(x)=
\frac{1}{\sqrt{2\pi}}\int^{\infty}_{-\infty}\frac{dk^1}{\sqrt{2k^0}}\,
\varepsilon(k^1)\,\Big[C(k^1)\,e^{\textstyle -ik\cdot x} +
C^{\dagger}(k^1)\,e^{\textstyle ik\cdot x}\Big].
\end{eqnarray}
Following the calculation of the commutators
$[c(k^1),c^{\dagger}(p^1)]$ and $[c(k^1),c(p^1)]$ represented in
(\ref{label4.7}) and (\ref{label4.8}) one can show that $C(k^1)$ and
$C^{\dagger}(p^1)$ are canonical quantum field operators obeying the
commutation relations
\begin{eqnarray}\label{label4.17}
&&[C(k^1),C(p^1)] = [C^{\dagger}(k^1),C^{\dagger}(p^1)] =
0,\nonumber\\ &&[C(k^1),C^{\dagger}(p^1)] = \delta(k^1 - p^1).
\end{eqnarray}
A one--parameter family of solutions of the equation of motion of the
massless Thirring model reads now 
\begin{eqnarray}\label{label4.18}
\psi(x) &=& e^{\textstyle -i\,\alpha\,S(x) -
i\,\beta\,\gamma^5\,P(x)}\,\Psi(x),\nonumber\\
\bar{\psi}(x)&=& \bar{\Psi}(x)\,e^{\textstyle i\,\alpha\,S(x) -
i\,\beta\,\gamma^5\,P(x)}.
\end{eqnarray}
Using the relations 
\begin{eqnarray}\label{label4.19}
J^{\mu}(x) &=& \frac{1}{\sqrt{\pi}}\,\partial^{\mu}S(x) = j^{\mu}(x)
=\frac{1}{\sqrt{\pi}}\,\partial^{\mu}s(x),\nonumber\\ J^{\mu}_5(x) &=&
\frac{1}{\sqrt{\pi}}\,\partial^{\mu}P(x) = j^{\mu}_5(x)
=\frac{1}{\sqrt{\pi}}\,\partial^{\mu}p(x)
\end{eqnarray}
and $j^{\mu}_5(x) = -\varepsilon^{\mu\nu}\,j_{\nu}(x)$ one can easily
show that the solution (\ref{label4.18}) obeys the equation of motion
(\ref{label2.1}) for $\alpha - \beta = g/\sqrt{\pi}$. Hence, the
equation of motion of the massless Thirring model possesses a
one--parameter family of solutions as claimed by Klaiber [4].

We would like to accentuate that this statement is only obtained for
the classical solutions of the classical equation of motion of the
massless Thirring model. In the next section we will show that the
quantum version of this hypothesis is not valid.

Now we would like to discuss the definition of annihilation operator
of antifermions with momentum $p^1$ given in Ref.[16]. According to
Eq.(2.38) of Ref.[16] in order to use Thirring's solution for a free
massless fermion field (\ref{label2.13}) with spinorial wave functions
(\ref{label2.30}) one can define the annihilation  operators of 
antifermions as follows
\begin{eqnarray}\label{label4.20}
b(p^1) \to \tilde{b}(p^1) = \left\{\begin{array}{r@{\quad,\quad}l}
+\,b(p^1) & p^1 > 0,\\
-\,b(p^1) & p^1 < 0.
\end{array}\right.
\end{eqnarray}
Formally, this relation can be defined in terms the sign function
$\varepsilon(p^1)$, i.e. $\tilde{b}(p^1) = \varepsilon(p^1)\,b(p^1)$.
The operator $b(p^1)$ should possess standard properties under the
parity transformation ${\cal P}b(p^1){\cal P}^{\dagger} = - b(-p^1)$
(\ref{label2.19}). In this case the expansion into plane waves of a
free massless fermion field $\Psi(x)$ should read
\begin{eqnarray}\label{label4.21}
\hspace{-0.3in}&&\Psi(x^0,x^1) = \int^{\infty}_{-\infty}
\frac{dp^1}{\sqrt{2\pi}}\, \frac{1}{\sqrt{
2p^0}}\,u(p^0,p^1)\,\Big[a(p^1)\,e^{\textstyle -ip^0x^0 + ip^1x^1} +
\tilde{b}^{\dagger}(p^1)\,e^{\textstyle ip^0x^0 -
ip^1x^1}\Big]=\nonumber\\
\hspace{-0.3in}&&= \int^{\infty}_{-\infty}\frac{dp^1}{\sqrt{2\pi}}\,
\frac{1}{\sqrt{ 2p^0}}\,u(p^0,p^1)\,\Big[a(p^1)\,e^{\textstyle
-ip^0x^0 + ip^1x^1} +
\varepsilon(p^1)\,b^{\dagger}(p^1)\,e^{\textstyle ip^0x^0 -
ip^1x^1}\Big].
\end{eqnarray}
Now let us make a parity transformation of the $\Psi$--field
(\ref{label4.21}) by using standard properties of the operators
$a(p^1)$ and $b^{\dagger}(p^1)$ (\ref{label2.19}) and the spinorial
function $u(p^0,p^1)$ (\ref{label2.18}). As a result we get
\begin{eqnarray}\label{label4.22}
\hspace{-0.3in}&&{\cal P}\Psi(x^0,x^1){\cal P}^{\dagger}
=\int^{\infty}_{-\infty} \frac{dp^1}{\sqrt{2\pi}}\, \frac{1}{\sqrt{
2p^0}}\,u(p^0,p^1)\,\Big[{\cal P}a(p^1){\cal
P}^{\dagger}\,e^{\textstyle -ip^0x^0 + ip^1x^1} +
\varepsilon(p^1)\nonumber\\
\hspace{-0.3in}&&\times\,{\cal P}b^{\dagger}(p^1){\cal
P}^{\dagger}\,e^{\textstyle ip^0x^0 - ip^1x^1}\Big] =
\int^{\infty}_{-\infty} \frac{dp^1}{\sqrt{2\pi}}\, \frac{1}{\sqrt{
2p^0}}\,\gamma^0 u(p^0,-p^1)\,\Big[a(-p^1)\,e^{\textstyle -ip^0x^0 + ip^1x^1}\nonumber\\
\hspace{-0.3in}&&- \varepsilon(p^1)\,b^{\dagger}(-p^1)\,e^{\textstyle
ip^0x^0 - ip^1x^1}\Big] = \gamma^0\Psi(x^0,- x^1),
\end{eqnarray}
where we have made a change of variables $p^1 \to -p^1$ and used the
relation $\varepsilon(-p^1) = - \varepsilon(p^1)$. This agrees with
standard properties of a free fermion field under parity
transformation [2,3] (see also Appendix A).

According to the prescription (\ref{label4.20}) the operators $c(k^1)$
and $c^{\dagger}(k^1)$ (\ref{label2.31}) should be replaced by the
operators $\tilde{c}(k^1)$ and $\tilde{c}^{\dagger}(k^1)$ defined by 
\begin{eqnarray}\label{label4.23}
\tilde{c}\,(k^1) &=&\frac{i}{\sqrt{k^0}}\int^{\infty}_{-\infty}dq^1\,
\Big\{\theta(k^1q^1 )\,[a^{\dagger}(q^1)\,a(k^1 + q^1)
-\varepsilon(q^1)\,\varepsilon(k^1+q^1)\,b^{\dagger}(q^1)\,b(k^1 +
q^1)]\nonumber\\ && + \theta(q^1(k^1 - q^1))\,\varepsilon(k^1 -
q^1)\,b(k^1 - q^1)\,a(q^1)\Big\},\nonumber\\ \tilde{c}^{\dagger}(k^1)
&=&-\frac{i}{\sqrt{k^0}}\int^{\infty}_{-\infty}dq^1\,\Big\{
\theta(k^1q^1)\,[a^{\dagger}(k^1 + q^1)\,a(q^1) -
\varepsilon(q^1)\,\varepsilon(k^1+q^1)\,b^{\dagger}(k^1 +
q^1)b(q^1)]\nonumber\\ && + \theta(q^1(k^1 - q^1))\,\varepsilon(k^1 -
q^1)\,a^{\dagger}(q^1)\,b^{\dagger}(k^1 - q^1)\Big\}.
\end{eqnarray}
One can show that these operators $\tilde{c}(k^1)$ and
$\tilde{c}^{\dagger}(k^1)$ are scalars under parity transformation
\begin{eqnarray}\label{label4.24}
{\cal P}\tilde{c}(k^1){\cal P}^{\dagger} &=&
\tilde{c}(-k^1),\nonumber\\ {\cal P}\tilde{c}^{\dagger}(k^1){\cal
P}^{\dagger} &=& \tilde{c}^{\dagger}(-k^1).
\end{eqnarray}
One can also show that the operators $\tilde{c}(k^1)$ and
$\tilde{c}^{\dagger}(k^1)$ obey canonical commutation relations
\begin{eqnarray}\label{label4.25}
&&[\tilde{c}(k^1),\tilde{c}^{\dagger}(p^1)] = \delta(k^1 -
p^1),\nonumber\\ &&[\tilde{c}(k^1),\tilde{c}(p^1)] =
[\tilde{c}^{\dagger}(k^1),\tilde{c}^{\dagger}(p^1)] = 0.
\end{eqnarray}
Since $\varepsilon(p^1)u(p^0,p^1) = v(p^0,p^1)$, where $v(p^0,p^1)$ is
defined by (\ref{label2.15}) the operators $\tilde{c}(k^1)$ and
$\tilde{c}^{\dagger}(k^1)$ introduced in Ref.[16] give other
representation of the operators $C(k^1)$ and $C^{\dagger}(k^1)$, which
we have defined in this section.

\section{Quantization of the massless Thirring model in \\
 Klaiber's operator formalism and quantum equation of motion} 
\setcounter{equation}{0}

\hspace{0.2in} Following Klaiber for the quantization of massless
Thirring fermion fields and the evaluation of correlation functions we
have to represent the classical solution of the classical equation of
motion (\ref{label4.18}) as follows
\begin{eqnarray}\label{label5.1}
\psi(x) &=& e^{\textstyle -i\,\alpha\,S^{(-)}(x) -
i\,\beta\,\gamma^5\,P^{(-)}(x)}\,\Psi(x)\,e^{\textstyle
-i\,\alpha\,S^{(+)}(x) - i\,\beta\,\gamma^5\,P^{(+)}(x)},\nonumber\\
\bar{\psi}(x)&=&e^{\textstyle + i\,\alpha\,S^{(-)}(x) -
i\,\beta\,\gamma^5\,P^{(-)}(x)}\,\bar{\Psi}(x)\,e^{\textstyle +
i\,\alpha\,S^{(+)}(x) - i\,\beta\,\gamma^5\,P^{(+)}(x)},
\end{eqnarray}
where $S^{(\pm)}(x)$ and $P^{(\pm)}(x)$ are positive and negative
frequency parts of the operators $S(x)$ and $P(x)$, respectively:
\begin{eqnarray}\label{label5.2}
S^{(+)}(x) &=&
\frac{1}{\sqrt{2\pi}}\int^{\infty}_{-\infty}\frac{dk^1}{\sqrt{2k^0}}\,
C(k^1)\,e^{\textstyle -ik\cdot x},\nonumber\\ S^{(-)}(x) &=&
\frac{1}{\sqrt{2\pi}}\int^{\infty}_{-\infty}\frac{dk^1}{\sqrt{2k^0}}\,
C^{\dagger}(k^1)\,e^{\textstyle + ik\cdot x},\nonumber\\ P^{(+)}(x)
&=&
\frac{1}{\sqrt{2\pi}}\int^{\infty}_{-\infty}\frac{dk^1}{\sqrt{2k^0}}\,
\varepsilon(k^1)\,C(k^1)\,e^{\textstyle -ik\cdot x},\nonumber\\
P^{(-)}(x) &=&
\frac{1}{\sqrt{2\pi}}\int^{\infty}_{-\infty}\frac{dk^1}{\sqrt{2k^0}}\,
\varepsilon(k^1)\,C^{\dagger}(k^1)\,e^{\textstyle + ik\cdot x}.
\end{eqnarray}
In analogy we define the positive and negative frequency parts
$\Psi^{(+)}(x)$ and $\Psi^{(-)}(x)$ of the $\Psi$--field
where $\Psi^{(+)}(y)$ is the positive frequency part of the
$\Psi$--field
\begin{eqnarray}\label{label5.3}
\Psi^{(+)}(y) &=&\frac{1}{\sqrt{2\pi}}\int^{\infty}_{-\infty}
\frac{dp^1}{\sqrt{2p^0}}\,u(p^0,p^1)\,a(p^1)\, e^{\textstyle -ip\cdot
y},\nonumber\\ \Psi^{(-)}(y) &=&
\frac{1}{\sqrt{2\pi}}\int^{\infty}_{-\infty}
\frac{dp^1}{\sqrt{2p^0}}\,v(p^0,p^1)\,b^{\dagger}(p^1)\, e^{\textstyle
+ip\cdot y}.
\end{eqnarray}
The operator relations (\ref{label5.1}) should be understood in matrix
form as follows
\begin{eqnarray}\label{label5.4}
\hspace{-0.5in}\psi_a(x) \!\!\!&=& \!\!\!
\Big(e^{\textstyle -i\,\alpha\,S^{(-)}(x) -
i\,\beta\,\gamma^5P^{(-)}(x)}\Big)_{cb}\Psi_c(x)
\Big(e^{\textstyle
-i\,\alpha\,S^{(+)}(x) -
i\,\beta\,\gamma^5P^{(+)}(x)}\Big)_{ba},\nonumber\\
\hspace{-0.5in}\bar{\psi}_a(x)\!\!\!&=& \!\!\!  \Big(e^{\textstyle +
i\,\alpha\,S^{(-)}(x) -
i\,\beta\,\gamma^5P^{(-)}(x)}\Big)_{cb}\bar{\Psi}_c(x)
\Big(e^{\textstyle + i\,\alpha\,S^{(+)}(x) - i\,\beta\,\gamma^5\,
P^{(+)}(x)}\Big)_{ba},
\end{eqnarray}
where Latin indices run over $a = 1,2$.

The operators $S^{(\pm)}(x)$ and $P^{(\pm)}(x)$ obey the $c$--valued
commutation relations
\begin{eqnarray}\label{label5.5}
&&[S^{(+)}(x),S^{(-)}(y)] = [P^{(+)}(x), P^{(-)}(y)] =
-i\,D^{(+)}(x-y),\nonumber\\ &&[S^{(-)}(x),S^{(+)}(y)] = [P^{(-)}(x),
P^{(+)}(y)] = -i\,D^{(-)}(x-y),\nonumber\\ &&[S^{(+)}(x),P^{(-)}(y)] =
-i\,D^{(+)}_5(x-y),\nonumber\\ &&[S^{(-)}(x),P^{(+)}(y)] =
-i\,D^{(-)}_5(x-y),
\end{eqnarray}
where the correlation functions in the r.h.s. of (\ref{label5.3}) are
determined by [4] (see also [16])
\begin{eqnarray}\label{label5.6}
\hspace{-0.5in}D^{(\pm)}(x) &=&\pm\,
\frac{i}{2\pi}\int^{\infty}_{-\infty} \frac{dk^1}{2k^0}\,e^{\textstyle
\mp ik\cdot x} = \mp \frac{i}{4\pi}\,{\ell n}[- \mu^2x^2 \pm
i\,0\cdot\varepsilon(x^0)],\nonumber\\
\hspace{-0.5in}D^{(\pm)}_5(x)
&=&\pm\,\frac{i}{2\pi}\int^{\infty}_{-\infty}
\frac{dk^1}{2k^0}\,\varepsilon(k^1)\,e^{\textstyle \mp ik\cdot x} =\mp\,\frac{i}{4\pi}\,{\ell n}\Bigg(\frac{x^0 + x^1 \mp i\,0}{x^0 - x^1
\mp i\,0}\Bigg),
\end{eqnarray}
where $\mu$ is an infrared cut--off and $x^2 = (x^0)^2 - (x^1)^2$. The
detailed calculation of the correlation functions $D^{(\pm)}(x)$ and
$D^{(\pm)}_5(x)$ we adduce in Appendix C. Under Lorentz
transformations
\begin{eqnarray}\label{label5.7}
x^0 \to \tilde{x}^0 &=& \frac{x^0 - v x^1}{\sqrt{1 - v^2}},\nonumber\\
x^1 \to \tilde{x}^1 &=& \frac{x^1 - v x^0}{\sqrt{1 - v^2}},
\end{eqnarray}
where $v$ is the velocity of a Lorentz frame, the correlation functions
$D^{(\pm)}(x)$ and $D^{(\pm)}_5(x)$ behave as follows
\begin{eqnarray}\label{label5.8}
D^{(\pm)}(x) \to D^{(\pm)}(\tilde{x}) &=& D^{(\pm)}(x),\nonumber\\
D^{(\pm)}_5(x) \to D^{(\pm)}_5(\tilde{x}) &=& D^{(\pm)}_5(x) \pm
\frac{i}{4\pi}\,{\ell n}\Big(\frac{1 + v}{1 - v}\Big).
\end{eqnarray}
Hence, the correlation function $D^{(\pm)}(x)$ is covariant and
$D^{(\pm)}_5(x)$ is non--covariant under Lorentz transformations,
respectively, since $x^2 \to \tilde{x}^2 = x^2$.

Now let us evaluate of the commutators $[S^{(\pm)}(x),\Psi(y)]$ and
$[P^{(\pm)}(x),\Psi(y)]$. The knowledge of these commutators is
important for the correct (i) derivation of the quantum equation of
motion for the massless Thirring model from the operator field
representation (\ref{label5.1}) and (ii) evaluation of correlation
functions of massless Thirring fermion fields.

Using (\ref{label5.2}) and (\ref{label2.13}) the evaluation of the
commutator $[S^{(+)}(x),\Psi(y)]$ runs in the way
\begin{eqnarray}\label{label5.9}
[S^{(+)}(x),\Psi(y)] &=&
\frac{1}{2\pi}\int^{\infty}_{-\infty}\frac{dk^1}{\sqrt{2k^0}}
\int^{\infty}_{-\infty}dp^1\Bigg[{\theta(+p^1)\choose
\theta(-p^1)}\,[C(k^1),a(p^1)]\,e^{\textstyle -ik\cdot x -ip\cdot
y}\nonumber\\ &&\hspace{1in} + {+\theta(+p^1)\choose
-\theta(-p^1)}\,[C(k^1),b^{\dagger}(p^1)]\,e^{\textstyle -ik\cdot x +
ip\cdot y}\Bigg].
\end{eqnarray}
According to the definition of the $C$--operator (\ref{label4.13}) and
the canonical anti--commutation relations (\ref{label2.14}) we get
\begin{eqnarray}\label{label5.10}
\hspace{-0.7in}&&[C(k^1),a(p^1)] =
-\frac{i}{\sqrt{k^0}}\,\theta(k^1p^1)\,a(k^1 + p^1),\nonumber\\
\hspace{-0.7in}&&[C(k^1),b^{\dagger}(p^1)] = -
\frac{i}{\sqrt{k^0}}\,\{\theta(k^1(p^1-k^1))\,b^{\dagger}(p^1-k^1)
+\varepsilon(k^1)\,\theta(p^1(k^1-p^1))\,a(k^1-p^1)\}.
\end{eqnarray}
Substituting (\ref{label5.10}) in (\ref{label5.9}), making the
corresponding changes of variables and using the properties of the
Heaviside functions we end up with the expression
\begin{eqnarray}\label{label5.11}
\hspace{-0.5in}&&[S^{(+)}(x),\Psi(y)] =\nonumber\\
\hspace{-0.5in}&&= -\,\sqrt{\pi}\,
\frac{i}{2\pi}\int^{\infty}_{-\infty} \frac{dk^1}{k^0}\,e^{\textstyle
-ik\cdot(x-y)}\int^{\infty}_{-\infty}\frac{dp^1}{\sqrt{2\pi}}\,
\Bigg[{\theta(+k^1)\theta(+p^1)\theta(p^1-k^1)\choose
\theta(-k^1)\theta(-p^1)\theta(k^1-p^1)}\,a(p^1)\,e^{\textstyle
-ip\cdot y}\nonumber\\
\hspace{-0.5in}&&+ {\theta(+k^1)\theta(+p^1)\theta(k^1 - p^1)\choose
\theta(-k^1)\theta(-p^1)\theta(p^1 - k^1)}\,a(p^1)\,e^{\textstyle
-ip\cdot y} + {+\theta(+k^1)\theta(+p^1)\choose
-\theta(-k^1)\theta(-p^1)}\,b^{\dagger}(p^1)\,e^{\textstyle + ip\cdot
y}\Bigg]=\nonumber\\
\hspace{-0.5in}&&= - \,\sqrt{\pi}\,
\frac{i}{2\pi}\int^{\infty}_{-\infty} \frac{dk^1}{k^0}\, e^{\textstyle
-ik\cdot(x-y)} \nonumber\\
\hspace{-0.5in}&&\times
\int^{\infty}_{-\infty}\frac{dp^1}{\sqrt{2\pi}}\,
\Bigg[{\theta(+k^1)\theta(+p^1)\choose
\theta(-k^1)\theta(-p^1)}\,a(p^1)\,e^{\textstyle -ip\cdot y} +
{+\theta(+k^1)\theta(+p^1)\choose
-\theta(-k^1)\theta(-p^1)}\,b^{\dagger}(p^1)\,e^{\textstyle + ip\cdot
y}\Bigg].
\end{eqnarray}
Due to the relations
\begin{eqnarray}\label{label5.12}
\theta(k^1)= \frac{1 + \varepsilon(k^1)}{2}\quad,\quad \theta(-k^1)=
\frac{1 - \varepsilon(k^1)}{2}
\end{eqnarray}
the r.h.s. of (\ref{label5.11}) can be recast into the form
\begin{eqnarray}\label{label5.13}
\hspace{-0.3in}&&[S^{(+)}(x),\Psi(y)] = - \sqrt{\pi} 
\frac{i}{2\pi}\int^{\infty}_{-\infty}
\frac{dk^1}{2k^0}\,e^{\textstyle
-ik\cdot(x-y)}\frac{1}{\sqrt{2\pi}}\int^{\infty}_{-\infty}
\frac{dp^1}{\sqrt{2p^0}}\Big[u(p^0,p^1)a(p^1)\,
e^{\textstyle -ip\cdot y}\nonumber\\
\hspace{-0.3in}&& + \varepsilon(k^1)\gamma^5u(p^0,p^1) a(p^1)
e^{\textstyle -ip\cdot y}+ v(p^0,p^1)\, b^{\dagger}(p^1) e^{\textstyle
+ ip\cdot y} +
\varepsilon(k^1)\gamma^5v(p^0,p^1)\,b^{\dagger}(p^1)\,e^{\textstyle +
ip\cdot y}\Big] \nonumber\\
\hspace{-0.3in}&&= -\,\sqrt{\pi}\,[D^{(+)}(x-y) +
\gamma^5D^{(+)}_5(x-y)]\,\Psi(y).
\end{eqnarray}
Thus, we have obtained
\begin{eqnarray}\label{label5.14}
[S^{(+)}(x),\Psi(y)] = -\sqrt{\pi}\,[D^{(+)}(x-y) +
\gamma^5D^{(+)}_5(x-y)]\,\Psi(y).
\end{eqnarray}
This is in agreement with Klaiber's commutator (see (IV.13) of
Ref.[4]).

In a similar way  we calculate the commutators
\begin{eqnarray}\label{label5.15}
\hspace{-0.5in}[S^{(-)}(x),\Psi(y)] &=& -\sqrt{\pi}\,[D^{(-)}(x-y) +
\gamma^5D^{(-)}_5(x-y)]\,\Psi(y),\nonumber\\
\hspace{-0.5in}[P^{(+)}(x),\Psi(y)] &=&-\sqrt{\pi}\,[D^{(+)}_5(x-y) +
\gamma^5D^{(+)}(x-y)]\,\Psi(y),\nonumber\\
\hspace{-0.5in}[P^{(-)}(x),\Psi(y)] &=& -\,\sqrt{\pi}\,[D^{(-)}_5(x-y)
+ \gamma^5D^{(-)}(x-y)]\,\Psi(y).
\end{eqnarray}
For the further analysis we need the commutation relations between the
scalar and pseudoscalar densities $S^{(\pm)}(x)$ and $P^{(\pm)}(x)$
and $\bar{\Psi}(y)$. These commutators read
\begin{eqnarray}\label{label5.16}
\hspace{-0.5in}[S^{(+)}(x),\bar{\Psi}(y)]
&=&+\,\sqrt{\pi}\,[D^{(+)}(x-y) -
\gamma^5D^{(+)}_5(x-y)]\,\bar{\Psi}(y),\nonumber\\
\hspace{-0.5in}[S^{(-)}(x),\bar{\Psi}(y)]
&=&+\,\sqrt{\pi}\,[D^{(-)}(x-y) -
\gamma^5D^{(-)}_5(x-y)]\,\bar{\Psi}(y),\nonumber\\
\hspace{-0.5in}[P^{(+)}(x),\bar{\Psi}(y)]
&=&+\,\sqrt{\pi}\,[D^{(+)}_5(x-y) -
\gamma^5D^{(+)}(x-y)]\,\bar{\Psi}(y),\nonumber\\
\hspace{-0.5in}[P^{(-)}(x),\bar{\Psi}(y)]
&=&+\,\sqrt{\pi}\,[D^{(-)}_5(x-y) -
\gamma^5D^{(-)}(x-y)]\,\bar{\Psi}(y).
\end{eqnarray}
For the evaluation of correlation functions one needs the relations
\begin{eqnarray}\label{label5.17}
&&e^{\textstyle +i\,\alpha\,S^{(+)}(x)\pm
i\,\beta\,\gamma^5\,P^{(+)}(x)}\,\Psi(y)\,e^{\textstyle
-i\,\alpha\,S^{(+)}(x) \mp i\,\beta\,\gamma^5\,P^{(+)}(x)}
=\nonumber\\ &&= e^{\textstyle -i\,\sqrt{\pi}\,(\alpha \pm
\beta)\,[D^{(+)}(x-y) + \gamma^5 D^{(+)}_5(x-y)]}\,\Psi(y),\nonumber\\
&&e^{\textstyle
+i\,\alpha\,S^{(-)}(x)\pm
i\,\beta\,\gamma^5\,P^{(-)}(x)}\,\Psi(y)\,e^{\textstyle
-i\,\alpha\,S^{(-)}(x) \mp i\,\beta\,\gamma^5\,P^{(-)}(x)}
=\nonumber\\ &&= e^{\textstyle -i\,\sqrt{\pi}\,(\alpha \pm
\beta)\,[D^{(-)}(x-y) + \gamma^5 D^{(-)}_5(x-y)]}\,\Psi(y),\nonumber\\ 
&&e^{\textstyle
+i\,\alpha\,S^{(+)}(x)\pm
i\,\beta\,\gamma^5\,P^{(+)}(x)}\,\bar{\Psi}(y)\,e^{\textstyle
-i\,\alpha\,S^{(+)}(x) \mp i\,\beta\,\gamma^5\,P^{(+)}(x)}
=\nonumber\\ &&= \bar{\Psi}(y)e^{\textstyle +i\,\sqrt{\pi}\,(\alpha
\mp \beta)\,[D^{(+)}(x-y) - \gamma^5 D^{(+)}_5(x-y)]},\nonumber\\
&&e^{\textstyle +i\,\alpha\,S^{(-)}(x)\pm
i\,\beta\,\gamma^5\,P^{(-)}(x)}\,\bar{\Psi}(y)\,e^{\textstyle
-i\,\alpha\,S^{(-)}(x) \mp i\,\beta\,\gamma^5\,P^{(-)}(x)}
=\nonumber\\ &&= \bar{\Psi}(y)\,e^{\textstyle +i\,\sqrt{\pi}\,(\alpha
\mp \beta)\,[D^{(-)}(x-y) - \gamma^5 D^{(-)}_5(x-y)]}.
\end{eqnarray}
Now we are able to derive the quantum equation of motion
for the massless Thirring model. For this aim we act with the operator
$i\gamma^{\mu}\partial_{\mu}$ on the operator of the $\psi(x)$--field
defined by (\ref{label5.1}).

We start the derivation of the quantum equation of motion from the
transformation of the $\psi$--field defined by (\ref{label5.1}). We
suggest to use the space--time splitting method suggested by Schwinger
[22]. Within Schwinger's splitting method we understand the
$\psi(x)$--field given by (\ref{label5.1}) as a limit
\begin{eqnarray}\label{label5.18}
\psi(x)= \lim_{\textstyle \epsilon \to 0}e^{\textstyle
-i\,Q^{(-)}(x + \epsilon)}\,\Psi(x) \,e^{\textstyle -i\, Q^{(+)}(x
- \epsilon)},
\end{eqnarray}
where $\epsilon$ is an infinitesimal space--like 2--vector, $\epsilon^2
< 0$ [4,21]. Then, for convenience we have introduced the notations:
$Q^{(\pm)}(x\mp \epsilon) = \alpha\,S^{(\pm)}(x \mp \epsilon) +
\beta\, \gamma^5 P^{(\pm)}(x \mp \epsilon)$.

Now we have to calculate the quantity
$i\gamma^{\mu}\partial_{\mu}\psi(x)$. The differentiation of
exponentials depending of operators $Q^{(\mp)}(x \pm\epsilon)$ we
would perform by using the formula
\begin{eqnarray}\label{label5.19}
\hspace{-0.3in}&&i\partial_{\mu}\Big(e^{\textstyle -iQ^{(\mp)}(x
\pm\epsilon)}\Big) =\int^1_0dt\,e^{\textstyle
-i\,(1-t)\,Q^{(\mp)}(x \pm\epsilon)}\,\partial_{\mu}Q^{(\mp)}(x
\pm\epsilon\,)e^{\textstyle -i\,t\,Q^{(\mp)}(x
\pm\epsilon)}.\nonumber\\ \hspace{-0.3in}&&
\end{eqnarray}
Applying (\ref{label5.19}) to the calculation of
$i\gamma^{\mu}\partial_{\mu}\psi(x)$ we obtain
\begin{eqnarray}\label{label5.20}
\hspace{-0.3in}&&i\gamma^{\mu}\partial_{\mu}\psi(x) = \lim_{\textstyle
\epsilon \to 0}\Big[ \gamma^{\mu}\int^1_0dt\,e^{\textstyle
\textstyle -i\,(1-t)\,Q^{(-)}(x +
\epsilon)}\,\partial_{\mu}Q^{(-)}(x +
\epsilon)\,\Psi(x)\,e^{\textstyle -\, i\,Q^{(+)}(x -
\epsilon)}\nonumber\\
\hspace{-0.3in}&&+ \gamma^{\mu}e^{\textstyle -\,i\,Q^{(-)}(x +
\epsilon)}\,\Psi(x)\int^1_0dt\,e^{\textstyle \textstyle
-i\,(1-t)\,Q^{(+)}(x - \epsilon)}\,\partial_{\mu}Q^{(+)}(x -
\epsilon)\,\Psi(x)\nonumber\\
\hspace{-0.3in}&&\times\,e^{\textstyle -\, i\,t\,Q^{(+)}(x -
\epsilon)}\Big] = \lim_{\textstyle
\epsilon \to 0}\Big[\gamma^{\mu}\,\partial_{\mu}Q^{(-)}(x +
\epsilon)\,e^{\textstyle \textstyle -i\,Q^{(-)}(x +
\epsilon)}\,\Psi(x)\, e^{\textstyle -\,i\, Q^{(+)}(x -
\epsilon)}\nonumber\\
\hspace{-0.3in}&&+ \gamma^{\mu}\,e^{\textstyle \textstyle
-i\,Q^{(-)}(x + \epsilon)}\,\Psi(x)\, e^{\textstyle -\,i\,Q^{(+)}(x
- \epsilon)}\,\partial_{\mu}Q^{(+)}(x - \epsilon)\Big] =
\partial_{\mu}Q^{(-)}(x)\,\gamma^{\mu}\psi(x)\nonumber\\
\hspace{-0.3in}&& + \gamma^{\mu}\psi(x)\,\partial_{\mu}Q^{(+)}(x) =
\sqrt{\pi}\,(\alpha- \beta)\,J^{(-)}_{\mu}(x)\,\gamma^{\mu}\psi(x) +
\sqrt{\pi}\,(\alpha -\beta)\,\gamma^{\mu}\psi(x)\,J^{(+)}_{\mu}(x) =
\nonumber\\
\hspace{-0.3in}&& =\sqrt{\pi}\,(\alpha-
\beta)\,:J_{\mu}(x)\gamma^{\mu}\psi(x):.
\end{eqnarray}
Here we have used the Wick definition of a normal--ordered product
[24,25]
\begin{eqnarray}\label{label5.21}
:\!J_{\mu}(x)\psi(x)\!: = J^{(-)}_{\mu}(x)\,\psi(x) +
\psi(x)\,J^{(+)}_{\mu}(x).
\end{eqnarray}
Thus, we have got 
\begin{eqnarray}\label{label5.22}
i\gamma^{\mu}\partial_{\mu}\psi(x) =\sqrt{\pi}\,(\alpha-
\beta)\!:\!J_{\mu}(x)\gamma^{\mu}\psi(x)\!:
\end{eqnarray}
In order to transform this equation to the equation of motion of the
massless Thirring model we have to find the relation between the
quantum currents $J_{\mu}(x)
=\,:\!\bar{\Psi}(x)\gamma_{\mu}\Psi(x)\!:$ and $j_{\mu}(x) =
\,:\!\bar{\psi}(x)\gamma_{\mu}\psi(x)\!:$. 

The normal--ordered current $j_{\mu}(x) =
\,:\!\bar{\psi}(x)\gamma_{\mu}\psi(x)\!:$ we define according to
Klaiber as follows
\begin{eqnarray}\label{label5.23}
j_{\mu}(x) = \,:\!\bar{\psi}(x)\gamma_{\mu}\psi(x)\!: =
\frac{1}{2}\lim_{\epsilon \to 0}(\gamma^{\mu})_{ab}[\bar{\psi}_a(x +
\epsilon)\psi_b(x) - \psi_b(x + \epsilon)\bar{\psi}_b(x)],
\end{eqnarray}
where $\epsilon$ is an infinitesimal 2--vector.

First, let us treat the quantity $\bar{\psi}(x +
\epsilon)\gamma^{\mu}\psi(x) = \bar{\psi}_a(x +
\epsilon)(\gamma^{\mu})_{ab}\psi_b(x)$:
\begin{eqnarray}\label{label5.24}
\hspace{-0.3in}\bar{\psi}_a(x +
\epsilon)(\gamma^{\mu})_{ab}\psi_b(x) &=& \Big(e^{\textstyle +
i\,\bar{Q}^{(-)}(x + \epsilon)}\Big)_{cd}\bar{\Psi}_c(x +
\epsilon) \Big(e^{\textstyle +i\,\bar{Q}^{(+)}(x +
\epsilon)}\Big)_{da}\nonumber\\
\hspace{-0.3in}&&\times\,(\gamma^{\mu})_{ab}\Big(e^{\textstyle
-i\,Q^{(-)}(x)}\Big)_{fe}\Psi_e(x)\Big(e^{\textstyle -i\,
Q^{(+)}(x)}\Big)_{fb},
\end{eqnarray}
where we have denoted $\bar{Q}^{(+)}(x + \epsilon) =
\alpha\,S^{(+)}(x+\epsilon) - \beta\,\gamma^5\,P^{(+)}(x +
\epsilon)$.

Using the commutation relations (\ref{label5.5}) and the relations
(\ref{label5.17}) we recast the r.h.s. of (\ref{label5.24}) into the
form
\begin{eqnarray}\label{label5.25}
\hspace{-0.3in}&&\bar{\psi}_a(x +
\epsilon)(\gamma^{\mu})_{ab}\psi_b(x)= \Big(e^{\textstyle +
i\,\bar{Q}^{(-)}(x + \epsilon) -
i\,\bar{Q}^{(-)}(x)}\Big)_{cd}\bar{\Psi}_c(x +
\epsilon)(\gamma^{\mu})_{da}\nonumber\\
\hspace{-0.3in}&&\times\,\Big(e^{\textstyle -i[(\alpha^2 +\beta^2)
+2\sqrt{\pi}(\alpha+\beta)]\,D{(+)}(\epsilon) - 2i[\alpha\beta +
\sqrt{\pi}(\alpha +
\beta)]\gamma^5D^{(+)}_5(\epsilon)}\Big)_{ab}\nonumber\\
\hspace{-0.3in}&&\times\,\Psi_e(x) \Big(e^{\textstyle +i\, Q^{(+)}(x +
\epsilon) - i\,Q^{(+)}(x)}\Big)_{eb}.
\end{eqnarray}
This expression contains the scalar, $D^{(+)}(\epsilon)$, and
pseudoscalar, $\gamma^5D^{(+)}_5(\epsilon)$, divergences at the limit
$\epsilon \to 0$. Since pseudoscalar divergences can be removed by
neither renormalization of wave functions of fermion fields nor chiral
rotations, we should impose the constraint
\begin{eqnarray}\label{label5.26}
\alpha\beta + \sqrt{\pi}(\alpha + \beta) = 0.
\end{eqnarray}
As we have shown in (\ref{label5.8}) the functions
$D^{(+)}_5(\epsilon)$ are non--covariant under Lorentz
transformations. Therefore, removal of contributions of these
functions is required too by Lorentz covariance. This agrees with
Klaiber's constraint making correlation functions covariant under
Lorentz transformations (see section V of Ref.[4]).

Due to the constraint (\ref{label5.26}) the relation (\ref{label5.25})
takes the form
\begin{eqnarray}\label{label5.27}
\hspace{-0.3in}&&\bar{\psi}_a(x +
\epsilon)(\gamma^{\mu})_{ab}\psi_b(x)= e^{\textstyle -i(\alpha -
\beta)^2 D^{(+)}(\epsilon)}\Big(e^{\textstyle + i\,\bar{Q}^{(-)}(x
+ \epsilon) - i\,\bar{Q}^{(-)}(x)}\Big)_{cd}\nonumber\\
\hspace{-0.3in}&&\times\,\bar{\Psi}_c(x +
\epsilon)(\gamma^{\mu})_{da}\Psi_e(x) \Big(e^{\textstyle +i\,
Q^{(+)}(x + \epsilon) - i\,Q^{(+)}(x)}\Big)_{ea}.
\end{eqnarray}
Now we have to rewrite the product $\bar{\Psi}_c(x +
\epsilon)\Psi_e(x)$ in the normal--ordered form. According to
Wick's theorem [24] we get
\begin{eqnarray}\label{label5.28}
\bar{\Psi}_c(x + \epsilon)\Psi_e(x) &=& :\!\bar{\Psi}_c(x +
\epsilon)\Psi_e(x)\!: +
\frac{1}{2\pi}\int^{\infty}_{-\infty}\frac{dk^1}{2k^0}\,
(\hat{k})_{ec}\,e^{\textstyle
-ik\cdot \epsilon}=\nonumber\\ &=&:\!\bar{\Psi}_c(x +
\epsilon)\Psi_e(x)\!: +
(\hat{\epsilon})_{ec}\,\delta(\epsilon^2) + \frac{1}{2\pi
i}\,\frac{(\hat{\epsilon})_{ec}}{\epsilon^2}.
\end{eqnarray}
Recall, that the momentum integral in (\ref{label5.28}) has been
calculated in [6] (see Eq.(6.6) of Ref.[6]).

Substituting (\ref{label5.5}) in (\ref{label5.27}), expanding the
exponentials in powers of $\epsilon$ and keeping only finite and
leading divergent contributions we arrive at the expression
\begin{eqnarray}\label{label5.29}
\hspace{-0.3in}&&\bar{\psi}(x + \epsilon)\gamma^{\mu}\psi(x) =
e^{\textstyle -i(\alpha - \beta)^2 D^{(+)}(\epsilon)}\,\Big[
\frac{1}{\pi i}\,\frac{\epsilon^{\,\mu}}{\epsilon^2} +
\epsilon^{\,\mu}\,\delta(\epsilon^2) +
:\!\bar{\Psi}(x)\gamma^{\mu}\Psi(x)\!: \nonumber\\
\hspace{-0.3in}&& +
\frac{1}{2\pi}\,\frac{\epsilon^{\,\nu}}{\epsilon^2}\,{\rm
tr}\{\hat{\epsilon}\gamma^{\mu}[\alpha\,\partial_{\nu}S(x) +
\beta\,\gamma^5\partial_{\nu}P(x)]\}\Big] = e^{\textstyle -i(\alpha -
\beta)^2 D^{(+)}(\epsilon)}\,\Big[\frac{1}{\pi
i}\,\frac{\epsilon^{\,\mu}}{\epsilon^2}  +
\epsilon^{\,\mu}\,\delta(\epsilon^2)\nonumber\\
\hspace{-0.3in}&& + J^{\mu}(x) + \frac{\alpha}{\sqrt{\pi}}\,
\frac{\epsilon^{\,\mu}\epsilon^{\,\nu}}{\epsilon^2}\,J_{\nu}(x) -
\frac{\beta}{\sqrt{\pi}}\,
\frac{\epsilon_{\lambda}\epsilon_{\nu}}{\epsilon^2}\,
\varepsilon^{\lambda\mu}\varepsilon^{\nu\rho}\,J_{\rho}(x)\Big].
\end{eqnarray}
Using then the identity
\begin{eqnarray}\label{label5.30}
\varepsilon^{\lambda\mu}\varepsilon^{\nu\rho} =
g^{\mu\nu}\,g^{\lambda\rho} - g^{\mu\rho}\,g^{\lambda\nu}
\end{eqnarray}
we get 
\begin{eqnarray}\label{label5.31}
\hspace{-0.7in}&&\bar{\psi}(x + \epsilon)\gamma^{\mu}\psi(x) = 
\frac{4i\epsilon^{\mu}}{(\alpha -
\beta)^2}\,\frac{d}{d\epsilon^2}\Big[e^{\textstyle -i(\alpha -
\beta)^2 D^{(+)}(\epsilon)}\Big] \nonumber\\
\hspace{-0.7in}&&+ e^{\textstyle -i(\alpha - \beta)^2
D^{(+)}(\epsilon)}\,\Big[\Big(1 +
\frac{\beta}{\sqrt{\pi}}\Big)\,J^{\mu}(x) + \frac{\alpha -
\beta}{\sqrt{\pi}}\,
\frac{\epsilon^{\,\mu}\epsilon^{\,\nu}}{\epsilon^2}\,J_{\nu}(x)\Big],
\end{eqnarray}
where we have used the relation 
\begin{eqnarray}\label{label5.32}
e^{\textstyle -i(\alpha - \beta)^2
D^{(+)}(\epsilon)}\,\Big[\frac{1}{\pi
i}\,\frac{\epsilon^{\,\mu}}{\epsilon^2} +
\epsilon^{\,\mu}\,\delta(\epsilon^2)\Big] =
\frac{4i\epsilon^{\mu}}{(\alpha -
\beta)^2}\,\frac{d}{d\epsilon^2}\Big[e^{\textstyle -i(\alpha -
\beta)^2 D^{(+)}(\epsilon)}\Big],
\end{eqnarray}
which follows from the definition of $D^{(+})(\epsilon)$ given by
(\ref{label5.5}).

In a similar way we can treat the second term in the r.h.s. of
(\ref{label5.23}). As result we get
\begin{eqnarray}\label{label5.33}
\hspace{-0.5in}(\gamma^{\mu})_{ab}\psi_b(x + \epsilon)\bar{\psi}_a(x)
&=& e^{\textstyle -i(\alpha - \beta)^2
D^{(+)}(\epsilon)}\Big(e^{\textstyle -i\,\bar{Q}^{(-)}(x + \epsilon) +
i\,\bar{Q}^{(-)}(x)}\Big)_{cd}\nonumber\\
\hspace{-0.5in}&\times&\Psi_c(x +
\epsilon)(\gamma^{\mu})_{db}\bar{\Psi}_a(x)\Big(e^{\textstyle
-i\,\bar{Q}^{(+)}(x + \epsilon) + i\,\bar{Q}^{(+)}(x)}\Big)_{ab}.
\end{eqnarray}
In the normal--ordered form the product $\Psi_c(x +
\epsilon)(\gamma^{\mu})_{db}\bar{\Psi}_a(x)$ is defined by
\begin{eqnarray}\label{label5.34}
\Psi_c(x + \epsilon)\bar{\Psi}_a(x) = -\!:\!\bar{\Psi}_a(x)\Psi_c(x +
\epsilon)\!: + (\hat{\epsilon})_{ca}\,\delta(\epsilon^2\,) +
\frac{1}{2\pi i}\,\frac{(\hat{\epsilon})_{ca}}{\epsilon^2}.
\end{eqnarray}
Substituting (\ref{label5.34}) in (\ref{label5.33}) and expanding in
powers of $\epsilon$, calculating traces over Dirac matrices and using
the identity (\ref{label5.30}) we obtain
\begin{eqnarray}\label{label5.35}
\hspace{-0.7in}&&(\gamma^{\mu})_{ab} \psi_b(x +
\epsilon)\bar{\psi}_a(x) = \frac{4i\epsilon^{\mu}}{(\alpha -
\beta)^2}\,\frac{d}{d\epsilon^2}\Big[e^{\textstyle -i(\alpha -
\beta)^2 D^{(+)}(\epsilon)}\Big] \nonumber\\
\hspace{-0.7in}&& - \,e^{\textstyle -i(\alpha - \beta)^2
D^{(+)}(\epsilon)}\,\Big[\Big(1 +
\frac{\beta}{\sqrt{\pi}}\Big)\,J^{\mu}(x) + \frac{\alpha -
\beta}{\sqrt{\pi}}\,
\frac{\epsilon^{\,\mu}\epsilon^{\,\nu}}{\epsilon^2}\,J_{\nu}(x)\Big].
\end{eqnarray}
Thus, the normal--ordered vector current $j^{\mu}(x)$ is defined by 
\begin{eqnarray}\label{label5.36}
\hspace{-0.3in}j^{\mu}(x) &=& :\!\bar{\psi}(x)\gamma^{\mu}\psi(x)\!:\,
=\nonumber\\
\hspace{-0.3in}&=& e^{\textstyle -i(\alpha - \beta)^2
D^{(+)}(\epsilon)}\!\Big[\Big(1 +
\frac{\beta}{\sqrt{\pi}}\Big)\,J^{\mu}(x) + \frac{\alpha -
\beta}{\sqrt{\pi}}\,
\frac{\epsilon^{\,\mu}\epsilon^{\,\nu}}{\epsilon^2}\,J_{\nu}(x)\Big].
\end{eqnarray}
In order to remove the non--covariant term proportional to
$\epsilon^{\,\mu}\epsilon^{\,\nu}/\epsilon^2$ we can follow the
standard point--splitting procedure [22,25] and average over all
directions of the $\epsilon^{\,\mu}$. However, Klaiber has invented
another method which leads, unfortunately, to violation of chiral
symmetry at intermediate steps.

In fact, Klaiber suggested to define a renormalized vector current
$K^{\mu}(x)$ as follows
\begin{eqnarray}\label{label5.37}
\hspace{-0.3in}&&K^{\mu}(x)=\lim_{\textstyle \delta \to
0}\lim_{\textstyle \epsilon \to 0}e^{\textstyle i(\alpha -
\beta)^2D^{(+)}(\epsilon)}\frac{1}{2}\,(\gamma_{\nu})_{ab}
\{\bar{\psi}_a(x+\epsilon)\psi_b(x)\,[g^{\mu\nu}
\Big(1-i\,g\,\epsilon^{\,\lambda}K_{\lambda}(x + \delta)\nonumber\\
\hspace{-0.3in}&& - i\,\sigma\,\epsilon^{\,\nu}K^{\mu}(x +
\delta)]-\psi_b(x + \epsilon)\bar{\psi}_a(x)\,[g^{\mu\nu}(1 +
i\,g\,\epsilon^{\,\lambda}K_{\lambda}(x + \delta) +
i\,\sigma\,\epsilon^{\,\nu}K^{\mu}(x + \delta)]\},
\end{eqnarray}
where $\sigma$ is a real parameter $\sigma \in \mathbb{R}^1$.

Using equations (\ref{label5.28}) and (\ref{label5.34}) we recast the
r.h.s. of (\ref{label5.37}) into the form
\begin{eqnarray}\label{label5.38}
K^{\mu}(x)=-\frac{\sigma}{\pi}\,K^{\mu}(x) + \Big(1 +
\frac{\beta}{\sqrt{\pi}}\Big)\,J^{\mu}(x) + \Big[\frac{\alpha -
\beta}{\sqrt{\pi}}\,J_{\nu}(x) - \frac{g}{\pi}\,K_{\nu}(x)\Big]
\lim_{\textstyle \epsilon \to
0}\frac{\epsilon^{\,\mu}\epsilon^{\,\nu}}{\epsilon^2}.
\end{eqnarray}
We would like to emphasize that the limit $\delta \to 0$ is not
singular and can be taken from the very beginning. So its presence is
rather formal. The requirement of the vanishing of the non--covariant
terms gives
\begin{eqnarray}\label{label5.39}
K_{\nu}(x) = \sqrt{\pi}\,\frac{\alpha -
\beta}{g}\,J_{\nu}(x).
\end{eqnarray}
Substituting (\ref{label5.39}) into (\ref{label5.22}) we get 
\begin{eqnarray}\label{label5.40}
i\gamma^{\mu}\partial_{\mu}\psi(x)
=g\!:\!K_{\mu}(x)\gamma^{\mu}\psi(x)\!:.
\end{eqnarray}
According to Klaiber this is the quantum equation of motion for the
massless Thirring model quantized in the chiral symmetric
phase. However, equation (\ref{label5.38}) together with
(\ref{label5.39}) leads to a relation between the parameters $\alpha$,
$\beta$ and $\sigma$
\begin{eqnarray}\label{label5.41}
1 + \frac{\beta}{\sqrt{\pi}} = \frac{\pi + \sigma}{g}\,\frac{\alpha -
\beta}{\sqrt{\pi}}.
\end{eqnarray}
Due to the relation (\ref{label5.25}) this leaves only one free
parameter in the approach confirming the one--parameter family of
solutions of the massless Thirring model [4].

Equation (\ref{label5.25}) can be fulfilled identically by [4]
\begin{eqnarray}\label{label5.42}
\alpha = \sqrt{\pi}\,(\rho - 1)\quad , \quad \beta =
\sqrt{\pi}\,\Big(\frac{1}{\rho} - 1\Big),
\end{eqnarray}
where $\rho \in \mathbb{R}^1$ and $\rho\neq 0$ [4].  The solution of
equation (\ref{label5.41}) with respect to $\rho$ is equal to
\begin{eqnarray}\label{label5.43}
\rho = \pm \sqrt{1 + \frac{g}{\pi + \sigma}}.
\end{eqnarray}
This agrees with Klaiber's result (see Eq.(X.13) of Ref.[4]).

However, we would like to emphasize that this method is not innocent
and leads to a breaking of chiral symmetry. In order to show this we
have to define the axial vector current $K^{\mu}_5(x)$ in a way
analogous to the vector current. First of all notice that the
$\delta$--limit is non--singular and, without loss of generality, we
can take the limit $\delta \to 0$ from the very beginning. This
reduces the r.h.s. of (\ref{label5.37}) to the form
\begin{eqnarray}\label{label5.44}
\hspace{-0.5in}K^{\mu}(x)&=&\lim_{\textstyle \epsilon
\to 0}e^{\textstyle i(\alpha -
\beta)^2D^{(+)}(\epsilon)}\frac{1}{2}\,(\gamma_{\nu})_{ab}
\{\bar{\psi}_a(x+\epsilon)\psi_b(x)\,[g^{\mu\nu}
(1-i\,g\,\epsilon^{\,\lambda}K_{\lambda}(x))\nonumber\\ 
\hspace{-0.5in}&& - i\,\sigma\,\epsilon^{\,\nu}K^{\mu}(x)]-\psi_b(x + \epsilon)\bar{\psi}_a(x)\,[g^{\mu\nu}(1 +
i\,g\,\epsilon^{\,\lambda}K_{\lambda}(x)) +
i\,\sigma\,\epsilon^{\,\nu}K^{\mu}(x)]\},
\end{eqnarray}
which can be rewritten as follows (see Appendix D)
\begin{eqnarray}\label{label5.45}
\hspace{-0.7in}&&K^{\mu}(x) = \lim_{\textstyle \epsilon \to
0}e^{\textstyle i(\alpha - \beta)^2D^{(+)}(\epsilon)}\nonumber\\
\hspace{-0.7in}&&\times\,\frac{1}{2}
\{\bar{\psi}_a(x+\epsilon)\,(\gamma_{\nu})_{ab}\,e^{\textstyle
- i\,g\int^{x+\epsilon}_x dz^{\lambda}K_{\lambda}(z)}\psi_b(x) -
i\,\sigma\,(\gamma_{\nu})_{ab}\bar{\psi}_a(x+\epsilon)\psi_b(x)\,
\epsilon^{\,\nu}K^{\mu}(x)]\nonumber\\
\hspace{-0.7in}&&\hspace{0.2in} - (\gamma_{\nu})_{ab}\psi_b(x +
\epsilon) \,e^{\textstyle +
i\,g\int^{x+\epsilon}_xdz^{\lambda}K_{\lambda}(z)}\bar{\psi}_a(x) +
i\,\sigma\,(\gamma_{\nu})_{ab}\psi_b(x+\epsilon)\bar{\psi}_a(x)\,\epsilon^{\,\nu}K^{\mu}(x)]\}.
\end{eqnarray}
The appearance of the vector field in the r.h.s. of (\ref{label5.45})
can be explained by the requirement of gauge invariance. The vector
current $j^{\mu}(x)=:\!\bar{\psi}(x)\gamma^{\mu}\psi(x)\!:$ and
the axial--vector current
$j^{\mu}_5(x)=:\!\bar{\psi}(x)\gamma^{\mu}\gamma^5\psi(x)\!:$ are
invariant under gauge transformations
\begin{eqnarray}\label{label5.46}
\psi(x)\to \psi\,'(x) &=& e^{\textstyle
-i\,\omega(x)}\psi(x),\nonumber\\ 
\bar{\psi}(x)\to \bar{\psi}\,'(x)
&=&\bar{\psi}\, e^{\textstyle +i\,\omega(x)},
\end{eqnarray}
where $\omega(x)$ is a gauge function. Gauge invariance breaks down
for the products $\bar{\psi}_a(x + \epsilon)\psi_b(x)$ and $\psi_b(x +
\epsilon)\bar{\psi}_a(x)$
\begin{eqnarray}\label{label5.47}
\hspace{-0.3in}&&\bar{\psi}_a(x + \epsilon)\psi_b(x) \to
\bar{\psi}\,'_a(x + \epsilon)\psi\,'_b(x) = \bar{\psi}_a(x +
\epsilon)\psi_b(x)\,e^{\textstyle + i\,\omega(x + \epsilon)
-i\,\omega(x) }, \nonumber\\ 
\hspace{-0.3in}&&\psi_b(x +\epsilon)\bar{\psi}_a(x) \to \psi\,'_b(x
+\epsilon)\bar{\psi}\,'_a(x) = \psi_b(x +\epsilon)\bar{\psi}_a(x)\,
e^{\textstyle - i\,\omega(x + \epsilon) +i\,\omega(x)}.
\end{eqnarray}
Since a difference of gauge functions $\omega(x + \epsilon) -
\omega(x)$ can be arbitrary large despite the infinitesimal magnitude
of $\epsilon$ [26], the violation of gauge invariance can be
arbitrarily strong. The presence of the vector field $K_{\lambda}(x)$
restores gauge invariance of the products (\ref{label5.47}):
\begin{eqnarray}\label{label5.48}
\hspace{-0.3in}&&\bar{\psi}_a(x + \epsilon)\,e^{\textstyle -
i\,g\int^{x+\epsilon}_x dz^{\lambda}K_{\lambda}(z)}\psi_b(x) =
\bar{\psi}\,'_a(x + \epsilon)\,e^{\textstyle - i\,g\int^{x+\epsilon}_x
dz^{\lambda}K'_{\lambda}(z)}\psi\,'_b(x),\nonumber\\
\hspace{-0.3in}&&\psi_b(x +\epsilon)\,e^{\textstyle +
i\,g\int^{x+\epsilon}_x dz^{\lambda}K_{\lambda}(z)}\bar{\psi}_a(x) \to
\psi\,'_b(x +\epsilon)\,e^{\textstyle + i\,g\int^{x+\epsilon}_x
dz^{\lambda}K'_{\lambda}(z)}\bar{\psi}\,'_a(x),
\end{eqnarray}
if the fields $K_{\lambda}(z)$ and $K'_{\lambda}(z)$ are related by 
\begin{eqnarray}\label{label5.49}
K'_{\lambda}(z) = K_{\lambda}(z) +
\frac{1}{g}\,\partial_{\lambda}\omega(z).
\end{eqnarray}
As a result gauge invariance of the vector current defined by the
expression (\ref{label5.44}) is broken only by non--covariant terms
proportional to $\sigma$. This violation is of order $O(\epsilon)$.

Now we are able to define the axial--vector current $K^{\mu}_5(x)$
\begin{eqnarray}\label{label5.50}
\hspace{-0.3in}&&K^{\mu}_5(x)=\lim_{\textstyle \epsilon \to 0}e^{\textstyle i(\alpha -
\beta)^2D^{(+)}(\epsilon)}\frac{1}{2}\,(\gamma_{\nu}\gamma^5)_{ab}
\{\bar{\psi}_a(x+\epsilon)\psi_b(x)\,[g^{\mu\nu}
(1-i\,g\,\epsilon^{\,\lambda}K_{\lambda}(x))\nonumber\\
\hspace{-0.3in}&& - i\,\sigma\,\epsilon^{\,\nu}K^{\mu}(x)]-\psi_b(x +
\epsilon)\bar{\psi}_a(x)\,[g^{\mu\nu}(1 +
i\,g\,\epsilon^{\,\lambda}K_{\lambda}(x)) +
i\,\sigma\,\epsilon^{\,\nu}K^{\mu}(x)]\}
\end{eqnarray}
and consider the divergence of the axial--vector current
$\partial_{\mu}K^{\mu}_5(x)$. The terms proportional to
$\epsilon^{\,\lambda}K_{\lambda}(x)$ are invented to retain gauge
invariance of the axial--vector current, and the non--covariant terms
proportional to $\epsilon^{\,\nu}K^{\mu}(x)$ provide a smooth
violation of gauge invariance of order $O(\epsilon)$. 

We should emphasize that the correct definition of the quantum vector
current $K^{\mu}(x)$ should give us the quantum axial--vector current
$K^{\mu}_5(x)$ related to $K^{\mu}(x)$ by the standard relation:
$K^{\mu}_5(x) = - \varepsilon^{\mu\nu}K_{\nu}(x)$.

According to the definition of the axial--vector current
(\ref{label5.50}) the divergence $\partial_{\mu}K^{\mu}_5(x)$ is
determined by
\begin{eqnarray}\label{label5.51}
\partial_{\mu}K^{\mu}_5(x)\!\!\!&=&\!\!\!\lim_{\textstyle \epsilon \to
0}e^{\textstyle i(\alpha - \beta)^2D^{(+)}(\epsilon)}(-i)\,g\,
\frac{1}{2}\,(\gamma^{\mu}\gamma^5)_{ab}
[\bar{\psi}_a(x+\epsilon)\psi_b(x) + \psi_b(x +
\epsilon)\bar{\psi}_a(x)]\nonumber\\ &&\times\,
\epsilon^{\,\lambda}\partial_{\mu}K_{\lambda}(x),
\end{eqnarray}
where we have taken into account the conservation of the vector
current $\partial_{\mu}K^{\mu}(x) = 0$.

Using the definition of the self--coupled fermionic fields $\psi$ in
terms of the free fermionic fields $\Psi$ given by (\ref{label5.18}),
the commutation relations (\ref{label5.5}) and
Eqs.(\ref{label5.13})--(\ref{label5.17}) we bring the r.h.s. of
(\ref{label5.51}) to the form
\begin{eqnarray}\label{label5.52}
\partial_{\mu}K^{\mu}_5(x)= \lim_{\textstyle \epsilon \to 0}
\frac{g}{2i}\,(\gamma^{\mu}\gamma^5)_{ab}
[\bar{\Psi}_a(x+\epsilon)\Psi_b(x) + \Psi_b(x +
\epsilon)\bar{\Psi}_a(x)]\,
\epsilon^{\,\lambda}\partial_{\mu}K_{\lambda}(x).
\end{eqnarray}
The product of the $\Psi$--fields is equal to 
\begin{eqnarray}\label{label5.53}
&&\bar{\Psi}_a(x + \epsilon)\Psi_b(x) + \Psi_b(x +
\epsilon)\bar{\Psi}_a(x) =\nonumber\\ &&= :\!\bar{\Psi}_a(x +
\epsilon)\Psi_b(x)\!: - :\!\bar{\Psi}_a(x)\Psi_b(x + \epsilon)\!: +
\frac{1}{\pi i}\,\frac{(\hat{\epsilon})_{ba}}{\epsilon^2},
\end{eqnarray}
where we have dropped an insignificant contribution of the
$\delta$--function $\delta(\epsilon^2\,)$.

Substituting (\ref{label5.53}) in (\ref{label5.52}) we obtain
\begin{eqnarray}\label{label5.54}
\partial_{\mu}K^{\mu}_5(x)=
\frac{g}{\pi}\,\partial_{\mu}K_{\lambda}(x)\lim_{\textstyle \epsilon
\to 0}\frac{1}{2}{\rm tr}(\gamma^5\gamma^{\mu}\hat{\epsilon})\,
\frac{\epsilon^{\,\lambda}}{\epsilon^2} =
\frac{g}{\pi}\,\partial_{\mu}K_{\lambda}(x)\,
\varepsilon^{\mu\nu}\lim_{\textstyle \epsilon \to 0}
\frac{\epsilon_{\nu}\,\epsilon^{\,\lambda}} {\epsilon^2}.
\end{eqnarray}
Hence, non--covariant terms are not canceled in the divergence of the
quantum axial--vector current that gives $\partial_{\mu}K^{\mu}_5(x)
\neq 0$. The former breaks explicitly chiral symmetry.

Chiral symmetry can be restored only after averaging over the
directions of the 2--vector $\epsilon$ [22,25]. This yields
\begin{eqnarray}\label{label5.55}
\partial_{\mu}K^{\mu}_5(x)=
\frac{g}{2\pi}\,\varepsilon^{\mu\nu}\partial_{\mu}K_{\nu}(x) =
\frac{\alpha -
\beta}{2\sqrt{\pi}}\,\varepsilon^{\mu\nu}\partial_{\mu}J_{\nu}(x).
\end{eqnarray}
The r.h.s. of (\ref{label5.55}) vanishes due to the conservation of
the axial--vector current of the free massless fermionic field $\Psi$,
$\varepsilon^{\mu\nu}\partial_{\mu}J_{\nu}(x) = 0$. 

Thereby, Klaiber's method does not provide a cancellation of
non--covariant terms in the divergence of the quantum axial--vector
current that breaks chiral symmetry explicitly.

Now let us show that the quantum axial--vector current defined by
(\ref{label5.50}) does not satisfy the standard relation $K^{\mu}_5(x)
= - \varepsilon^{\mu\nu}K_{\nu}(x)$. Making use of the sequence of
actions that we have made for the derivation of the quantum vector
current $K^{\mu}(x)$ we reduce the r.h.s. of (\ref{label5.50}) to the
form
\begin{eqnarray}\label{label5.56}
K^{\mu}_5(x) = J^{\mu}_5(x) + \frac{\beta}{\sqrt{\pi}}\,
\frac{\epsilon^{\,\mu}\epsilon^{\,\nu}}{\epsilon^2}\, J_{5\nu}(x) +
\varepsilon^{\mu\nu}\frac{\epsilon_{\nu}
\epsilon_{\lambda}}{\epsilon^2}\,\Big[-\frac{\alpha}
{\sqrt{\pi}}\,J^{\lambda}(x) + \frac{g}{\pi}\,K^{\lambda}(x)\Big].
\end{eqnarray}
With the relation (\ref{label5.39}) the quantum
axial--vector current (\ref{label5.56}) takes the form
\begin{eqnarray}\label{label5.57}
K^{\mu}_5(x) = J^{\mu}_5(x) + \frac{\beta}{\sqrt{\pi}}\,
\frac{\epsilon^{\,\mu}\epsilon^{\,\nu}}{\epsilon^2}\, J_{5\nu}(x) -
\frac{\beta} {\sqrt{\pi}}\,\varepsilon^{\mu\nu}\frac{\epsilon_{\nu}
\epsilon_{\lambda}}{\epsilon^2}\,\,J^{\lambda}(x).
\end{eqnarray}
Thus, Klaiber's method for the removal of non--covariant contributions
fails for the covariant definition of the quantum axial--vector
current. Averaging over all directions of a 2--vector $\epsilon$ we
arrive at the expression
\begin{eqnarray}\label{label5.58}
K^{\mu}_5(x) = \Big(1+ \frac{\beta}{\sqrt{\pi}}\Big)\, J^{\mu}_5(x),
\end{eqnarray}
which does not obey the standard relation $K^{\mu}_5(x) =
-\varepsilon^{\mu\nu}K_{\nu}(x)$ if the quantum vector current
$K_{\nu}(x)$ is given by
\begin{eqnarray}\label{label5.59}
K^{\mu}(x)= \Big[\Big(1 +
\frac{\beta}{\sqrt{\pi}}\Big)-\frac{\sigma}{g}\,\frac{\alpha -
\beta}{\sqrt{\pi}}\Big]\,J^{\mu}(x).
\end{eqnarray}
In order to fulfill the relation $K^{\mu}_5(x) =
-\varepsilon^{\mu\nu}K_{\nu}(x)$ one should set $\sigma = 0$. Compared
to Klaiber's discussion this condition leads to a substantial
narrowing of the class of solutions of the massless Thirring model.

Hence, due to the impossibility to remove non--covariant terms
simultaneously in the vector and axial--vector currents and the
explicit breakdown of chiral symmetry Klaiber's parameterization based
on the definition of the quantum vector current (\ref{label5.37})
seems to be incorrect. Therefore, it cannot be accepted even if for
$\sigma = 0$, the case that Klaiber adduced finally in [4], and that
has been used then by Coleman for his proof of equivalence between the
massive Thirring model and the sine--Gordon model [18].

In turn, if we follow the standard point--splitting prescription
[22,25] and average over all directions of a 2--vector $\epsilon$ in
(\ref{label5.36}) and (\ref{label5.56}), the former, of course,
without the contribution of $K_{\lambda}(x)$, one arrives at the
definition of the quantum currents
\begin{eqnarray}\label{label5.60}
\hspace{-0.3in}j^{\mu}(x) &=& e^{\textstyle -i(\alpha - \beta)^2
D^{(+)}(\epsilon)}\!\Big(1 + \frac{\alpha +
\beta}{\sqrt{\pi}}\,\Big)\,J_{\nu}(x),\nonumber\\
\hspace{-0.3in}j^{\mu}_5(x) &=& e^{\textstyle -i(\alpha - \beta)^2
D^{(+)}(\epsilon)}\!\Big(1 + \frac{\alpha +
\beta}{\sqrt{\pi}}\,\Big)\,J^{\mu}_5(x).
\end{eqnarray}
which agree completely with the standard relation $j^{\mu}_5(x) = -
\varepsilon^{\mu\nu}j_{\nu}(x)$.

Inserting (\ref{label5.60}) in (\ref{label5.22}) we obtain
\begin{eqnarray}\label{label5.61}
i\gamma^{\mu}\partial_{\mu}\psi(x) = e^{\textstyle i(\alpha - \beta)^2
D^{(+)}(\epsilon)}\;\frac{\sqrt{\pi}\,(\alpha - \beta)}{\displaystyle
1 + \frac{\alpha + \beta}{2\sqrt{\pi}}}\!:\!
j^{\mu}(x)\gamma_{\mu}\psi(x)\!:.
\end{eqnarray}
If we define the renormalized quantum vector and axial--vector
currents as follows [4]
\begin{eqnarray}\label{label5.62}
\hspace{-0.3in}{\cal J}^{\mu}(x) &=& e^{\textstyle +i(\alpha - \beta)^2
D^{(+)}(\epsilon)}\,j^{\mu}(x) = \Big(1 + \frac{\alpha +
\beta}{\sqrt{\pi}}\,\Big)\,J_{\nu}(x),\nonumber\\
\hspace{-0.3in}{\cal J}^{\mu}_5(x)&=& e^{\textstyle +i(\alpha - \beta)^2
D^{(+)}(\epsilon)}\,j^{\mu}_5(x) = \Big(1 + \frac{\alpha +
\beta}{\sqrt{\pi}}\,\Big)\,J^{\mu}_5(x),
\end{eqnarray}
we arrive at the quantum equation of motion
\begin{eqnarray}\label{label5.63}
i\gamma^{\mu}\partial_{\mu}\psi(x) = \frac{\sqrt{\pi}\,(\alpha -
\beta)}{\displaystyle 1 + \frac{\alpha + \beta}{2\sqrt{\pi}}}:\!
{\cal J}^{\mu}(x)\gamma_{\mu}\psi(x)\!:.
\end{eqnarray}
In terms of the parameter $\rho$ (\ref{label5.42}) the equation of
motion (\ref{label5.63}) reads
\begin{eqnarray}\label{label5.64}
i\gamma^{\mu}\partial_{\mu}\psi(x) = 2\pi\,\frac{\rho^2 -1}{\rho^2 +
1}:\!{\cal J}^{\mu}(x)\gamma_{\mu}\psi(x)\!:.
\end{eqnarray}
Since the constant factor in front of the product $:\!{\cal
J}^{\mu}(x)\gamma_{\mu}\psi(x)\!:$ should be equal to $g$, we obtain
the equation
\begin{eqnarray}\label{label5.65}
\frac{2\pi}{g}\,\frac{\rho^2 - 1}{\rho^2 + 1} = 1.
\end{eqnarray}
The solution of this equation defines the parameter $\rho$:
\begin{eqnarray}\label{label5.66}
\rho = \sqrt{\frac{2\pi + g}{2\pi - g}}.
\end{eqnarray}
This confirms Klaiber's statement [27] concerning the presence of
singularities of the massless Thirring model at $g = \pm 2\pi$ in the
$g$--plane. 

We would like to emphasize that in our definition of quantum vector
and axial--vector currents the parameter $\rho$ should be only
positive. Indeed, for negative $\rho$ in the limit of a free fermionic
field theory the renormalized vector and axial--vector currents given
by (\ref{label5.62}) vanish. Therefore, for real $\rho$ the coupling
constant $g$ should range between $-2\pi < g < 2\pi$.

The main conclusion of this result is the absence of a one--parameter
family of solutions for the massless Thirring model with fermionic
fields quantized in the chiral symmetric phase. This statement is
valid for both Klaiber's parameterization in terms of the parameter
$\sigma$ fixed to the value $\sigma = 0$ and ours given by
(\ref{label5.66}).

In order to retain the existence of a one--parameter family of
solutions of the quantized version of the massless Thirring model we
suggest (i) to renormalize the wave function of the fermionic field $\psi(x) \to Z^{1/2}_2\,\psi(x)$
\begin{eqnarray}\label{label5.67}
i\gamma^{\mu}\partial_{\mu}\psi(x) = e^{\textstyle i(\alpha - \beta)^2
D^{(+)}(\epsilon)}\;\frac{\sqrt{\pi}\,(\alpha - \beta)}{\displaystyle
1 + \frac{\alpha + \beta}{2\sqrt{\pi}}}\!:\!
j^{\mu}(x)\gamma_{\mu}\psi(x)\!:.
\end{eqnarray}
and denote
\begin{eqnarray}\label{label5.68}
g = Z_2\,e^{\textstyle i(\alpha - \beta)^2
D^{(+)}(\epsilon)}\;\frac{\sqrt{\pi}\,(\alpha - \beta)}{\displaystyle
1 + \frac{\alpha + \beta}{2\sqrt{\pi}}}.
\end{eqnarray}
This yields the quantum equation of motion
\begin{eqnarray}\label{label5.69}
i\gamma^{\mu}\partial_{\mu}\psi(x) =g:\!
j^{\mu}(x)\gamma_{\mu}\psi(x)\!:
\end{eqnarray}
without any constraints on the parameters. Moreover, by using the
definition of the parameters $\alpha$ and $\beta$ in terms of the
parameter $\rho$ one can easily show that the $\rho$--depending factor
can be absorbed by a redefinition of the infrared cut--off
regularizing the function $D^{(+)}(\epsilon)$. Thus, this derivation
of the quantum equation of motion of the massless Thirring model
introduces neither new arbitrary parameters nor additional
constraints. The parameters $\alpha$ and $\beta$ defining a
one--parameter family of classical solutions (\ref{label4.18}) are
related by $\alpha - \beta = g/\sqrt{\pi}$. In turn, for the quantum
solutions (\ref{label5.1}), the parameters $\alpha$ and $\beta$ obey
only the relation (\ref{label5.26}).

In section 11 we show that Klaiber's operator approach applied to the
evaluation of correlation functions of massless Thirring fermion
fields is inconsistent with non--perturbative ultra--violet
renormalizability of the massless Thirring model for any parameters
$\alpha$ and $\beta$ related to each other in the operator formalism.
Non--perturbative renormalizability of the massless Thirring model
demands contributions induced by local chiral rotations in terms of
chiral Jacobians entering with an arbitrary parameter related to a
regularization procedure [11].

\section{Correlation functions of left(right)handed fermion densities.
Path--integral approach } 
\setcounter{equation}{0}

\hspace{0.2in} For the derivation of an equivalence between the
massive Thirring model and the sine--Gordon model Coleman analysed the
following set of $n$--point correlation functions of left(right)handed
fermion densities $\bar{\psi}(x)(1\pm\gamma^5)\psi(x)$ [17]
\begin{eqnarray}\label{label6.1}
\Big\langle 0\Big|{\rm
T}\Big(\prod^n_{i=1}\bar{\psi}(x_i)\Big(\frac{1-\gamma^5}{2}\Big)
\psi(x_i)\,\bar{\psi}(y_i)\Big(\frac{1+\gamma^5}{2}\Big)
\psi(y_i)\Big)\Big|0\Big\rangle.
\end{eqnarray}
This set of correlation functions can be obtained by using the
generating functional (\ref{label1.23}) and the generating functional
(\ref{label1.32}). This yields
\begin{eqnarray}\label{label6.2}
\hspace{-0.5in}&&\Big\langle 0\Big|{\rm
T}\Big(\prod^n_i\bar{\psi}(x_i)\Big(\frac{1-\gamma^5}{2}\Big)
\psi(x_i)\,\bar{\psi}(y_i)\Big(\frac{1+\gamma^5}{2}\Big)
\psi(y_i)\Big)\Big|0\Big\rangle =\nonumber\\
\hspace{-0.5in}&&= \prod^n_{i=1}\frac{1}{i}\frac{\delta}{\delta
J(x_i)}\frac{1}{i}\frac{\delta}{\delta
\bar{J}(x_i)}\frac{1}{i}\frac{\delta}{\delta
J(y_i)}\frac{1}{i}\frac{\delta}{\delta \bar{J}(y_i)}Z_{\rm
Th}[J,\bar{J}]\Big|_{\textstyle J = \bar{J} = 0}=\nonumber\\
\hspace{-0.5in}&=&\prod^n_{i=1}\frac{1}{i}\,\frac{\delta}{\delta
\sigma_+(x_i)}\,\frac{1}{i}\,\frac{\delta}{\delta \sigma_-(y_i)}Z_{\rm
Th}[\sigma_-,\sigma_+]\Big|_{\textstyle \sigma_- = \sigma_+ = 0}.
\end{eqnarray}
As a result the $n$--point correlation functions (\ref{label6.1}) can
be represented by the path integral
\begin{eqnarray}\label{label6.3}
&&\Big\langle 0\Big|{\rm
T}\Big(\prod^n_i\bar{\psi}(x_i)\Big(\frac{1-\gamma^5}{2}\Big)
\psi(x_i)\,\bar{\psi}(y_i)\Big(\frac{1+\gamma^5}{2}\Big)
\psi(y_i)\Big)\Big|0\Big\rangle =\nonumber\\ &&=\int {\cal D}\Psi{\cal
D} \bar{\Psi}\,e^{\textstyle \,i\int
d^2x\,\bar{\Psi}(x)i\gamma^{\mu}\partial_{\mu}\Psi(x)}\,
\prod^n_{i=1}\bar{\Psi}(x_i)
\Big(\frac{1-\gamma^5}{2}\Big)
\Psi(x_i)\,\bar{\Psi}(y_i)\Big(\frac{1+\gamma^5}{2}\Big)
\Psi(y_i)\,\nonumber\\ &&\times\,\int {\cal D}\xi\,e^{\textstyle
-\,i\,\frac{1}{2}\int d^2x\,\partial_{\mu}\xi(x)\partial^{\mu}
\xi(x)}\prod^n_{k=1}e^{\textstyle \,2\,i\,\lambda(\alpha_J)\,\xi(x_k) -
\,2\,i\,\lambda(\alpha_J)\,\xi(y_k)}.
\end{eqnarray}
For the integration over fermionic degrees of freedom we suggest to
consider the following generating functional
\begin{eqnarray}\label{label6.4}
Z_{\rm F}[J,\bar{J}]=\int {\cal D}\Psi{\cal D}
\bar{\Psi}\,e^{\textstyle \,i\int
d^2x\,[\bar{\Psi}(x)i\gamma^{\mu}\partial_{\mu}\Psi(x) +
\bar{J}(x)\Psi(x) + \bar{\Psi}(x)J(x)]},
\end{eqnarray}
where $J(x)$ and $\bar{J}(x) = J^{\dagger}(x)\gamma^0$ are column
and row matrices of external sources of fermion fields $\bar{\Psi}(x)$
and $\Psi(x)$, respectively, with components $J(x) = (J_1(x), J_2(x))$
and $\bar{J}(x) = (J^{\dagger}_2(x), J^{\dagger}_1(x))$. The
fermionic field $\Psi(x)$ is a column matrix with components $\Psi(x) =
(\Psi_1(x), \Psi_2(x))$. It is convenient to rewrite the r.h.s. of
(\ref{label6.4}) in terms of the fermionic field components
\begin{eqnarray}\label{label6.5}
\hspace{-0.7in}&&Z_{\rm F}[J,\bar{J}] =\nonumber\\
\hspace{-0.7in}&&=\int {\cal D}\Psi_1{\cal D} \Psi^{\dagger}_1\exp
i\!\!\int d^2x\Big[\Psi^{\dagger}_1(x)i \Big(\frac{\partial}{\partial
x^0} - \frac{\partial}{\partial x^1}\Big)\Psi_1(x) +
\Psi^{\dagger}_1(x)J_2(x) + J^{\dagger}_2(x)\Psi_1(x)\Big]\nonumber\\
\hspace{-0.7in}&&\times\int {\cal D}\Psi_2{\cal D}\Psi^{\dagger}_2\exp
i\!\!\int d^2x \Big[ \Psi^{\dagger}_2(x)i
\Big(\frac{\partial}{\partial x^0} + \frac{\partial}{\partial
x^1}\Big)\Psi_2(x) + \Psi^{\dagger}_2(x)J_1(x) +
J^{\dagger}_1(x)\Psi_2(x)\Big].
\end{eqnarray}
Integrating out the fermion fields we obtain
\begin{eqnarray}\label{label6.6}
\hspace{-0.3in}Z_{\rm F}[J,\bar{J}]=e^{\textstyle \,i\!\int\!\!\!\int
d^2x\,d^2y\,J^{\dagger}_2(x)S_{11}(x - y)J_2(y)}e^{\textstyle
\,i\!\int\!\!\!\int d^2x\,d^2y\,J^{\dagger}_1(x)S_{22}(x - y)J_1(y)},
\end{eqnarray}
where $S_{11}(x-y)$ and $S_{22}(x-y)$ are the Green functions
\begin{eqnarray}\label{label6.7}
S_{11}(x-y)&=&i\langle 0|{\rm
T}(\Psi_1(x)\Psi^{\dagger}_1(y))|0\rangle = \nonumber\\
&=&\frac{1}{2\pi}\, \frac{1}{v - V - i\,0\cdot\varepsilon(x^0-y^0)} =
\frac{1}{2\pi}\, \frac{u - U}{(x-y)^2 - i\,0},\nonumber\\
S_{22}(x-y)&=&i\langle 0|{\rm
T}(\Psi_2(x)\Psi^{\dagger}_2(y))|0\rangle =\nonumber\\ &=&
\frac{1}{2\pi}\, \frac{1}{u - U - i\,0\cdot\varepsilon(x^0-y^0)} =
\frac{1}{2\pi}\, \frac{v - V}{(x-y)^2 - i\,0},
\end{eqnarray}
where $v = x^0 - x^1$, $V = y^0 - y^1$, $u = x^0 + x^1$ and $U = y^0 +
y^1$, and $\varepsilon(x^0 - y^0)$ is a sign function. 

In terms of $S_{11}(x-y)$ and $S_{22}(x-y)$ the covariant Green
function $S_{\rm F}(x-y)$ is defined by
\begin{eqnarray}\label{label6.8}
S_{\rm F}(x-y) &=&i\langle 0|{\rm T}(\psi(x)\bar{\psi}(y))|0\rangle
=\nonumber\\ &=&\left(\begin{array}{cc} 0 & S_{11}(x-y)\\ S_{22}(x-y)
& 0
\end{array}\right) = \frac{1}{2\pi}\,\frac{1}{(x-y)^2 - i\,0}\,
\left(\begin{array}{cc}
0 & u - U\\ v - V & 0
\end{array}\right) = \nonumber\\
&=& \frac{1}{2\pi}\, \frac{1}{(x-y)^2 - i\,0}\,[\gamma^0(x^0-y^0) -
\gamma^1(x^1 - y^1)] = \frac{1}{2\pi}\, \frac{\hat{x} -
\hat{y}}{(x-y)^2 - i\,0}.
\end{eqnarray}
For the evaluation of the Green functions (\ref{label6.7}) we have
used expansions of the fermion fields $\Psi_1(x)$ and $\Psi_2(x)$ into
plane waves [6]:
\begin{eqnarray}\label{label6.9}
\hspace{-0.5in}\Psi_1(x) &=&\int^{\infty}_{-\infty}
\frac{dk^1}{\sqrt{2\pi}}\, \theta(+ k^1)\Big[a(k^1)\,e^{\textstyle
-ik^0x^0 + ik^1x^1} + b^{\dagger}(k^1)\,e^{\textstyle ik^0x^0 -
ik^1x^1}\Big],\nonumber\\
\hspace{-0.5in}\Psi_2(x)
&=&\int^{\infty}_{-\infty}\frac{dk^1}{\sqrt{2\pi}}\, \theta(-
k^1)\Big[a(k^1)\,e^{\textstyle -ik^0x^0 + ik^1x^1} -
b^{\dagger}(k^1)\,e^{\textstyle ik^0x^0 - ik^1x^1}\Big],
\end{eqnarray}
where $a(k^1)(a^{\dagger}(k^1))$ and $b(k^1)(b^{\dagger}(k^1))$ are
annihilation (creation) operators of fermions and anti--fermions with
momentum $k^1$.

The evaluation of the Green functions (\ref{label6.7}) runs in the way
\begin{eqnarray}\label{label6.10}
&&S_{11}(x-y) = i\langle 0|{\rm
T}(\Psi_1(x)\Psi^{\dagger}_1(y))|0\rangle =\nonumber\\
&&=i\,\theta(x^0 - y^0)\,\langle
0|\Psi_1(x)\Psi^{\dagger}_1(y)|0\rangle - i\,\theta(y^0 -
x^0)\,\langle 0|\Psi^{\dagger}_1(y)\Psi_1(x)|0\rangle =\nonumber\\ &&=
i\,\theta(x^0 - y^0)\int^{\infty}_{-\infty}\frac{dk^1}{2\pi}\,
\theta(+k^1)\,e^{\textstyle -ik^0(x^0 - y^0) + ik^1(x^1 -
y^1)}\nonumber\\ && - i\,\theta(y^0 -
x^0)\int^{\infty}_0\frac{dk^1}{2\pi}\, \theta(+k^1)\,e^{\textstyle
ik^0(x^0 - y^0) - ik^1(x^1 - y^1)}=\nonumber\\
&&= i\,\theta(x^0 -
y^0)\int^{\infty}_0\frac{dk^1}{2\pi}\,e^{\textstyle -ik^0(x^0 - y^0) +
ik^1(x^1 - y^1)}\nonumber\\
&& - i\,\theta(y^0 -
x^0)\int^{\infty}_0\frac{dk^1}{2\pi}\, e^{\textstyle ik^0(x^0 - y^0) -
ik^1(x^1 - y^1)}=\nonumber\\ &&=\theta(x^0 -
y^0)\,\frac{1}{2\pi}\,\frac{1}{v - V -i0} + \theta(y^0 -
x^0)\,\frac{1}{2\pi}\,\frac{1}{v - V +i0} = \nonumber\\ && =\theta(x^0
- y^0)\,\frac{1}{2\pi}\,\Big({\rm P}\frac{1}{v - V} +\pi i\,\delta(v
-V)\Big)\nonumber\\ && + \theta(y^0 - x^0)\,\frac{1}{2\pi}\,\Big({\rm
P}\frac{1}{v - V} - \pi i\,\delta(v -V)\Big)=\nonumber\\ && =
\frac{1}{2\pi}\,\Big({\rm P}\frac{1}{v - V} +\pi
i\,\varepsilon(x^0-y^0)\,\delta(v - V)\Big) =
\frac{1}{2\pi}\,\frac{1}{v - V - i0\cdot
\varepsilon(x^0-y^0)},\nonumber\\ &&S_{22}(x-y) = i\langle 0|{\rm
T}(\Psi_2(x)\Psi^{\dagger}_2(y))|0\rangle =\nonumber\\
&&=i\,\theta(x^0 - y^0)\,\langle
0|\Psi_2(x)\Psi^{\dagger}_2(y)|0\rangle - i\,\theta(y^0 -
x^0)\,\langle 0|\Psi^{\dagger}_2(y)\Psi_2(x)|0\rangle =\nonumber\\ &&=
i\,\theta(x^0 - y^0)\int^{\infty}_{-\infty}\frac{dk^1}{2\pi}\,
\theta(-k^1)\,e^{\textstyle -ik^0(x^0 - y^0) + ik^1(x^1 -
y^1)}\nonumber\\ && - i\,\theta(y^0 -
x^0)\int^{\infty}_0\frac{dk^1}{2\pi}\, \theta(-k^1)\,e^{\textstyle
ik^0(x^0 - y^0) - ik^1(x^1 - y^1)}=\nonumber\\ &&= i\,\theta(x^0 -
y^0)\int^{\infty}_0\frac{dk^1}{2\pi}\,e^{\textstyle -ik^0(x^0 - y^0) -
ik^1(x^1 - y^1)}\nonumber\\ && - i\,\theta(y^0 -
x^0)\int^{\infty}_0\frac{dk^1}{2\pi}\, e^{\textstyle ik^0(x^0 - y^0) +
ik^1(x^1 - y^1)}=\nonumber\\ &&=\theta(x^0 -
y^0)\,\frac{1}{2\pi}\,\frac{1}{u - U -i0} + \theta(y^0 -
x^0)\,\frac{1}{2\pi}\,\frac{1}{u - U +i0} = \nonumber\\ && =\theta(x^0
- y^0)\,\frac{1}{2\pi}\,\Big({\rm P}\frac{1}{u - U} +\pi i\,\delta(u
-U)\Big)\nonumber\\ && + \theta(y^0 - x^0)\,\frac{1}{2\pi}\,\Big({\rm
P}\frac{1}{u - U} - \pi i\,\delta(u -U)\Big)=\nonumber\\ && =
\frac{1}{2\pi}\,\Big({\rm P}\frac{1}{u - U} +\pi
i\,\varepsilon(x^0-y^0)\,\delta(u - U)\Big) =
\frac{1}{2\pi}\,\frac{1}{u - U - i0\cdot \varepsilon(x^0-y^0)},
\end{eqnarray}
where the symbol ${\rm P}$ denotes the principle value and
$\varepsilon(x^0-y^0) = \theta(x^0 - y^0) - \theta(y^0 - x^0)$.

The Green functions $S_{11}(x-y)$ and $S_{22}(x-y)$ obey the equations
\begin{eqnarray}\label{label6.11}
&&i\Big(\frac{\partial}{\partial x^0} + \frac{\partial }{\partial
x^1}\Big)\,S_{11}(x-y) = -\,\delta^{(2)}(x-y),\nonumber\\
&&i\Big(\frac{\partial}{\partial x^0} - \frac{\partial }{\partial
x^1}\Big)\,S_{22}(x-y) = -\,\delta^{(2)}(x-y).
\end{eqnarray}
One can show that direct solutions of these equations leads to the
expressions given by (\ref{label6.10}).

The correlation function (\ref{label6.3}) can be rewritten as follows
\begin{eqnarray}\label{label6.12}
&&\Big\langle 0\Big|{\rm
T}\Big(\prod^n_i\bar{\psi}(x_i)\Big(\frac{1-\gamma^5}{2}\Big)
\psi(x_i)\,\bar{\psi}(y_i)\Big(\frac{1+\gamma^5}{2}\Big)
\psi(y_i)\Big)\Big|0\Big\rangle = \int {\cal D}\Psi_1{\cal
D}\Psi^{\dagger}_1{\cal D}\Psi_2{\cal D}\Psi^{\dagger}_2\nonumber\\
&&\times\,\exp \,i\int d^2x\,\Big[\Psi^{\dagger}_1(x)i
\Big(\frac{\partial}{\partial x^0} - \frac{\partial}{\partial
x^1}\Big)\Psi_1(x) +\Psi^{\dagger}_2(x)i \Big(\frac{\partial}{\partial
x^0} + \frac{\partial}{\partial x^1}\Big)\Psi_2(x)\Big]\nonumber\\
&&\times\,\prod^n_{i=1}\Psi^{\dagger}_1(x_i)\Psi_2(x_i)\,
\Psi^{\dagger}_2(y_i)\Psi_1(y_i)\int {\cal D}\xi\,e^{\textstyle
-\,i\,\frac{1}{2}\int d^2x\,\partial_{\mu}\xi(x)\partial^{\mu}
\xi(x)}\nonumber\\ &&\times\,\prod^n_{k=1}e^{\textstyle
\,2\,i\,\lambda(\alpha_J)\,\xi(x_k) -
\,2\,i\,\lambda(\alpha_J)\,\xi(y_k)}.
\end{eqnarray}
Using the generating functional (\ref{label6.6}) the integration over
fermionic degrees of freedom gives (see (8.35) of Ref.[6])
\begin{eqnarray}\label{label6.13}
&& \int {\cal D}\Psi_1{\cal D}\Psi^{\dagger}_1{\cal D}\Psi_2{\cal
D}\Psi^{\dagger}_2\,\exp \,i\int d^2x\,\Big[\Psi^{\dagger}_1(x)i
\Big(\frac{\partial}{\partial x^0} - \frac{\partial}{\partial
x^1}\Big)\Psi_1(x)\nonumber\\ && + \Psi^{\dagger}_2(x)i
\Big(\frac{\partial}{\partial x^0} + \frac{\partial}{\partial
x^1}\Big)\Psi_2(x)\Big]\,\prod^n_{i=1}\Psi^{\dagger}_1(x_i)\Psi_2(x_i)\,
\Psi^{\dagger}_2(y_i)\Psi_1(y_i)=\nonumber\\
&&=\frac{1}{(2\pi)^{2n}}\, \frac{\displaystyle \prod^{n}_{j <k}[-(x_j
- x_k)^2 + i\,0\,]\prod^{n}_{j <k}[-(y_j - y_k)^2 +
i\,0\,]}{\displaystyle\prod^{n}_{j = 1}\prod^{n}_{k = 1}[-(x_j -
y_k)^2 + i\,0\,]}.
\end{eqnarray}
Now the correlation function (\ref{label6.12}) is equal to
\begin{eqnarray}\label{label6.14}
&&\Big\langle 0\Big|{\rm
T}\Big(\prod^n_i\bar{\psi}(x_i)\Big(\frac{1-\gamma^5}{2}\Big)
\psi(x_i)\,\bar{\psi}(y_i)\Big(\frac{1+\gamma^5}{2}\Big)
\psi(y_i)\Big)\Big|0\Big\rangle =\nonumber\\ &&=
\frac{1}{(2\pi)^{2n}}\, \frac{\displaystyle \prod^{n}_{j <k}[-(x_j -
x_k)^2 + i\,0\,]\prod^{n}_{j <k}[-(y_j - y_k)^2 + i\,0\,
]}{\displaystyle\prod^{n}_{j = 1}\prod^{n}_{k = 1}[-(x_j - y_k)^2 +
i\,0\,]}\nonumber\\ &&\times \int {\cal D}\xi\,e^{\textstyle
-\,i\,\frac{1}{2}\int d^2x\,\partial_{\mu}\xi(x)\partial^{\mu}
\xi(x)}\,\prod^n_{k=1}e^{\textstyle
\,2\,i\,\lambda(\alpha_J)\,\xi(x_k) -
\,2\,i\,\lambda(\alpha_J)\,\xi(y_k)}.
\end{eqnarray}
For the evaluation of the path integral over the $\xi$--field we
apply the infrared regularization suggested by Coleman [17] and
discussed then in [8] (see also [6]):
\begin{eqnarray}\label{label6.15}
\hspace{-0.5in}&&\int {\cal D}\xi\,e^{\textstyle -\,i\,\frac{1}{2}\int
d^2x\,\partial_{\mu}\xi(x)\partial^{\mu}
\xi(x)}\,\prod^n_{k=1}e^{\textstyle i\,\lambda(\alpha_J)\,\xi(x_k) -
i\,\lambda(\alpha_J)\,\xi(y_k)}\nonumber\\ \hspace{-0.5in}&& \to \lim_{\mu \to
0}\int {\cal D}\xi\,e^{\textstyle -\,i\,\frac{1}{2}\int
d^2x\,[\partial_{\mu}\xi(x)\partial^{\mu} \xi(x) -
\mu^2\,\xi^2(x)]}\,\prod^n_{k=1}e^{\textstyle
i\,\lambda(\alpha_J)\,\xi(x_k) -
i\,\lambda(\alpha_J)\,\xi(y_k)}=\nonumber\\
\hspace{-0.5in}&& \to \lim_{\mu \to 0}\int {\cal D}\xi\,e^{\textstyle
\,i\,\frac{1}{2}\int d^2x\,\xi(x)(\Box +
\mu^2)\,\xi(x)}\,\prod^n_{k=1}e^{\textstyle
\,2\,i\,\lambda(\alpha_J)\,\xi(x_k) -
\,2\,i\,\lambda(\alpha_J)\,\xi(y_k)}.
\end{eqnarray}
The result of the integration over the $\xi$--field reads (see
(2.6) of Ref.[6])
\begin{eqnarray}\label{label6.16}
&&\int {\cal D}\xi\,e^{\textstyle \,i\,\frac{1}{2}\int
d^2x\,\xi(x)(\Box + \mu^2)\,\xi(x)}\,\prod^n_{k=1}e^{\textstyle
2\,i\,\lambda(\alpha_J)\,\xi(x_k) - 2\,i\,\lambda(\alpha_J)\,\xi(y_k)}=\nonumber\\ &&=
\exp\Big\{ -\,n\,4\,\lambda^2(\alpha_J)\,i\Delta(0)\Big\}\,\exp\Big\{
-4\,\lambda^2(\alpha_J)\sum^{n}_{j <k}i\Delta(x_j - x_k)\nonumber\\
\hspace{-0.3in}&& - 4\,\lambda^2(\alpha_J)\sum^{n}_{j <k}i\Delta(y_j -
y_k) + 4\,\lambda^2(\alpha_J)\sum^{n}_{j = 1}\sum^{n}_{k =
1}i\Delta(x_j - y_k)\Big\},
\end{eqnarray}
where the Green function $\Delta(x-y)$ is determined by 
\begin{eqnarray}\label{label6.17}
\Delta(x-y) = i\langle 0|{\rm T}(\xi(x)\xi(y))|0\rangle
\end{eqnarray}
and obeys the equation 
\begin{eqnarray}\label{label6.18}
(\Box + \mu^2)\,\Delta(x-y) = \delta^{(2)}(x - y).
\end{eqnarray}
In the limit $\mu \to 0$ the Green function $\Delta(x-y)$ is given by
the expression (for a detailed calculation see Appendix C)
\begin{eqnarray}\label{label6.19}
i\Delta(x-y) =\frac{1}{4\pi}\,{\ell n}[-\mu^2(x-y)^2 + i\,0\,].
\end{eqnarray}
For $(x-y) = 0$ the Green function $i\Delta(0)$ amounts to
\begin{eqnarray}\label{label6.20}
i\Delta(0) = -\,\frac{1}{4\pi}\,{\ell
n}\Big(\frac{\Lambda^2}{\mu^2}\Big),
\end{eqnarray}
where $\Lambda$ is an ultra--violet cut--off.

It is important to emphasize that the path integrals over the
$\xi$--field are ill--defined in the infrared region
\begin{eqnarray}\label{label6.21}
&&\int {\cal D}\xi\,e^{\textstyle \,i\,\frac{1}{2}\int
d^2x\,\xi(x)(\Box +
\mu^2)\,\xi(x)}\,\prod^p_{k=1}\prod^n_{j=1}e^{\textstyle
\,2\,i\,\lambda(\alpha_J)\,\xi(x_k) -
\,2\,i\,\lambda(\alpha_J)\,\xi(y_j)}=\nonumber\\ &&= \exp\Big\{
-\frac{1}{2}\,(p+n)\,4\,\lambda^2(\alpha_J)\,i\Delta(0)\Big\}\,\exp\Big\{
-4\,\lambda^2(\alpha_J)\sum^{p}_{j <k}i\Delta(x_j - x_k)\nonumber\\
\hspace{-0.3in}&& - 4\,\lambda^2(\alpha_J)\sum^{n}_{j <k}i\Delta(y_j -
y_k) + 4\,\lambda^2(\alpha_J)\sum^{p}_{j = 1}\sum^{n}_{k =
1}i\Delta(x_j - y_k)\Big\} \nonumber\\ && \sim (\mu^2)^{\textstyle
-(p-n)^2\lambda^2(\alpha_J)/2\pi}
\end{eqnarray}
and diverge in the limit $\mu^2 \to 0$ for $p\not= n$. Such a
behaviour of correlation functions with $p\not= n$ is related to the
ghost--nature of the $\xi$--field having an incorrect sign of the
kinetic term.

After the integration over the $\xi$--field the correlation function
(\ref{label6.1}) reads
\begin{eqnarray}\label{label6.22}
\hspace{-0.7in}&&\Big\langle 0\Big|{\rm
T}\Big(\prod^n_i\bar{\psi}(x_i)\Big(\frac{1-\gamma^5}{2}\Big)
\psi(x_i)\,\bar{\psi}(y_i)\Big(\frac{1+\gamma^5}{2}\Big)
\psi(y_i)\Big)\Big|0\Big\rangle =
\Bigg(\frac{\Lambda}{2\pi}\Bigg)^{2n}\nonumber\\
\hspace{-0.7in}&&\times\,\frac{\displaystyle \prod^{n}_{j
<k}[-\Lambda^2(x_j - x_k)^2 + i\,0\,]^{\textstyle 1 + d_{\textstyle
\bar{\psi}\psi}(g,\alpha_J)}\prod^{n}_{j <k}[-\Lambda^2(y_j - y_k)^2 +
i\,0\,]^{\textstyle 1 + d_{\textstyle
\bar{\psi}\psi}(g,\alpha_J)}}{\displaystyle\prod^{n}_{j =
1}\prod^{n}_{k = 1}[-\Lambda^2(x_j - y_k)^2 + i\,0\,]^{\textstyle 1 +
d_{\textstyle \bar{\psi}\psi}(g,\alpha_J)}},
\end{eqnarray}
where $d_{\textstyle \bar{\psi}\psi}(g,\alpha_J)$ are dynamical
dimensions of the left(right)handed fermion densities $\bar{\psi}(x)(1
\pm\gamma^5)\psi(x)$. According to Wilson's analysis [28]
$d_{\textstyle \bar{\psi}\psi}(g,\alpha_J)$ is equal to
\begin{eqnarray}\label{label6.23}
d_{\textstyle \bar{\psi}\psi}(g,\alpha_J) = - \frac{\displaystyle
g/\pi}{\displaystyle 1 + \alpha_J\,g/\pi}.
\end{eqnarray}
The dependence of $d_{\textstyle \bar{\psi}\psi}(g,\alpha_J)$ on the
arbitrary parameter $\alpha_J$ agrees with a one--parameter family of
solutions of the massless Thirring model.

For the two--point correlation function we obtain
\begin{eqnarray}\label{label6.24}
\hspace{-0.5in}&&\Big\langle 0\Big|{\rm
T}\Big(\bar{\psi}(x)\Big(\frac{1-\gamma^5}{2}\Big)
\psi(x)\,\bar{\psi}(y)\Big(\frac{1+\gamma^5}{2}\Big)
\psi(y)\Big)\Big|0\Big\rangle = \nonumber\\
\hspace{-0.5in}&&=\frac{\Lambda^2}{4\pi^2}\, [-\Lambda^2(x - y)^2 +
i\,0\,]^{\textstyle - 1 - d_{\textstyle \bar{\psi}\psi}(g,\alpha_J)}.
\end{eqnarray}
We would like to emphasize that the correlation functions
(\ref{label6.22}) evaluated within the path--integral approach do not
depend on an infrared cut--off. This is unlike the correlation
functions evaluated within Klaiber's operator formalism. In sections 8
and 9 we consider the evaluation of $2n$--point Green functions and
$n$--point correlation functions of the left(right)handed fermion
densities within Klaiber's operator formalism and discuss the
dependence of these expressions on the infrared cut--off.

\section{Bosonization of generating functionals 
$Z_{\rm Th}[\sigma_-,\sigma_+]$ and  $Z_{\rm Th}[\sigma]$ }
\setcounter{equation}{0}

\hspace{0.2in} The expression (\ref{label6.22}) for the $n$--point
correlation function (\ref{label6.1}) can be obtained using the
generating functional $Z_{\rm Th}[\sigma_-,\sigma_+]$ in
(\ref{label1.32}). In this section we show that the fermionic degrees
of freedom in the generating functionals $Z_{\rm
Th}[\sigma_-,\sigma_+]$ and $Z_{\rm Th}[\sigma]$ can be fully replaced
by bosonic degrees. This means that these generating functionals admit
full bosonization.

For this aim we suggest to expand $Z_{\rm Th}[\sigma_-,\sigma_+]$ 
in powers of $\sigma_{\mp}$. This yields
\begin{eqnarray}\label{label7.1}
\hspace{-0.5in}&&Z_{\rm Th}[\sigma_-,\sigma_+]
=\sum^{\infty}_{n=1}\sum^{\infty}_{m=1}\frac{i^n}{n!}\,
\frac{i^m}{m!}\prod^n_{i=1}\prod^m_{j=1}\int \!\!\!\int d^2x_i
d^2y_j\,\sigma_+(x_i)\sigma_-(y_j)\,\nonumber\\
\hspace{-0.5in}&&\times\int {\cal
D}\Psi {\cal D}\bar{\Psi} {\cal D}\xi\, \exp\,i\!\!\int
d^2z\,\Bigg\{\bar{\Psi}(z)i\gamma^{\mu}\partial_{\mu}\Psi(z) -
\frac{1}{2}\,\partial_{\mu}\xi(z)\partial^{\mu}\xi(z)\Bigg\}\nonumber\\
\hspace{-0.5in}&&\times\,\Bigg[\bar{\Psi}(x_i)
\Bigg(\frac{1-\gamma^5}{2}\Bigg)\Psi(x_i)\Bigg]
\Bigg[\bar{\Psi}(y_j)\Bigg(\frac{1+\gamma^5}{2}\Bigg)
\Psi(y_j)\Bigg]\,e^{\textstyle
-2i\lambda(\alpha_J)\,\xi(x_i)}\,e^{\textstyle +
2i\lambda(\alpha_J)\,\xi(y_j)} =\nonumber\\
\hspace{-0.5in}&&=\sum^{\infty}_{n=1}\sum^{\infty}_{m=1}\frac{i^n}{n!}\,
\frac{i^m}{m!}\prod^n_{i=1}\prod^m_{j=1}\int \!\!\!\int d^2x_i
d^2y_j\,\sigma_+(x_i)\sigma_-(y_j)\,\nonumber\\
\hspace{-0.5in}&&\times\,\Bigg\langle 0\Bigg|{\rm
T}\Bigg(\Bigg[\bar{\Psi}(x_i)
\Bigg(\frac{1-\gamma^5}{2}\Bigg)\Psi(x_i)\Bigg]
\Bigg[\bar{\Psi}(y_j)\Bigg(\frac{1+\gamma^5}{2}\Bigg)
\Psi(y_j)\Bigg]\Bigg)\Bigg|0\Bigg\rangle\nonumber\\
\hspace{-0.5in}&&\times\int {\cal D}\xi\, \exp\,\Bigg\{- i\int
d^2z\,\frac{1}{2}\,\partial_{\mu}\xi(z)\partial^{\mu}\xi(z) -
2i\lambda(\alpha_J)\,\xi(x_i) + 2i\lambda(\alpha_J)\,\xi(y_j)\Bigg\}.
\end{eqnarray}
The time--ordered vacuum expectation value of products of free fermion
fields $\Psi$ and $\bar{\Psi}$ can be represented in terms of a
time--ordered vacuum expectation value of products of a massless
pseudoscalar (or scalar) field $\vartheta$. Using Eq.(2.10) of Ref.[6]
we get
\begin{eqnarray}\label{label7.2}
\hspace{-0.3in}&&\Bigg\langle 0\Bigg|{\rm
T}\Bigg(\prod^n_{i=1}\prod^m_{j=1}\Bigg[\bar{\Psi}(x_i)\Bigg(\frac{1 -
\gamma^5}{2}\Bigg)\Psi(x_i)\Bigg]\Bigg[\bar{\Psi}(y_j)\Bigg(\frac{1 +
\gamma^5}{2}\Bigg)\Psi(y_j)\Bigg]\Bigg) \Bigg|0\Bigg\rangle
=\nonumber\\
\hspace{-0.3in}&& = \delta_{n m}\,\frac{1}{(2\pi)^{n + m}}
\frac{\displaystyle \prod^{n}_{i <k}[-(x_i - x_k)^2]\prod^{m}_{j
<k}[-(y_j - y_k)^2]}{\displaystyle\prod^{n}_{i = 1}\prod^{m}_{j =
1}[-(x_i - y_j)^2]} = \nonumber\\
\hspace{-0.3in}&& = \prod^n_{i=1}\prod^m_{j=1}\Bigg\langle 0\Bigg|{\rm
T}\Bigg(\frac{\Lambda}{2\pi}\,e^{\textstyle
i\sqrt{4\pi}\,\vartheta(x_i)}\,\frac{\Lambda}{2\pi}\,e^{\textstyle -
i\sqrt{4\pi}\,\vartheta(y_j)}\Bigg)\Bigg|0\Bigg\rangle,
\end{eqnarray}
where $\delta_{n m}$ is the Kronecker symbol and $\Lambda$ is an
ultra--violet cut--off.

Substituting (\ref{label7.2}) and (\ref{label7.1}) we get
\begin{eqnarray}\label{label7.3}
\hspace{-0.5in}&&Z_{\rm Th}[\sigma_-,\sigma_+]
=\sum^{\infty}_{n=1}\sum^{\infty}_{m=1}\frac{i^n}{n!}\,
\frac{i^m}{m!}\prod^n_{i=1}\prod^m_{j=1}\int \!\!\!\int d^2x_i
d^2y_j\,\sigma_+(x_i)\sigma_-(y_j)\,\nonumber\\
\hspace{-0.5in}&&\times\,\Bigg\langle 0\Bigg|{\rm
T}\Bigg(\frac{\Lambda}{2\pi}\,e^{\textstyle
i\sqrt{4\pi}\,\vartheta(x_i)}\,\frac{\Lambda}{2\pi}\,e^{\textstyle -
i\sqrt{4\pi}\,\vartheta(y_j)}\Bigg)\Bigg|0\Bigg\rangle\nonumber\\
\hspace{-0.5in}&&\times\int {\cal D}\xi\, \exp\,\Bigg\{- i\int
d^2z\,\frac{1}{2}\,\partial_{\mu}\xi(z)\partial^{\mu}\xi(z) -
2i\lambda(\alpha_J)\,\xi(x_i) + 2i\lambda(\alpha_J)\,\xi(y_j)\Bigg\}.
\end{eqnarray}
Since the vacuum expectation value can be represented in terms of a
path integral over the $\vartheta$--field (see Eq.(2.5) of Ref.[6]),
the generating functional (\ref{label7.3}) takes the form
\begin{eqnarray}\label{label7.4}
\hspace{-0.5in}&&Z_{\rm Th}[\sigma_-,\sigma_+]
=\sum^{\infty}_{n=1}\sum^{\infty}_{m=1}\frac{i^n}{n!}\,
\frac{i^m}{m!}\prod^n_{i=1}\prod^m_{j=1}\int \!\!\!\int d^2x_i
d^2y_j\,\sigma_+(x_i)\sigma_-(y_j)\,\nonumber\\
\hspace{-0.5in}&&\times\int {\cal D}\vartheta {\cal D}\xi\,
\exp\,i\!\!\int
d^2z\,\Bigg\{\frac{1}{2}\,\partial_{\mu}\vartheta(z)
\partial^{\mu}\vartheta(z)
-
\frac{1}{2}\,\partial_{\mu}\xi(z)\partial^{\mu}\xi(z)\Bigg\}
\nonumber\\
\hspace{-0.5in}&&\times\,
\frac{\Lambda}{2\pi}\,e^{\textstyle
i\sqrt{4\pi}\,\vartheta(x_i)}\,\frac{\Lambda}{2\pi}\,e^{\textstyle -
i\sqrt{4\pi}\,\vartheta(y_j)}\,e^{\textstyle
-2i\lambda(\alpha_J)\,\xi(x_i) + 2i\lambda(\alpha_J)\,\xi(y_j)}.
\end{eqnarray}
Summing up the series we get
\begin{eqnarray}\label{label7.5}
\hspace{-0.5in}&&Z_{\rm Th}[\sigma_-,\sigma_+] =\int {\cal D}\vartheta
{\cal D}\xi\, \exp\,i\!\!\int
d^2z\,\Bigg\{\frac{1}{2}\,\partial_{\mu}\vartheta(z)
\partial^{\mu}\vartheta(z) -
\frac{1}{2}\,\partial_{\mu}\xi(z)\partial^{\mu}\xi(z)\nonumber\\
\hspace{-0.5in}&& + \sigma_-(z)\,\frac{\Lambda}{2\pi}\,e^{\textstyle -
i\sqrt{4\pi}\,\vartheta(z) +2i\lambda(\alpha_J)\,\xi(z)} +
\sigma_+(z)\,\frac{\Lambda}{2\pi}\,e^{\textstyle +
i\sqrt{4\pi}\,\vartheta(z) - 2i\lambda(\alpha_J)\,\xi(z)}\Bigg\}.
\end{eqnarray}
In order to remove unphysical degrees of freedom from the path
integral (\ref{label7.5}) we suggest to make a change of variables
\begin{eqnarray}\label{label7.6}
\Theta(x) &=& \frac{\sqrt{4\pi}\,\vartheta(x) -
2\lambda(\alpha_J)\,\xi(x)} {\sqrt{4\pi -
4\lambda^2(\alpha_J)}},\nonumber\\ \Phi(x) &=&
\frac{2\lambda(\alpha_J)\,\vartheta(x) -
\sqrt{4\pi}\,\xi(x)}{\sqrt{4\pi - 4\lambda^2(\alpha_J)}}
\end{eqnarray}
Now the generating functional (\ref{label7.5}) reads 
\begin{eqnarray}\label{label7.7}
\hspace{-0.7in}Z_{\rm Th}[\sigma_-,\sigma_+] &=&\int {\cal
D}\Theta\,{\cal D}\Phi\,\exp\,i\int
d^2z\,\Bigg\{\frac{1}{2}\,\partial_{\mu}\Theta(z)
\partial^{\mu}\Theta(z) -
\frac{1}{2}\,\partial_{\mu}\Phi(z)\partial^{\mu}\Phi(z)\nonumber\\
\hspace{-0.7in}&& +
\sigma_-(z)\,\frac{\Lambda}{2\pi}\,\exp\,\Big(-i\sqrt{4\pi\,(1 +
d_{\textstyle \bar{\psi}\psi}(g,\alpha_J))}\,\Theta(z)\Big)
\nonumber\\
\hspace{-0.7in}&& +
\sigma_+(z)\,\frac{\Lambda}{2\pi}\,\exp\,\Big(+i\sqrt{4\pi\,(1 +
d_{\textstyle \bar{\psi}\psi}(g,\alpha_J))}\,\Theta(z)\Big)\Bigg\},
\end{eqnarray}
where we have used the relation
\begin{eqnarray}\label{label7.8}
\sqrt{4\pi - 4\lambda^2(\alpha_J)} = \sqrt{4\pi\,(1 +
d_{\textstyle \bar{\psi}\psi}(g,\alpha_J))}
\end{eqnarray}
with $d_{\textstyle \bar{\psi}\psi}(g,\alpha_J)$ given by 
(\ref{label6.23}).

Integrating out the $\Phi$--field we define the generating functional
$Z_{\rm Th}[\sigma_-,\sigma_+]$ by the path integral over the
$\Theta$--field only
\begin{eqnarray}\label{label7.9}
\hspace{-0.7in}Z_{\rm Th}[\sigma_-,\sigma_+] &=&\int {\cal
D}\Theta\,\exp\,i\int
d^2z\,\Bigg\{\frac{1}{2}\,\partial_{\mu}\Theta(z)
\partial^{\mu}\Theta(z)\nonumber\\
\hspace{-0.7in}&& +
\sigma_-(z)\,\frac{\Lambda}{2\pi}\,\exp\,\Big(-i\sqrt{4\pi\,(1 +
d_{\textstyle \bar{\psi}\psi}(g,\alpha_J))}\,\Theta(z)\Big)
\nonumber\\
\hspace{-0.7in}&& +
\sigma_+(z)\,\frac{\Lambda}{2\pi}\,\exp\,\Big(+i\sqrt{4\pi\,(1 +
d_{\textstyle \bar{\psi}\psi}(g,\alpha_J))}\,\Theta(z)\Big)\Bigg\}
\end{eqnarray}
It is important to accentuate that the generating functional
(\ref{label7.9}) is in terms of physical degrees of freedom only. It
is easy to verify that substituting the generating functional
(\ref{label7.9}) and integrating over the $\Theta$--field one obtains
the result (\ref{label6.22}).

Setting $\sigma_-(z) = \sigma_+(z) = \sigma(z)$ in (\ref{label7.9}) we
obtain the bosonized version of the generating functional $Z_{\rm
Th}[\sigma]$. It reads
\begin{eqnarray}\label{label7.10}
\hspace{-0.3in}Z_{\rm Th}[\sigma] &=& \int {\cal D}\Theta\,
\exp\,i\!\!\int d^2z\Bigg\{\frac{1}{2}\partial_{\mu}\Theta(z)
\partial^{\mu}\Theta(z)\nonumber\\
\hspace{-0.3in}&&+ \sigma(z)\,\frac{\Lambda}{\pi}\,
\cos\Big(\sqrt{4\pi\,(1 +
d_{\textstyle \bar{\psi}\psi}(g,\alpha_J))}\,\Theta(z)\Big) \Bigg\}. 
\end{eqnarray}
Using the generating functional (\ref{label7.10}) we are able to
evaluate the two--point correlation function of scalar fermion
densities
\begin{eqnarray}\label{label7.11}
\hspace{-0.3in}&&\langle 0|{\rm T}(\bar{\psi}(x)
\psi(x)\bar{\psi}(y)\psi(y))|0\rangle =
\frac{1}{i}\,\frac{\delta}{\delta
\sigma(x)}\,\frac{1}{i}\,\frac{\delta}{\delta \sigma(y)}Z_{\rm
Th}[\sigma]\Big|_{\textstyle \sigma = 0} =\nonumber\\
\hspace{-0.3in}&&=\frac{\Lambda^2}{\pi^2}\int {\cal D}\Theta\,
\exp\,i\int
d^2z\,\Bigg\{\frac{1}{2}\,\partial_{\mu}\Theta(z)
\partial^{\mu}\Theta(z)\Bigg\}\nonumber\\
\hspace{-0.3in}&&\times\,\cos\Big(\sqrt{4\pi\,(1 +
d_{\textstyle \bar{\psi}\psi}(g,\alpha_J))}\,\Theta(x)\Big)\,
\cos\Big(\sqrt{4\pi\,(1 +
d_{\textstyle \bar{\psi}\psi}(g,\alpha_J))}\,\Theta(y)\Big).
\end{eqnarray}
Integrating over the $\Theta$--field we get 
\begin{eqnarray}\label{label7.12}
\hspace{-0.5in}\langle 0|{\rm T}(\bar{\psi}(x)\psi(x)
\bar{\psi}(y)\psi(y))|0\rangle =
\frac{\Lambda^2}{2\pi^2}\,[-\Lambda^2(x-y)^2 + i\,0\,]^{\textstyle - 1
- d_{\textstyle \bar{\psi}\psi}(g,\alpha_J)}.
\end{eqnarray}
The same result for the two--point correlation function
(\ref{label7.12}) can be derived from (\ref{label6.24}).

Setting $\sigma(z) = -m$, where $m$ has the meaning of the mass of
Thirring fermions, the generating functional
(\ref{label7.10}) reduces to the partition function of the
sine--Gordon (SG) model
\begin{eqnarray}\label{label7.13}
\hspace{-0.7in}&&Z_{\rm Th}[\sigma]\Big|_{\sigma = - m} = Z_{\rm SG} =
\nonumber\\ \hspace{-0.7in}&&= \int {\cal D}\Theta \exp\,i\!\!\int
d^2z\Bigg\{\frac{1}{2}\partial_{\mu}\Theta(z) \partial^{\mu}\Theta(z)
- m \,\frac{\Lambda}{\pi}\,\cos\Big(\sqrt{4\pi\,(1 +
d_{\textstyle \bar{\psi}\psi}(g,\alpha_J))}\,\Theta(z)\Big) \Bigg\}.
\end{eqnarray}
Recall that the sine--Gordon model is described by the Lagrangian
(\ref{label1.28}). A positive sign for the interaction in the
r.h.s. of (\ref{label7.13}) one can get by the shift 
\begin{eqnarray}\label{label7.14}
\Theta(z) \to \Theta(z) + \pi\,\sqrt{\frac{1}{4\pi}\, \frac{1}{1 +
d_{\textstyle \bar{\psi}\psi}(g,\alpha_J)}}.
\end{eqnarray}
Matching the Lagrangian of the sine--Gordon model
(\ref{label1.29}) with the effective Lagrangian in the exponent of the
r.h.s. of the generating functional (\ref{label7.13}) we obtain the
relations between the sine--Gordon parameters $\bar{\alpha}$ and
$\bar{\beta}$ and the Thirring model parameters
\begin{eqnarray}\label{label7.15}
\frac{\bar{\beta}^2}{4\pi}&=& 1 + d_{\textstyle
\bar{\psi}\psi}(g,\alpha_J),\nonumber\\ \bar{\alpha} &=& 4\,
m\,\Lambda\,\frac{\bar{\beta}^2}{4\pi}= 4\,m\,\Lambda\,(1 +
d_{\textstyle \bar{\psi}\psi}(g,\alpha_J)).
\end{eqnarray}
Substituting the dynamical dimension $d_{\textstyle
\bar{\psi}\psi}(g,\alpha_J)$ given by (\ref{label6.22}) we get
\begin{eqnarray}\label{label7.16}
\frac{\bar{\beta}^2}{4\pi}&=&\frac{\displaystyle1 + (\alpha_J -
1)\,g/\pi}{\displaystyle 1 + \alpha_J g/\pi}\Bigg|_{\textstyle
\alpha_J= 1} = \frac{1}{\displaystyle 1 + g/\pi},\nonumber\\
\bar{\alpha} &=& 4\,m\,\Lambda\,\frac{\displaystyle 1 + (\alpha_J -
1)\,g/\pi}{\displaystyle 1 + \alpha_J g/\pi}\Bigg|_{\textstyle
\alpha_J = 1} = \frac{4\,m\,\Lambda}{\displaystyle 1 + g/\pi}.
\end{eqnarray}
Our results obtained in this section agree with Na${\acute{\rm o}}$n's
calculations [8] up to the uncertainties defined by the parameter
$\alpha_J$ and related to regularizations of chiral Jacobians.

\section{Two--point Green function of the massless Thirring fermion 
field. The path--integral approach} 
\setcounter{equation}{0}

\hspace{0.2in} In this section we discuss the evaluation of the
two--point Green function of the Thirring fermionic field $S_{\rm
F}(x,y) = i\langle 0|{\rm T}(\psi(x)\bar{\psi}(y))|0\rangle $ in the
path--integral approach suggested in Refs.[7,8].

In terms of the generating functional of Green functions given by
(\ref{label1.11}) the two--point Green function $S_{\rm F}(x,y)$ is
determined by
\begin{eqnarray}\label{label8.1}
S_{\rm F}(x,y) =i\,\langle 0|{\rm T}(\psi(x)\bar{\psi}(y))|0\rangle =
i\,\frac{\delta}{\delta \bar{J}(x)}\,\frac{\delta}{\delta J(y)}Z_{\rm
Th}[J,\bar{J}]\Big|_{\textstyle J = \bar{J} = 0}.
\end{eqnarray}
Using the path--integral representation of the generating functional
(\ref{label1.23}) the two--point Green function $S_{\rm F}(x,y)$ is
given by 
\begin{eqnarray}\label{label8.2}
\hspace{-0.5in}&&S_{\rm F}(x,y) = i\,\langle 0|{\rm
T}(\psi(x)\bar{\psi}(y))|0\rangle = i\int {\cal D}\Psi{\cal D}
\bar{\Psi}{\cal D}\eta\,{\cal D}\xi\,\nonumber\\
\hspace{-0.5in}&&\times\,\exp\,i\int
d^2z\,\Big[\bar{\Psi}(z)i\gamma^{\mu}\partial_{\mu}\Psi(z) +
\frac{1}{2}\,\partial_{\mu}\eta(z)\partial^{\mu}\eta(z) -
\frac{1}{2}\,\partial_{\mu}\xi(z)\partial^{\mu} \xi(z)\Big]\nonumber\\
\hspace{-0.5in}&&\Big[\Big(\frac{1+\gamma^5}{2}\Big)
\Psi(x)e^{\textstyle i\sqrt{g}\,\eta(x) + i \lambda(\alpha_J)\, \xi(x) }
+ \Big(\frac{1-\gamma^5}{2}\Big)\Psi(x)e^{\textstyle i
\sqrt{g}\,\eta(x) - i \lambda(\alpha_J)\, \xi(x)}\Big]\nonumber\\
\hspace{-0.5in}&&\times\,
\Big[\bar{\Psi}(y)\Big(\frac{1+\gamma^5}{2}\Big)\,e^{\textstyle - i\sqrt{g}\,\eta(y) + i
\lambda(\alpha_J)\, \xi(y)} +
\bar{\Psi}(y)\Big(\frac{1-\gamma^5}{2}\Big)\,e^{\textstyle -
i\sqrt{g}\,\eta(y) - i \lambda(\alpha_J)\, \xi(y)}\Big].\nonumber\\
\hspace{-0.5in}&&
\end{eqnarray}
In a more convenient form the r.h.s. of (\ref{label8.2}) can be
written as
\begin{eqnarray}\label{label8.3}
\hspace{-0.5in}S_{\rm F}(x,y) &=&\int {\cal D}\eta\,{\cal
D}\xi\,\exp\,\Big\{\,i\int d^2z\,\Big[
\frac{1}{2}\,\partial_{\mu}\eta(z)\partial^{\mu}\eta(z) -
\frac{1}{2}\,\partial_{\mu}\xi(z)\partial^{\mu}
\xi(z)\Big]\Big\}\nonumber\\
\hspace{-0.5in}&&\hspace{0.1in}\times\Bigg\{i\Big\langle 0\Big|{\rm
T}\Big(\Big(\frac{1+\gamma^5}{2}\Big)\Psi(x)\bar{\Psi}(y)
\Big(\frac{1+\gamma^5}{2}\Big)\Big)\Big|0\Big\rangle \nonumber\\
\hspace{-0.5in}&&\hspace{0.3in}\times\, e^{\textstyle +
i\sqrt{g}\,\eta(x) + i \lambda(\alpha_J)\, \xi(x)- i\sqrt{g}\,\eta(y)
+ i \lambda(\alpha_J)\, \xi(y)}\nonumber\\
\hspace{-0.5in}&&\hspace{0.1in} + i\Big\langle 0\Big|{\rm
T}\Big(\Big(\frac{1-\gamma^5}{2}\Big)\Psi(x)\bar{\Psi}(y)
\Big(\frac{1-\gamma^5}{2}\Big)\Big)\Big|0\Big\rangle \nonumber\\
\hspace{-0.5in}&&\hspace{0.3in}\times\,e^{\textstyle +
i\sqrt{g}\,\eta(x) - i \lambda(\alpha_J)\, \xi(x)- i\sqrt{g}\,\eta(y)
- i \lambda(\alpha_J)\, \xi(y)}\nonumber\\
\hspace{-0.5in}&&\hspace{0.1in} + i\,\Big\langle 0\Big|{\rm
T}\Big(\Big(\frac{1+\gamma^5}{2}\Big)\Psi(x)\bar{\Psi}(y)
\Big(\frac{1-\gamma^5}{2}\Big)\Big)\Big|0\Big\rangle\nonumber\\
\hspace{-0.5in}&&\hspace{0.3in}\times\, e^{\textstyle +
i\sqrt{g}\,\eta(x) + i \lambda(\alpha_J)\, \xi(x)- i\sqrt{g}\,\eta(y)
- i \lambda(\alpha_J)\, \xi(y)}\nonumber\\
\hspace{-0.5in}&&\hspace{0.1in} + i\,\Big\langle 0\Big|{\rm
T}\Big(\Big(\frac{1-\gamma^5}{2}\Big)\Psi(x)\bar{\Psi}(y)
\Big(\frac{1+\gamma^5}{2}\Big)\Big)\Big|0\Big\rangle\nonumber\\
\hspace{-0.5in}&&\hspace{0.3in}\times\, e^{\textstyle +
i\sqrt{g}\,\eta(x) - i \lambda(\alpha_J)\, \xi(x)- i\sqrt{g}\,\eta(y)
+ i \lambda(\alpha_J)\, \xi(y)}\Bigg\}.
\end{eqnarray}
The Green function of the free massless fermionic field $\Psi(x)$ is
defined by (\ref{label1.10}). By using this expression the r.h.s. of
(\ref{label8.3}) can be recast into the form
\begin{eqnarray}\label{label8.4}
\hspace{-0.5in}&&S_{\rm F}(x,y) = \int {\cal D}\eta\,{\cal
D}\xi\,\exp\,\Big\{\,i\int d^2z\,\Big[
\frac{1}{2}\,\partial_{\mu}\eta(z)\partial^{\mu}\eta(z) -
\frac{1}{2}\,\partial_{\mu}\xi(z)\partial^{\mu}
\xi(z)\Big]\Big\}\nonumber\\
\hspace{-0.5in}&&\times\Bigg\{\Big(\frac{1+\gamma^5}{2}\Big)\frac{1}{2\pi}\,\frac{\hat{x}
- \hat{y}}{(x - y)^2 - i\,0}\,e^{\textstyle + i\sqrt{g}\,\eta(x) + i
\lambda(\alpha_J)\, \xi(x)- i\sqrt{g}\,\eta(y) - i \lambda(\alpha_J)\,
\xi(y)}\nonumber\\ \hspace{-0.5in}&&+
\Big(\frac{1-\gamma^5}{2}\Big)\frac{1}{2\pi}\,\frac{\hat{x} -
\hat{y}}{(x - y)^2 - i\,0}\,e^{\textstyle + i\sqrt{g}\,\eta(x) - i
\lambda(\alpha_J)\, \xi(x)- i\sqrt{g}\,\eta(y) + i \lambda(\alpha_J)\,
\xi(y)}\Bigg\}.
\end{eqnarray}
Since the integrands are identical up to the transformation $\xi \to
-\,\xi$, the two--point Green function is defined by
\begin{eqnarray}\label{label8.5}
\hspace{-0.5in}&&S_{\rm F}(x,y) = i\,\langle 0|{\rm
T}(\psi(x)\bar{\psi}(y))|0\rangle = \frac{1}{2\pi}\,\frac{\hat{x} -
\hat{y}}{(x - y)^2 - i\,0}\nonumber\\
\hspace{-0.5in}&&\times \int {\cal D}\eta\,{\cal
D}\xi\,\exp\,\Big\{\,i\int d^2z\,\Big[
\frac{1}{2}\,\partial_{\mu}\eta(z)\partial^{\mu}\eta(z) -
\frac{1}{2}\,\partial_{\mu}\xi(z)\partial^{\mu} \xi(z)\Big]\nonumber\\
\hspace{-0.5in}&& + i\sqrt{g}\,\eta(x) + i \lambda(\alpha_J)\, \xi(x)-
i\sqrt{g}\,\eta(y) - i \lambda(\alpha_J)\, \xi(y)\Big\}
\end{eqnarray}
The integration over $\eta$ and $\xi$ fields leads to
\begin{eqnarray}\label{label8.6}
&&\int {\cal D}\eta\,\exp\,\Big\{\,i\int d^2z\,
\frac{1}{2}\,\partial_{\mu}\eta(z)\partial^{\mu}\eta(z) +
i\sqrt{g}\,\eta(x) - i\sqrt{g}\,\eta(y) \Big\}\nonumber\\ && \to \int
{\cal D}\eta\,\exp\Big\{-\,i\int d^2z\,\frac{1}{2}\,\eta(z)(\Box +
\mu^2)\eta(z) + i\sqrt{g}\,\eta(x) - i\sqrt{g}\,\eta(y) \Big\}
=\nonumber\\ &&=\exp\Big\{\,g\,i\,\Delta(0) -
\,g\,i\,\Delta(x-y)\Big\} = \exp\Big\{-\frac{g}{4\pi}\,{\ell
n}\frac{\Lambda^2}{\mu^2} - \frac{g}{4\pi}\,{\ell n}[-\mu^2(x-y)^2 +
i\,0\,]\Big\}=\nonumber\\ &&= \exp\Big\{- \frac{g}{4\pi}\,{\ell
n}[-\Lambda^2(x-y)^2 + i\,0\,]\Big\},\nonumber\\ &&\int {\cal
D}\xi\,\exp\,\Big\{-\,i\int d^2z\,
\frac{1}{2}\,\partial_{\mu}\xi(z)\partial^{\mu}\xi(z) +
i\,\lambda(\alpha_J)\,\xi(x) -
i\,\lambda(\alpha_J)\,\xi(y)\Big\}\nonumber\\ && \to \int {\cal
D}\xi\,\exp\,\Big\{\,i\int d^2z\,\frac{1}{2}\,\xi(z)(\Box +
\mu^2)\xi(z) + i\,\lambda(\alpha_J)\,\xi(x) -
i\,\lambda(\alpha_J)\,\xi(y) \Big\} =\nonumber\\
&&=\exp\Big\{-\,\lambda^2(\alpha_J)\,i\,\Delta(0)
+\,\lambda^2(\alpha_J)\,i\,\Delta(x-y)\Big\} =\nonumber\\
&&=\exp\Big\{ \frac{\lambda^2(\alpha_J)}{4\pi}\,{\ell
n}[-\Lambda^2(x-y)^2 + i\,0]\Big\}.
\end{eqnarray}
Thus, the two--point Green function of the Thirring fermionic field is
equal to
\begin{eqnarray}\label{label8.7}
\hspace{-0.5in}&&S_{\rm F}(x,y) = i\,\langle 0|{\rm
T}(\psi(x)\bar{\psi}(y))|0\rangle = \frac{1}{2\pi}\,\frac{\hat{x} -
\hat{y}}{(x - y)^2 - i\,0}\nonumber\\
\hspace{-0.5in}&&\times\,\exp\Big\{
\frac{\lambda^2(\alpha_J)}{4\pi}\,{\ell n}[-\Lambda^2(x-y)^2 + i\,0] -
\frac{g}{4\pi}\,{\ell n}[-\Lambda^2(x-y)^2 + i\,0\,]\Big\}=\nonumber\\
\hspace{-0.5in}&&= \frac{1}{2\pi}\,\frac{\hat{x} -
\hat{y}}{\displaystyle (x - y)^2 - i\,0}\, [-\Lambda^2(x - y)^2 +
i\,0\,]^{\textstyle -(g-\lambda^2(\alpha_J))/4\pi} = \nonumber\\
\hspace{-0.5in}&&= \frac{1}{2\pi}\,\frac{\hat{x} -
\hat{y}}{\displaystyle (x - y)^2 - i\,0}\,[-\Lambda^2(x - y)^2 +
i\,0\,]^{\textstyle - d_{\psi}(g,\alpha_J)},
\end{eqnarray}
where $d_{\psi}(g,\alpha_J)$, the dynamical dimension of the massless
Thirring fermionic field [21]
\begin{eqnarray}\label{label8.8}
d_{\psi}(g,\alpha_J) = \frac{g -\lambda^2(\alpha_J)}{4\pi} =
\frac{g^2}{4\pi^2}\frac{\alpha_J}{\displaystyle 1 +
\alpha_J\,\frac{g}{\pi}}.
\end{eqnarray}
It depends on $\alpha_J$ as well as the dynamical dimension of the
left(right)handed fermion densities (\ref{label6.23}).

The two--point Green function (\ref{label8.7}) describes the Green
function of massless fermion fields quantized in the chiral symmetric
phase. We emphasize that the expression (\ref{label8.7}) does not
depend on the infrared cut--off.

\section{$2n$--point Green functions of fermion fields. 
Path--integral approach and Klaiber's operator formalism}
\setcounter{equation}{0}

\hspace{0.2in} In the massless Thirring model there are two
irreducible $2n$--point Green functions. They read
\begin{eqnarray}\label{label9.1}
G^{(1)}(x_1,\ldots,x_n;y_1,\ldots,y_n) &=&\langle 0|{\rm
T}(\psi_1(x_1)\ldots\psi_1(x_n)\psi^{\dagger}_1(y_1)
\ldots\psi^{\dagger}_1(y_n))|0\rangle,\nonumber\\
G^{(2)}(x_1,\ldots,x_n;y_1,\ldots,y_n) &=&\langle 0|{\rm
T}(\psi_2(x_1)\ldots\psi_2(x_n)\psi^{\dagger}_2(y_1)
\ldots\psi^{\dagger}_2(y_n))|0\rangle.
\end{eqnarray}
For the evaluation of these Green functions we apply the generating
functional (\ref{label1.11}) given by
\begin{eqnarray}\label{label9.2}
Z_{\rm Th}[J,\bar{J}] &=& \int {\cal D}\psi_1{\cal
D}\psi^{\dagger}_1{\cal D}\psi_2{\cal D}\psi^{\dagger}_2\nonumber\\
&\times&\exp\,i\int
d^2x\,\Big[\bar{\psi}(x)i\gamma^{\mu}\partial_{\mu}\psi(x) -
\frac{1}{2}\,g\,\bar{\psi}(x)\gamma^{\mu}\psi(x)\bar{\psi}(x)
\gamma_{\mu}\psi(x)\nonumber\\ &+& \psi^{\dagger}_1(x)J_2(x) +
\psi^{\dagger}_2(x)J_1(x) + J^{\dagger}_2(x)\psi_1(x) +
J^{\dagger}_1(x)\psi_2(x)\Big],
\end{eqnarray}
After the set of transformations given by (\ref{label1.12}) --
(\ref{label1.20}) the generating functional (\ref{label9.2}) acquires
the form
\begin{eqnarray}\label{label9.3}
\hspace{-0.5in}&&Z_{\rm Th}[J,\bar{J}] = \int {\cal D}\Psi_1{\cal
D}\Psi^{\dagger}_1{\cal D}\Psi_2{\cal D}\Psi^{\dagger}_2{\cal
D}\eta\,{\cal D}\xi\,\nonumber\\
\hspace{-0.5in}&&\times \exp\,i\int
d^2x\,\Big\{\bar{\Psi}(x)i\gamma^{\mu}\partial_{\mu}\Psi(x) +
\frac{1}{2}\,\partial_{\mu}\eta(x)\partial^{\mu}\eta(x) -
\frac{1}{2}\,\partial_{\mu}\xi(x)\partial^{\mu}
\xi(x)\nonumber\\
\hspace{-0.5in}&& + \Psi^{\dagger}_1(x)J_2(x)\,
e^{\textstyle - i\sqrt{g}\,\eta(x) - i \lambda(\alpha_J)\,\xi(x)} +\Psi^{\dagger}_2(x)J_1(x)\,e^{\textstyle - i\sqrt{g}\,\eta(x) + i
\lambda(\alpha_J)\, \xi(x)}\nonumber\\
\hspace{-0.5in}&& + J^{\dagger}_2(x)\Psi_1(x)\, e^{\textstyle
i\sqrt{g}\,\eta(x) + i \lambda(\alpha_J)\, \xi(x)} +
J^{\dagger}_1(x)\Psi_2(x)\,e^{\textstyle i \sqrt{g}\,\eta(x) - i
\lambda(\alpha_J)\, \xi(x)}\Big\}.
\end{eqnarray}
In terms of the generating functional $Z_{\rm Th}[J,\bar{J}]$ the
Green functions (\ref{label9.1}) are defined by
\begin{eqnarray}\label{label9.4}
\hspace{-0.5in}&&G^{(1)}(x_1,\ldots,x_n;y_1,\ldots,y_n) =\langle 0|{\rm
T}(\psi_1(x_1)\ldots\psi_1(x_n)\psi^{\dagger}_1(y_1)
\ldots\psi^{\dagger}_1(y_n))|0\rangle =\nonumber\\
\hspace{-0.5in}&&=\frac{\delta}{\delta
J^{\dagger}_2(x_1)}\ldots\frac{\delta}{\delta
J^{\dagger}_2(x_n)}\,\frac{\delta}{\delta
J_2(y_1)}\,\ldots\,\frac{\delta}{\delta J_2(y_n)}Z_{\rm
Th}[J,\bar{J}]\Big|_{\textstyle J_1 = J_2 = J^{\dagger}_1 =
J^{\dagger}_2 = 0},\nonumber\\
\hspace{-0.5in}&&G^{(2)}(x_1,\ldots,x_n;y_1,\ldots,y_n) =\langle 0|{\rm
T}(\psi_2(x_1)\,\ldots\,\psi_2(x_n)\psi^{\dagger}_2(y_1)
\ldots\psi^{\dagger}_2(y_n))|0\rangle =\nonumber\\
\hspace{-0.5in}&&=\frac{\delta}{\delta
J^{\dagger}_1(x_1)}\,\ldots\,\frac{\delta}{\delta
J^{\dagger}_1(x_n)}\,\frac{\delta}{\delta
J_1(y_1)}\,\ldots\,\frac{\delta}{\delta J_1(y_n)}Z_{\rm
Th}[J,\bar{J}]\Big|_{\textstyle J_1 = J_2 = J^{\dagger}_1 =
J^{\dagger}_2 = 0}.
\end{eqnarray}
Substituting (\ref{label9.3}) in (\ref{label9.4}) we get
\begin{eqnarray}\label{label9.5}
\hspace{-0.5in}&&G^{(1)}(x_1,\ldots,x_n;y_1,\ldots,y_n) =\langle 0|{\rm
T}(\Psi_1(x_1)\ldots\Psi_1(x_n)\Psi^{\dagger}_1(y_1)
\ldots\Psi^{\dagger}_1(y_n))|0\rangle \nonumber\\
\hspace{-0.5in}&&\times \int {\cal D}\eta\,{\cal
D}\xi\,\exp\,i\int d^2x\,\Big\{
\frac{1}{2}\,\partial_{\mu}\eta(x)\partial^{\mu}\eta(x) -
\frac{1}{2}\,\partial_{\mu}\xi(x)\partial^{\mu} \xi(x)\Big\}\nonumber\\
\hspace{-0.5in}&&\times\,\prod^n_{j=1}
e^{\textstyle i\sqrt{g}\,\eta(x_j) +
i\lambda(\alpha_J)\,\xi(x_j)}\,e^{\textstyle -i\sqrt{g}\,\eta(y_j) -
i\lambda(\alpha_J)\,\xi(y_j)},\nonumber\\
\hspace{-0.5in}&&G^{(2)}(x_1,\ldots,x_n;y_1,\ldots,y_n) =\langle 0|{\rm
T}(\Psi_2(x_1)\ldots\Psi_2(x_n)\Psi^{\dagger}_2(y_1)
\ldots\Psi^{\dagger}_2(y_n))|0\rangle \nonumber\\
\hspace{-0.5in}&&\times \int {\cal D}\eta\,{\cal
D}\xi\,\exp\,i\int d^2x\,\Big\{
\frac{1}{2}\,\partial_{\mu}\eta(x)\partial^{\mu}\eta(x) -
\frac{1}{2}\,\partial_{\mu}\xi(x)\partial^{\mu} \xi(x)\Big\}\nonumber\\
\hspace{-0.5in}&&\times\,\prod^n_{j=1}
e^{\textstyle i\sqrt{g}\,\eta(x_j) -
i\lambda(\alpha_J)\,\xi(x_j)}\,e^{\textstyle -i\sqrt{g}\,\eta(y_j) +
i\lambda(\alpha_J)\,\xi(y_j)}
\end{eqnarray}
Integration over $\eta$ and $\xi$ fields gives
\begin{eqnarray}\label{label9.6}
\hspace{-0.5in}&&G^{(1)}(x_1,\ldots,x_n;y_1,\ldots,y_n) =
\langle 0|{\rm
T}(\Psi_1(x_1)\ldots\Psi_1(x_n)\Psi^{\dagger}_1(y_1)
\ldots\Psi^{\dagger}_1(y_n))|0\rangle \nonumber\\
\hspace{-0.5in}&&\times \frac{\displaystyle
\prod^{n}_{j <k}[-\Lambda^2(x_j - x_k)^2 + i\,0\,]^{\textstyle
(g-\lambda^2(\alpha_J))/4\pi}\prod^{n}_{j <k}[-\Lambda^2(y_j -
y_k)^2 + i\,0\,]^{\textstyle (g-\lambda^2(\alpha_J))/4\pi}}
{\displaystyle\prod^{n}_{j =
1}\prod^{n}_{k = 1}[-\Lambda^2(x_j - y_k)^2 + i\,0\,]^{\textstyle
(g - \lambda^2(\alpha_J))/4\pi}},\nonumber\\
\hspace{-0.5in}&&G^{(2)}(x_1,\ldots,x_n;y_1,\ldots,y_n) =\langle 0|{\rm
T}(\Psi_2(x_1)\ldots\Psi_2(x_n)\Psi^{\dagger}_2(y_1)
\ldots\Psi^{\dagger}_2(y_n))|0\rangle \nonumber\\
\hspace{-0.5in}&&\times \frac{\displaystyle \prod^{n}_{j
<k}[-\Lambda^2(x_j - x_k)^2 + i\,0\,]^{\textstyle (g -
\lambda^2(\alpha_J))/4\pi}\prod^{n}_{j <k}[-\Lambda^2(y_j -
y_k)^2 + i\,0\,]^{\textstyle
(g-\lambda^2(\alpha_J))/4\pi}}{\displaystyle\prod^{n}_{j =
1}\prod^{n}_{k = 1}[-\Lambda^2(x_j - y_k)^2 + i\,0\,]^{\textstyle
(g-\lambda^2(\alpha_J))/4\pi}}.
\end{eqnarray}
In order to compare the expressions (\ref{label9.8}) with Klaiber's
result [4] we suggest to represent them in the form
\begin{eqnarray}\label{label9.7}
\hspace{-0.5in}&&G^{(1)}(x_1,\ldots,x_n;y_1,\ldots,y_n) =\langle 0|{\rm
T}(\Psi_1(x_1)\ldots\Psi_1(x_n)\Psi^{\dagger}_1(y_1)
\ldots\Psi^{\dagger}_1(y_n))|0\rangle\,e^{\textstyle
iF(x,y)},\nonumber\\
\hspace{-0.5in}&&G^{(2)}(x_1,\ldots,x_n;y_1,\ldots,y_n) =\langle
0|{\rm T}(\Psi_2(x_1)\ldots\Psi_2(x_n)\Psi^{\dagger}_2(y_1)
\ldots\Psi^{\dagger}_2(y_n))|0\rangle\,e^{\textstyle iF(x,y)},
\end{eqnarray}
where the exponential $e^{\textstyle iF(x,y)}$ is determined by
\begin{eqnarray}\label{label9.8}
\hspace{-0.5in}&&e^{\textstyle iF(x,y)} = e^{\textstyle
-i\,n\,\frac{\textstyle g - \lambda^2(\alpha_J)}{\textstyle
4}}\,\Bigg(\frac{1}{\Lambda}\Bigg)^{\textstyle n\,
\frac{\textstyle g - \lambda^2(\alpha_J)}{\textstyle 2\pi}}\nonumber\\
\hspace{-0.5in}&&\times\prod^n_{j<k} (v_j - v_k
-i\,0\cdot\varepsilon(x^0_j - x^0_k))^{\textstyle \frac{\textstyle g -
\lambda^2(\alpha_J)}{\textstyle 4\pi}} (u_j - u_k -i\,0\cdot\varepsilon(x^0_j - x^0_k))^{\textstyle
\frac{\textstyle g }{\textstyle 4\pi}}\nonumber\\
\hspace{-0.5in}&&\times \prod^n_{j<k}(V_j - V_k
-i\,0\cdot\varepsilon(y^0_j - y^0_k))^{\textstyle \frac{\textstyle g -
\lambda^2(\alpha_J)}{\textstyle 4\pi}} (U_j - U_k
-i\,0\cdot\varepsilon(y^0_j - y^0_k))^{\textstyle \frac{\textstyle g -
\lambda^2(\alpha_J)}{\textstyle 4\pi}}\nonumber\\ \hspace{-0.5in}
&&\times\prod^n_{j = 1}\prod^n_{k = 1}\Bigg(\frac{1}{v_j - V_k -
i\,0\cdot\varepsilon(x^0_j - y^0_k)}\Bigg)^{\textstyle
\frac{\textstyle g - \lambda^2(\alpha_J)}{\textstyle
4\pi}}\Bigg(\frac{1}{u_j - U_k - i\,0\cdot\varepsilon(x^0_j -
y^0_k)}\Bigg)^{\textstyle \frac{\textstyle g - \lambda^2(\alpha_J)
}{\textstyle 4\pi}}.\nonumber\\
\hspace{-0.5in}&&
\end{eqnarray}
In Klaiber's approach this contribution reads (see Eq.(VII.2) of
Ref.[4])
\begin{eqnarray}\label{label9.9}
\hspace{-0.5in}&&e^{\textstyle iF(x,y)_{\rm K}} = e^{\textstyle
-i\,n\,\frac{\textstyle a+b }{\textstyle
4}}\,\Bigg(\frac{1}{\mu}\Bigg)^{\textstyle n\,\frac{\textstyle a+b
}{\textstyle 2\pi}}\nonumber\\
\hspace{-0.5in}&&\times \prod^n_{j<k}(v_j - v_k
-i\,0\cdot\varepsilon(x^0_j - x^0_k))^{\textstyle \frac{\textstyle a+b -
2\lambda }{\textstyle 4\pi}}(u_j - u_k -i\,0\cdot\varepsilon(x^0_j - x^0_k))^{\textstyle
\frac{\textstyle a+b+2\lambda }{\textstyle 4\pi}}\nonumber\\
\hspace{-0.5in}&&\times \prod^n_{j<k}(V_j - V_k
-i\,0\cdot\varepsilon(y^0_j - y^0_k))^{\textstyle \frac{\textstyle
a+b-2\lambda }{\textstyle 4\pi}}(U_j - U_k -i\,0\cdot\varepsilon(y^0_j
- y^0_k))^{\textstyle \frac{\textstyle a+b+2\lambda }{\textstyle
4\pi}}\nonumber\\ \hspace{-0.5in}&&\times\prod^n_{j = 1}\prod^n_{k =
1}\Bigg(\frac{1}{v_j - V_k - i\,0\cdot\varepsilon(x^0_j -
y^0_k)}\Bigg)^{\textstyle \frac{\textstyle a+b -2\lambda }{\textstyle
4\pi}}\Bigg(\frac{1}{u_j - U_k - i\,0\cdot\varepsilon(x^0_j -
y^0_k)}\Bigg)^{\textstyle \frac{\textstyle a+b+2\lambda }{\textstyle
4\pi}},\nonumber\\ 
\hspace{-0.5in}&
\end{eqnarray}
where $\mu$ is an infrared cut--off and the parameters $a$, $b$ and
$\lambda$ are defined by (see Eq.(IV.25) of Ref.[4])
\begin{eqnarray}\label{label9.10}
a &=&\alpha^2 +2\,\sqrt{\pi}\,\alpha,\nonumber\\ b &=&\beta^2
+2\,\sqrt{\pi}\,\beta,\nonumber\\ \lambda &=& \alpha\,\beta +
\sqrt{\pi}\,\alpha +\sqrt{\pi}\,\beta.
\end{eqnarray}
Invariance of the correlation function (\ref{label9.10}) under Lorentz
transformations imposes the constraint $\lambda = 0$ [4] (see also the
discussion in section 5). This reduces the correlation function
(\ref{label9.9}) to the form
\begin{eqnarray}\label{label9.11}
\hspace{-0.5in}&&e^{\textstyle iF(x,y)_{\rm K}} = e^{\textstyle
-i\,n\,\frac{\textstyle a+b }{\textstyle
4}}\,\Bigg(\frac{1}{\mu}\Bigg)^{\textstyle n\,\frac{\textstyle a+b
}{\textstyle 2\pi}}\nonumber\\
\hspace{-0.5in}&&\times \prod^n_{j<k}(v_j - v_k
-i\,0\cdot\varepsilon(x^0_j - x^0_k))^{\textstyle \frac{\textstyle a+b
}{\textstyle 4\pi}}(u_j - u_k -i\,0\cdot\varepsilon(x^0_j -
x^0_k))^{\textstyle \frac{\textstyle a+b }{\textstyle
4\pi}}\nonumber\\
\hspace{-0.5in}&&\times \prod^n_{j<k}(V_j - V_k
-i\,0\cdot\varepsilon(y^0_j - y^0_k))^{\textstyle \frac{\textstyle a+b
}{\textstyle 4\pi}}(U_j - U_k -i\,0\cdot\varepsilon(y^0_j -
y^0_k))^{\textstyle \frac{\textstyle a+b }{\textstyle
4\pi}}\nonumber\\
\hspace{-0.5in}&&\times \prod^n_{j = 1}\prod^n_{k =
1}\Bigg(\frac{1}{v_j - V_k - i\,0\cdot\varepsilon(x^0_j -
y^0_k)}\Bigg)^{\textstyle \frac{\textstyle a+b }{\textstyle
4\pi}}\Bigg(\frac{1}{u_j - U_k - i\,0\cdot\varepsilon(x^0_j -
y^0_k)}\Bigg)^{\textstyle \frac{\textstyle a+b }{\textstyle 4\pi}},
\nonumber\\
\hspace{-0.5in}&&
\end{eqnarray}
where $a+b = (\alpha - \beta)^2$ valid at $\lambda = 0$. For further
discussion it is convenient to define the exponential (\ref{label9.11})
relative to the ultra--violate cut--off
\begin{eqnarray}\label{label9.12}
\hspace{-0.5in}&&e^{\textstyle iF(x,y)_{\rm K}} = e^{\textstyle
-i\,n\,\frac{\textstyle a+b }{\textstyle
4}}\,\Bigg(\frac{\Lambda}{\mu}\Bigg)^{\textstyle n\,\frac{\textstyle a+b
}{\textstyle 2\pi}}\nonumber\\
\hspace{-0.5in}&&\times \prod^n_{j<k}[\Lambda(v_j - v_k
-i\,0\cdot\varepsilon(x^0_j -
x^0_k))]^{\textstyle \frac{\textstyle a+b }{\textstyle
4\pi}}[\Lambda(u_j - u_k -i\,0\cdot\varepsilon(x^0_j -
x^0_k))]^{\textstyle \frac{\textstyle a+b
}{\textstyle 4\pi}}\nonumber\\
\hspace{-0.5in}&&\times \prod^n_{j<k}[\Lambda(V_j - V_k
-i\,0\cdot\varepsilon(y^0_j - y^0_k))]^{\textstyle \frac{\textstyle
a+b }{\textstyle 4\pi}}[\Lambda(U_j - U_k -i\,0\cdot\varepsilon(y^0_j
- y^0_k))]^{\textstyle \frac{\textstyle a+b }{\textstyle
4\pi}}\nonumber\\ \hspace{-0.5in}&&\times \prod^n_{j = 1}\prod^n_{k =
1}\Bigg(\frac{1}{[\Lambda(v_j - V_k - i\,0\cdot\varepsilon(x^0_j -
y^0_k))]}\Bigg)^{\textstyle \frac{\textstyle a+b }{\textstyle
4\pi}}\Bigg(\frac{1}{[\Lambda(u_j - U_k - i\,0\cdot\varepsilon(x^0_j -
y^0_k))]}\Bigg)^{\textstyle \frac{\textstyle a+b }{\textstyle
4\pi}}.\nonumber\\ \hspace{-0.5in}&&
\end{eqnarray}
It is seen that Klaiber's correlation function is singular in the
infrared limit $\mu \to 0$.

The dynamical dimension of the massless fermionic field in terms of
Klaiber's parameters is determined by
\begin{eqnarray}\label{label9.13}
d_{\textstyle \psi}(g,a,b) = \frac{a+b}{4\pi}.
\end{eqnarray}
Vacuum expectation values of the time--ordered products of the fields
$\Psi_1$ and $\Psi_2$ can be evaluated by means of the generating
functional $Z_{\rm F}[J,\bar{J}]$ (\ref{label6.6}). The results read
\begin{eqnarray}\label{label9.14}
&&\langle 0|{\rm T}(\Psi_1(x_1)\ldots\Psi_1(x_n)\Psi^{\dagger}_1(y_1)
\ldots\Psi^{\dagger}_1(y_n))|0\rangle =\nonumber\\ &&= \frac{1}{(2\pi
i)^n}\,\frac{\displaystyle \prod^n_{j<k}(u_j - u_k -i\,0\cdot
\varepsilon(x^0_j - x^0_k))\prod^n_{j<k}(U_j - U_k -i\,0\cdot
\varepsilon(y^0_j - y^0_k))}{\prod^n_{j = 1} \prod^n_{k=1}(u_j - U_k -
i\,0\cdot \varepsilon(x^0_j - y^0_k))},\nonumber\\ &&\langle 0|{\rm
T}(\Psi_2(x_1)\ldots\Psi_2(x_n) \Psi^{\dagger}_2(y_1)
\ldots\Psi^{\dagger}_2(y_n))|0\rangle =\nonumber\\ &&=\frac{1}{(2\pi
i)^n}\,\frac{\displaystyle \prod^n_{j<k}(v_j - v_k -i\,0\cdot
\varepsilon(x^0_j - x^0_k))\prod^n_{j<k}(V_j - V_k -i\,0\cdot
\varepsilon(y^0_j - y^0_k))}{\prod^n_{j = 1} \prod^n_{k=1}(v_j - V_k -
i\,0\cdot \varepsilon(x^0_j - y^0_k))}.
\end{eqnarray}
Recall that $u = x^0 + x^1$, $v = x^0 - x^1$, $U = y^0 + y^1$ and 
$V =
y^0 - y^1$ (\ref{label6.7}).

Thus the $2n$--point Green functions (\ref{label9.1}) are equal to
\begin{eqnarray}\label{label9.15}
\hspace{-0.5in}&&G^{(1)}(x_1,\ldots,x_n;y_1,\ldots,y_n) =\langle 0|{\rm
T}(\psi_1(x_1)\ldots\psi_1(x_n)\psi^{\dagger}_1(y_1)
\ldots\psi^{\dagger}_1(y_n))|0\rangle =\nonumber\\
\hspace{-0.5in}&&\times\,(-1)^{\textstyle n\,d_{\textstyle
\psi}(g,\alpha_J)}\Bigg(\frac{\Lambda}{2\pi i}\Bigg)^n \prod^n_{j<k}
[\Lambda(v_j - v_k -i\,0\cdot \varepsilon(x^0_j - x^0_k))]^{\textstyle
d_{\textstyle \psi}(g,\alpha_J)}\nonumber\\
\hspace{-0.5in}&&\times\, [\Lambda(u_j - u_k -i\,0\cdot
\varepsilon(x^0_j - x^0_k))]^{\textstyle 1 + d_{\textstyle
\psi}(g,\alpha_J)}\nonumber\\
\hspace{-0.5in}&&\times \prod^n_{j<k}[\Lambda(V_j - V_k -i\,0\cdot
\varepsilon(y^0_j - y^0_k))]^{\textstyle d_{\textstyle
\psi}(g,\alpha_J) } [\Lambda (U_j - U_k -i\,0\cdot \varepsilon(y^0_j -
y^0_k))]^{\textstyle 1 + d_{\textstyle \psi}(g,\alpha_J)}\nonumber\\
\hspace{-0.5in} &&\times\prod^n_{j = 1}\prod^n_{k =
1}\Bigg(\frac{1}{[\Lambda(v_j - V_k - i\,0\cdot \varepsilon(x^0_j -
y^0_k))]}\Bigg)^{\textstyle d_{\textstyle
\psi}(g,\alpha_J)}\nonumber\\
\hspace{-0.5in}&&\times\,\Bigg(\frac{1}{[\Lambda(u_j - U_k - i\,0\cdot
\varepsilon(x^0_j - y^0_k))]}\Bigg)^{\textstyle 1 + d_{\textstyle
\psi}(g,\alpha_J)},\nonumber\\
\hspace{-0.5in}&&G^{(2)}(x_1,\ldots,x_n;y_1,\ldots,y_n) = \langle
0|{\rm T}(\psi_2(x_1)\ldots\psi_2(x_n)\psi^{\dagger}_2(y_1)
\ldots\psi^{\dagger}_2(y_n))|0\rangle=\nonumber\\
\hspace{-0.5in}&&\times\,(-1)^{\textstyle n\,d_{\textstyle
\psi}(g,\alpha_J)}\Bigg(\frac{\Lambda}{2\pi i}\Bigg)^n\prod^n_{j<k}
[\Lambda(v_j - v_k -i\,0\cdot \varepsilon(x^0_j - x^0_k))]^{\textstyle
1+ d_{\textstyle \psi}(g,\alpha_J)}\nonumber\\
\hspace{-0.5in}&&\times\, [\Lambda(u_j - u_k -i\,0\cdot
\varepsilon(x^0_j - x^0_k))]^{\textstyle d_{\textstyle
\psi}(g,\alpha_J)}\nonumber\\
\hspace{-0.5in}&&\times \prod^n_{j<k}[\Lambda(V_j - V_k -i\,0\cdot
\varepsilon(y^0_j - y^0_k))]^{\textstyle 1 + d_{\textstyle
\psi}(g,\alpha_J) } [\Lambda (U_j - U_k -i\,0\cdot \varepsilon(y^0_j -
y^0_k))]^{\textstyle d_{\textstyle \psi}(g,\alpha_J)}\nonumber\\
\hspace{-0.5in} &&\times\prod^n_{j = 1}\prod^n_{k =
1}\Bigg(\frac{1}{[\Lambda(v_j - V_k - i\,0\cdot \varepsilon(x^0_j -
y^0_k))]}\Bigg)^{\textstyle 1 + d_{\textstyle
\psi}(g,\alpha_J)}\nonumber\\
\hspace{-0.5in}&&\times\,\Bigg(\frac{1}{[\Lambda(u_j - U_k - i\,0\cdot
\varepsilon(x^0_j - y^0_k))]}\Bigg)^{\textstyle d_{\textstyle
\psi}(g,\alpha_J)},
\end{eqnarray}
where $d_{\textstyle \psi}(g,\alpha_J)$ is the dynamical dimension of
the massless Thirring fermionic field defined by (\ref{label8.8}).

We can reduce the number of bosonic degrees of freedom defining
$2n$--point Green function. Making $O(1,1)$ rotations in the
$(\eta,\xi)$ functional space
\begin{eqnarray}\label{label9.16}
\Theta(x) = \frac{\sqrt{g}\,\eta(x) +
\lambda(\alpha_J)\,\xi(x)}{\sqrt{g - \lambda^2(\alpha_J)}}&,&
\Phi(x) = \frac{\lambda(\alpha_J)\,\eta(x) +
\sqrt{g}\,\xi(x)}{\sqrt{g - \lambda^2(\alpha_J)}},\nonumber\\
\Theta(x) = \frac{\sqrt{g}\,\eta(x) -
\lambda(\alpha_J)\,\xi(x)}{\sqrt{g - \lambda^2(\alpha_J)}}&,&
\Phi(x) = \frac{\lambda(\alpha_J)\,\eta(x) -
\sqrt{g}\,\xi(x)}{\sqrt{g - \lambda^2(\alpha_J)}}
\end{eqnarray}
for $G^{(1)}(x_1,\ldots,x_n;y_1,\ldots,y_n)$ and
$G^{(2)}(x_1,\ldots,x_n;y_1,\ldots,y_n)$, respectively, we arrive at
the expressions
\begin{eqnarray}\label{label9.17}
\hspace{-0.5in}&&G^{(1)}(x_1,\ldots,x_n;y_1,\ldots,y_n) =\langle 0|{\rm
T}(\Psi_1(x_1)\ldots\Psi_1(x_n)\Psi^{\dagger}_1(y_1)
\ldots\Psi^{\dagger}_1(y_n))|0\rangle \nonumber\\
\hspace{-0.5in}&&\times \int {\cal D}\Theta \,{\cal D}\Phi\,
e^{\textstyle\,i\int d^2x\,
\Big[\frac{1}{2}\,\partial_{\mu}\Theta(x)\partial^{\mu} \Theta(x) -
\frac{1}{2}\,\partial_{\mu}\Phi(x)\partial^{\mu}
\Phi(x)\Big]}\nonumber\\
\hspace{-0.5in}&&\times\,\prod^n_{j=1} 
e^{\textstyle \,i\sqrt{4\pi\,d_{\textstyle \psi}(g,\alpha_J)}\,
\Theta(x_j)}\,\prod^n_{j=1} e^{\textstyle
-\,i\sqrt{4\pi\,d_{\textstyle \psi}(g,\alpha_J)}\, 
\Theta(y_j)},\nonumber\\
\hspace{-0.5in}&&G^{(2)}(x_1,\ldots,x_n;y_1,\ldots,y_n) =
\langle 0|{\rm
T}(\Psi_2(x_1)\ldots\Psi_2(x_n)\Psi^{\dagger}_2(y_1)
\ldots\Psi^{\dagger}_2(y_n))|0\rangle \nonumber\\
\hspace{-0.5in}&&\times \int {\cal D}\Theta\,{\cal
D}\Phi\,e^{\textstyle \,i\int
d^2x\,\Big[\frac{1}{2}\,\partial_{\mu}\Theta(x)\partial^{\mu}
\Theta(x) - \frac{1}{2}\,\partial_{\mu}\Phi(x)\partial^{\mu}
\Phi(x)\Big] }\nonumber\\
\hspace{-0.5in}&&\times\,\prod^n_{j=1} e^{\textstyle
\,i\sqrt{4\pi\,d_{\textstyle
\psi}(g,\alpha_J)}\,\Theta(x_j)}\,\prod^n_{j=1} e^{\textstyle
-\,i\sqrt{4\pi\,d_{\textstyle \psi}(g,\alpha_J)}\, \Theta(y_j)}.
\end{eqnarray}
The $\Phi$--field is decoupled from the system and can be integrated
out. This yields
\begin{eqnarray}\label{label9.18}
\hspace{-0.5in}&&G^{(1)}(x_1,\ldots,x_n;y_1,\ldots,y_n) =\langle
0|{\rm T}(\Psi_1(x_1)\ldots\Psi_1(x_n)\Psi^{\dagger}_1(y_1)
\ldots\Psi^{\dagger}_1(y_n))|0\rangle \nonumber\\
\hspace{-0.5in}&&\times \int {\cal D}\Theta\,e^{\textstyle\,i\int
d^2x\, \frac{1}{2}\,\partial_{\mu}\Theta(x)
\partial^{\mu}\Theta(x)}\,\prod^n_{j=1} e^{\textstyle \,i\sqrt{4\pi\,d_{\textstyle \psi}(g,\alpha_J)}\,[\Theta(x_j) - \Theta(y_j)]},\nonumber\\
\hspace{-0.5in}&&G^{(2)}(x_1,\ldots,x_n;y_1,\ldots,y_n) =\langle 0|{\rm
T}(\Psi_2(x_1)\ldots\Psi_2(x_n)\Psi^{\dagger}_2(y_1)
\ldots\Psi^{\dagger}_2(y_n))|0\rangle \nonumber\\
\hspace{-0.5in}&&\times \int {\cal D}\Theta\,e^{\textstyle \,i\int
d^2x\, \frac{1}{2}\,\partial_{\mu}\Theta(x)\partial^{\mu}\Theta(x)}
\,\prod^n_{j=1} e^{\textstyle \,i\sqrt{4\pi\,d_{\textstyle
\psi}(g,\alpha_J)}\,[\Theta(x_j) -\Theta(y_j)]}.
\end{eqnarray}
The integration over the $\Theta$--field returns us to the expressions
(\ref{label9.6}).

Thus, the $2n$--point Green function evaluated within the
path--integral approach agree with Klaiber's expressions up to an
infrared renormalization removing infrared singularities of Klaiber's
expression in the limit $\mu \to 0$. However, as we will show in
section 11, this agreement is only superficial. The correlation
functions evaluated in Klaiber's approach do not admit
non--perturbative ultra--violet renormalization, whereas the
correlation functions obtained within the path--integral approach do.

\section{\hspace{-0.1in}Correlation functions of left(right)handed 
fermion densities in Klaiber's operator formalism} 
\setcounter{equation}{0}

\hspace{0.2in} In this section we analyse correlation functions of
left(right)handed fermion densities within Klaiber's operator
approach. In order to understand the peculiarities of correlation
functions of left(right)handed fermion densities of Klaiber's approach
it is sufficient to deal with the two--point correlation function
\begin{eqnarray}\label{label10.1}
\hspace{-0.5in}&&\Big\langle 0\Big|{\rm
T}\Big(\bar{\psi}(x)\Big(\frac{1-\gamma^5}{2}\Big)
\psi(x)\,\bar{\psi}(y)\Big(\frac{1+\gamma^5}{2}\Big)
\psi(y)\Big)\Big|0\Big\rangle.
\end{eqnarray}
Then, it is convenient to treat the correlation function
(\ref{label10.1}) as a limit of four--point correlation function
\begin{eqnarray}\label{label10.2}
\hspace{-0.3in}&&\Big\langle 0\Big|{\rm
T}\Big(\bar{\psi}(x)\Big(\frac{1-\gamma^5}{2}\Big)
\psi(x)\,\bar{\psi}(y)\Big(\frac{1+\gamma^5}{2}\Big)
\psi(y)\Big)\Big|0\Big\rangle 
\end{eqnarray}
at $x_1 \to x$ and $y_1 \to y$ [17].

Following Klaiber's prescription and using equations ((IV.12) and
(IV.13) of Ref.[4]) for the four--point correlation function in the
r.h.s. of (\ref{label10.2}) we obtain
\begin{eqnarray}\label{label10.3}
&&\Big\langle 0\Big|{\rm
T}\Big(\bar{\psi}(x)\Big(\frac{1-\gamma^5}{2}\Big)
\psi(x_1)\,\bar{\psi}(y)\Big(\frac{1+\gamma^5}{2}\Big)
\psi(y_1)\Big)\Big|0\Big\rangle = \nonumber\\ &&=\Big\langle
0\Big|{\rm T}\Big(\bar{\Psi}(x)\Big(\frac{1-\gamma^5}{2}\Big)
\Psi(x_1)\,\bar{\Psi}(y)\Big(\frac{1+\gamma^5}{2}\Big)
\Psi(y_1)\Big)\Big|0\Big\rangle\nonumber\\
&&\times\,\exp\{-(a-b)\,i\Delta(x - x_1) - (a-b)\,i\Delta(y - y_1) +
(a-b)\,i\Delta(x - y)\nonumber\\ && + (a-b)\,i\Delta(x_1 - y_1) -
(a+b)\,i\Delta(x - y_1) - (a+b)\,i\Delta(x_1 - y)\}.
\end{eqnarray}
The parameters $a$ and $b$ are defined by (\ref{label9.10}) or
(\ref{label9.13}) in terms of Klaiber's parameters $\alpha$ and
$\beta$ or $\sigma$, respectively.

Using (\ref{label6.19}) for the definition of the Green function
$i\Delta(x-y)$ we get
\begin{eqnarray}\label{label10.4}
&&\Big\langle 0\Big|{\rm
T}\Big(\bar{\psi}(x)\Big(\frac{1-\gamma^5}{2}\Big)
\psi(x_1)\,\bar{\psi}(y)\Big(\frac{1+\gamma^5}{2}\Big)
\psi(y_1)\Big)\Big|0\Big\rangle = \nonumber\\ &&=\Big\langle
0\Big|{\rm T}\Big(\bar{\Psi}(x)\Big(\frac{1-\gamma^5}{2}\Big)
\Psi(x_1)\,\bar{\Psi}(y)\Big(\frac{1+\gamma^5}{2}\Big)
\Psi(y_1)\Big)\Big|0\Big\rangle\nonumber\\
&&\times\,\frac{\displaystyle [-\mu^2(x - y)^2 +
i\,0\,]^{\frac{\textstyle a-b}{\textstyle 4\pi}}[-\mu^2(x_1 - y_1)^2 +
i\,0\,]^{\frac{\textstyle a-b}{\textstyle 4\pi}}}{\displaystyle
[-\mu^2(x - x_1)^2 + i\,0\,]^{\frac{\textstyle a-b}{\textstyle
4\pi}}[-\mu^2(y - y_1)^2 + i\,0\,]^{\frac{\textstyle a-b}{\textstyle
4\pi}}}\nonumber\\ &&\times\,\frac{1}{\displaystyle [-\mu^2(x - y_1)^2
+ i\,0\,]^{\frac{\textstyle a+b}{\textstyle 4\pi}}[-\mu^2(x_1 - y)^2 +
i\,0\,]^{\frac{\textstyle a+b}{\textstyle 4\pi}}},
\end{eqnarray}
where $\mu$ is infrared cut--off. Finally it should be taken in the
limit $\mu \to 0$. The correlation function (\ref{label10.4}) agrees
completely with Klaiber's expression (VII.5).

Setting then $x_1 = x$ and $y_1 = y$ in (\ref{label10.3}) we obtain the
two--point correlation function
\begin{eqnarray}\label{label10.5}
&&\Big\langle 0\Big|{\rm
T}\Big(\bar{\psi}(x)\Big(\frac{1-\gamma^5}{2}\Big)
\psi(x)\,\bar{\psi}(y)\Big(\frac{1+\gamma^5}{2}\Big)
\psi(y)\Big)\Big|0\Big\rangle = \nonumber\\ &&=\Big\langle 0\Big|{\rm
T}\Big(\bar{\Psi}(x)\Big(\frac{1-\gamma^5}{2}\Big)
\Psi(x)\,\bar{\Psi}(y)\Big(\frac{1+\gamma^5}{2}\Big)
\Psi(y)\Big)\Big|0\Big\rangle\nonumber\\
&&\times\,\exp\{-2\,(a-b)\,i\Delta(0) - 4\,b\,i\Delta(x
-y)\}=\nonumber\\ &&= \Bigg(\frac{\Lambda}{\mu}\Bigg)^{\textstyle
4\,\frac{\textstyle a + b}{\textstyle
4\pi}}\,\frac{\Lambda^2}{4\pi^2}\,[-\Lambda^2(x-y)^2 +
i\,0\,]^{\textstyle - 1 - b/\pi}.
\end{eqnarray}
From (\ref{label10.5}) we conclude that \\
\noindent (i) the dynamical dimension of
the left(right)handed fermion densities $\bar{\psi}(x)(1 \pm
\gamma^5)\psi(x)$ is equal to
\begin{eqnarray}\label{label10.6}
d_{\textstyle \bar{\psi}\psi}(g,a,b) = \frac{b}{\pi} ,
\end{eqnarray}
\noindent (ii) in the limit $\mu \to 0$ due to the factor
$(\Lambda/\mu)^{\textstyle (a+b)/\pi}$ Klaiber's expressions for
correlation functions of $\bar{\psi}(x)(1 \pm \gamma^5)\psi(x)$ are
either singular or vanish in accordance with the sign of $(a+b)$,\\
\noindent (iii) for the $n$--point correlation function of the
left(right)handed fermion densities the transition to the
ultra--violet cut--off would be accompanied by the appearance of the
factor
\begin{eqnarray}\label{label10.7}
\Bigg(\frac{\Lambda}{\mu}\Bigg)^{\textstyle
4n\,\frac{\textstyle a + b}{\textstyle
4\pi}} = \Bigg(\frac{\Lambda}{\mu}\Bigg)^{\textstyle
4n\,d_{\textstyle \psi}(g,a,b)},
\end{eqnarray}
where $d_{\textstyle \psi}(g,a,b)$ is defined by (\ref{label9.13}).

Now we suggest to evaluate the four--point correlation function
(\ref{label10.3}) within the path--integral approach. For this aim it
is convenient to introduce the generating functional
\begin{eqnarray}\label{label10.8}
&&Z_{\rm Th}[J_-,J_+,\bar{J}_-,\bar{J}_+] =\nonumber\\ &&=\int {\cal
D}\psi{\cal D}\bar{\psi}\,\exp\,i\int
d^2x\,\Big\{\bar{\psi}(x)i\gamma^{\mu}\partial_{\mu}\psi(x) -
\frac{1}{2}\,g\,\bar{\psi}(x)\gamma^{\mu}\psi(x)\bar{\psi}(x)
\gamma_{\mu}\psi(x)\nonumber\\ \hspace{-0.5in}&& +
\bar{\psi}(x)\Big(\frac{1+\gamma^5}{2}\Big)J_-(x) +
\bar{\psi}(x)\Big(\frac{1-\gamma^5}{2}\Big)J_+(x)\nonumber\\ &&+
\bar{J}_-(x)\Big(\frac{1-\gamma^5}{2}\Big)\psi(x) +
\bar{J}_+(x)\Big(\frac{1+\gamma^5}{2}\Big)\psi(x)\Big\},
\end{eqnarray}
where $J_{\pm}(x)$ and $\bar{J}_{\pm}(x)$ are external sources of the
corresponding fermion densities.

After the set of transformations (\ref{label1.12}) --
(\ref{label1.22}) the generating functional (\ref{label10.8}) acquires
the form
\begin{eqnarray}\label{label10.9}
&&Z_{\rm Th}[J_-,J_+,\bar{J}_-,\bar{J}_+] = \int {\cal D}\Psi{\cal
D}\bar{\Psi}{\cal D}\eta {\cal D}\xi\nonumber\\
&&\times\,\exp\,i\int
d^2x\,\Big\{\bar{\Psi}(x)i\gamma^{\mu}\partial_{\mu}\Psi(x) +
\frac{1}{2}\,\partial_{\mu}\eta(x) \partial^{\mu}\eta(x) -
\frac{1}{2}\,\partial_{\mu}\xi(x) \partial^{\mu}\xi(x)\nonumber\\ &&
+ \bar{\Psi}(x)\Bigg(\frac{1+\gamma^5}{2}\Bigg)J_-(x)\,e^{\textstyle -
i\sqrt{g}\,\eta(x) + i \lambda(\alpha_J)\,\xi(x)}\nonumber\\ && +
\bar{\Psi}(x)\Bigg(\frac{1-\gamma^5}{2}\Bigg)J_+(x)\, e^{\textstyle -
i\sqrt{g}\,\eta(x) - i \lambda(\alpha_J)\, \xi(x)}\nonumber\\ && +
\bar{J}_-(x)\Bigg(\frac{1-\gamma^5}{2}\Bigg)\Psi(x)\,e^{\textstyle
i\sqrt{g}\,\eta(x) - i \lambda(\alpha_J)\, \xi(x)}\nonumber\\ && +
\bar{J}_+(x)\Bigg(\frac{1+\gamma^5}{2}\Bigg)\Psi(x)\,e^{\textstyle
i\sqrt{g}\,\eta(x) + i \lambda(\alpha_J)\, \xi(x)}\Big\}.
\end{eqnarray}
In terms of $Z_{\rm Th}[J_-,J_+,\bar{J}_-,\bar{J}_+]$ the correlation
function (\ref{label10.3}) is defined by
\begin{eqnarray}\label{label10.10}
\hspace{-0.5in}&&\Big\langle 0\Big|{\rm
T}\Big(\bar{\psi}(x)\Big(\frac{1-\gamma^5}{2}\Big)
\psi(x_1)\,\bar{\psi}(y)\Big(\frac{1+\gamma^5}{2}\Big)
\psi(y_1)\Big)\Big|0\Big\rangle  = \nonumber\\
\hspace{-0.5in}&& = \frac{\delta}{\delta
J_+(x)}\frac{\delta}{\delta
\bar{J}_-(x_1)}\frac{\delta}{\delta
J_-(y)}\frac{\delta}{\delta \bar{J}_+(y_1)}Z_{\rm
Th}[J_-,J_+,\bar{J}_-,\bar{J}_+]\Big|_{\textstyle J_-=J_+
=\bar{J}_-=\bar{J}_+ = 0}=\nonumber\\
\hspace{-0.5in}&& = \Big\langle 0\Big|{\rm
T}\Big(\bar{\Psi}(x)\Big(\frac{1-\gamma^5}{2}\Big)
\Psi(x_1)\,\bar{\Psi}(y)\Big(\frac{1+\gamma^5}{2}\Big)
\Psi(y_1)\Big)\Big|0\Big\rangle\nonumber\\
\hspace{-0.5in}&&\times\int{\cal D}\eta {\cal D}\xi\,\exp\,i\int
d^2x\,\Big\{\bar{\Psi}(x)i\gamma^{\mu}\partial_{\mu}\Psi(x) +
\frac{1}{2}\,\partial_{\mu}\eta(x) \partial^{\mu}\eta(x) -
\frac{1}{2}\,\partial_{\mu}\xi(x) \partial^{\mu}\xi(x)\nonumber\\
\hspace{-0.5in}&&\times\,e^{\textstyle -i\sqrt{g}\,[\eta(x) -
\eta(x_1) + \eta(y) - \eta(y_1)]}\,e^{\textstyle
-i\lambda(\alpha_J)\,[\xi(x) + \xi(x_1) - \xi(y) - \xi(y_1)]}.
\end{eqnarray}
Integrating over $\eta$ and $\xi$ fields we obtain
\begin{eqnarray}\label{label10.11}
\hspace{-0.5in}&&\Big\langle 0\Big|{\rm
T}\Big(\bar{\psi}(x)\Big(\frac{1-\gamma^5}{2}\Big)
\psi(x_1)\,\bar{\psi}(y)\Big(\frac{1+\gamma^5}{2}\Big)
\psi(y_1)\Big)\Big|0\Big\rangle  = \nonumber\\
\hspace{-0.5in}&& =  \Big\langle 0\Big|{\rm
T}\Big(\bar{\Psi}(x)\Big(\frac{1-\gamma^5}{2}\Big)
\Psi(x_1)\,\bar{\Psi}(y)\Big(\frac{1+\gamma^5}{2}\Big)
\Psi(y_1)\Big)\Big|0\Big\rangle\nonumber\\
\hspace{-0.5in}&&\times\,\exp\{g\,[2i\Delta(0) - i\Delta(x-x_1) +
i\Delta(x-y) - i\Delta(x-y_1)\nonumber\\
\hspace{-0.5in}&& - i\Delta(x_1-y) + i\Delta(x_1-y_1) -
i\Delta(y-y_1)\}\nonumber\\
\hspace{-0.5in}&&\times\exp\{-\lambda^2(\alpha_J)\,[2i\Delta(0) +
i\Delta(x-x_1) - i\Delta(x-y) - i\Delta(x-y_1)\nonumber\\
\hspace{-0.5in}&& - i\Delta(x_1-y) - i\Delta(x_1-y_1) +
i\Delta(y-y_1)\}=\nonumber\\
\hspace{-0.5in}&&=\Big\langle 0\Big|{\rm
T}\Big(\bar{\Psi}(x)\Big(\frac{1-\gamma^5}{2}\Big)
\Psi(x_1)\,\bar{\Psi}(y)\Big(\frac{1+\gamma^5}{2}\Big)
\Psi(y_1)\Big)\Big|0\Big\rangle\nonumber\\
\hspace{-0.5in}&&\times\,\exp\{2\,(g -\lambda^2(\alpha_J))
\,i\Delta(0)- (g +\lambda^2(\alpha_J))\,i\Delta(x-x_1) + (g
+\lambda^2(\alpha_J))\,i\Delta(x-y)\nonumber\\
\hspace{-0.5in}&&- (g -\lambda^2(\alpha_J))\,i\Delta(x-y_1) - (g
-\lambda^2(\alpha_J))\,i\Delta(x_1-y) + (g
+\lambda^2(\alpha_J))\,i\Delta(x_1-y_1)\nonumber\\
\hspace{-0.5in}&&- (g
+\lambda^2(\alpha_J))\,i\Delta(y-y_1)\}.
\end{eqnarray}
Using (\ref{label6.19}) and (\ref{label6.20}) we get
\begin{eqnarray}\label{label10.12}
\hspace{-0.5in}&&\Big\langle 0\Big|{\rm
T}\Big(\bar{\psi}(x)\Big(\frac{1-\gamma^5}{2}\Big)
\psi(x_1)\,\bar{\psi}(y)\Big(\frac{1+\gamma^5}{2}\Big)
\psi(y_1)\Big)\Big|0\Big\rangle  = \nonumber\\
\hspace{-0.5in}&& =  \Big\langle 0\Big|{\rm
T}\Big(\bar{\Psi}(x)\Big(\frac{1-\gamma^5}{2}\Big)
\Psi(x_1)\,\bar{\Psi}(y)\Big(\frac{1+\gamma^5}{2}\Big)
\Psi(y_1)\Big)\Big|0\Big\rangle\nonumber\\
\hspace{-0.5in}&&\times\, \frac{\displaystyle [-\Lambda^2(x -
y)^2 + i\,0\,]^{\frac{\textstyle g + \lambda^2(\alpha_J)}{\textstyle
4\pi}}[-\Lambda^2(x_1 - y_1)^2 + i\,0\,]^{\frac{\textstyle g +
\lambda^2(\alpha_J)}{\textstyle 4\pi}}}{\displaystyle [-\Lambda^2(x -
x_1)^2 + i\,0\,]^{\frac{\textstyle g + \lambda^2(\alpha_J)}{\textstyle
4\pi}}[-\Lambda^2(y - y_1)^2 + i\,0\,]^{\frac{\textstyle g +
\lambda^2(\alpha_J)}{\textstyle 4\pi}}}\nonumber\\
&&\times\,\frac{1}{\displaystyle [-\Lambda^2(x -
y_1)^2 + i\,0\,]^{\frac{\textstyle g - \lambda^2(\alpha_J)}{\textstyle
4\pi}}[-\Lambda^2(x_1 - y)^2 + i\,0\,]^{\frac{\textstyle g -
\lambda^2(\alpha_J)}{\textstyle 4\pi}}}.
\end{eqnarray}
Matching (\ref{label10.12}) with (\ref{label10.4}) up to a
redefinition of powers
\begin{eqnarray}\label{label10.13}
\frac{\textstyle g + \lambda^2(\alpha_J)}{\textstyle 4\pi} \to \frac{a
- b}{4\pi}\quad,\quad \frac{\textstyle g -
\lambda^2(\alpha_J)}{\textstyle 4\pi} \to \frac{a + b}{4\pi}
\end{eqnarray}
the expression (\ref{label10.12}) agrees with (\ref{label10.4}). The
most important distinction is the dependence of Klaiber's expression
(\ref{label10.4}) on the infrared cut--off $\mu$, whereas the
correlation function (\ref{label10.12}) evaluated within the
path--integral approach is infrared stable and depends only on the
ultra--violet cut--off. This confirms our statement that Klaiber's
correlation functions demand infrared regularization. After infrared
regularization they should be compatible to the correlations functions
evaluated within the path--integral approach. The problem of infrared
regularization of Klaiber's correlation functions we solve in the next
section. But it is important to emphasize that Klaiber's correlation
functions are not subject to non--perturbative ultra--violet
renormalization.

\section{Non--perturbative renormalization of the massless Thirring 
model in the chiral symmetric phase}
\setcounter{equation}{0}

\hspace{0.2in} As has been shown above a comparison of the correlation
functions evaluated within Klaiber's operator formalism with the
results obtained in the path--integral technique testifies the
necessity of infrared renormalization of Klaiber's expressions making
them infrared convergent. This infrared renormalization is related to
full extent to the non--perturbative renormalization of the massless
Thirring model in the chiral symmetric phase. According to the
standard procedure of renormalization in quantum field theory [29] we
would understand the renormalizability of the massless Thirring model
as a possibility to remove all infrared and ultra--violet divergences
by renormalization of the wave function of the massless Thirring
fermion field $\psi(x)$ and the coupling constant $g$.

Let us rewrite the Lagrangian (\ref{label1.1}) in terms of {\it bare}
quantities
\begin{eqnarray}\label{label11.1}
{\cal L}_{\rm Th}(x) =
\bar{\psi}_0(x)i\gamma^{\mu}\partial_{\mu}\psi_0(x) -
\frac{1}{2}\,g_0\,\bar{\psi}_0(x)\gamma^{\mu}\psi_0(x)\bar{\psi}_0(x)
\gamma_{\mu}\psi_0(x),
\end{eqnarray}
where $\psi_0(x)$, $\bar{\psi}_0(x)$ are {\it bare} fermionic field
operators and $g_0$ is a {\it bare} coupling constant.

The renormalized Lagrangian ${\cal L}(x)$ of the massless Thirring
model should then read [29]
\begin{eqnarray}\label{label11.2}
{\cal L}_{\rm Th}(x) &=&
\bar{\psi}(x)i\gamma^{\mu}\partial_{\mu}\psi(x) -
\frac{1}{2}\,g\,\bar{\psi}(x)\gamma^{\mu}\psi(x)\bar{\psi}(x)
\gamma_{\mu}\psi(x)\nonumber\\ &+&(Z_2 -
1)\,\bar{\psi}(x)i\gamma^{\mu}\partial_{\mu}\psi(x) -
\frac{1}{2}\,g\,(Z_1 -
1)\,\bar{\psi}(x)\gamma^{\mu}\psi(x)\bar{\psi}(x)\gamma_{\mu}\psi(x)
=\nonumber\\ &=&Z_2\,\bar{\psi}(x)i\gamma^{\mu}\partial_{\mu}\psi(x) -
\frac{1}{2}\,g\,Z_1\,\bar{\psi}(x) \gamma^{\mu}\psi(x)\bar{\psi}(x)
\gamma_{\mu}\psi(x),
\end{eqnarray}
where $Z_1$ and $Z_2$ are the renormalization constants of the
coupling constant and the wave function of the fermion field.

The renormalized fermionic field operator $\psi(x)$ and the coupling
constant $g$ are related to {\it bare} quantities by the relations
[29]
\begin{eqnarray}\label{label11.3}
\psi_0(x) &=& Z^{1/2}_2\,\psi(x),\nonumber\\
g_0 &=& Z_1Z^{-2}_2\,g.
\end{eqnarray}
For the correlation functions of massless Thirring fermions the
renormalizability of the massless Thirring model means the possibility
to replace the infrared cut--off $\mu$ or the ultra--violet cut--off
$\Lambda$ by another finite scale $M$ by means of the renormalization
constants $Z_1$ and $Z_2$.

According to the general theory of renormalizations [29] the
renormalization constants $Z_1$ and $Z_2$ depend on the renormalized
quantities $g$, the infrared scale $\mu$, the ultra--violet scale
$\Lambda$ and the finite scale $M$:
\begin{eqnarray}\label{label11.4}
Z_1 &=& Z_1(g, M;\mu,\Lambda),\nonumber\\
Z_2 &=& Z_2(g, M;\mu,\Lambda).
\end{eqnarray}
Since the correlation functions evaluated within the path--integral
approach do not depend on the infrared cut--off $\mu$, let us, first,
consider the renormalization of them.  In this a little simpler case
the renormalization constants should be written as
\begin{eqnarray}\label{label11.5}
Z_1 &=& Z_1(g, M;\Lambda),\nonumber\\ Z_2 &=& Z_2(g, M;\Lambda).
\end{eqnarray}
For the analysis of the feasibility of the replacement $\Lambda \to M$
it is convenient to introduce the following notations
\begin{eqnarray}\label{label11.6}
\hspace{-0.5in}G^{(i)}_0(x_1,\ldots,x_n;y_1,\ldots,y_n) &=&
\Lambda^n\,G^{(i)}_0(d_{\textstyle
\psi}(g_0,\alpha_J);\Lambda x_1,\ldots,\Lambda x_n;
\Lambda y_1,\ldots,\Lambda y_n),\nonumber\\
\hspace{-0.5in}G^{(\pm)}_0(x_1,\ldots,x_n;y_1,\ldots,y_n) &=&
\Lambda^{2n}\,G^{(\pm)}_0(d_{\textstyle
\bar{\psi}\psi}(g_0,\alpha_J);\Lambda x_1,\ldots,\Lambda x_n;\Lambda
y_1,\ldots,\Lambda y_n),
\end{eqnarray}
where $G^{(i)}_0(x_1,\ldots,x_n;y_1,\ldots,y_n)$ and
$G^{(\pm)}_0(x_1,\ldots,x_n;y_1,\ldots,y_n)$ are the $2n$--point Green
function (\ref{label8.6}) and the $n$--point correlation function
(\ref{label6.22}).

The transition to a finite scale $M$ changes the Green functions
(\ref{label11.6}) as follows
\begin{eqnarray}\label{label11.7}
\hspace{-0.5in}&&G^{(i)}_0(x_1,\ldots,x_n;y_1,\ldots,y_n) 
=\nonumber\\
\hspace{-0.5in}&&=\Bigg(\frac{\Lambda}{M}\Bigg)^{\textstyle 
- 2nd_{\textstyle
\psi}(g_0,\alpha_J)}\,M^n\,G^{(i)}_0(d_{\textstyle
\psi}(g_0,\alpha_J);M x_1,\ldots,M x_n;
M y_1,\ldots,M y_n),\nonumber\\
\hspace{-0.5in}&&G^{(\pm)}_0(x_1,\ldots,x_n;y_1,\ldots,y_n) 
=\nonumber\\
\hspace{-0.5in}&&=\Bigg(\frac{\Lambda}{M}\Bigg)^{\textstyle 
- 2nd_{\textstyle
\bar{\psi}\psi}(g_0,\alpha_J)}\,M^{2n}\,G^{(\pm)}_0(d_{\textstyle
\bar{\psi}\psi}(g_0,\alpha_J);
M x_1,\ldots,M x_n;M y_1,\ldots,M y_n).
\end{eqnarray}
The renormalized correlation functions are related to the {\it bare}
ones by the relations [29]:
\begin{eqnarray}\label{label11.8}
\hspace{-0.5in}&&G^{(i)}(x_1,\ldots,x_n;y_1,\ldots,y_n) 
=Z^{-n}_2\,G^{(i)}_0(x_1,\ldots,x_n;y_1,\ldots,y_n)=\nonumber\\
\hspace{-0.5in}&&=Z^{-n}_2\Bigg(\frac{\Lambda}{M}\Bigg)^{\textstyle\! -
2nd_{\textstyle \psi}(g_0,\alpha_J)}\!M^n\,G^{(i)}_0(d_{\textstyle
\psi}(g_0,\alpha_J);M x_1,\ldots,M x_n; M y_1,\ldots,M
y_n),\nonumber\\
\hspace{-0.5in}&&G^{(\pm)}(x_1,\ldots,x_n;y_1,\ldots,y_n) 
=Z^{-2n}_2G^{(\pm)}_0(x_1,\ldots,x_n;y_1,\ldots,y_n)=\nonumber\\
\hspace{-0.5in}&&=Z^{-2n}_2\Bigg(\frac{\Lambda}{M}\Bigg)^{\textstyle\! -
2nd_{\textstyle
\bar{\psi}\psi}(g_0,\alpha_J)}\!M^{2n}\,G^{(\pm)}_0(d_{\textstyle
\bar{\psi}\psi}(g_0,\alpha_J); M x_1,\ldots,M x_n;M y_1,\ldots,M y_n).
\nonumber\\
\hspace{-0.5in}
\end{eqnarray}
Renormalizability demands the relations
\begin{eqnarray}\label{label11.9}
\hspace{-0.3in}G^{(i)}(x_1,\ldots,x_n;y_1,\ldots,y_n) &=&
M^n\,G^{(i)}(d_{\textstyle
\psi}(g,\alpha_J);M x_1,\ldots,M x_n;M y_1,\ldots,M y_n),\nonumber\\
\hspace{-0.3in}G^{(\pm)}(x_1,\ldots,x_n;y_1,\ldots,y_n) &=&
M^{2n}\,G^{(\pm)}(d_{\textstyle
\bar{\psi}\psi}(g,\alpha_J);M x_1,\ldots,M x_n;M
y_1,\ldots,M y_n),\nonumber\\
\hspace{-0.3in}&&
\end{eqnarray}
which impose constraints on the dynamical dimensions and
renormalization constants
\begin{eqnarray}\label{label11.10}
d_{\textstyle \psi}(g,\alpha_J) &=& d_{\textstyle \psi}(Z_1Z^{-2}_2
g,\alpha_J),\nonumber\\ d_{\textstyle \bar{\psi}\psi}(g,\alpha_J)
&=& d_{\textstyle \bar{\psi}\psi}(Z_1Z^{-2}_2
g,\alpha_J),\nonumber\\
Z^{-1}_2\,\Bigg(\frac{\Lambda}{M}\Bigg)^{\textstyle - 2d_{\textstyle
\psi}(g,\alpha_J)} &=&
Z^{-1}_2\,\Bigg(\frac{\Lambda}{M}\Bigg)^{\textstyle - d_{\textstyle
\bar{\psi}\psi}(g,\alpha_J)} = 1.
\end{eqnarray}
Using the exact expressions for the dynamical dimensions
(\ref{label6.22}) and (\ref{label8.8}) one can show that the first two
constraints on the dynamical dimensions are fulfilled only if the
renormalization constants are related by
\begin{eqnarray}\label{label11.11}
Z_1 = Z^2_2.
\end{eqnarray}
The important consequence of this relation is that the coupling
constant $g$ of the massless Thirring model is unrenormalized, i.e.
\begin{eqnarray}\label{label11.12}
g_0 = g.
\end{eqnarray}
This also implies that the Gell--Mann--Low $\beta$--function, defined
by [29]
\begin{eqnarray}\label{label11.13}
M\frac{dg}{dM} = \beta(g, M),
\end{eqnarray}
should vanish, since $g$ is equal to $g_0$ which does not depend on
$M$, i.e. $\beta(g, M) = 0$ in the massless Thirring model. Our
observation concerning the unrenormalizability of the coupling
constant, $g_0 = g$, is supported by the analysis of renormalizability
of the massive Thirring model carried out by Mueller and Trueman [19]
and Gomes and Lowenstein [20].

The fulfillment of the third constraint in (\ref{label11.10}) demands
a relation between the dynamical dimensions
\begin{eqnarray}\label{label11.14}
d_{\textstyle \bar{\psi}\psi}(g,\alpha_J) = 2\,d_{\textstyle
\psi}(g,\alpha_J).
\end{eqnarray}
In order to satisfy this relation we can use the arbitrariness of the
parameter $\alpha_J$. Solving the equation (\ref{label11.14}) with
respect to $\alpha_J$ we find
\begin{eqnarray}\label{label11.15}
\alpha_J = - \frac{2\pi}{g}.
\end{eqnarray}
The dynamical dimensions are equal to
\begin{eqnarray}\label{label11.16}
d_{\textstyle \bar{\psi}\psi}(g,\alpha_J) = 2\,d_{\textstyle
\psi}(g,\alpha_J)\Big|_{\textstyle \alpha_J = -2\pi/g} =
\frac{g}{\pi}.
\end{eqnarray}
The renormalization constants $Z_1$ and $Z_2$ read then
\begin{eqnarray}\label{label11.17}
Z_1(g, M;\Lambda) = \Big(\frac{M}{\Lambda}\Big)^{\textstyle
2g/\pi} \quad,\quad  Z_2(g, M;\Lambda) =
\Big(\frac{M}{\Lambda}\Big)^{\textstyle g/\pi}.
\end{eqnarray}
Hence, for example, the {\it bare} two--point Green function and the
renormalized one are related by
\begin{eqnarray}\label{label11.18}
S^{(0)}_{\rm F}(x - y) &=& i\langle 0|{\rm
T}(\psi_0(x)\bar{\psi}_0(y))|0\rangle = \frac{1}{2\pi}\,\frac{\hat{x}
- \hat{y}}{(x-y)^2 - i\,0\,}\, [-\Lambda^2(x-y)^2 +
i\,0\,]^{\textstyle - g_0/2\pi} \to\nonumber\\ S_{\rm F}(x - y)
&=& i\langle 0|{\rm T}(\psi(x)\bar{\psi}(y))|0\rangle =
\frac{1}{2\pi}\,\frac{\hat{x} - \hat{y}}{(x-y)^2 - i\,0\,}\, [-
M^2(x-y)^2 + i\,0\,]^{\textstyle - g/2\pi}.\nonumber\\ &&
\end{eqnarray}
The Fourier transform of the two--point Green function reads (see
Appendix E):
\begin{eqnarray}\label{label11.19}
S_{\rm F}(p) = -\,\frac{\displaystyle g\,e^{\textstyle
-ig/2}}{\displaystyle
2\,\sin\Bigg(\frac{g}{2}\Bigg)}\,\frac{1}{\displaystyle
\Gamma^2\Bigg(1 + \frac{g}{2\pi}\Bigg)}\,\frac{\hat{p}}{p^2 +
i\,0}\,\Bigg(\frac{p^2 + i\,0}{\bar{M}^2}\Bigg)^{\displaystyle
\frac{g}{2\pi}},
\end{eqnarray}
where $\bar{M} = 2M$. The Fourier transform of the two--point Green
function possesses singularities in the $g$--plane at $g = \pm
2\pi$ in agreement with Klaiber's analysis [27]. The Green function
(\ref{label11.19}) agrees well with Glaser's analysis of the massless
Thirring model [30,31] and the spectral analysis of two--point
correlation functions developed by Schroer [32].

Now let us show that infrared singularities of Klaiber's correlation
functions can be removed by renormalization of the wave function of
the massless fermionic field in the case of the unrenormalized coupling
constant $g_0 = g$.

In Klaiber's formalism the correlation functions are defined by 
\begin{eqnarray}\label{label11.20}
\hspace{-0.5in}G^{(i)}_0(x_1,\ldots,x_n;y_1,\ldots,y_n)_{\rm K} &=&
\mu^n\,G^{(i)}_0(d_{\textstyle \psi}(g_0,a,b);\mu
x_1,\ldots,\mu x_n;\mu y_1,\ldots,\mu y_n)_{\rm K},\nonumber\\
\hspace{-0.5in}G^{(\pm)}_0(x_1,\ldots,x_n;y_1,\ldots,y_n)_{\rm K} &=&
\mu^{2n}\,G^{(\pm)}_0(d_{\textstyle
\bar{\psi}\psi}(g_0,a,b);\mu x_1,\ldots,\mu x_n;\mu
y_1,\ldots,\mu y_n)_{\rm K},\nonumber\\
\hspace{-0.5in}&&
\end{eqnarray}
where $\mu$ is an infrared cut--off. The dynamical dimensions
$d_{\textstyle \psi}(g_0,a,b)$ and $d_{\textstyle
\bar{\psi}\psi}(g_0,a,b)$ are defined by (\ref{label9.13}) and
(\ref{label10.6}), respectively.

In order to compare Klaiber's expressions with those evaluated within
the path integral approach we should consider the transition of the
infrared into an ultra--violet scale, $\mu \to \Lambda$. This changes
the Green functions (\ref{label11.20}) as follows
\begin{eqnarray}\label{label11.21}
\hspace{-0.5in}&&G^{(i)}_0(x_1,\ldots,x_n;y_1,\ldots,y_n)_{\rm K} 
=\nonumber\\
\hspace{-0.5in}&&=\Bigg(\frac{\mu}{\Lambda}\Bigg)^{\textstyle 
- 2nd_{\textstyle
\psi}(g_0,a,b)}\!\Lambda^n\,G^{(i)}_0(d_{\textstyle
\psi}(g_0,a,b);\Lambda x_1,\ldots,\Lambda x_n;
\Lambda y_1,\ldots,\Lambda y_n)_{\rm K},\nonumber\\
\hspace{-0.5in}&&G^{(\pm)}_0(x_1,\ldots,x_n;y_1,\ldots,y_n)_{\rm K} 
=\nonumber\\
\hspace{-0.5in}&&=\Bigg(\frac{\mu}{\Lambda}\Bigg)^{\textstyle -
4nd_{\textstyle
\psi}(g_0,a,b)}\!\Lambda^{2n}\,G^{(\pm)}_0(d_{\textstyle
\bar{\psi}\psi}(g_0,a,b); \Lambda x_1,\ldots,\Lambda x_n;\Lambda
y_1,\ldots,\Lambda y_n)_{\rm K}.\nonumber\\
\hspace{-0.5in}&&
\end{eqnarray}
Here we have taken into account Eq.(\ref{label10.5}) testifying the
appearance of the factor (\ref{label10.7}) for the rescaling $\mu \to
\Lambda$.

The infrared--renormalized correlation functions are related to the
{\it bare} ones by the relations [29]:
\begin{eqnarray}\label{label11.22}
\hspace{-0.5in}&&G^{(i)}(x_1,\ldots,x_n;y_1,\ldots,y_n)_{\rm K} 
=z^{-n}_2\,G^{(i)}_0(x_1,\ldots,x_n;y_1,\ldots,y_n)_{\rm K}=\nonumber\\
\hspace{-0.5in}&&=
z^{-n}_2\,\Bigg(\frac{\mu}{\Lambda}\Bigg)^{\textstyle - 2nd_{\textstyle
\psi}(g_0,a,b)}\,M^n\,G^{(i)}_0(d_{\textstyle
\psi}(g_0,a,b);\Lambda x_1,\ldots,\Lambda x_n; \Lambda
y_1,\ldots,\Lambda y_n)_{\rm K},\nonumber\\
\hspace{-0.5in}&&G^{(\pm)}(x_1,\ldots,x_n;y_1,\ldots,y_n)_{\rm K} 
=z^{-2n}_2\,G^{(\pm)}_0(x_1,\ldots,x_n;y_1,\ldots,y_n)_{\rm K}=\nonumber\\
\hspace{-0.5in}&&=z^{-2n}_2\,\Bigg(\frac{\mu}{\Lambda}\Bigg)^{\textstyle
- 4nd_{\textstyle
\bar{\psi}}(g_0,a,b)}\,\Lambda^{2n}\,G^{(\pm)}_0(d_{\textstyle
\bar{\psi}\psi}(g_0,a,b); \Lambda x_1,\ldots,\Lambda x_n; \Lambda
y_1,\ldots, \Lambda y_n)_{\rm K}.\nonumber\\
\hspace{-0.5in}&&
\end{eqnarray}
We denote the renormalization constants providing the infrared
renormalization of correlation functions with small letters $z_1$ and
$z_2$, respectively. They depend on $g$, $\Lambda$ and $\mu$
\begin{eqnarray}\label{label11.23}
z_1 &=& z_1(g,\Lambda; \mu),\nonumber\\
z_2 &=& z_2(g,\Lambda; \mu).
\end{eqnarray}
Infrared renormalizability, i.e. independence on the infrared cut--off
$\mu$, assumes the relations
\begin{eqnarray}\label{label11.24}
\hspace{-0.3in}G^{(i)}(x_1,\ldots,x_n;y_1,\ldots,y_n) &=&
\Lambda^n\,G^{(i)}(d_{\textstyle \psi}(g,a,b); \Lambda
x_1,\ldots, \Lambda x_n; \Lambda y_1,\ldots, \Lambda y_n)_{\rm K},
\nonumber\\
\hspace{-0.3in}G^{(\pm)}(x_1,\ldots,x_n;y_1,\ldots,y_n)_{\rm K} &=&
\Lambda^{2n}\,G^{(\pm)}(d_{\textstyle \bar{\psi}\psi}(g,a,b);
\Lambda x_1,\ldots, \Lambda x_n; \Lambda y_1,\ldots, \Lambda y_n)_{\rm
K},\nonumber\\
\hspace{-0.3in}&&
\end{eqnarray}
which impose the relations
\begin{eqnarray}\label{label11.25}
d_{\textstyle \psi}(g,a,b) &=& d_{\textstyle \psi}(z_1z^{-2}_2
g,a,b),\nonumber\\ d_{\textstyle \bar{\psi}\psi}(g,a,b) &=&
d_{\textstyle \bar{\psi}\psi}(z_1z^{-2}_2 g,a,b),\nonumber\\
z^{-1}_2\,\Bigg(\frac{\mu}{\Lambda}\Bigg)^{\textstyle - 2d_{\textstyle
\psi}(g,a,b)} &=&1.
\end{eqnarray}
This gives the following relations. $g_0 = g$, $z_1 = z^2_2$  and 
\begin{eqnarray}\label{label11.26}
z_2 = \Bigg(\frac{\mu}{\Lambda}\Bigg)^{\textstyle
2d_{\textstyle \psi}(g,a,b)}.
\end{eqnarray}
Hence, we have shown that the infrared singularities of Klaiber's
correlation functions can be removed by regularization of the wave
functions of the massless fermionic field at the unrenormalized
coupling constant $g_0 = g$.

The ultra--violet renormalizability of the massless Thirring model 
in Klaiber's formalism imposes the constraint
\begin{eqnarray}\label{label11.27}
a=b
\end{eqnarray}
which leads to the dynamical dimension of the massless fermionic field
equal to
\begin{eqnarray}\label{label11.28}
d_{\textstyle \psi}(g_0,a,b) = \frac{b}{2\pi}.
\end{eqnarray}
The dynamical dimension $d_{\textstyle 2n\psi}$ of any $2n$--fermion
correlation function is merely the product $d_{\textstyle 2n\psi} =
2nd_{\textstyle \psi}(g_0,a,b) = nb/\pi$. 

In order to show that the constraint (\ref{label11.27}) leads to a
quantum field theory of a free massless fermionic field we suggest to
use the parameterization of the parameters $a$ and $b$ in terms of
$\rho$. This yield
\begin{eqnarray}\label{label11.29}
a = \pi\,(\rho^2 -1)\quad,\quad b = \pi\,\Big(\frac{1}{\rho^2} -
1\Big).
\end{eqnarray}
Thereby, the constraint (\ref{label11.27}) entails that $\rho = \pm
1$. This gives
\begin{eqnarray}\label{label11.30}
a = b = 0
\end{eqnarray}
and corresponds to a quantum field theory of a free massless fermionic
field. In fact, in our parameterization (\ref{label5.66}) we have
$\rho = 1$ that provides $\alpha = \beta = 0$ (see
(\ref{label5.25})). Setting $\alpha = \beta = 0$ in (\ref{label4.18})
and (\ref{label5.1}) we get $\psi(x) = \Psi(x)$.

It is interesting to notice that in Klaiber's parameterization there
are two free massless fermionic fields $\psi(x)$ defined for $\alpha =
\beta = 0$ at $\rho = 1$ and for $\alpha = \beta = - 2\sqrt{\pi}$ at
$\rho = - 1$.

\section{Fermion condensate and instability of 
chiral symmetric phase under spontaneous breaking of chiral symmetry}
\setcounter{equation}{0}

\hspace{0.2in} As has been shown in Ref.[6] in the massless Thirring
model the chirally broken phase is energetically preferable with
respect to the chiral symmetric one. In this section we show that the
solution of the massless Thirring model in the chiral symmetric phase
is unstable under spontaneous breaking of chiral symmetry. For this
aim we treat a causal two--point Green function (\ref{label11.19}) and
show the existence of a non--vanishing value of the fermion
condensate.

The two--point Green function $S_{\rm F}(x) = i\langle 0|{\rm
T}(\psi(x)\bar{\psi}(0))|0\rangle$ can be represented in the form
\cite{[silva]}
\begin{eqnarray}\label{label12.1}
\hspace{-0.3in}&&S_{\rm F}(x) = i\langle 0|{\rm
T}(\psi(x)\bar{\psi}(0))|0\rangle = \frac{\displaystyle
g^2\,e^{\textstyle - ig/2}}{\displaystyle
\bar{M}^2\sin\Bigg(\frac{g}{2}\Bigg)}\,\frac{1}{\displaystyle
\Gamma^2\Bigg(1 + \frac{g}{2\pi}\Bigg)}\nonumber\\
\hspace{-0.3in}&&\times\int\frac{d^2p}{(2\pi)^2}\,e^{\textstyle
-ip\cdot x}\int\frac{d^2k}{(2\pi)^2i}\,\frac{\hat{k}}{k^2 +
i\,0}\,\Bigg(\frac{(p-k)^2 + i\,0}{\bar{M}^2}\Bigg)^{\textstyle -1 +
g/2\pi}.
\end{eqnarray}
The Fourier transform is defined by 
\begin{eqnarray}\label{label12.2}
\hspace{-0.3in}S_{\rm F}(p) = \frac{\displaystyle g^2\,e^{\textstyle -
i g/2}}{\displaystyle
\bar{M}^2\sin\Bigg(\frac{g}{2}\Bigg)}\,\frac{1}{\displaystyle
\Gamma^2\Bigg(1 +
\frac{g}{2\pi}\Bigg)}\int\frac{d^2k}{(2\pi)^2i}\,\frac{\hat{k}}{k^2 +
i\,0}\,\Bigg(\frac{(p-k)^2 + i\,0}{\bar{M}^2}\Bigg)^{\textstyle -1 +
g/2\pi}.
\end{eqnarray}
Now let us proceed to Euclidean momentum space [16]
\begin{eqnarray}\label{label12.3}
S_{\rm F}(p_{\rm E}) = -\,e^{\textstyle -
ig/2}\,\frac{C(g)}{\bar{M}^2}\int\frac{d^2k_{\rm
E}}{(2\pi)^2}\,\frac{\hat{k}_{\rm E}}{k^2_{\rm E} -
i\,0}\,\Bigg(\frac{- (p_{\rm E}-k_{\rm E})^2 +
i\,0}{\bar{M}^2}\Bigg)^{\textstyle -1 + g/2\pi},
\end{eqnarray}
where we have denoted
\begin{eqnarray}\label{label12.4}
C(g) = \frac{\displaystyle g^2}{\displaystyle
\sin\Bigg(\frac{g}{2}\Bigg)}\,\frac{1}{\displaystyle \Gamma^2\Bigg(1 +
\frac{g}{2\pi}\Bigg)}.
\end{eqnarray}
Using the integral representation
\begin{eqnarray}\label{label12.5}
&&\Bigg(\frac{- (p_{\rm E}-k_{\rm E})^2 +
i\,0}{\bar{M}^2}\Bigg)^{\textstyle -1 + g/2\pi} =\nonumber\\ &&=
\frac{\displaystyle -\,i\,e^{\textstyle ig/4}}{\displaystyle
\Gamma\Bigg(1 - \frac{g}{2\pi}\Bigg)}\int^{\infty}_0dt\, t^{\textstyle
- g/2\pi}\,\exp\Bigg\{-i\Bigg(\frac{(p_{\rm E} - k_{\rm E})^2
-i\,0}{\bar{M}^2}\Bigg)\,t\Bigg\},
\end{eqnarray}
substituting (\ref{label12.5}) in (\ref{label12.3}) and integrating
over directions of 2--Euclidean vector $k_{\rm E}$ we obtain
\begin{eqnarray}\label{label12.6}
\hspace{-0.3in}S_{\rm F}(p_{\rm E}) &=& \frac{C(g)}{2\pi\bar{M}^2}\,
\frac{\displaystyle i\, e^{\textstyle - ig/4}}{\displaystyle
\Gamma\Bigg(1 - \frac{g}{2\pi}\Bigg)}\frac{\hat{p}_{\rm E}}{p_{\rm
E}}\int^{\infty}_0dt\, t^{\textstyle -
g/2\pi}\exp\Bigg(-i\,\frac{p^2_{\rm E}t}{\bar{M}^2}\Bigg)\nonumber\\
&&\times\int^{\infty}_0dk_{\rm
E}\,J_1\Big(\frac{2t}{\bar{M}^2}\,p_{\rm E}k_{\rm
E}\Big)\,\exp\Bigg(-i\,\frac{k^2_{\rm E}t}{\bar{M}^2}\Bigg),
\end{eqnarray}
where $J_1(z)$ is the Bessel function \cite{[abr]}. Integrating over
$k_{\rm E}$ we obtain \cite{[grad]}
\begin{eqnarray}\label{label12.7}
\hspace{-0.3in}\int^{\infty}_0dk_{\rm
E}\,J_1\Big(\frac{2t}{\bar{M}^2}\,p_{\rm E}k_{\rm
E}\Big)\exp\Bigg(-i\,\frac{k^2_{\rm E}t}{\bar{M}^2}\Bigg) =
\frac{\bar{M}^2}{itp_{\rm E}}\,\sin\Bigg(\frac{p^2_{\rm
E}t}{2\bar{M}^2}\Bigg)\exp\Bigg(+i\,\frac{p^2_{\rm
E}t}{2\bar{M}^2}\Bigg).
\end{eqnarray}
Substituting (\ref{label12.7}) in (\ref{label12.6}) we arrive at the
expression
\begin{eqnarray}\label{label12.8}
\hspace{-0.3in}S_{\rm F}(p_{\rm E})= -\,\frac{C(g)}{4\pi}\,
\frac{\displaystyle e^{\textstyle -ig/4}}{\displaystyle \Gamma\Bigg(1
- \frac{g}{2\pi}\Bigg)}\,\frac{i}{\hat{p}_{\rm
E}}\int^{\infty}_0\frac{dt}{\displaystyle t^{\textstyle 1 +
g/2\pi}}\Bigg[1 - \exp\Bigg(-i \frac{p^2_{\rm
E}t}{\bar{M}^2}\Bigg)\Bigg].
\end{eqnarray}
In order to regularize the integral over $t$ we introduce the infrared
cut--off $\mu$
\begin{eqnarray}\label{label12.9}
\hspace{-0.3in}S_{\rm F}(p_{\rm
E})=-\frac{C(g)}{4\pi}\frac{\displaystyle e^{\textstyle
-ig/4}}{\displaystyle \Gamma\Bigg(1 -
\frac{g}{2\pi}\Bigg)}\,\frac{i}{\hat{p}_{\rm
E}}\int^{\infty}_0\frac{dt}{\displaystyle t^{\textstyle 1 +
g/2\pi}}\Bigg[\exp\!\Bigg(-i\frac{\mu^2t}{\bar{M}^2}\Bigg) -
\exp\!\Bigg(-i\frac{p^2_{\rm E}t}{\bar{M}^2}\Bigg)\Bigg].
\end{eqnarray}
The result of the integration over $t$ reads
\begin{eqnarray}\label{label12.10}
\hspace{-0.3in}S_{\rm F}(p_{\rm E}) = \frac{\displaystyle
g}{\displaystyle 2\sin\Bigg(\frac{g}{2}\Bigg)}\frac{1}{\displaystyle
\Gamma^2\Bigg(1 + \frac{g}{2\pi}\Bigg)}\,\frac{i}{\hat{p}_{\rm
E}}\,\Bigg[\Bigg(\frac{\mu^2}{\bar{M}^2}\Bigg)^{\textstyle g/2\pi} -
\Bigg(\frac{p^2_{\rm E}}{\bar{M}^2}\Bigg)^{\textstyle g/2\pi}\Bigg].
\end{eqnarray}
Taking the limit $\mu \to 0$ we obtain
\begin{eqnarray}\label{label12.11}
\hspace{-0.3in}S_{\rm F}(p_{\rm E}) = - \frac{\displaystyle
g}{\displaystyle 2\sin\Bigg(\frac{g}{2}\Bigg)}\frac{1}{\displaystyle
\Gamma^2\Bigg(1 + \frac{g}{2\pi}\Bigg)}\frac{i}{\hat{p}_{\rm
E}}\Bigg(\frac{p^2_{\rm E}}{\bar{M}^2}\Bigg)^{\textstyle g/2\pi}.
\end{eqnarray}
This agrees with (\ref{label11.19}). Then, the fermion condensate is
defined by
\begin{eqnarray}\label{label12.12}
\langle 0|\bar{\psi}(0)\psi(0)|0\rangle = - \int\frac{d^2p_{\rm
E}}{(2\pi)^2}\,{\rm tr}\{S_{\rm F}(p_{\rm E})\}.
\end{eqnarray}
Since the integrand is singular at $\hat{p}_{\rm E} = 0$, we suggest
to use the following integral representation
\begin{eqnarray}\label{label12.13}
\frac{i}{\hat{p}_{\rm E}} = \int^{\infty}_0d\tau\,e^{\textstyle
i\hat{p}_{\rm E}\tau} = \int^{\infty}_0d\tau\,\Big[\cos(p_{\rm E}\tau)
+\,i\,\frac{\hat{p}_{\rm E}}{p_{\rm E}}\,\sin(p_{\rm E}\tau)\Big]
\end{eqnarray}
valid for $p_{\rm E}\neq 0$. In fact, the direct calculation of the
integral over $\tau$ gives
\begin{eqnarray}\label{label12.14}
\int^{\infty}_0d\tau\,\Big[\cos(p_{\rm E}\tau)
+\,i\,\frac{\hat{p}_{\rm E}}{p_{\rm E}}\,\sin(p_{\rm E}\tau)\Big] =
\pi\,\delta(p_{\rm E}) + \frac{i}{\hat{p}_{\rm E}}.
\end{eqnarray}
Since the l.h.s. of (\ref{label12.13}) is well--defined for $p_{\rm
E}\neq 0$, the $\delta$--function vanishes and we obtain
$i/\hat{p}_{\rm E}$.

For the fermion condensate we obtain
\begin{eqnarray}\label{label12.15}
\hspace{-0.3in}\langle 0|\bar{\psi}(0)\psi(0)|0\rangle
=\frac{\displaystyle g}{\displaystyle
\sin\Bigg(\frac{g}{2}\Bigg)}\frac{\bar{M}}{\displaystyle
\Gamma^2\Bigg(1 + \frac{g}{2\pi}\Bigg)}
\int^{\infty}_0\!\frac{d\tau}{2\pi}\!\int^{\infty}_0\!dp_{\rm
E}\cos(p_{\rm E}\tau)\Bigg(\frac{p_{\rm E}}{\bar{M}}\Bigg)^{\textstyle
1 + g/\pi}\!.
\end{eqnarray}
The regularization procedure implies the interchangeability of the
integrations over $\tau$ and $p_{\rm E}$. The r.h.s. of
(\ref{label12.15}) does not vanish and gives a non-trivial value for
the fermion condensate. The appearance of a non--vanishing fermion
condensate testifies instability of the chiral symmetric phase under
spontaneous breaking of chiral symmetry.

The calculation of the integral over $p_{\rm E}$ should be carried out
in the meaning of generalized functions \cite{[gel]}. This yields
\begin{eqnarray}\label{label12.16}
\int^{\infty}_0dp_{\rm E}\cos(p_{\rm E}\tau)\Bigg(\frac{p_{\rm
E}}{\bar{M}}\Bigg)^{\textstyle 1 + g/\pi} =
-\cos\Bigg(\frac{g}{2}\Bigg)\,\Gamma\Bigg(2 +
\frac{g}{\pi}\Bigg)\,\tau^{\textstyle -2 - g/\pi}.
\end{eqnarray}
Substituting (\ref{label12.16}) in (\ref{label12.15}), regularizing
the integral over $\tau$ by the ultra--violet parameter $\Lambda$ and
renormalizing the fermion fields $\psi(x) \to
Z^{1/2}_2(g,\Lambda,\bar{M})\,\psi(x)$ we obtain
\begin{eqnarray}\label{label12.17}
\hspace{-0.3in}\langle 0|\bar{\psi}(0)\psi(0)|0\rangle = -
\frac{M(g)}{g},
\end{eqnarray}
where $M(g)$, a dynamical mass of Thirring fermions in the chirally
broken phase, we define following [6]
\begin{eqnarray}\label{label12.18}
M(g) = \frac{\Lambda}{\displaystyle \sqrt{e^{\textstyle 2\pi/g} - 1}}
\end{eqnarray}
and $Z_2(g,\Lambda,\bar{M})$, the renormalization constant of the
wave function of Thirring fermion fields, is equal to
\begin{eqnarray}\label{label12.19}
Z_2(g,\Lambda,\bar{M}) = \Bigg(e^{\textstyle 2\pi/g} -
1\Bigg)^{1/2}\frac{g^2}{2\pi}\,\frac{\displaystyle {\rm
ctg}\Bigg(\frac{g}{2}\Bigg)}{\displaystyle \Gamma\Bigg(1 +
\frac{g}{2\pi}\Bigg)}\Bigg(\frac{\Lambda}{\bar{M}}\Bigg)^{\textstyle
g/\pi}.
\end{eqnarray}
Introducing the normalization scale $\bar{\mu}$ defined by
\begin{eqnarray}\label{label12.20}
\bar{\mu} =\Bigg(\frac{\Lambda^2}{\bar{M}}\Bigg)
\left[\Bigg(e^{\textstyle 2\pi/g} -
1\Bigg)^{1/2}\frac{g^2}{2\pi}\,\frac{\displaystyle {\rm
ctg}\Bigg(\frac{g}{2}\Bigg)}{\displaystyle \Gamma\Bigg(1 +
\frac{g}{2\pi}\Bigg)}\right]^{\textstyle \pi/g}.
\end{eqnarray}
we transform the renormalization constant $Z_2(g,\Lambda,\bar{M})$
given by (\ref{label12.19}) to the form
\begin{eqnarray}\label{label12.21}
Z_2(g,\Lambda,\bar{\mu}) =
\Bigg(\frac{\bar{\mu}}{\Lambda}\Bigg)^{\textstyle g/\pi}
\end{eqnarray}
in agreement with the definition (\ref{label11.17}).

The result (\ref{label12.17}) confirms qualitatively our conclusion
concerning the instability of the chiral symmetric phase of the
massless Thirring model under spontaneous breaking of chiral symmetry
pointed out in Ref.[6].

\section*{Conclusion}

\hspace{0.2in} We have analysed the solution of the massless Thirring
model with fermion fields quantized in the chiral symmetric phase. In
the literature the solution of the massless Thirring model in the form
of $n$--point correlation functions has been carried out within the
operator formalism developed by Klaiber [4] and by the path--integral
technique initiated by Furuya, Gamboa Saravi and Schaposnik [7] and
Na${\acute{\rm o}}$n [8]. We have discussed these two independent
approaches. We have found that Klaiber's operator formalism suffers
from the following problems: (i) unusual properties of composite
fermionic field operators caused by the use of a solution of the Dirac
equation contradicting to standard parity properties of fermionic
fields, (ii) the definition of quantum vector and axial--vector
fermion currents within the point--splitting technique leading to
explicit breaking of chiral symmetry and (iii) the absence of a
one--parameter family of solutions of quantum equations of motion of
the massless Thirring model. In fact, a simultaneous evaluation of
quantum vector and axial--vector fermion currents within Klaiber's
approach supplemented by the requirement of the fulfillment of the
standard relation $j^{\mu}_5(x) = -\varepsilon^{\mu\nu}j_{\nu}(x)$,
caused by the properties of Dirac matrices in 1+1--dimensional
space--time, fixes the parameter $\sigma$ to $\sigma = 0$. Therefore,
unlike Klaiber's claim none relation between Klaiber's [4] and
Johnson's solutions of the massless Thirring model
\cite{[john],[wess]} exists. Taking into account that at intermediate
steps Klaiber's definition of quantum vector and axial--vector
currents leads to an explicit breakdown of chiral symmetry
(\ref{label5.54}), it is obvious that Klaiber's parameterization
(\ref{label5.43}) is not applicable at all. On this way Klaiber's
parameterization should be replaced by ours given by
(\ref{label5.66}). In fact, this parameterization has been introduced
within a correct definition of quantum vector and axial--vector
current without violation of chiral symmetry and the standard relation
$j^{\mu}_5(x) = -\varepsilon^{\mu\nu}j_{\nu}(x)$.

Unlike Klaiber's operator formalism the existence of a
one--parameter family of solutions of the massless Thirring model is
supported within the path--integral approach to the solution of the
massless Thirring model. We have evaluated $n$--point correlation
functions of left(right)handed fermion densities and $2n$--point Green
functions of fermion fields and found dynamical dimensions of these
quantities depending on an arbitrary parameter $\alpha_J$ caused by
ambiguities of the evaluation of chiral Jacobians quoted by Christos
[11].

Such a free parameter in the solution of the massless Thirring model
has turned out to be of use for the non--perturbative renormalization
of the model. We have analysed the non--perturbative renormalization
of the massless Thirring model in the standard meaning of
renormalization as a possibility to replace the dependence of
correlation functions and Green functions on the ultra--violet
cut--off $\Lambda$ by any finite scale $M$, $\Lambda \to M$, by means
of the renormalization of the wave function of fermion fields and the
coupling constant. We have shown that the coupling constant of the
model $g$ is unrenormalized that corresponds to the vanishing
Gell--Mann--Low function, $\beta(g,M) = 0$. Our result on the
unrenormalized coupling constant $g$ is supported by similar
conclusions which have been obtained by Mueller and Trueman [19] and
Gomes and Lowenstein [20] who investigated the renormalizability of
the massive Thirring model. The requirement of the complete
replacement $\Lambda \to M$ in the expressions for $n$--point
correlation functions of left(right)handed fermion field densities
$\bar{\psi}(x)(1\pm\gamma^5)\psi(x)$and $2n$--point Green function has
allowed to fix the parameter $\alpha_J$ to the value $\alpha_J =
-2\pi/g$. According to this result the dynamical dimensions of any
$n$--fermion correlation function $d_{\textstyle n\psi}$ are equal to
$d_{\textstyle n\psi} = nd_{\textstyle \psi}$, where $d_{\textstyle
\psi} = g/2\pi$.

We would like to emphasize that for $\alpha_J = -2\pi/g$ the kinetic
term of the pseudoscalar field $\xi(x)$ acquires a correct sign. Due
to this the symmetry group of the massless scalar and pseudoscalar
fields $(\eta(x),\xi(x))$ becomes equal to the compact $O(2)$ group
instead of the former non--compact $O(1,1)$ group.

We have noted that $n$--point correlation functions evaluated with
Klaiber's operator formalism contain infrared divergences. This is
unlike the path--integral approach giving correlations functions and
Green functions regular in the infrared limit. We have shown that by
means of the renormalization of the wave functions of fermion fields
the infrared cut--off can be replaced by an ultra--violet cut--off.
This confirms the infrared stability of the massless Thirring model.
Unfortunately, unlike the path--integral approach Klaiber's operator
formalism has turned out to be inconsistent with a non--perturbative
ultra--violet renormalization of the massless Thirring model. We have
shown that Klaiber's definition of dynamical dimensions of $n$--point
correlation functions of left(right)handed fermion densities
$\bar{\psi}(x)(1\pm \gamma^5)\psi(x)$ and $2n$--point fermion Green
functions are not invariant under the replacement of the ultra--violet
cut-off $\Lambda$ by any finite scale $M$. After a change $\Lambda \to
M$ multiplicative factors depending on $\Lambda$ are left which cannot
be removed by a renormalization of wave functions of fermion fields.

One has to confess that Klaiber's operator formalism is not flexible
enough for the non--perturbative ultra--violet renormalizability of
the massless Thirring model. The dependence of the scalar and
pseudoscalar fermion densities $S(x)$ and $P(x)$, related to vector
and axial vector currents $J_{\mu}(x) =\partial_{\mu}S(x)/\sqrt{\pi}$
and $J_{5\mu}(x) =\partial_{\mu}P(x)/\sqrt{\pi}$, on the operators of
creation $C^{\dagger}(k^1)$ and annihilation $C(k^1)$ leads to the
appearance of Lorentz non--covariant contributions to correlation
functions of fermion fields (\ref{label9.9}) which can be removed only
by means of the constraint (\ref{label5.26}). Due to these constraint
Klaiber's parameters $\alpha$ and $\beta$ become expressed in terms of
only one parameter $\rho$ (\ref{label5.42}). This rules out the
possibility for these parameters to be equal to each other.  In fact,
the requirement of an ultra--violet non--perturbative
renormalizability of the massless Thirring model (\ref{label11.27})
leads to equality $\rho = \pm 1$.

Taking into account that the derivation of the quantum equation of
motion of the massless Thirring model fixes all parameters of the
canonical transformation suggested by Klaiber, (\ref{label4.18}) and
(\ref{label5.1}), the possibility of the realization of a
one--parameter family of solutions should be attributed to
non--canonical contributions like chiral Jacobians in the
path--integral approach. Hence, we argue that Klaiber's operator
approach is not complete and should be supplemented by non--canonical
contributions which are determined by chiral Jacobians, for example.
We are planning to carry out this program in our forthcoming
publications.

The transition from Klaiber's parameterization to ours changes
Coleman's coupling constant relation (\ref{label1.30}) between the
coupling constants of the Thirring model and the sine--Gordon
model. Now according to our parameterization carried out within
Klaiber's operator formalism it should read
$$
\frac{4\pi}{\bar{\beta}^2} = \frac{\displaystyle 1 +
\frac{g}{2\pi}}{\displaystyle 1 - \frac{g}{2\pi}}.\eqno(\Pi.1)
$$
In turn, within the path--integral definition of the dynamical
dimension of the fermion field densities
$\bar{\psi}(x)(1\pm\gamma^5)\psi(x)$, $d_{\textstyle \bar{\psi}\psi} =
g/\pi$, Coleman's relation between coupling constants $g$ and
$\bar{\beta}$ should read
$$
\frac{\bar{\beta}^2}{4\pi} = 1 + \frac{g}{\pi}.\eqno(\Pi.2)
$$
This relation is preferable with respect to $(\Pi.1)$, since it is
obtained in the massless Thirring model subject to all requirements of
non--perturbative renormalizability. The non--perturbative
renormalizability of the sine--Gordon model we demonstrate in Appendix
F by deriving a renormalized generating functional of Green functions.

We would like to emphasize that in the strong coupling limit $(\Pi.2)$
agrees with the relation obtained in Ref.[6] for the massive Thirring
model quantized in the chirally broken phase. Indeed, as has been
found in Ref.[6] in the chirally broken phase the relation between the
coupling constants $g$ and $\beta$ reads
$$
\frac{8\pi}{\bar{\beta}^2} = 1 - e^{\textstyle - 2\pi/g}.\eqno(\Pi.3)
$$
In the strong coupling limit $g \gg\pi$ the relations $(\Pi.2)$ and
$(\Pi.3)$ have the same asymptotic behaviour
$$
\frac{8\pi}{\bar{\beta}^2} = \frac{2\pi}{g} + O(g^{-2}).\eqno(\Pi.4)
$$
This implies the inequality $\bar{\beta}^2 \gg 8\pi$ that leads to the
population of the 1+1--dimensional world with solitons having quantum
numbers of Thirring fermions, see also Ref.[6]. It is important to
underscore that according to Coleman's relation (\ref{label1.30}) the
coupling constant $\bar{\beta}$ tends to zero, $\bar{\beta} \to 0$, at
$g \to \infty$. This suppresses the production of soliton states with
respect to the production of single quanta of the sine--Gordon field.
The former is hardly comprehensible from a physical point of view,
since in the strong coupling limit collective phenomena should play an
important role. In Appendix F we have carried out a non--perturbative
renormalization of the sine--Gordon model. We have shown that the
sine--Gordon model is not asymptotically free. This is in complete
agreement with the equivalence between the Thirring model and the
sine--Gordon model. The unrenormalizability of the coupling constant
$\bar{\beta}$ leading to the vanishing Gell--Mann--Low function can be
easily understood following the similarity between $\bar{\beta}$ and
$\hbar$ which has been drawn in Ref.[6]. As has been shown in Ref.[6]
the limits $\bar{\beta} \to 0$ and $\bar{\beta} \to \infty$
distinguish {\it classical} and {\it quantum} regimes of the
sine--Gordon model. Within such an understanding of the coupling
constant $\bar{\beta}$ its unrenormalizability is justified by the
unrenormalizability of $\hbar$.

Now we would like to make some remarks concerning Coleman's
perturbation theory developed with respect to the mass $m$ of the
Thirring fermion field which was used for the derivation of
equivalence between the massive Thirring model and the sine--Gordon
model.  An expansion of the generating functional of correlation
functions in powers of $m$ assumes the integration over fermionic
degrees of freedom described by a free massless fermion field.
However, as we have discussed in section 2 due to, say, the
non--standard properties under parity transformations breaking the
relation $v(p^0, p^1) = - \gamma^0v(p^0, - p^1)$ (see
(\ref{label2.18})), Klaiber's solution for a free massless fermion
field cannot be obtained as the massless limit of a free massive
fermion field. In fact, a free massive fermion field with mass $m$
possesses standard properties under parity transformations obeying the
relation $v(p^0, p^1) = - \gamma^0v(p^0, - p^1)$ (see Appendix A ({\rm
A}.15)). Therefore, the quantum field theory of a free massless
fermion field obtained in the limit $m \to 0$ differs from the quantum
field theory of a free massless fermion field in Klaiber's
approach. Thus, we think that Coleman's perturbation theory developed
with respect to $m$ does not exist from the point of view of
constructive quantum field theory.

Finally, we would like to note that we have shown that the chiral
symmetric phase of the massless Thirring model is not stable, and
fermion condensation can occur even if the fermion system evolves from
the chiral symmetric phase. Taking the Fourier transform of the
two--point Green function of a massless Thirring fermion field
evaluated in the chiral symmetric phase we have shown that fermion
condensation can occur dynamically. The former confirms our statement
in Ref.[6] concerning the energetic advantage of the chirally broken
phase relative to the chiral symmetric one.

\section*{Acknowledgement}

\hspace{0.2in} This work was supported in part by Fonds zur
F\"orderung der Wissenschaftlichen Forschung P13997-TPH.

\newpage

\section*{Appendix A. Parity properties of fermion fields}

\hspace{0.2in} In this Appendix we recall the properties of fermion
fields under parity transformation.

First, let us discuss fermion fields in 3+1--dimensional
space--time. A free massive Dirac field of mass $m$ obeys the
equation
$$
(i\,\gamma^{\mu}\partial_{\mu} - m)\,\psi(x) = 0,\eqno({\rm A}.1)
$$
where $\gamma^{\mu} = (\gamma^0,\vec{\gamma}\,)$ are Dirac matrices in
the Dirac representation and $\psi(x) = \psi(t,\vec{x}\,)$. The
solution of ({\rm A}.1) can be represented in the standard form of the
expansion into plane waves
$$
\psi(t,\vec{x}\,) =
\int\frac{d^3p}{(2\pi)^{3/2}}\,\frac{1}{\sqrt{2p^0}}
$$
$$
\times\sum_{\sigma =
\pm 1/2}\Big[u(p^0,\vec{p},\sigma)\,a(\vec{p},\sigma)\,e^{\textstyle -ip^0t
+ i\vec{p}\cdot \vec{x}} +
v(p^0,\vec{p},\sigma)\,b^{\dagger}(\vec{p},\sigma)\,e^{\textstyle
ip^0t - i\vec{p}\cdot \vec{x}}\Big],\eqno({\rm A}.2)
$$
where $a(\vec{p},\sigma)$ and $b(\vec{p},\sigma)$ are annihilation
operators of fermions and antifermions with momentum $\vec{p}$,
energy $p^0 = \sqrt{\vec{p}^{\,2} + m^2}$ and polarization
$\sigma$. The Dirac bispinorial functions $u(p^0,\vec{p},\sigma)$ and
$v(p^0,\vec{p},\sigma)$ have the standard form
$$
u(p^0,\vec{p},\sigma) = \sqrt{p^0 + m}{\displaystyle
\left(\begin{array}{c}{\displaystyle \varphi_{\sigma}} \\
{\displaystyle \frac{\vec{\sigma}\cdot \vec{p}}{p^0 +
m}\,\varphi_{\sigma}}
\end{array}\right)}\quad,\quad v(p^0,\vec{p},\sigma)= 
\sqrt{p^0 + m}{\displaystyle \left(\begin{array}{c}{\displaystyle
\frac{\vec{\sigma}\cdot \vec{p}}{p^0 + m}\,\chi_{\sigma}}
\\{\displaystyle \chi_{\sigma}}
\end{array}\right)},\eqno({\rm A}.3)
$$
where $\vec{\sigma} = (\sigma_1,\sigma_2,\sigma_3)$ are Pauli
matrices, $\varphi_{\sigma}$ and $\chi_{\sigma}$ are two--component
spinors. The Dirac bispinorial functions ({\rm A}.3) are normalized to
$$
u^{\dagger}(p^0,\vec{p},\sigma)\,u(p^0,\vec{p},\sigma'\,) =
v^{\dagger}(p^0,\vec{p},\sigma)\,v(p^0,\vec{p},\sigma'\,) =
2p^0\,\delta_{\sigma\,\sigma'},
$$
$$
\bar{u}(p^0,\vec{p},\sigma)\,u(p^0,\vec{p},\sigma'\,) =-
\bar{v}(p^0,\vec{p},\sigma)\,v(p^0,\vec{p},\sigma'\,)
2m\,\delta_{\sigma\,\sigma'},
$$
$$
u^{\dagger}(p^0,\vec{p},\sigma)\,v(p^0,\vec{p},\sigma'\,) =
v^{\dagger}(p^0,\vec{p},\sigma)\,u(p^0,\vec{p},\sigma'\,) = 0
\eqno({\rm A}.4)
$$
and satisfy the matrix relations
$$
\sum_{\sigma = \pm
1/2}u(p^0,\vec{p},\sigma)\,\bar{u}(p^0,\vec{p},\sigma) =
\left(\begin{array}{cc} p^0 + m & -\vec{\sigma}\cdot \vec{p} \\
\vec{\sigma}\cdot \vec{p} & p^0 + m
\end{array} \right) = \hat{p} + m,
$$
$$
\sum_{\sigma = \pm
1/2}v(p^0,\vec{p},\sigma)\,\bar{v}(p^0,\vec{p},\sigma) =
\left(\begin{array}{cc} p^0 - m & -\vec{\sigma}\cdot \vec{p} \\
\vec{\sigma}\cdot \vec{p} & - p^0 - m
\end{array} \right) = \hat{p} - m.\eqno({\rm A}.5)
$$
The Dirac bispinorial functions ({\rm A}.3) are normalized by the
conditions ({\rm A}.4) and obey the identities
$$
u(p^0,-\vec{p},\sigma) = \gamma^0 u(p^0,\vec{p},\sigma)\quad,\quad
v(p^0,-\vec{p},\sigma) = -\gamma^0 v(p^0,\vec{p},\sigma).\eqno({\rm
A}.6)
$$
Under parity transformation the fermionic field ({\rm A}.1) transforms
in the standard way [2,3]
$$
{\cal P}\,\psi(t,\vec{x}\,)\,{\cal P}^{\dagger} =
\gamma^0\,\psi(t,-\vec{x}\,),\eqno({\rm A}.7)
$$
where we have dropped an insignificant phase factor. This relation
follows directly from the scalar properties of the Lagrangian [3]
$$
{\cal P}\,{\cal L}(t,\vec{x}\,)\,{\cal P}^{\dagger} = {\cal
L}(t,-\vec{x}\,).\eqno({\rm A}.8)
$$
Using the properties of the Dirac bispinorial functions ({\rm A}.6)
one can show that the operators $a(\vec{p},\sigma)$,
$a^{\dagger}(\vec{p},\sigma)$, $b(\vec{p},\sigma)$ and
$b^{\dagger}(\vec{p},\sigma)$ transform as follows
$$
{\cal P}\,a(\vec{p},\sigma)\,{\cal P}^{\dagger} =
a(-\vec{p},\sigma)\quad,\quad {\cal
P}\,b(\vec{p},\sigma)\,{\cal P}^{\dagger} = -
b(-\vec{p},\sigma),
$$
$$
{\cal P}\,a^{\dagger}(\vec{p},\sigma)\,{\cal P}^{\dagger} =
a^{\dagger}(-\vec{p},\sigma)\quad,\quad {\cal
P}\,b^{\dagger}(\vec{p},\sigma)\,{\cal P}^{\dagger} = -
b^{\dagger}(-\vec{p},\sigma).\eqno({\rm A}.9)
$$
The properties of the fermionic field discussed above do not depend on
the dimension of space--time and should be retained in
1+1--dimensional space--time too. We should only recall that in
1+1--dimensional space--time fermion fields have no spin and the
dependence on $\sigma$, the projection of the spin onto the fermion
momentum, vanishes.

In 1+1--dimensional space--time a solution of the Dirac equation for a
free massive fermionic field expanded into plane waves reads [5,16]
$$
\psi(x) = \int^{\infty}_{-\infty}\frac{dp^1}{\sqrt{2\pi}}\,
\frac{1}{\sqrt{ 2p^0}}\,\Big[u(p^0,p^1)a(p^1)\,e^{\textstyle -ip\cdot
x} + v(p^0,p^1)b^{\dagger}(p^1)\,e^{\textstyle ip\cdot x}\Big],
$$
$$
\bar{\psi}(x) =\psi^{\dagger}(x)\gamma^0 =
\int^{\infty}_{-\infty}\frac{dp^1}{\sqrt{2\pi}}\,
\frac{1}{\sqrt{2p^0}}\Big[\bar{u}(p^0,p^1)a^{\dagger}(p^1)
\,e^{\textstyle ip\cdot x} + \bar{v}(p^0,p^1)b(p^1)\,e^{\textstyle
-ip\cdot x}\Big],\eqno({\rm A}.10)
$$
where $p\cdot x = p^0x^0 - p^1x^1$. The creation
$a^{\dagger}(p^1)\,(b^{\dagger}(p^1))$ and annihilation
$a(p^1)\,(b(p^1))$ operators of fermions (antifermions) with momentum
$p^1$ and energy $p^0 = \sqrt{(p^1)^2 + m^2}$ obey the anticommutation
relations
$$
\{a(p^1), a^{\dagger}(q^1)\} = \{b(p^1), b^{\dagger}(q^1)\} =
\delta(p^1-q^1),
$$
$$
\{a(p^1), a(q^1)\} = \{a^{\dagger}(p^1), a^{\dagger}(q^1)\} = 
\{b(p^1), b(q^1)\} =
\{b^{\dagger}(p^1), b^{\dagger}(q^1)\} = 0.\eqno({\rm A}.11)
$$
The wave functions $u(p^0,p^1)$ and $v(p^0,p^1)$ are the
solutions of the Dirac equation in the momentum representation for
positive and negative energies, respectively. They are defined by
$$
u(p^0,p^1) = {\displaystyle
\left(\begin{array}{c}\sqrt{p^0 + p^1} \\ \sqrt{p^0 - p^1}
\end{array}\right)}\;,\;\bar{u}(p^0,p^1) = (\sqrt{p^0 - p^1},
\sqrt{p^0 + p^1})
$$
$$
v(p^0,p^1) = {\displaystyle
\left(\begin{array}{c}\sqrt{p^0 + p^1} \\ -\sqrt{p^0 - p^1}
\end{array}\right)}\;,\;
\bar{v}(p^0,p^1) = (- \sqrt{p^0 - p^1}, \sqrt{p^0 + p^1})\eqno({\rm
A}.12)
$$
at $p^0 = \sqrt{(p^1)^2 + m^2}$ and normalized to 
$$
u^{\dagger}(p^0,p^1)u(p^0,p^1) = v^{\dagger}(p^0,p^1)v(p^0,p^1) = 2p^0,
$$
$$
\bar{u}(p^0,p^1)u(p^0,p^1) = - \bar{v}(p^0,p^1)v(p^0,p^1) = 2m,
$$
$$
\bar{u}(p^0,p^1)v(p^0,p^1) = \bar{v}(p^0,p^1)u(p^0,p^1) = 0.\eqno({\rm
A}.13)
$$
The functions $u(p^0,p^1)$ and $v(p^0,p^1)$ satisfy the following matrix
relations
$$
u(p^0,p^1)\bar{u}(p^0,p^1) = {\displaystyle {\sqrt{p^0 + p^1}\choose
\sqrt{p^0 - p^1}}}(\sqrt{p^0 - p^1}, \sqrt{p^0 + p^1}) =
$$
$$
= \left(\begin{array}{cc}\sqrt{(p^0)^2 - (p^1)^2} & p^0 + p^1 \\ p^0 -
p^1 & \sqrt{(p^0)^2 - (p^1)^2}
\end{array} \right) =  \left(\begin{array}{cc} m  & p^0 + p^1 \\
p^0 - p^1 &  m 
\end{array} \right) = 
$$
$$
= \gamma^0p^0 - \gamma^1p^1 + m = \hat{p} + m,
$$
$$
v(p^0,p^1)\bar{v}(p^0,p^1) = {\displaystyle
{\sqrt{p^0 + p^1}\choose -\sqrt{p^0 - p^1}}}(-\sqrt{p^0 - p^1}, 
\sqrt{p^0 +
p^1}) = 
$$
$$
= \left(\begin{array}{cc}-\sqrt{(p^0)^2 - (p^1)^2} & p^0 + p^1 \\ 
p^0 - p^1 &
-\sqrt{(p^0)^2 - (p^1)^2}
\end{array} \right) =  \left(\begin{array}{cc} - m  & p^0 + p^1 \\
p^0 - p^1 & - m
\end{array} \right) = 
$$
$$
= \gamma^0p^0 - \gamma^1p^1 - m = \hat{p} - m\eqno({\rm A}.14)
$$
and the identities
$$
u(p^0, -p^1) = \gamma^0u(p^0,p^1)\quad,\quad v(p^0, - p^1) =
-\gamma^0v(p^0,p^1).\eqno({\rm A}.15)
$$
Under parity transformations a fermionic field in 1+1--dimensional
space--time transforms in the usual way
$$
{\cal P}\,\psi(t,x)\,{\cal P}^{\dagger} = \gamma^0 \psi(t,-x).
\eqno({\rm A}.15)
$$
By virtue of the identities ({\rm A}.15) the operators of creation and
annihilation of fermions and antifermions behave under parity
transformation in the standard way
$$
{\cal P}\,a(p^1)\,{\cal P}^{\dagger} =
a(- p^1)\quad,\quad {\cal
P}\,b(p^1)\,{\cal P}^{\dagger} = -
b(-p^1),
$$
$$
{\cal P}\,a^{\dagger}(p^1)\,{\cal P}^{\dagger} = a^{\dagger}(-
p^1)\quad,\quad {\cal P}\,b^{\dagger}(p^1)\,{\cal P}^{\dagger} = -
b^{\dagger}(- p^1).\eqno({\rm A}.16)
$$
In the massless limit the wave functions $u(p^0,p^1)$ and $v(p^0,p^1)$
can be written in terms of Heaviside functions and read
$$
u(p^0,p^1) = \sqrt{2p^0}\,{\displaystyle
\left(\begin{array}{c}\theta(+p^1) \\ \theta(-p^1)
\end{array}\right)}\;,\;\bar{u}(p^0,p^1) = \sqrt{2p^0}\,(\theta(-p^1),
\theta(+p^1))
$$
$$
v(p^0,p^1) = \sqrt{2p^0}\,{\displaystyle
\left(\begin{array}{c} \theta(+p^1)\ \\ -\theta(-p^1)
\end{array}\right)}\;,\;
\bar{v}(p^0,p^1) = \sqrt{2p^0}\,(- \theta(-p^1),
\theta(+p^1)),\eqno({\rm A}.17)
$$
where $\theta(\pm p^1)$ are Heaviside functions.

In his seminal paper [1] Thirring and then Klaiber [4] used the
solution for a free massless fermionic field in 1+1--dimensional
space--time with identical spinorial functions $u(p^0,p^1)$ and
$v(p^0,p^1)$
$$
u(p^0,p^1) = v(p^0,p^1) = \sqrt{2p^0}\,{\displaystyle
\left(\begin{array}{c}\theta(+p^1) \\ \theta(-p^1)
\end{array}\right)}.\eqno({\rm A}.18)
$$
Since the wave function $v(p^0,p^1)$ taken in the form ({\rm
A}.17) breaks the identity ({\rm A}.14) that entails, in turn, a
contradiction with the standard properties of fermion fields under
parity transformations
$$
{\cal P}\,a(p^1)\,{\cal P}^{\dagger} =
a(- p^1)\quad,\quad {\cal
P}\,b(p^1)\,{\cal P}^{\dagger} = 
b(-p^1),
$$
$$
{\cal P}\,a^{\dagger}(p^1)\,{\cal P}^{\dagger} = a^{\dagger}(-
p^1)\quad,\quad {\cal P}\,b^{\dagger}(p^1)\,{\cal P}^{\dagger} =
b^{\dagger}(- p^1).\eqno({\rm A}.19)
$$
This violates the smooth transition from massive to massless
fermionic fields in the zero--mass limit.

\section*{Appendix B. Chiral Jacobian}

\hspace{0.2in} In this Appendix we adduce the calculation of the
Jacobian induced by chiral rotations. We follow the procedure
formulated in Refs.[7--14]. For the calculation of the chiral Jacobian
we start with the Lagrangian defined by
$$
{\cal L}_{\psi}(x) = \bar{\psi}(x)i\gamma^{\mu}(\partial_{\mu} -
i\partial_{\mu}\eta(x) -
i\varepsilon_{\mu\nu}\partial^{\nu}\xi(x))\psi(x) =
\bar{\psi}(x)D(x;0)\psi(x),\eqno({\rm B}.1)
$$
where $D(x;0)$ is the Dirac operator given by 
$$
D(x;0) = i\gamma^{\mu}(\partial_{\mu} - i\partial_{\mu}\eta(x) -
i\varepsilon_{\mu\nu}\partial^{\nu}\xi(x)).\eqno({\rm B}.2)
$$
By  a chiral rotation 
$$
\psi(x) = e^{\textstyle i\eta(x) + i\,\alpha\,\gamma^5\xi(x)
}\,\chi(x),
$$
$$
\bar{\psi}(x) = \bar{\chi}(x)\,e^{\textstyle - i\eta(x) +
i\,\alpha\,\gamma^5\xi(x)},\eqno({\rm B}.3)
$$
where $0\le \alpha \le 1$, we reduce the Lagrangian $({\rm B}.1)$ to
the form
$$
{\cal L}_{\chi}(x) = \bar{\chi}(x\,)D(x,\alpha)\,\chi(x).\eqno({\rm
B}.4)
$$
The Dirac operator $D(x;\alpha)$ reads
$$
D(x;\alpha) = i\gamma^{\mu}\partial_{\mu} + (1 -
\alpha)\,\gamma^{\mu}\gamma^5\,\partial_{\mu}\xi(x).\eqno({\rm
B}.5)
$$
At $\alpha = 1$ we obtain the Lagrangian
$$
{\cal L}_{\chi}(x) = \bar{\chi}(x)\,D(x,1)\,\chi(x) =
\bar{\chi}(x)\,i\gamma^{\mu}\partial_{\mu}\chi(x).\eqno({\rm B}.6)
$$
Due to the chiral rotation ({\rm B}.3) the fermionic measure changes
as follows
$$
{\cal D}\psi\,{\cal D}\bar{\psi} = J[\xi]\,{\cal D}\chi\,{\cal
D}\bar{\chi},\eqno({\rm B}.7)
$$
For the calculation $J[\vartheta]$ we follow Fujikawa's procedure
[7--14] and introduce eigenfunctions $\varphi_n(x;\alpha)$ and
eigenvalues $\lambda_n(\alpha)$ of the Dirac operator $D(x;\alpha)$:
$$
D(x;\alpha)\,\varphi_n(x;\alpha) =
\lambda_n(\alpha)\,\varphi_n(x;\alpha).\eqno({\rm B}.8)
$$
The operator $D^2(x;\alpha)$ reads then
$$
D^2(x;\alpha) = - \Box -
2\,i\,(1-\alpha)\,\gamma^{\mu}\gamma^{\nu}\gamma^5\,
\partial_{\mu}\xi(x)\partial_{\nu} + i\,(1-\alpha)\,\gamma^5\,\Box\xi
- (1-\alpha)^2\,\partial_{\mu}\xi(x)\partial^{\mu}\xi(x).\eqno({\rm
B}.9)
$$
In terms of the eigenfunctions and eigenvalues of the Dirac operator
$D(x;\alpha)$ the Jacobian $J[\xi]$ is defined by [9,10]
$$
J[\xi] = \exp\,2\,i\!\!\int^1_0d\alpha\,w[\xi;\alpha],\eqno({\rm
B}.10)
$$
where the functional $w[\xi;\alpha]$ is given by [9,10]
$$
w[\xi;\alpha] = -\lim_{\Lambda_{\rm F} \to
\infty}\sum_n\varphi^{\dagger}_n(x;\alpha)\,
\gamma^5\,\xi(x)\,e^{\textstyle i\lambda^2(\alpha_J)_n/\Lambda^2_{\rm F}}
\varphi_n(x;\alpha)=
$$
$$
=-\lim_{\Lambda_{\rm F} \to \infty} \int
d^2x\,\xi(x)\int\frac{d^2k}{(2\pi)^2}\,{\rm
tr}\Big\{\gamma^5\Big\langle x\Big|e^{\textstyle
iD^2(x;\alpha)/\Lambda^2_{\rm F}}\Big|x\Big\rangle\Big\}.\eqno({\rm
B}.10)
$$
For the calculation of the matrix element $\langle x|\ldots|x\rangle$
we use plane waves [7--14] and get
$$
\Big\langle x\Big|e^{\textstyle iD^2(x;\alpha)/\Lambda^2_{\rm
F}}\Big|x\Big\rangle =
$$
$$
= \exp\Big\{\frac{i}{\Lambda^2_{\rm F}}\,[k^2 +
2\,(1-\alpha)\,\gamma^{\mu}\gamma^{\nu}\gamma^5\,
\partial_{\mu}\xi(x)\,k_{\nu} + i\,(1-\alpha)\,\gamma^5\,\Box\xi -
(1-\alpha)^2\,\partial_{\mu}\xi(x)\partial^{\mu}\xi(x)\Big\}.
\eqno({\rm B}.11)
$$
Substituting $({\rm B}.11)$ in $({\rm B}.10)$ we obtain
$$
\lim_{\Lambda_{\rm F} \to \infty} \int\frac{d^2k}{(2\pi)^2}\,{\rm
tr}\Big\{\gamma^5\Big\langle x\Big|e^{\textstyle
iD^2(x;\alpha)/\Lambda^2_{\rm F}}\Big|x\Big\rangle\Big\} =
-\,\frac{1}{2\pi}\,(1-\alpha)\,\Box\xi(x).  \eqno({\rm B}.12)
$$
The functional $w[\xi,\alpha]$ is then given by
$$
w[\xi,\alpha] = \frac{1}{2\pi}\,(1-\alpha)\int
d^2x\,\xi(x)\,\Box\xi(x) .\eqno({\rm B}.13)
$$
Inserting $w[\xi,\alpha]$ into $({\rm B}.9)$ and integrating over
$\alpha$ we get the Jacobian
$$
J[\xi] = \exp\Big\{-\,i\!\!\int d^2x\,\frac{1}{2\pi}\,
\partial^{\mu}\xi(x)\partial_{\mu}\xi(x)\Big\}.\eqno({\rm B}.14)
$$
For the derivation of the exponent we have integrated by parts and
dropped the surface contributions.

According to Refs.[10--12] the exponent in ({\rm B}.14) is
ill--defined and depends on the regularization procedure. Therefore,
the chiral Jacobian can be written as
$$
J[\xi] = \exp\Big\{-\,i\!\!\int d^2x\,\frac{\alpha_J}{2\pi}\,
\partial^{\mu}\xi(x)\partial_{\mu}\xi(x)\Big\},\eqno({\rm B}.15)
$$
where $\alpha_J$ is an arbitrary parameter.

In fact, following Christos [10] the functional $w[\xi;\alpha]$ can be
represented in the form
$$
w[\xi;\alpha] =
$$
$$
= -\lim_{\Lambda_{\rm F} \to
\infty}\sum_n\varphi^{\dagger}_n(x;\alpha)\,
\gamma^5\,\xi(x)\,\Bigg[\alpha_J\,
e^{\textstyle i\lambda^2(\alpha_J)_n/\Lambda^2_{\rm F}} 
+ (1 - \alpha_J)\,e^{\textstyle i
(i\hat{\partial})^2/\Lambda^2_{\rm F}}\Bigg]
\varphi_n(x;\alpha)=
$$
$$
=-\lim_{\Lambda_{\rm F} \to \infty} \int
d^2x\,\xi(x)\int\frac{d^2k}{(2\pi)^2}\,{\rm
tr}\Big\{\gamma^5\Big\langle x\Big|\alpha_J\,e^{\textstyle
iD^2(x;\alpha)/\Lambda^2_{\rm F}} + (1 - \alpha_J)\,e^{\textstyle
i(i\hat{\partial})^2/\Lambda^2_{\rm
F}}\Big|x\Big\rangle\Big\}.\eqno({\rm B}.16)
$$
Since the contribution of the last term vanishes, the functional
$w[\xi;\alpha]$ becomes equal to
$$
w[\xi,\alpha] = \frac{\alpha_J}{2\pi}\,(1-\alpha)\int
d^2x\,\xi(x)\,\Box\xi(x) .\eqno({\rm B}.17)
$$
This leads to the chiral Jacobian given by ({\rm B}.15).

In turn, in the massless Thirring model with fermionic fields
quantized in the chirally broken phase the chiral Jacobian is
well--defined and is equal to unity. In fact, as has been shown in [6]
following Fujikawa's procedure chiral Jacobian can obtained in the
form
$$
J[\vartheta] = \exp\,i\!\!\int d^2x\,\Big[\frac{1}{8\pi}\,
\partial^{\mu}\vartheta(x)\partial_{\mu}\vartheta(x) 
+ \frac{M^2}{4\pi}\,(\cos2\vartheta(x) - 1)\Big],\eqno({\rm B}.18)
$$
where $\vartheta(x)$ is a pseudoscalar field. According to Christos'
remark the exponent in the r.h.s. of (\ref{label11.19}) should enter
with an arbitrary parameter $\alpha_J$, i.e.
$$
J[\vartheta] = \exp\,i\,\frac{\alpha_J}{4\pi}\int
d^2x\,\Big[\frac{1}{2}\,
\partial^{\mu}\vartheta(x)\partial_{\mu}\vartheta(x) +
M^2\,(\cos2\vartheta(x) - 1)\Big].\eqno({\rm B}.19)
$$
The term proportional to $M^2$ gives a contribution to the effective
potential breaking the rotational symmetry of the effective potential
under global rotations $\vartheta(x)\to \vartheta(x) + \theta$, where
$\theta$ is an arbitrary constant. The effective potential has been
evaluated explicitly [6] without chiral rotations and found invariant
under global rotations $\vartheta(x)\to \vartheta(x) +
\theta$. Therefore, the requirement of rotational symmetry of all
contributions to the effective potential gives $\alpha_J = 0$. This
yields $J[\vartheta] = 1$ [6].

\section*{Appendix C. Two--point correlation functions of scalar and 
pseudoscalar fermionic field densities $S^{(\pm)}(x)$ and
$P^{(\pm)}(y)$}

\hspace{0.2in} In this Appendix we calculate the correlation functions
$D^{(\pm)}(x)$ and $D^{(\pm)}_5(x)$ defining the commutators of scalar
and pseudoscalar fermionic field densities $S^{(\pm)}(x)$ and
$P^{(\pm)}(y)$ given by (\ref{label5.5}) and (\ref{label5.6}).

Since the momentum integrals defining $D^{(\pm)}(x)$ and
$D^{(\pm)}_5(x)$ are divergent in the infrared region, we have to
regularize them. The simplest regularization is related to the
inclusion of a non--zero mass $\mu$ which finally should be taken in
the limit $\mu \to 0$. The correlation functions $D^{(\pm)}(x)$ and
$D^{(\pm)}_5(x)$ read
$$
D^{(\pm)}(x) \to D^{(\pm)}(x;\mu) =
\pm\,\frac{i}{4\pi}\int^{\infty}_{-\infty}\frac{dk^1}{\sqrt{(k^1)^2 +
\mu^2}}\,e^{\textstyle \mp i\sqrt{(k^1)^2 + \mu^2}\,x^0 \pm i\,k^1x^1},
$$
$$
D^{(\pm)}_5(x) \to D^{(\pm)}_5(x;\mu) =
\pm\,\frac{i}{4\pi}\int^{\infty}_{-\infty}\frac{dk^1}{\sqrt{(k^1)^2 +
\mu^2}}\,\varepsilon(k^1)\,e^{\textstyle \mp i\sqrt{(k^1)^2 +
\mu^2}\,x^0 \pm i\,k^1x^1}.\eqno({\rm C}.1)
$$
By a change of variables $k^1=\mu\,\sinh\varphi$ we recast the
integral over $k^1$ into the form
$$
D^{(\pm)}(x) \to D^{(\pm)}(x;\mu) =
\pm\,\frac{i}{4\pi}\int^{\infty}_{-\infty}d\varphi\,e^{\textstyle
-i\,\mu\,(\pm\,x^0)\cosh\varphi \pm i\,\mu\,x^1\sinh\varphi},
$$
$$
D^{(\pm)}_5(x) \to D^{(\pm)}_5(x;\mu) =
\pm\,\frac{i}{4\pi}\int^{\infty}_{-\infty}d\varphi\,
\varepsilon(\varphi)\,e^{\textstyle -i\,\mu\,(\pm\,x^0)\cosh\varphi
\pm i\,\mu\,x^1\sinh\varphi}.\eqno({\rm C}.2)
$$
For the convergence of the integrals over $\varphi$ we should
introduce an infinitesimal imaginary part for $x^0$: $\pm\,x^0 \to
\pm\,x^0 - i\,0$. This yields
$$
D^{(\pm)}(x) \to D^{(\pm)}(x;\mu) =
\pm\,\frac{i}{4\pi}\int^{\infty}_{-\infty}d\varphi\,e^{\textstyle
-i\,\mu\,(\pm\,x^0 - i\,\,0)\cosh\varphi \pm i\,\mu\,x^1\sinh\varphi},
$$
$$
D^{(\pm)}_5(x) \to D^{(\pm)}_5(x;\mu) =
\pm\,\frac{i}{4\pi}\int^{\infty}_{-\infty}d\varphi\,
\varepsilon(\varphi)\,e^{\textstyle -i\,\mu\,(\pm\,x^0 -
i\,0)\cosh\varphi \pm i\,\mu\,x^1\sinh\varphi}.\eqno({\rm C}.3)
$$
Then, the calculation of $D^{(\pm)}(x;\mu)$ runs in the way
$$
D^{(\pm)}(x;\mu) =
\pm\,\frac{i}{4\pi}\int^{\infty}_{-\infty}d\varphi\,e^{\textstyle
-\,\mu\,\sqrt{- (x^0)^2 -(x^1)^2 \pm i\,0\cdot x^0}\cosh(\varphi -
\varphi_0)} =
$$
$$
=\pm\,\frac{i}{4\pi}\int^{\infty}_{-\infty}d\varphi\,e^{\textstyle
-\,\mu\,\sqrt{ - x^2 \pm i0\cdot\varepsilon(x^0)}\cosh\varphi}
=\pm\,\frac{i}{2\pi}\,K_0(\mu\sqrt{- x^2 \pm \,i\,0\cdot
\varepsilon(x^0)}),\eqno({\rm C}.4)
$$
where $x^2 = (x^0)^2 - (x^1)^2$, $K_0(z)$ is the McDonald function
\cite{[abr]} and we have replaced $i\,0\,x^0 \to i\,0\cdot
\varepsilon(x^0)$. For the calculation of $D^{(\pm)}(x;\mu)$ in ({\rm
C}.4) we have denoted
$$
\cosh\varphi_0 = \frac{\pm\,x^0 - \,i\,0}{\sqrt{x^2 \mp i\,0\cdot
\varepsilon(x^0)}}\quad,\quad \sinh\varphi_0 =
\frac{\pm\,x^1}{\sqrt{x^2 \mp i\,0\cdot \varepsilon(x^0)}}.\eqno({\rm
C}.5)
$$
and made a shift $\varphi - \varphi_0 \to \varphi$. 

In the limit $\mu \to 0$ the correlation function $D^{(\pm)}(x;\mu)$
is defined by
$$
D^{(\pm)}(x;\mu) = \mp \frac{i}{4\pi}\,{\ell n}[- \mu^2x^2 \pm
\,i\,0\cdot\varepsilon(x^0)],\eqno({\rm C}.6)
$$
where we have used the asymptotic behaviour of the McDonald function
$K_0(z)$ at $z \to 0$ \cite{[abr]}. Our result ({\rm C}.6) agrees with that
obtained in Ref.[16] (see Eqs.(2.6) and (2.11) of Ref.[16]).

For the correlation function $D^{(\pm)}_5(x; \mu)$ we get
$$
D^{(\pm)}_5(x;\mu) =
\pm\,\frac{i}{4\pi}\int^{\infty}_{-\infty}d\varphi\,
\varepsilon(\varphi)\,e^{\textstyle -i\,\mu\,(\pm\,x^0 -\,
i\,0)\cosh\varphi \pm i\,\mu\,x^1\sinh\varphi} =
$$
$$
= \pm\,\frac{i}{4\pi}\Bigg[\int^0_{-\infty}d\varphi\,
\varepsilon(\varphi)\,e^{\textstyle - \mu\,\sqrt{-x^2
\pm\,i\,0\cdot\varepsilon(x^0)}\,\cosh(\varphi - \varphi_0)}
$$
$$
+\int^{\infty}_0d\varphi\, \varepsilon(\varphi)\,e^{\textstyle
- \mu\,\sqrt{- x^2 \pm\,i\,0\cdot\varepsilon(x^0)}\cosh(\varphi -
\varphi_0)}\Bigg]=
$$
$$
= \pm\,\frac{i}{4\pi}\Bigg[-\int^{\infty}_0d\varphi\, e^{\textstyle
-\mu\,\sqrt{- x^2 \pm\,i\,0\cdot\varepsilon(x^0)}\,\cosh(\varphi +
\varphi_0)}
$$
$$
+\int^{\infty}_0d\varphi\,e^{\textstyle -\mu\,\sqrt{- x^2
\pm\,i\,0\cdot\varepsilon(x^0)}\cosh(\varphi - \varphi_0)}\Bigg]=
$$
$$
=\mp\,\frac{i}{4\pi}\int^{\,\varphi_0}_{-\varphi_0}d\varphi\,
e^{\textstyle - \mu\,\sqrt{- x^2 \pm
\,i\,0\cdot\varepsilon(x^0)}\cosh\varphi} =
\mp\,\frac{i}{2\pi}\,\varphi_0 + O(\mu).\eqno({\rm C}.7)
$$
Taking into account that
$$
\varphi_0 = \frac{1}{2}\,{\ell n}\Bigg(\frac{x^0 + x^1 \mp \,i\,0}{x^0 -
x^1 \mp i\,0}\Bigg)\eqno({\rm C}.8)
$$
we obtain
$$
D^{(\pm)}_5(x) = \lim_{\textstyle \mu \to 0}D^{(\pm)}_5(x; \mu) =
\mp\,\frac{i}{4\pi}\,{\ell n}\Bigg(\frac{x^0 + x^1 \mp \,i\,0}{x^0 -
x^1 \mp \,i\,0}\Bigg).\eqno({\rm C}.9)
$$
Within the operator approach we apply these correlation functions
to the derivation of quantum equations of motion of the massless
Thirring model and the evaluation of $n$--point correlation functions
of left(right)handed fermionic densities $\bar{\psi}(x)(1 \pm
\gamma^5)\psi(x)$ of Coleman's proof of equivalence between the
massive Thirring model and the sine--Gordon model [18].

The causal Green function $\Delta(x) = i\langle 0|{\rm
T}(S(x)S(0))|0\rangle = i\langle 0|{\rm T}(P(x)P(0))|0\rangle$ is
equal to
$$
\Delta(x) = \theta(x^0)\,D^{(+)}(x) -\theta(-x^0)\,D^{(-)}(x) =
$$
$$
= -\frac{i}{4\pi}\,\theta(+ x^0)\,{\ell n}[- \mu^2x^2
+\,i\,0\cdot\varepsilon(x^0)] - \frac{i}{4\pi}\,\theta(-x^0)\,{\ell
n}[- \mu^2x^2 - \,i\,0\cdot\varepsilon(x^0)]=
$$
$$
= -\frac{i}{4\pi}\,{\ell n}(- \mu^2x^2 +\,i\,0).\eqno({\rm C}.10)
$$
The same result can be obtained by using the momentum representation of
$\Delta(x)$ [6,16]
$$
\Delta(x) = i\int \frac{d^2p}{(2\pi)^2i}\,\frac{\textstyle
e^{\textstyle -ip\cdot x}}{\mu^2 - p^2 - i\,0} =
i\int^{\infty}_0\frac{dp_{\rm E}}{(2\pi)^2}\,\frac{p_{\rm E}}{\mu^2 +
p^2_{\rm E}}\int^{2\pi}_0d\varphi\,e^{\textstyle ip_{\rm E}\sqrt{-x^2
+ i\,0}\,\cos\varphi} =
$$
$$
= i\int^{\infty}_0\frac{dp_{\rm E}}{2\pi}\,\frac{p_{\rm E}J_0(p_{\rm
E}\sqrt{-x^2 + i\,0})}{\mu^2 + p^2_{\rm E}} =
\frac{i}{2\pi}\,K_0(\mu\sqrt{-x^2 + i\,0}) = -\frac{i}{4\pi}\,{\ell
n}(- \mu^2x^2 +\,i\,0).\eqno({\rm C}.11)
$$
By the Wick rotation $p^0 \to i\,p_2$ we have passed to Euclidean
momentum space and used the integral representations for the Bessel
and McDonald functions $J_0(z)$ and $K_0(z)$ (see p. 360, eq.(9.1.18)
and p.488, eq.(11.4.44) of Ref.\cite{[abr]})).

\section*{\hspace{-0.1in}Appendix D. On Klaiber's ansatz for the
definition of the quantum vector fermionic current $K^{\mu}(x)$}

\hspace{0.2in} In section 5 we have discussed the definition of the
fermion vector current $K^{\mu}(x)$ determined by (\ref{label5.37})
and the analogous definition of the axial--vector current
$K^{\mu}_5(x)$ given by (\ref{label5.50}). In this Appendix we would
like to give some hints for our interpretation of Klaiber's ansatz
(\ref{label5.37}). For this aim let us transcribe the Lagrangian
(\ref{label1.1}) as follows
$$
{\cal L}_{\rm Th}(x) =
\bar{\psi}(x)i\gamma^{\mu}(\partial_{\mu}
+i\,g\,K_{\mu}(x))\psi(x) + \frac{1}{2}\,g\,K_{\mu}(x)K^{\mu}(x).
\eqno({\rm D}.1)
$$
The equations of motion for this Lagrangian are equal to
$$
\frac{\delta {\cal L}_{\rm Th}(x)}{\delta \bar{\psi}(x)} =
i\gamma^{\mu}(\partial_{\mu} + i\,g\,K_{\mu}(x))\psi(x) = 0,
$$
$$
\frac{\delta {\cal L}_{\rm Th}(x)}{\delta K^{\mu}(x)} =
-g\,\bar{\psi}(x)\gamma^{\mu}\psi(x) + g\,K_{\mu}(x) = 0.
\eqno({\rm D}.2)
$$
Thus, the equations of motion read
$$
i\gamma^{\mu}(\partial_{\mu} + i\,g\,K_{\mu}(x))\psi(x) = 0 \eqno({\rm
D}.3)
$$
and 
$$
K_{\mu}(x) = \bar{\psi}(x)\gamma^{\mu}\psi(x).\eqno({\rm D}.4)
$$
Substituting ({\rm D}.4) in ({\rm D}.3) we arrive at the equation of
motion of the massless Thirring model
$$
i\gamma^{\mu}\partial_{\mu}\psi(x) =
g\,\gamma_{\mu}\psi(x)\bar{\psi}(x)\gamma^{\mu}\psi(x).\eqno({\rm
D}.5)
$$
The general form of the solution of the equation of motion ({\rm D}.3)
can be given in the form
$$
\psi(x) =
\exp\,\Big\{i\,g\int^x_{-\infty}dz^{\lambda}\,K_{\lambda}(z)\Big\}\,
\Psi(x), \eqno({\rm D}.6)
$$
where the fermion field $\Psi(x)$ obeys a free Dirac equation
$$
i\gamma^{\mu}\partial_{\mu}\Psi(x) = 0.\eqno({\rm D}.7)
$$
The quantum fermionic vector current $K^{\mu}(x)$ should be defined in 
the normal--ordered form
$$
K^{\mu}(x) = :\!\bar{\psi}(x)\gamma^{\mu}\psi(x)\!:,\eqno({\rm D}.7)
$$
which could be understood within the point--splitting technique as
follows
$$
K^{\mu}(x) =\lim_{\textstyle \epsilon \to
0}\frac{1}{2}\,(\gamma^{\mu})_{ab}[\bar{\psi}_a(x + \epsilon)\psi_b(x) -
\psi_b(x + \epsilon)\bar{\psi}_a(x)].\eqno({\rm D}.8)
$$
Using the solution ({\rm D}.6) of the equation of motion ({\rm D}.3)
and substituting it into ({\rm D}.8) we obtain the quantum vector
current $K^{\mu}(x)$ in the form
$$
K^{\mu}(x) =\lim_{\textstyle \epsilon \to
0}\frac{1}{2}\,(\gamma^{\mu})_{ab}[\bar{\Psi}_a(x +
\epsilon)\,e^{\textstyle -i\,g\int^{x + \epsilon}_x
dz^{\lambda}\,K_{\lambda}(z)}\,\Psi_b(x)
$$
$$
 - \Psi_b(x + \epsilon)\,e^{\textstyle +i\,g\int^{x +
\epsilon}_x dz^{\lambda}\,K_{\lambda}(z)}\,\bar{\Psi}_a(x)].
\eqno({\rm D}.9)
$$
Expanding the r.h.s. of ({\rm D}.9) in powers of $\epsilon$ we get
$$
K^{\mu}(x) =
$$
$$
= \lim_{\textstyle \epsilon \to
0}\frac{1}{2}\,(\gamma^{\mu})_{ab}\{\bar{\Psi}_a(x +
\epsilon)\,\Psi_b(x)\,[1 -i\,g\,\epsilon^{\lambda}\,K_{\lambda}(x)] -
\Psi_b(x + \epsilon)\,\bar{\Psi}_a(x)\,[1 +
i\,g\,\epsilon^{\lambda}\,K_{\lambda}(x)]\}=
$$
$$
= \lim_{\textstyle \epsilon \to
0}\frac{1}{2}\,(\gamma^{\mu})_{ab}[\bar{\Psi}_a(x +
\epsilon)\,\Psi_b(x) - \Psi_b(x + \epsilon)\,\bar{\Psi}_a(x)] 
$$
$$
+ \lim_{\textstyle \epsilon \to
0}\frac{1}{2}\,(\gamma^{\mu})_{ab}\{\bar{\Psi}_a(x +
\epsilon)\,\Psi_b(x) + \Psi_b(x +
\epsilon)\,\bar{\Psi}_a(x)\}\,(-ig)\,\epsilon^{\lambda}\,K_{\lambda}(x).
\eqno({\rm D}.10)
$$
Using (\ref{label5.28}) and (\ref{label5.34}) we recast the r.h.s. of
({\rm D}.10) into the form
$$
K^{\mu}(x) =:\!\bar{\Psi}(x)\gamma^{\mu}\Psi(x)\!: -
\frac{g}{\pi}\,\frac{\epsilon^{\,\mu}\epsilon^{\,\lambda}}
{\epsilon^2}\,K_{\lambda}(x).  \eqno({\rm D}.11)
$$
Averaging over the directions of the 2--vector $\epsilon$ and solving
the obtained relation with respect to $K^{\mu}(x)$ we get
$$
K^{\mu}(x) =\frac{1}{\displaystyle 1 +
\frac{g}{2\pi}}:\!\bar{\Psi}(x)\gamma^{\mu}\Psi(x)\!:.  \eqno({\rm D}.12)
$$
The quantum axial--vector current $K^{\mu}_5(x)$ can be defined in an
analogous way
$$
K^{\mu}_5(x) =\lim_{\textstyle \epsilon \to
0}\frac{1}{2}\,(\gamma^{\mu}\gamma^5)_{ab}[\bar{\psi}_a(x + \epsilon)\psi_b(x) -
\psi_b(x + \epsilon)\bar{\psi}_a(x)]=
$$
$$
=\lim_{\textstyle \epsilon \to
0}\frac{1}{2}\,(\gamma^{\mu}\gamma^5)_{ab}[\bar{\Psi}_a(x +
\epsilon)\,e^{\textstyle -i\,g\int^{x + \epsilon}_x
dz^{\lambda}\,K_{\lambda}(z)}\,\Psi_b(x)
$$
$$
 - \Psi_b(x + \epsilon)\,e^{\textstyle +i\,g\int^{x +
\epsilon}_x dz^{\lambda}\,K_{\lambda}(z)}\,\bar{\Psi}_a(x)]=
$$
$$
= \lim_{\textstyle \epsilon \to
0}\frac{1}{2}\,(\gamma^{\mu}\gamma^5)_{ab}\{\bar{\Psi}_a(x +
\epsilon)\,\Psi_b(x)\,[1 -i\,g\,\epsilon^{\lambda}\,K_{\lambda}(x)] -
\Psi_b(x + \epsilon)\,\bar{\Psi}_a(x)\,[1 +
i\,g\,\epsilon^{\lambda}\,K_{\lambda}(x)]\}=
$$
$$
= \lim_{\textstyle \epsilon \to
0}\frac{1}{2}\,(\gamma^{\mu}\gamma^5)_{ab}[\bar{\Psi}_a(x +
\epsilon)\,\Psi_b(x) - \Psi_b(x + \epsilon)\,\bar{\Psi}_a(x)] 
$$
$$
+ \lim_{\textstyle \epsilon \to
0}\frac{1}{2}\,(\gamma^{\mu}\gamma^5)_{ab}\{\bar{\Psi}_a(x +
\epsilon)\,\Psi_b(x) + \Psi_b(x +
\epsilon)\,\bar{\Psi}_a(x)\}\,(-ig)\,\epsilon^{\lambda}\,K_{\lambda}(x)=
$$
$$
=:\!\bar{\Psi}(x)\gamma^{\mu}\gamma^5\Psi(x)\!: +\,
\varepsilon^{\mu\nu}\,\frac{g}{\pi}\,
\frac{\epsilon_{\nu}\epsilon_{\lambda}}{\epsilon^2}\,K^{\lambda}(x).
\eqno({\rm D}.13)
$$
Averaging again over the directions of the 2--vector $\epsilon$, using
the identity $\varepsilon^{\mu\nu}\gamma_{\nu} = -
\gamma^{\mu}\gamma^5$ and solving the result with respect to
$K^{\mu}_5(x)$ we obtain
$$
K^{\mu}_5(x) = \frac{1}{\displaystyle 1 +
\frac{g}{2\pi}}:\!\bar{\Psi}(x)\gamma^{\mu}\gamma^5\Psi(x)\!:.  \eqno({\rm
D}.14)
$$
One can see that the quantum vector and axial--vector currents
$K^{\mu}(x)$ and $K^{\mu}_5(x)$ defined by ({\rm D}.12) and ({\rm
D}.14), respectively, are related by the standard relation
$K^{\mu}_5(x) = - \varepsilon^{\mu\nu}\,K_{\nu}(x)$.

Since according to the solution ({\rm D}.6)) we are able to define
$\Psi(x)$ is terms of $\psi(x)$
$$
\Psi(x) =
\exp\,\Big\{-\,i\,g\int^x_{-\infty}dz^{\lambda}\,K_{\lambda}(z)\Big\}\,
\psi(x), \eqno({\rm D}.15)
$$
the quantum vector and axial--vector fermionic currents $K^{\mu}(x)$
and $K^{\mu}_5(x)$ can be determined in terms of the self--coupled
field $\psi(x)$
$$
K^{\mu}(x) =\frac{1}{\displaystyle 1 +
\frac{g}{2\pi}}:\!\bar{\psi}(x)\gamma^{\mu}\psi(x)\!:\quad,\quad
K^{\mu}_5(x) =\frac{1}{\displaystyle 1 + \frac{g}{2\pi}}
:\!\bar{\psi}(x)\gamma^{\mu}\gamma^5\psi(x)\!:.  \eqno({\rm D}.16)
$$
Making a finite renormalization of the wave function of the
$\psi$--field, $\psi(x) \to Z^{1/2}_2\psi(x)$ we can remove the common
factor in the definition of quantum vector and axial--vector current
and obtain
$$
K^{\mu}(x) =:\!\bar{\psi}(x)\gamma^{\mu}\psi(x)\!:\quad,\quad K^{\mu}_5(x) =
:\!\bar{\psi}(x)\gamma^{\mu}\gamma^5\psi(x)\!:.  \eqno({\rm D}.17)
$$
This testifies the self--consistency of the approach.

In turn, the inclusion of gauge dependent terms violates the
self--consistency. Indeed, with such terms the definition of the
vector current $K^{\mu}(x)$ given by ({\rm D}.9) should read (see
(\ref{label5.45}))
$$
K^{\mu}(x) =\!\lim_{\textstyle \epsilon \to
0}\frac{1}{2}\,(\gamma^{\mu})_{ab}[\bar{\Psi}_a(x +
\epsilon)\,e^{\textstyle -i\,g\int^{x + \epsilon}_x
dz^{\lambda}\,K_{\lambda}(z)}\,\Psi_b(x) 
$$
$$
 - \Psi_b(x + \epsilon)\,e^{\textstyle +i\,g\int^{x +
\epsilon}_x dz^{\lambda}\,K_{\lambda}(z)}\,\bar{\Psi}_a(x)]
$$
$$
-i\,\sigma\!\lim_{\textstyle \epsilon \to
0}\frac{1}{2}\,(\hat{\epsilon})_{ab}[\bar{\Psi}_a(x +
\epsilon)\,\Psi_b(x) + \Psi_b(x +
\epsilon)\,\bar{\Psi}_a(x)]\,K^{\mu}(x)=
$$
$$
=:\!\bar{\Psi}(x)\gamma^{\mu}\Psi(x)\!: -
\frac{g}{\pi}\,\frac{\epsilon^{\,\mu}\epsilon^{\,\lambda}}
{\epsilon^2}\,K_{\lambda}(x) - \frac{\sigma}{\pi}\,K^{\mu}(x).
\eqno({\rm D}.18)
$$
Thus, unlike Klaiber stated, the contribution of the terms
violating gauge invariance does not remove non--covariant term and an
average over directions of the 2--vector $\epsilon$ is demanded. This
yields
$$
K^{\mu}(x) = :\!\bar{\Psi}(x)\gamma^{\mu}\Psi(x)\!: -\Bigg(
\frac{g}{2\pi} + \frac{\sigma}{\pi}\Bigg)\,K^{\mu}(x).\eqno({\rm
D}.19)
$$
Solving this relation with respect to $K^{\mu}(x)$ we get a new
expression for the quantum vector current $K^{\mu}(x)$
$$
K^{\mu}(x) =\frac{1}{\displaystyle 1 + \frac{g}{2\pi} +
\frac{\sigma}{\pi} }:\!\bar{\Psi}(x)\gamma^{\mu}\Psi(x)\!:.
\eqno({\rm D}.20)
$$
In order to underscore the inconsistency of this extended definition
of a quantum vector fermionic current we have to define a quantum
axial--vector current in analogues way
$$
K^{\mu}_5(x) =\!\lim_{\textstyle \epsilon \to
0}\frac{1}{2}\,(\gamma^{\mu}\gamma^5)_{ab}[\bar{\Psi}_a(x +
\epsilon)\,e^{\textstyle -i\,g\int^{x + \epsilon}_x
dz^{\lambda}\,K_{\lambda}(z)}\,\Psi_b(x) 
$$
$$
 - \Psi_b(x + \epsilon)\,e^{\textstyle +i\,g\int^{x +
\epsilon}_x dz^{\lambda}\,K_{\lambda}(z)}\,\bar{\Psi}_a(x)]
$$
$$
-i\,\sigma\!\lim_{\textstyle \epsilon \to
0}\frac{1}{2}\,(\hat{\epsilon}\gamma^5)_{ab}[\bar{\Psi}_a(x +
\epsilon)\,\Psi_b(x) + \Psi_b(x +
\epsilon)\,\bar{\Psi}_a(x)]\,K^{\mu}(x).  \eqno({\rm D}.21)
$$
The presence of the $\gamma^5$--matrix thwarts the appearance of
contributions proportional to $\sigma$ and restores fully
the former result ({\rm D}.13). After averaging over directions of the
2--vector $\epsilon$ we get again ({\rm D}.14). Due to the necessity
to satisfy the standard relation $K^{\mu}_5(x) =
-\varepsilon^{\mu\nu}\,K_{\nu}(x)$ the only solution for the parameter
$\sigma$ is $\sigma = 0$. This confirms our statement below
(\ref{label5.59}).

\section*{Appendix E. Fourier transform of the
two--point Green function of the massless Thirring fermion field}

\hspace{0.2in} In this Appendix we calculate the Fourier transform of
the two--point renormalized Green function of the massless Thirring
fermion field defined by (\ref{label11.18})
$$
S_{\rm F}(p) = \int d^2x\,\frac{M^2}{2\pi}\,\frac{\textstyle -
\,\hat{x}\,e^{\textstyle +\,i\,p\cdot x}}{(-M^2x^2 + i\,0\,)^{\textstyle
\lambda}},\eqno({\rm E}.1)
$$
where we have denoted $\lambda = 1 + g/2\pi$.

The r.h.s. of ({\rm E}.1) we transcribe in equivalent form
$$
S_{\rm F}(p) =-\, \frac{M^2}{2\pi
i}\,\gamma^{\mu}\frac{\partial}{\partial p^{\mu}}\int
d^2x\,\frac{\textstyle e^{\textstyle +\,i\,p\cdot x}}{(-M^2x^2 + i\,0\,)^{\textstyle \lambda}} =
\gamma^{\mu}\frac{\partial}{\partial p^{\mu}}\,\Delta
(p),\eqno({\rm E}.2)
$$
where $\Delta(p^0,p^1)$ is defined by
$$
\Delta(p) = -\, \frac{M^2}{2\pi i}\int d^2x\,\frac{\textstyle
e^{\textstyle +\,i\,p\cdot x}}{(-M^2x^2 + i\,0\,)^{\textstyle
\lambda}}.\eqno({\rm E}.4)
$$
For the denominator we suggest to use the following integral
representation
$$
\frac{1}{(-M^2x^2 + i\,0)^{\textstyle
\lambda}}=\frac{(-i)^{\textstyle \lambda}}{\Gamma(\lambda)}
\int^{\infty}_0dt\,t^{\lambda - 1}\,e^{\textstyle +i\,(-M^2x^2
+\,i\,0)\,t}. \eqno({\rm E}.5)
$$
Substituting ({\rm E}.5) in ({\rm E}.4) and integrating over $x$
with the help of 
$$
\int d^2x\,e^{\textstyle -\,i\,M^2x^2t + i\,p\cdot x} =
\frac{\pi}{M^2t}\,e^{\textstyle +i(p^2 +i\,0)/4M^2t}. \eqno({\rm E}.6)
$$
Thus, we define the function $\Delta(p)$ in terms of the integral over
$t$
$$
\Delta(p) = - \frac{1}{2i}\,\frac{(-i)^{\textstyle
\lambda}}{\Gamma(\lambda)}\int^{\infty}_0dt\,t^{\lambda -
2}\,e^{\textstyle +i(p^2 +i\,0)/4M^2t}. \eqno({\rm E}.7)
$$
Integrating over $t$ we obtain
$$
\Delta(p) = - \frac{1}{2}\,e^{\textstyle
-i\pi\lambda}\,\frac{\Gamma(1-\lambda)}{\Gamma(\lambda)}\,
\Bigg(\frac{\bar{M}^2}{p^2 + i\,0}\Bigg)^{\textstyle 1-\lambda},
\eqno({\rm E}.8)
$$
where $\bar{M} = 2M$. 

Substituting ({\rm E}.8) in ({\rm E}.2) we derive the Fourier
transform of the Green function ({\rm E}.1)
$$
S_{\rm F}(p) = e^{\textstyle -i\pi\lambda}\,\frac{\hat{p}}{p^2 +
i\,0}\,
\frac{\Gamma(2-\lambda)}{\Gamma(\lambda)}\,\Bigg(\frac{\bar{M}^2}{p^2
+ i\,0}\Bigg)^{\textstyle 1-\lambda}, \eqno({\rm E}.9)
$$
Setting $\lambda = 1 + g/2\pi$ we obtain the Fourier transform of the
 causal two--point Green function of massless Thirring fermion field
$$
S_{\rm F}(p) = - \,e^{\textstyle -ig/2}\frac{\hat{p}}{p^2 + i\,0}\,
\frac{\Gamma(1-g/2\pi)}{\Gamma(1 + g/2\pi)}\,\Bigg(\frac{p^2 +
i\,0}{\bar{M}^2}\Bigg)^{\textstyle g/2\pi}. \eqno({\rm E}.10)
$$
Using the properties of $\Gamma$--functions \cite{[abr]} we reduce
$S_{\rm F}(p)$ to the form
$$
S_{\rm F}(p) = -\,\frac{\displaystyle g\,e^{\textstyle
-ig/2}}{\displaystyle
2\,\sin\Bigg(\frac{g}{2}\Bigg)}\,\frac{1}{\displaystyle
\Gamma^2\Bigg(1 + \frac{g}{2\pi}\Bigg)}\,\frac{\hat{p}}{p^2 +
i\,0}\,\Bigg(\frac{p^2 + i\,0}{\bar{M}^2}\Bigg)^{\displaystyle
\frac{g}{2\pi}}. \eqno({\rm E}.11)
$$
In the limit $g \to 0$ we get the Green function of a free massless
fermion field: $S_{\rm F}(p) =\hat{p}/(- p^2 - i\,0)$.

The obtained result agrees well with Glaser's analysis of the massless
Thirring model [30,31] and spectral analysis of two--point correlation
functions in 1+1--dimensional quantum field theories developed by
Schroer [32].

For the calculation of the fermion condensate it is convenient to use
the double--integral representation for the two--point Green function
$S_{\rm F}(x)$ \cite{[silva]}:
$$
S_{\rm F}(x) = i\langle 0|{\rm T}(\psi(x)\bar{\psi}(0))|0\rangle =
\frac{\displaystyle g^2\,e^{\textstyle -ig/2}}{\displaystyle
\bar{M}^2\sin\Bigg(\frac{g}{2}\Bigg)}\,\frac{1}{\displaystyle
\Gamma^2\Bigg(1 + \frac{g}{2\pi}\Bigg)}
$$
$$
\times\int\frac{d^2p}{(2\pi)^2}\,e^{\textstyle -ip\cdot
x}\int\frac{d^2k}{(2\pi)^2i}\,\frac{\hat{k}}{k^2 +
i\,0}\,\Bigg(\frac{(p-k)^2 + i\,0}{\bar{M}^2}\Bigg)^{\textstyle -1 +
g/2\pi}. \eqno({\rm E}.12)
$$
Let us prove this representation. Making a shift of variables $p\to k
+ q$ we bring up the r.h.s. of ({\rm E}.12) to the form
$$
S_{\rm F}(x) = \frac{\displaystyle g^2e^{\textstyle
-ig/2}}{\displaystyle
\bar{M}^2\sin\Bigg(\frac{g}{2}\Bigg)}\,\frac{1}{\displaystyle
\Gamma^2\Bigg(1 +
\frac{g}{2\pi}\Bigg)}\int\frac{d^2k}{(2\pi)^2i}\,\frac{\hat{k}}{k^2 +
i\,0}\,e^{\textstyle -ik\cdot x}
$$
$$
\times\int\frac{d^2q}{(2\pi)^2}\,\Bigg(\frac{q^2 +
i\,0}{\bar{M}^2}\Bigg)^{\textstyle -1 + g/2\pi}\,e^{\textstyle
-iq\cdot x}. \eqno({\rm E}.13)
$$
The integral over $k$ is equal to
$$
\int\frac{d^2k}{(2\pi)^2i}\,\frac{\hat{k}}{k^2 + i\,0}\,e^{\textstyle
-ik\cdot x} = - \gamma^{\mu}\frac{\partial }{\partial
x^{\mu}}\Delta(x) =\frac{i}{2\pi}\,\frac{\hat{x}}{x^2 - i\,0}.
\eqno({\rm E}.14)
$$
For the integration over $q$ we would use the representation analogous
to ({\rm E}.5) and change the order of integrations. This yields
$$
\int\frac{d^2q}{(2\pi)^2}\,\Bigg(\frac{q^2 +
i\,0}{\bar{M}^2}\Bigg)^{\textstyle -1 + g/2\pi}\,e^{\textstyle
-iq\cdot x} = \frac{\displaystyle -i\,e^{\textstyle
ig/4}}{\displaystyle \Gamma\Bigg(1 -
\frac{g}{2\pi}\Bigg)}\int^{\infty}_0\frac{dt}{\displaystyle
t^{\textstyle g/2\pi}}
$$
$$
\times\int\frac{d^2q}{(2\pi)^2}\,e^{\textstyle it(q^2
+i\,0)/\bar{M}^2}\,e^{\textstyle -iq\cdot x} = \frac{\displaystyle
-i\,e^{\textstyle ig/4}}{\displaystyle \Gamma\Bigg(1 -
\frac{g}{2\pi}\Bigg)}\int^{\infty}_0\frac{dt}{\displaystyle
t^{\textstyle g/2\pi}}\frac{M^2}{\pi t}\,e^{\textstyle i(-M^2x^2 +
i\,0)/t}=
$$
$$
=-i\,\,e^{\textstyle ig/2}\,\frac{M^2}{\pi}\,\frac{\displaystyle
\Gamma\Bigg(\frac{g}{2\pi}\Bigg)}{\displaystyle \Gamma\Bigg(1 -
\frac{g}{2\pi}\Bigg)}\,(-M^2x^2 + i\,0)^{\textstyle - g/2\pi}.
\eqno({\rm E}.15)
$$
Substituting ({\rm E}.15) and ({\rm E}.14) in ({\rm E}.12) and using
the properties of $\Gamma$--functions \cite{[abr]} we obtain
$$
S_{\rm F}(x) = \frac{1}{2\pi}\,\frac{\hat{x}}{x^2 - i\,0}\,(-M^2x^2 +
i\,0)^{\textstyle - g/2\pi}.  \eqno({\rm E}.16)
$$
By the change $x \to x - y$ we bring up the r.h.s. of the
expression ({\rm E}.16) to the form of the Green function
(\ref{label11.18}). We would like to accentuate that this agreement
has been obtained due to shifts of variables -- the procedure which is
rather sensitive to convergence of integrals. For the derivation of
the final expression ({\rm E}.16) we have tacitly assumed that all
integrals we have dealt with are convergent even if in the meaning of
generalized functions \cite{[gel]}.

\section*{\hspace{-0.1in}Appendix F. Non--perturbative 
renormalizability of the sine--Gordon model}

\hspace{0.2in} As has been shown in section 7 the massless Thirring
model is equivalent to the sine--Gordon (SG) model, when the mass of
Thirring fermion fields $m$ is considered as an external source
$\sigma(x) = - m$ of the scalar fermion density
$\bar{\psi}(x)\psi(x)$. Therefore, the properties of non--perturbative
renormalizability of the massless Thirring model investigated in
section 11 should be fully extended to the SG model.

The generating functional of Green functions in the SG model we define
as
$$
Z_{\rm SG}[J] = \int {\cal D}\vartheta\,\exp\,i\int
d^2x\,\Big\{\frac{1}{2}\,\partial_{\mu}\vartheta(x)\partial^{\mu}
\vartheta(x) +
\frac{\bar{\alpha}}{\bar{\beta}^2}\,(\cos\bar{\beta}\vartheta(x) - 1)
+ \vartheta(x)J(x)\Big\},\eqno({\rm F}.1)
$$
where $J(x)$ is an external source of the $\vartheta(x)$--field.

The Lagrangian of the SG model is invariant under the transformations
$$
\vartheta(x) \to \vartheta\,'(x) = \vartheta(x) + \frac{2\pi
n}{\bar{\beta}},\eqno({\rm F}.2)
$$
where $n$ is an integer number running over $n = 0,\pm 1,\pm
2,\ldots$.  In order to get the generating functional $Z_{\rm SG}[J]$
invariant under the transformations ({\rm F}.2) it is sufficient to
restrict the class of functions describing the external source of the
$\vartheta$--field and impose the constraint
$$
\int d^2x\,J(x) = 0.\eqno({\rm F}.3)
$$
Formally, for the pseudoscalar SG field $\vartheta(x)$, that we really
have [6,18], the constraint ({\rm F}.3) is fulfilled
automatically. Indeed, due to the conservation of parity the external
source of the SG field $\vartheta(x)$ should be a pseudoscalar
quantity $J(x^0,x^1) = - J(x^0,-x^1)$. In the case of the scalar SG
field $\vartheta(x)$ the situation with the fulfillment of the
constraint ({\rm F}.3) is a little bit more complicated but not
crucial for the validity of it. Below we show that the constraint
({\rm F}.3) will be a great deal of importance for the infrared
renormalizability of the SG model.

Non--perturbative renormalizability of the SG model we understand as a
possibility to remove all divergences by renormalizing the coupling
constant $\bar{\alpha}$. Indeed, since the coupling constant
$\bar{\beta}$ is related to the coupling constant of the Thirring
model $g$ which is unrenormalized $g_0 = g$, so the coupling constant
$\bar{\beta}$ should possesses the same property, i.e. $\bar{\beta}_0
= \bar{\beta}$. Hence, only the coupling constant $\alpha$ should
undergo renormalization.

The Lagrangian of the SG model written in terms of {\it bare}
quantities reads
$$
{\cal L}_{\rm SG}(x) =
\frac{1}{2}\partial_{\mu}\vartheta_0(x)\partial^{\mu}\vartheta_0(x) +
\frac{\bar{\alpha}_0}{\bar{\beta}^2_0}\,(\cos\bar{\beta}_0\vartheta_0(x)
- 1).\eqno({\rm F}.4)
$$
Since $\bar{\beta}$ is the unrenormalized coupling constant, the field
$\vartheta(x)$ should be also unrenormalized, $\vartheta_0(x) =
\vartheta(x)$. This means that there is no renormalization of the wave
function of the $\vartheta$--field.  As a result the Lagrangian ${\cal
L}_{\rm SG}(x)$ of the SG model in terms of renormalized quantities
can be written by
$$
{\cal L}_{\rm SG}(x) =
\frac{1}{2}\partial_{\mu}\vartheta(x)\partial^{\mu}\vartheta(x) +
\frac{\bar{\alpha}}{\bar{\beta}^2}\,(\cos\bar{\beta}\vartheta(x) - 1)
+ (Z_1 -
1)\,\frac{\bar{\alpha}}{\bar{\beta}^2}\,(\cos\bar{\beta}\vartheta(x) -
1) =
$$
$$
= \frac{1}{2}\partial_{\mu}\vartheta(x)\partial^{\mu}\vartheta(x) +
Z_1\,\frac{\bar{\alpha}}{\bar{\beta}^2}\,(\cos\bar{\beta}\vartheta(x)
- 1),\eqno({\rm F}.5)
$$
where $Z_1$ is the renormalization constant of the coupling constant
$\bar{\alpha}$. The renormalized coupling constant $\bar{\alpha}$ is
related to the {\it bare} one by the relation [29]
$$
\bar{\alpha} = Z^{-1}_1\,\bar{\alpha}_0.\eqno({\rm F}.6)
$$
Renormalizability of the SG model as well as the Thirring model we
understand as the possibility to replace the infrared cut--off $\mu$
and the ultra--violet cut--off $\Lambda$ by another finite scale $M$
by means of the renormalization constant $Z_1$. According to the
general theory of renormalizations [29] $Z_1$ should be a function of
the coupling constants $\bar{\beta}$, $\bar{\alpha}$, the infrared
cut--off $\mu$, the ultra--violet cut--off $\Lambda$ and a finite
scale $M$:
$$
Z_1 = Z_1(\bar{\beta},\bar{\alpha}, M; \mu, \Lambda).\eqno({\rm
F}.7)
$$
Now let us proceed to the evaluation of the generating functional
({\rm F}.1). For this aim, we first replace $\bar{\alpha} \to
\bar{\alpha}_0$ and expand the integrand in powers of
$\bar{\alpha}_0$. This gives
$$
Z_{\rm SG}[J] = \lim_{\textstyle \mu \to 0}e^{\textstyle -i\int
d^2x\,{\displaystyle
\frac{\bar{\alpha}_0}{\bar{\beta}^2}}}\sum^{\infty}_{n=0}
\frac{i^n}{n!}\,\Bigg(\frac{\bar{\alpha}_0}{\bar{\beta}^2}\Bigg)^n
\prod^n_{i=1}\int d^2x_i
$$
$$
\times \int {\cal
D}\vartheta\prod^n_{i=1}\cos\bar{\beta}\vartheta(x_i)\,\exp\, i\int
d^2x\,\Big\{\frac{1}{2}\,\partial_{\mu}\vartheta(x)
\partial^{\mu}\vartheta(x) - \frac{1}{2}\,\mu^2\vartheta^2(x) +
\vartheta(x)J(x)\Big\}.\eqno({\rm F}.8)
$$
The integration over the $\vartheta$--field can be carried out
explicitly and we get
$$
Z_{\rm SG}[J] = \lim_{\textstyle \mu \to 0}\sum^{\infty}_{n=0}
\frac{i^n}{n!}\,\Bigg(\frac{\bar{\alpha}_0}{2\bar{\beta}^2}\Bigg)^n\sum^n_{p=0}\frac{n!}{(n-p)!\,p!}\prod^{n-p}_{j=1}\prod^p_{k=1}\int
d^2x_j d^2y_k
$$
$$
\times\,\exp\,\Big\{\frac{1}{2}\,n\,\bar{\beta}^2 i \Delta(0) +
\bar{\beta}^2\sum^{n-p}_{j < k}i\Delta(x_j - x_k) +
\bar{\beta}^2\sum^{p}_{j < k}i \Delta(y_j - y_k) -
\bar{\beta}^2\sum^{n-p}_{j = 1}\sum^{p}_{k = 1}i\Delta(x_j -
y_k)\Big\}
$$
$$
\times\,\exp\,\Big\{\int d^2x\,\bar{\beta}\,[\sum^{n-p}_{j =
1}i\Delta(x_j - x) - \sum^{p}_{k = 1}i\Delta(y_j - x)]\,J(x) +
\int\!\!\!\int d^2x\,d^2y\,\frac{1}{2}\,J(x)\,i\Delta(x -
y)\,J(y)\Big\},\eqno({\rm F}.9)
$$
where the Green functions $i\Delta(x-y)$ and $i\Delta(0)$ are defined
by (\ref{label6.19}) and (\ref{label6.20}).

Taking the limit $\mu \to 0$ we reduce the r.h.s. of ({\rm F}.6) to
the form
$$
Z_{\rm SG}[J] = \sum^{\infty}_{p=0} \frac{(-1)^p}{(p!)^2}\,
\Bigg(\frac{\bar{\alpha}_0}{2\bar{\beta}^2}\Bigg)^{2p}
\prod^{p}_{j=1}\int\!\!\!\int d^2x_j d^2y_j\,
\Bigg[\Bigg(\frac{M}{\Lambda}\Bigg)^{\textstyle
\bar{\beta}^2/2\pi}\Bigg]^{\!2p}
$$
$$
\times\exp\Big\{\frac{\bar{\beta}^2}{4\pi}\sum^p_{j < k} \Big({\ell
n}[ - M^2(x_j - x_k)^2\,] + {\ell n}[ - M^2(y_j - y_k)^2\,]\Big) -
\frac{\bar{\beta}^2}{4\pi}\sum^p_{j=1}\sum^p_{k=1}{\ell n}[ - M^2(x_j
- y_k)^2\,]\Big\}
$$
$$
\times\exp\Big\{\frac{\bar{\beta}^2}{4\pi}\int d^2x\,\sum^p_{j = 1}
{\ell n}\Bigg[\frac{(x_j - x)^2}{(y_j - x)^2}\Bigg]\,J(x) +
\frac{1}{8\pi}\int\!\!\!\int d^2x\,d^2y\,J(x)\,{\ell
n}[-M^2(x-y)^2\,]\,J(y)\Big\}
$$
$$
\times\lim_{\textstyle \mu \to 0} \exp\Big\{ -\frac{1}{4\pi}\,{\ell
n}\Big(\frac{M}{\mu}\Big) \Big(\int
d^2x\,J(x)\Big)^{\!\!2}\Big\}.\eqno({\rm F}.10)
$$
Due to the constraint ({\rm F}.3) the generating functional $Z_{\rm
SG}[J]$ does not depend on the infrared cut--off.  Using ({\rm F}.3)
we get
$$
Z_{\rm SG}[J] = \sum^{\infty}_{p=0} \frac{(-1)^p}{(p!)^2}\,
\Bigg(\frac{\bar{\alpha}_0}{2\bar{\beta}^2}\Bigg)^{2p}
\prod^{p}_{j=1}\int\!\!\!\int d^2x_j d^2y_j\,
\Bigg[\Bigg(\frac{M}{\Lambda}\Bigg)^{\textstyle
\bar{\beta}^2/2\pi}\Bigg]^{\!2p}
$$
$$
\times\exp\Big\{\frac{\bar{\beta}^2}{4\pi}\sum^p_{j < k} \Big({\ell
n}[ - M^2(x_j - x_k)^2\,] + {\ell n}[ - M^2(y_j - y_k)^2\,]\Big) -
\frac{\bar{\beta}^2}{4\pi}\sum^p_{j=1}\sum^p_{k=1}{\ell n}[ - M^2(x_j
- y_k)^2\,]\Big\}
$$
$$
\times\exp\Big\{\frac{\bar{\beta}^2}{4\pi}\int d^2x\,\sum^p_{j = 1}
{\ell n}\Bigg[\frac{(x_j - x)^2}{(y_j - x)^2}\Bigg]\,J(x) +
\frac{1}{8\pi}\int\!\!\!\int d^2x\,d^2y\,J(x)\,{\ell
n}[-M^2(x-y)^2\,]\,J(y)\Big\}.\eqno({\rm F}.11)
$$
Passing to a renormalized constant $\bar{\alpha}$ we recast the
r.h.s. of ({\rm F}.11) into the form
$$
Z_{\rm SG}[J] = \sum^{\infty}_{p=0}\Bigg[\frac{1}{Z_1}\,
\Bigg(\frac{M}{\Lambda}\Bigg)^{ \!\textstyle
\bar{\beta}^2/2\pi}\Bigg]^{\!2p} \frac{(-1)^p}{(p!)^2}\,
\Bigg(\frac{\bar{\alpha}}{2\bar{\beta}^2}\Bigg)^{2p}
\prod^{p}_{j=1}\int\!\!\!\int d^2x_j d^2y_j\,
$$
$$
\times\exp\Big\{\frac{\bar{\beta}^2}{4\pi}\sum^p_{j < k} \Big({\ell
n}[ - M^2(x_j - x_k)^2\,] + {\ell n}[ - M^2(y_j - y_k)^2\,]\Big) -
\frac{\bar{\beta}^2}{4\pi}\sum^p_{j=1}\sum^p_{k=1}{\ell n}[ - M^2(x_j
- y_k)^2\,]\Big\}
$$
$$
\times\exp\Big\{\frac{\bar{\beta}^2}{4\pi}\int d^2x\,\sum^p_{j = 1}
{\ell n}\Bigg[\frac{(x_j - x)^2}{(y_j - x)^2}\Bigg]\,J(x) +
\frac{1}{8\pi}\int\!\!\!\int d^2x\,d^2y\,J(x)\,{\ell
n}[-M^2(x-y)^2\,]\,J(y)\Big\}.\eqno({\rm F}.12)
$$
Setting 
$$
Z_1 = \Bigg(\frac{M}{\Lambda}\Bigg)^{ \!\textstyle
\bar{\beta}^2/2\pi}\eqno({\rm F}.13)
$$
we remove the dependence of the generating functional $Z_{\rm SG}[J]$
on the ultra--violet cut--off $\Lambda$
$$
Z_{\rm SG}[J] = \sum^{\infty}_{p=0}\frac{(-1)^p}{(p!)^2}\,
\Bigg(\frac{\bar{\alpha}}{2\bar{\beta}^2}\Bigg)^{2p}
\prod^{p}_{j=1}\int\!\!\!\int d^2x_j d^2y_j\,
$$
$$
\times\exp\Big\{\frac{\bar{\beta}^2}{4\pi}\sum^p_{j < k} \Big({\ell
n}[ - M^2(x_j - x_k)^2\,] + {\ell n}[ - M^2(y_j - y_k)^2\,]\Big) -
\frac{\bar{\beta}^2}{4\pi}\sum^p_{j=1}\sum^p_{k=1}{\ell n}[ - M^2(x_j
- y_k)^2\,]\Big\}
$$
$$
\times\exp\Big\{\frac{\bar{\beta}^2}{4\pi}\int d^2x\,\sum^p_{j = 1}
{\ell n}\Bigg[\frac{(x_j - x)^2}{(y_j - x)^2}\Bigg]\,J(x) +
\frac{1}{8\pi}\int\!\!\!\int d^2x\,d^2y\,J(x)\,{\ell
n}[-M^2(x-y)^2\,]\,J(y)\Big\}.\eqno({\rm F}.14)
$$
However, for the evaluation of correlation functions there appears the
causal Green function $\Delta(x-y; M)$ taken at $x=y$. For the
definition of $\Delta(0;M)$ one can use dimensional or analytical
regularization procedures allowing to set $\Delta(0; M) = 0$ [6]. As a
result no divergences appear for the evaluation of any correlation
function of the SG model.

The generating functional ({\rm F}.14) is expressed in terms of the
renormalized constant $\bar{\alpha}$, the constant $\bar{\beta}$ and
the finite scale $M$. Now it can be applied to the evaluation of any
renormalized correlation function of the SG model. This testifies the
complete non--perturbative renormalizability of the SG model.

We would like to emphasize that due to unrenormalizability of the
coupling constant, $\bar{\beta}_0 = \bar{\beta}$, the Gell--Mann--Low
$\beta$--function vanishes, $\beta(\bar{\beta}, M) = 0$. This means
that the SG model is not an asymptotically free theory that would
require non--zero negative value for the Gell--Mann--Low function,
$\beta(\bar{\beta}, M) < 0$. Our result disagrees with that obtained
within perturbation theory \cite{[sg]}.

The unrenormalizability of the coupling constant $\bar{\beta}$ leading
to the vanishing Gell--Mann--Low function can be easily understood
following the similarity between $\bar{\beta}$ and $\hbar$ which has
been drawn in Ref.[6]. As has been shown in Ref.[6] the limits
$\bar{\beta} \to 0$ and $\bar{\beta} \to \infty$ distinguish {\it
classical} and {\it quantum} regimes of the sine--Gordon
model. Within such an understanding of the coupling constant
$\bar{\beta}$ its unrenormalizability is justified by the
unrenormalizability of $\hbar$.

The same procedure which we have applied to the renormalization of the
SG model model can be implemented to the formulation of a quantum
field theory of a free massless (pseudo)scalar field in
1+1--dimensional space--time free of the {\it infrared} problem
pointed out by Klaiber [4] and Coleman \cite{[col]}. This program we
are planning to realize in forthcoming publications.

\end{document}